\documentclass{article}
\usepackage[margin=.8in,left=.8in]{geometry}

\usepackage{amsthm}
\usepackage{amssymb}
\usepackage{amsmath}
\usepackage{mathtools}
\usepackage{adjustbox}
\usepackage{bbding}
\usepackage{xcolor}
\usepackage{stmaryrd}    
\usepackage{hyperref}
\hypersetup{
  colorlinks = true,
  allcolors=.
}
\usepackage[all]{xy}
\usepackage{rotating}
\usepackage{xfrac}

\usepackage{color, colortbl}

\usepackage{enumitem} \setlist{nosep}

\usepackage{tikz}
\usetikzlibrary{decorations.pathmorphing}
\usetikzlibrary{arrows}
\usetikzlibrary{cd}

\newcommand{\longsquiggly}{\xymatrix{{}\ar@{~>}[r]&{}}}

\newcommand{\Id}{\mathrm{Id}}
\newcommand{\id}{\mathrm{id}}
\newcommand{\defneq}{\equiv}

\newcommand{\underoverset}[3]{\underset{#1}{\overset{#2}{#3}}}

\newcommand*\circled[1]{\tikz[baseline=(char.base)]{
            \node[shape=circle,draw,inner sep=1pt] (char) {#1};}}

\usepackage[titletoc]{appendix}

\usepackage[safe]{tipa} 

\definecolor{darkblue}{rgb}{0.05,0.25,0.65}
\definecolor{greenii}{RGB}{20,140,10}
\definecolor{darkgreen}{rgb}{0.00,0.85,0.1}
\definecolor{lightgray}{rgb}{0.9,0.9,0.9}
\definecolor{orangeii}{RGB}{200,100,5}
\definecolor{darkyellow}{rgb}{.91,.91,0}

\usepackage{multirow}

\DeclareMathAlphabet{\mathpzc}{OT1}{pzc}{m}{it} 
\newcommand\mathscr[1]{\scalebox{1.1}{$\mathpzc{#1}$}}

\newcommand{\Bit}{\mathrm{Bit}}

\usepackage{mathptmx}
\usepackage{amsmath}
\usepackage{graphicx}
\DeclareRobustCommand{\coprod}{\mathop{\text{\fakecoprod}}}
\newcommand{\fakecoprod}{%
  \sbox0{$\prod$}%
  \smash{\raisebox{\dimexpr.9625\depth-\dp0}{\scalebox{1}[-1]{$\prod$}}}%
  \vphantom{$\prod$}%
}

\newcommand{\Sets}{
  \mathrm{Set}
}

\newcommand{\Types}{\mathrm{Type}}
\newcommand{\Propositions}{\mathrm{Prop}}
\newcommand{\Objects}{\mathrm{Obj}}

\newcommand{\isTruncatedType}[1]{\mathrm{isType}_{\leq{#1}}}
\newcommand{\TruncatedTypes}[1]
  { \Types_{\leq{#1}} }

\newcommand{\Bits}{\mathrm{Bit}}

\newcommand{\Spaces}{
  \mathrm{Spc}
}

\newcommand{\SimplicialSets}{
  \simplicial\Sets
}

\newcommand{\NeutralElement}{
  \mathrm{e}
}

\newcommand{\Groups}{
  \mathrm{Grp}
}

\newcommand{\PointedConnectedOneTypes}{
  \Types
    _{0 < \bullet \leq 1}
    ^{\ast\!\raisebox{-1pt}{\scalebox{.7}{$/$}}}
}

\newcommand{\Rings}{\mathrm{Ring}}

\newcommand{\Torsors}{\mathrm{Tors}}

\newcommand{\AbelianGroups}{
  \mathrm{Ab}\Groups
}

\newcommand{\NaturalNumbers}{
  \mathbb{N}
}

\newcommand{\Integers}{
  \mathbb{Z}
}

\newcommand{\RationalNumbers}{
  \mathbb{Q}
}

\newcommand{\RealNumbers}{
  \mathbb{R}
}

\newcommand{\ImaginaryUnit}{
  \mathrm{i}
}

\newcommand{\ComplexNumbers}{\mathbb{C}}

\newcommand{\CyclicGroup}[1]{\mathbb{Z}_{#1}}

\newcommand{\UnitaryGroup}{
  \mathrm{U}
}

\newcommand{\SpecialUnitaryLieAlgebra}[1]{\mathfrak{su}_{#1}}
\newcommand{\suTwo}{\SpecialUnitaryLieAlgebra{2}}

\newcommand{\suAffine}[2]{\widehat{\SpecialUnitaryLieAlgebra{#1}}^{\raisebox{-2.5pt}{\scalebox{.73}{\hspace{-1pt}$#2$}}}}

\newcommand{\suTwoAffine}[1]{\suAffine{2}{#1}}

\newcommand{\CircleGroup}{
  \UnitaryGroup(1)
}

\newcommand{\Hausdorff}{
  \mathrm{Haus}
}

\newcommand{\topological}{
  \mathrm{Top}
}

\newcommand{\simplicial}{
  \Delta
}

\newcommand{\Groupoids}[1]{
  \mathrm{Grpd}_{#1}
}

\newcommand{\ComplexPlane}{\ComplexNumbers}

\newcommand{\ConfigurationSpace}[1]{  \underset{
    \scalebox{.65}{$
      \{1,\cdots,#1\}
    $}
  }
  {\mathrm{Conf}}
}

\newcommand{\weight}{\mathrm{w}}

\newcommand{\level}{k}
\newcommand{\Level}{\level}
\newcommand{\ShiftedLevel}{\kappa}

\newcommand{\TopologicalSpaces}{
  \topological\Spaces
}

\newcommand{\TopologicalSpace}{
  \mathrm{X}
}

\newcommand{\kTopologicalSpaces}{
  \mathrm{k}\TopologicalSpaces
}

\newcommand{\InfinityGroupoids}{
  \Groupoids{\infty}
}

\newcommand{\Fibrations}{
  \mathrm{Fib}
}

\newcommand{\KanFibrations}{
  \mathrm{Kan}\Fibrations
}

\newcommand{\shape}{
  \raisebox{1pt}{\rm\normalfont\textesh}
}

\newcommand{\SingularSimplicialComplex}{
  \mathrm{Pth}
}

\newcommand{\kHausdorffSpaces}{
  \mathrm{k}\HausdorffSpaces
}

\newcommand{\proofstep}[1]{
  \mbox{\small #1}
}

\newcommand{\HausdorffSpaces}{
  \Hausdorff\Spaces
}

\newcommand{\Actions}[1]{
  {#1}\,\mathrm{Act}
}

\newcommand{\Truncation}[3]{
  #1[
    #3
  #1]_{#2}
}

\def\AmbientCategory{\mathbf{H}}

\newcommand{\Topos}{
  \AmbientCategory
}

\newcommand{\Maps}[3]{
  \mathrm{Map}
  #1(
    #2
    ,\,
    #3
  #1)
}

\newcommand{\fiber}[2]{
  \mathrm{fib}_{#2}\left({#1}\right)
}

\newcommand{\SliceMaps}[4]{
  \Maps{#1}{#3}{#4}_{\!\scalebox{.7}{$#2$}}
}

\newcommand{\PointsMaps}[3]{
  \Topos
  #1(
    #2
    ,\,
    #3
  #1)
}

\newcommand{\HomotopyQuotient}[2]{
  #1 \!\sslash\! #2
}

\newcommand{\Modules}{\mathrm{Mod}}
\newcommand{\Lines}{\mathrm{Line}}

\usepackage[new]{old-arrows}   

\newdir{> }{{}*!/10pt/@{>}}

\usepackage{amssymb}

\def\acts{\raisebox{1.2pt}{\;\rotatebox[origin=c]{90}{$\curvearrowright$}}\hspace{.5pt}}

\def\rightacts{\raisebox{-1pt}{\rotatebox[origin=c]{180}{$\acts$}}}

\makeatletter
\newif\if@sup
\newtoks\@sups
\def\append@sup#1{\edef\act{\noexpand\@sups={\the\@sups #1}}\act}%
\def\reset@sup{\@supfalse\@sups={}}%
\def\mk@scripts#1#2{\if #2/ \if@sup ^{\the\@sups}\fi \else%
  \ifx #1_ \if@sup ^{\the\@sups}\reset@sup \fi {}_{#2}%
  \else \append@sup#2 \@suptrue \fi%
  \expandafter\mk@scripts\fi}
\def\tensor#1#2{\reset@sup#1\mk@scripts#2_/}
\def\multiscripts#1#2#3{\reset@sup{}\mk@scripts#1_/#2%
  \reset@sup\mk@scripts#3_/}
\makeatother

\makeatletter
\newbox\slashbox \setbox\slashbox=\hbox{$/$}
\def\itex@pslash#1{\setbox\@tempboxa=\hbox{$#1$}
  \@tempdima=0.5\wd\slashbox \advance\@tempdima 0.5\wd\@tempboxa
  \copy\slashbox \kern-\@tempdima \box\@tempboxa}
\def\slash{\protect\itex@pslash}
\makeatother

\def\clap#1{\hbox to 0pt{\hss#1\hss}}
\def\mathllap{\mathpalette\mathllapinternal}
\def\mathrlap{\mathpalette\mathrlapinternal}
\def\mathclap{\mathpalette\mathclapinternal}
\def\mathllapinternal#1#2{\llap{$\mathsurround=0pt#1{#2}$}}
\def\mathrlapinternal#1#2{\rlap{$\mathsurround=0pt#1{#2}$}}
\def\mathclapinternal#1#2{\clap{$\mathsurround=0pt#1{#2}$}}

\let\oldroot\root
\def\root#1#2{\oldroot #1 \of{#2}}
\renewcommand{\sqrt}[2][]{\oldroot #1 \of{#2}}

\DeclareSymbolFont{symbolsC}{U}{txsyc}{m}{n}
\SetSymbolFont{symbolsC}{bold}{U}{txsyc}{bx}{n}
\DeclareFontSubstitution{U}{txsyc}{m}{n}

\DeclareSymbolFont{stmry}{U}{stmry}{m}{n}
\SetSymbolFont{stmry}{bold}{U}{stmry}{b}{n}

\DeclareFontFamily{OMX}{MnSymbolE}{}
\DeclareSymbolFont{mnomx}{OMX}{MnSymbolE}{m}{n}
\SetSymbolFont{mnomx}{bold}{OMX}{MnSymbolE}{b}{n}
\DeclareFontShape{OMX}{MnSymbolE}{m}{n}{
    <-6>  MnSymbolE5
   <6-7>  MnSymbolE6
   <7-8>  MnSymbolE7
   <8-9>  MnSymbolE8
   <9-10> MnSymbolE9
  <10-12> MnSymbolE10
  <12->   MnSymbolE12}{}

\usepackage{cleveref}

\crefformat{section}{\S#2#1#3} 
\crefformat{subsection}{\S#2#1#3}
\crefformat{subsubsection}{\S#2#1#3}

\theoremstyle{italics}
\newtheorem{theorem}{Theorem}[section]
\newtheorem{lemma}[theorem]{Lemma}

\newtheorem{proposition}[theorem]{Proposition}

\theoremstyle{definition}
\newtheorem{definition}[theorem]{Definition}
\newtheorem{notation}[theorem]{Notation}
\newtheorem{literature}[theorem]{Literature}
\newtheorem{example}[theorem]{Example}

\newtheorem{remark}[theorem]{Remark}

\usepackage{amsfonts}
\usepackage{colortbl}

\renewcommand{\emph}{\textit}

\newcommand{\rbraid}{
  \begin{scope}[yscale=.5]
  \draw
  [line width=2pt]
    (-1,1)
    .. controls (.1,1) and (-.1,-1) ..
    (1,-1);

  \draw[white, line width=6pt]
    (-1,-1)
    .. controls (.1,-1) and (-.1,+1) ..
    (1,+1);
  \draw
  [line width=2pt]
    (-1,-1)
    .. controls (.1,-1) and (-.1,+1) ..
    (1,+1);
  \end{scope}
}
\newcommand{\lbraid}{
  \begin{scope}[yscale=-1]
    \rbraid
  \end{scope}
}
\newcommand{\strand}
{
  \draw
  [line width=2pt]
    (-1,0) to (1,0);
}

\newcommand{\dprod}[1]{\left(#1\right)\to}           
\newcommand{\dsum}[1]{\left(#1\right) \times }  

\tikzset{>=Straight Barb
}    

\newcommand{\descriptionbox}[3]{
  {
    \rotatebox{#1}{
      \scalebox{#2}{
        \color{orangeii}
        \bf
        \def\arraystretch{.9}
        \begin{tabular}{c}
          #3
        \end{tabular}
      }
    }
  }
}
\newcommand{\descriptionboxdefault}[1]{\descriptionbox{0}{.8}{#1}}

\begin{document}

\title{
  Topological Quantum Gates in
  Homotopy Type Theory
}

\author{
  David Jaz Myers${}^\ast$
  \and
  Hisham Sati${}^{\ast, \dagger}$
  \and
  Urs Schreiber${}^\ast$
}
\date{}
\maketitle

\begin{abstract}

Despite the evident necessity of topological protection for realizing scalable quantum computers, the conceptual
underpinnings of topological quantum logic gates had arguably remained shaky, both regarding their physical realization
as well as their information-theoretic nature.

\smallskip
Building on recent results on defect branes in string/M-theory \cite{SS22AnyonicDefectBranes} and on their
holographically dual anyonic defects in condensed matter theory \cite{SS22AnyonictopologicalOrder}, here we explain
(as announced in \cite{SS22TQC}) how the specification of realistic topological quantum gates,
operating by anyon defect braiding in topologically ordered quantum materials, has a surprisingly slick formulation in parameterized
point-set topology, which is so fundamental that it lends itself to certification in modern homotopically typed programming languages,
such as cubical {\tt Agda}.

\smallskip
We propose that this remarkable confluence of concepts may jointly kickstart the development of topological quantum
programming proper as well as of real-world application of homotopy type theory, both of which have arguably been
falling behind their high expectations; in any case, it provides a powerful paradigm for simulating and verifying
topological quantum computing architectures with high-level certification languages
aware of the actual physical principles of realistic topological quantum hardware.

\smallskip
In a companion article \cite{QPinLHOTT} (announced in \cite{Schreiber22}), we explain how further passage to ``dependent linear''
homotopy data types naturally extends this scheme to a full-blown quantum
programming/certification language in which our topological quantum gates may be compiled to verified quantum circuits,
complete with quantum measurement gates and classical control.

\end{abstract}

\begin{center}
{\small
  {\it  In Memoriam} of Yuri Manin
  \\
  who introduced quantum computation
  \\
  as well as Gauss-Manin connections
  \\
  unknowing of their close relationship.
  }
\end{center}

\tableofcontents

\vfill

\hrule
\vspace{5pt}

{
\footnotesize
\noindent
\def\arraystretch{1}
\tabcolsep=0pt
\begin{tabular}{ll}
${}^*$\,
&
Mathematics, Division of Science; and
\\
&
Center for Quantum and Topological Systems,
\\
&
NYUAD Research Institute,
\\
&
New York University Abu Dhabi, UAE.
\end{tabular}
\hfill
\adjustbox{raise=-15pt}{
\includegraphics[width=3cm]{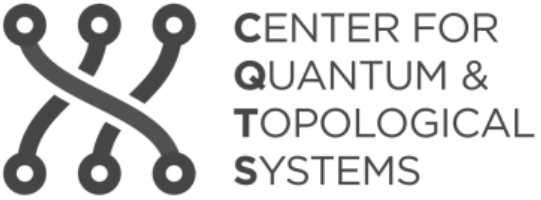}
}

\vspace{1mm}
\noindent ${}^\dagger$The Courant Institute for Mathematical Sciences, NYU, NY

\vspace{.2cm}

\noindent
The authors acknowledge the support by {\it Tamkeen} under the
{\it NYU Abu Dhabi Research Institute grant} {\tt CG008}.
}

\newpage

\section{Introduction}

\noindent
{\bf The need for topology in quantum computation.} While the hopes connected with the idea of quantum computation (Lit. \ref{LiteratureQuantumComputation}) are hard to overstate, experts are well-aware (Lit. \ref{NeedForTopologicalProtection}) that practically useful quantum computation beyond the presently existing NISQ machines (Lit. \ref{LiteratureOnNISQMachines}) will require the development of profound stabilization mechanisms to protect quantum data against decoherence. This might be achievable at the software level by implementing enough redundancy within existing quantum hardware paradigms (``quantum error correction'') but more likely it will (in addition) require error protection right at the hardware level, utilizing quantum materials whose quantum states are stabilized by fundamental physical principles broadly known as {\it topological} (Lit. \ref{LiteratureTopologyAndHomotopyTheory}), specifically  by {\it topological order} appearing in {\it topological phases} of quantum materials (Lit. \ref{TopologicalQuantumMaterials}).

\medskip

\noindent
{\bf The open problem of topological quantum gates.}
Inevitable as the path of {\it topological quantum computing} (Lit. \ref{LiteratureTopologicalQuantumComputation}) may thus be in the long run,
its theoretical underpinnings had arguably remained shaky (cf. \cite{Valera21}\cite[p. 2]{SS22AnyonictopologicalOrder}, Lit. \ref{MathematicsOfAnyonsLiterature}), despite considerable
interest and in contrast to the impression one may glean from a cursory perusal of the literature. This might have in part contributed to the
apparent failure of the only attempt to date at implementing topologically protected qbits in the laboratory (Lit. \ref{LiteratureMajoranaZeroModes}). Generally, the original and still most promising idea of topological quantum logic gates operating on topologically ordered ground states
(Lit. \ref{TopologicalQuantumMaterials})
by {\it adiabatic} (Lit. \ref{TQCAsAdiabaticQuantumComputing})
{\it braiding} (Lit. \ref{LiteratureBraiding})
of {\it anyonic} (Lit. \ref{AnyonLiterature}) defect worldlines had shifted out of the community's focus: Experimentalists have been focusing on topological states which, even if detected, would be intrinsically immobile and hence un-braidable (Lit. \ref{LiteratureMajoranaZeroModes}); while theorists have been exploring anyonic braiding in a generality remote from considerations of physical realizability (cf. Lit. \ref{MathematicsOfAnyonsLiterature} and p. \pageref{Outlook}).

\medskip

\noindent
{\bf A solution gleaned from high energy physics...}
But in a recent re-analysis of defects in topological quantum materials \cite{SS22AnyonictopologicalOrder} (Lit. \ref{TopologicalQuantumMaterials}) -- following analogous (``dual'') considerations for stable {\it defect branes} in
string/M-theory \cite{SS22AnyonicDefectBranes} (Lit. \ref{LiteratureIntersectingBraneModels}) -- we found a detailed
realistic model for adiabatic anyon braiding,
showing how the established classification of topological phases of matter (Lit. \ref{TopologicalQuantumMaterials}) by
{\it topological K-theory} (Lit. \ref{LiteratureKTheoryClassificationOfTopologicalPhases}) extends to describe topologically ordered ground states supporting braid group statistics (Lit. \ref{AnyonLiterature}).
This description of anyon braiding turns out to flow naturally from just fundamental constructions in {\it parameterized homotopy theory} (discussed \cref{ViaParameterizedPointSetTopology}), revealing a deeper purely homotopy-theoretic nature (Lit. \ref{LiteratureTopologyAndHomotopyTheory}) of topological quantum gates than has been appreciated before.

\medskip

\noindent
{\bf ...lending itself to certified quantum programming.}
This formulation of topological quantum gates in parameterized homotopy theory
is noteworthy also in view of a remarkable modern development in certified (``typed'') programming language theory (Lit. \ref{VerificationLiterature}): where strict adherence to the principle of assigning {\it data types} to all data, in particular also to  certificates of identification of pairs of other data, leads to these data types behaving just as the {\it homotopy types} of parameterized homotopy theory. In effect, the fundamental certification language now called {\it homotopy type theory} (``HoTT'', Lit. \ref{LiteratureHomotopyTypeTheory}) serves at once as a general-purpose programming language as well as a proof language for constructions in parameterized homotopy theory.

\medskip

\noindent
{\bf Claim and broad Outline.} In summary, this suggests that {\it homotopy-typed programming languages} (Lit. \ref{LiteratureHomotopyTypeTheory}, exposition in \cref{HoTTIdea}) naturally serve
for encoding (simulating) and formally verifying realistic topological quantum logic gates (Lit. \ref{LiteratureTopologicalQuantumComputation}),
providing a natural theoretical basis for simulation and certification of realistic topological quantum computing platforms.
We had briefly announced this result in \cite{SS22TQC}; here we introduce and explain in detail:

\smallskip
{\bf in \cref{ViaParameterizedPointSetTopology}} the relevant parameterized point-set topology as understood from traditional algebraic topology literature (Lit. \ref{LiteratureTopologyAndHomotopyTheory});

{\bf in \cref{ViaDependentHomotopyTypeTheory}} the relevant dependent homotopy type theory, as operational in the programming language {\tt Agda} (Lit. \ref{LiteratureAgda});

{\bf in \cref{KZConnectionsInHomotopyTypeTheory}} the construction of the homotopy data structure (Def. \ref{DataStructureOfConformalBlocks}) which encodes topological quantum gates: Theorem \ref{TheTheorem}.

\medskip

\noindent

\noindent
{\bf More technical outline.}
Before concluding, here to outline our construction/claim in a little more technical detail:

Our starting point is the following two facts which are
separately ``well-known'' to their respective experts, but whose striking conjunction does not seem to have been appreciated before:

\smallskip
\begin{itemize}[leftmargin=.8cm]
 \item[\bf (1)]
 \cite[\S 5]{SS22AnyonicDefectBranes}:
  Plausible future topological quantum computation
  hardware realizes (only) those anyon braid quantum gates (Lit. \ref{LiteratureBraiding}) which act on quantum states by
  {monodromy braid representations}
  (Lit. \ref{LiteratureBraidRepresentations})
  of Knizhnik-Zamolodchikov (KZ) connections
  (Lit. \ref{KZConnectionsOnConformalBlocksReferences})
  on ``$\suTwoAffine{\Level}$-{conformal blocks}''  from conformal field theory.
  At level $k = 2$ these include the popular Majorana/Ising anyons, and for $k = 3$ the universal ``Fibonacci anyons'' (Lit. \ref{LiteratureAnyonSpecies}).

  The point is that anyon braiding is often broadly hypothesized (cf. \cite{Valera21})
  to be described by any braid group representation or
  any unitary braided fusion category. However, plausible physical realizations are expected to be much more specifically given by
  modular tensor categories of affine $\SpecialUnitaryLieAlgebra{2}$- representations, these
  being quantum states in Chern-Simons theory hence correlators in (chiral) current algebra (i.e: WZW) conformal field theory (Lit. \ref{KZConnectionsOnConformalBlocksReferences}).

\smallskip
  \item[\bf (2)]
  \cite[\S 2]{SS22AnyonicDefectBranes}:
  A technical result known as the ``{\it hypergeometric integral construction}'' (Lit. \ref{HypergeometricIntegralReferences}) of
  KZ-solutions serves to show that
  just these $\suTwoAffine{\Level}$-monodromy braid representations have a natural construction in algebraic topology (Lit. \ref{LiteratureTopologyAndHomotopyTheory}), where they
  are given by the monodromy of canonical {Gauss-Manin connections}
  (Lit. \ref{LiteratureGaussManinConnections})
  on fiberwise twisted cohomology groups (Lit. \ref{LiteratureCohomology})
  of configurations spaces of points in the plane (Lit. \ref{LiteratureConfigurationSpaces}).

\end{itemize}

\noindent
This is striking, because the first item above, when taken at face value, invokes a fairly long and intricate sequence of constructions from
conformal field theory and representation theory of affine Lie algebras, while the second item invokes only the most basic concepts of algebraic topology.
It is via this translation from conformal quantum field theory to plain algebraic topology (discussed in \cref{ViaParameterizedPointSetTopology})
that topological quantum gates can be fully grasped by a homotopically typed programming language (as discussed in
\cref{ViaDependentHomotopyTypeTheory}).

More in detail, the structure of our construction and proof is indicated in the following flow diagram:

\vspace{-.2cm}
\begin{center}
\small
\begin{tikzcd}[row sep=37pt,
  column sep=20pt,
  decoration=snake
]
  \mathllap{
    \scalebox{.8}{
      Lit. \ref{LiteratureBraiding}
      \;\;
    }
  }
  \fcolorbox{black}{lightgray}{
    \hspace{-.3cm}
    \def\arraystretch{.9}
    \color{darkblue}
    \begin{tabular}{c}
      \color{darkblue}
      Monodromy
      \\
      braid representation
      \\
    \end{tabular}
    \hspace{-.2cm}
  }
  \ar[
    rrrr,
    Latex-Latex,
    "{
      \fcolorbox{gray}{white}{
        \cref{TopologicalQuantumGates}
      }
    }"
  ]
  &[-20pt]
  {}
  &
  {}
  \ar[
    d,
    phantom,
    shift right=32pt,
    "\scalebox{.7}{
      Lit. \ref{LiteratureLocalSystems}
    }"{pos=.4}
  ]
  &
  {}
  &[-10pt]
  \fcolorbox{black}{lightgray}{
    \hspace{-.3cm}
    \def\arraystretch{.9}
    \color{darkblue}
    \begin{tabular}{c}
      \color{darkblue}
      Specified topological
      \\
      quantum logic gates
      \\
    \end{tabular}
  }
  \mathrlap{
    \scalebox{.8}{
      \;
      Lit. \ref{LiteratureTopologicalQuantumComputation}
     }
  }
  \\
  \mathllap{
    \scalebox{.8}{
      Lit. \ref{KZConnectionsOnConformalBlocksReferences}
      \,
    }
  }
  \fcolorbox{black}{lightgray}{
    \hspace{-.3cm}
    \def\arraystretch{.9}
    \color{darkblue}
    \begin{tabular}{c}
      \color{darkblue}
      Knizhnik-Zamol.
      \\
      connection
    \end{tabular}
    \hspace{-.2cm}
  }
  \ar[
    u,
    decorate,
    shorten >= -1.5pt,
    "{
      \scalebox{.75}{
        \def\arraystretch{.9}
        \color{greenii}
        \begin{tabular}{c}
        \bf
        holonomy
        \end{tabular}
      }
    }"{xshift=3pt},
  ]
  \ar[
    rrrr,
    Latex-Latex
  ]
  &
  {}
  &
  {}
  &{}
  &
  \underset{
    \scalebox{.7}{
      \hspace{30pt}
      Def. \ref{DataStructureOfConformalBlocks}
    }
  }{
  \fcolorbox{black}{lightgray}{
    \hspace{-.3cm}
    \def\arraystretch{.9}
    \color{darkblue}
    \begin{tabular}{c}
      \color{darkblue}
      Transport of 0-truncated
      \\
      types of dependent functions
      \\
      to twisted higher $\ComplexNumbers$-deloopings
    \end{tabular}
    \hspace{-.2cm}
  }
  }
  \ar[
    u,
    decorate,
    shorten >= -1.5pt,
    "{
      \scalebox{.75}{
        \def\arraystretch{.9}
        \color{greenii}
        \begin{tabular}{c}
         \bf  transport
        \end{tabular}
      }
    }"{xshift=+3pt},
    "{
      \scalebox{.85}{
        \;\;
        Thm.
        \ref{TheTheorem}
      }
    }"{swap}
  ]
  \\[+6pt]
  \mathllap{
    \scalebox{.8}{
      Lit. \ref{LiteratureGaussManinConnections}
      \;
    }
  }
  \fcolorbox{black}{lightgray}{
    \hspace{-.3cm}
    \def\arraystretch{.9}
    \color{darkblue}
    \begin{tabular}{c}
      Gauss-Manin
      \\
      connections
    \end{tabular}
    \hspace{-.2cm}
  }
  \ar[
    dr,
    decorate,
    shorten <=-3pt,
    shorten >=-1.5pt,
    "{
      \scalebox{.7}{
        \color{greenii}
    \bf     on
      }
    }"{sloped, yshift=2.3pt, pos=.4}
  ]
  \ar[
    dr,
    decorate,
    shorten <=-3pt,
    shorten >=-1.5pt,
    "{
      \scalebox{.7}{
        \color{greenii}
     \bf    (fiberwise)
      }
    }"{sloped, swap, pos=.4, yshift=-2pt}
  ]
  \ar[
    r,
    phantom,
    "{
      \scalebox{.7}{among}
    }"
  ]
  \ar[
    u,
    decorate,
    shorten >= -1.5pt,
    "{
      \scalebox{.75}{
        \def\arraystretch{.9}
        \begin{tabular}{c}
          \color{greenii} \bf
          over configuration
          \\
          \color{greenii} \bf
          spaces of points
          \\
          Lit.
          \ref{LiteratureConfigurationSpaces}
        \end{tabular}
      }
    }"{swap, xshift=-3pt},
    "{
      \scalebox{.85}{
        Ex. \ref{TheKZConnection}
        \;\;
      }
    }"
  ]
  &
  \underset{
    \hspace{5pt}
    \scalebox{.7}{
      Lit. \ref{LiteratureLocalSystems}
    }
  }{
  \fcolorbox{black}{lightgray}{
    \hspace{-.3cm}
    \def\arraystretch{.9}
    \color{darkblue}
    \begin{tabular}{c}
      Flat
      \\
      connections
    \end{tabular}
    \hspace{-.2cm}
  }
  }
  \ar[
    rrr,
    Latex-Latex
  ]
  &
  {}
  &
  {}
  &
  \underset{
    \raisebox{-1pt}{
      \scalebox{.7}{
      Def. \ref{TypeTheoreticGaussManinConnection}
      }
    }
  }{
  \fcolorbox{black}{lightgray}{
    \hspace{-.3cm}
    \def\arraystretch{.9}
    \color{darkblue}
    \begin{tabular}{c}
      Transport of 0-truncated
      \\
      dependent function types
    \end{tabular}
    \hspace{-.2cm}
  }
  }
  \ar[
    u,
    decorate,
    shorten >= -1.5pt,
    "{
      \scalebox{.75}{
        \def\arraystretch{.9}
        \begin{tabular}{c}
          \color{greenii} \bf
          over deloopings
          \\
          \color{greenii} \bf
          of braid groups
          \\
          Lit. \ref{LiteratureBraiding}
        \end{tabular}
      }
    }"{xshift=+0pt},
    "{
      \scalebox{.7}{
        Lem. \ref{AssigningPhasesToPureArtinGenerators}
      }
    }"{swap, xshift=2pt}
  ]
  \ar[
    uu,
    bend right=38,
    shift right=48pt,
    decorate,
    "{
      \scalebox{.95}{
        \fcolorbox{gray}{white}{
          \cref{KZConnectionsInHomotopyTypeTheory}
        }
      }
    }"{description}
  ]
  \\
  &
  \underset{
    \mathclap{
    \scalebox{.7}{
      \begin{tabular}{c}
        Lit \ref{LiteratureCohomology}
      \end{tabular}
     }
    }
  }{
  \fcolorbox{black}{lightgray}{
    \hspace{-.3cm}
    \def\arraystretch{.9}
    \color{darkblue}
    \hspace{-5pt}
    \begin{tabular}{c}
      Twisted generalized
      \\
      cohomology
    \end{tabular}
    \hspace{-5pt}
  }
  }
  \ar[
    rrr,
    Latex-Latex
  ]
  &&&
  \underset{
    \raisebox{-1pt}{
    \scalebox{.7}{
      \hspace{00pt}
      \eqref{TwistedCohomologyFirstVersion}
    }
    }
  }{
  \fcolorbox{black}{lightgray}{
    \hspace{-.3cm}
    \def\arraystretch{.9}
    \color{darkblue}
    \begin{tabular}{c}
      0-truncations of
      \\
      dependent function types
    \end{tabular}
    \hspace{-.2cm}
  }
  }
  \ar[
    u,
    decorate
  ]
  \\[-3pt]
  \mathllap{
    \scalebox{.8}{
      Lit. \ref{LiteratureTopologyAndHomotopyTheory}
      \;
    }
  }
  \fcolorbox{black}{lightgray}{
    \hspace{-.3cm}
    \def\arraystretch{.9}
    \color{orangeii}
    \begin{tabular}{c}
      Parameterized
      \\
      homotopy theory
    \end{tabular}
    \hspace{-.2cm}
  }
  \ar[
    uuuu,
    decorate,
    shift left=35pt,
    bend left=33,
    "{
      \scalebox{1}{
        \fcolorbox{gray}{white}{
        \cref{ViaParameterizedPointSetTopology}
        }
      }
    }"{description}
  ]
  \ar[
    uu,
    decorate,
    "{
      \scalebox{.75}{
        \def\arraystretch{.9}
        \hspace{-14pt}
        \begin{tabular}{c}
          \color{greenii} \bf
          fiberwise homotopy
          \\
          \color{greenii} \bf
          classes of slice maps
        \end{tabular}
        \hspace{-14pt}
      }
    }"{xshift=-14pt, pos=.55, description},
    "{
      \scalebox{.85}{
         Thm.
         \ref{GaussManinConnectionInTwistedGeneralizedCohomologyViaMappingSpaces}
      }
    }"{pos=.8, xshift=-5pt}
  ]
  \ar[
    ur,
    decorate,
    shorten >=-6pt,
    "{
      \scalebox{.75}{
        \def\arraystretch{.9}
        \color{greenii} \bf
        \begin{tabular}{c}
          homotopy
          \\
          classes of
          \\
          slice maps
        \end{tabular}
      }
    }"{sloped, yshift=1pt}
  ]
  \ar[
    from=rrrr,
    -Latex,
    shift right=10-2pt,
    "{
      \scalebox{1}{
        \color{orangeii}
        \bf
        semantics
      }
    }"{swap, yshift=2pt}
  ]
  \ar[
    rrrr,
    -Latex,
    shift right=10+2pt,
    "{
      \scalebox{1}{
        \color{orangeii}
        \bf
        syntax
      }
    }"{swap, yshift=-2pt}
  ]
  \ar[
    rrrr,
    shift right=2pt,
    phantom,
    "{
      \fbox{
      \hspace{-3pt}
      \scalebox{.9}{
        \cref{HoTTIdea}
      }
      \hspace{-3pt}
      }
    }"
  ]
  \ar[
    rr,
    -Latex,
    shorten >=18pt,
    shift right=10+2pt,
  ]
  &&
  \ar[
    from=rr,
    -Latex,
    shorten >=-34pt,
    shift right=10-2pt,
  ]
  &&
  \fcolorbox{black}{lightgray}{
    \hspace{-.3cm}
    \color{orangeii}
    \def\arraystretch{.9}
    \begin{tabular}{c}
      Homotopy type
      \\
      theory (HoTT)
    \end{tabular}
    \hspace{-.2cm}
  }
  \mathrlap{
    \scalebox{.8}{
      \;
      Lit. \ref{LiteratureHomotopyTypeTheory}
    }
  }
  \ar[
    uu,
    decorate,
    shift right=38pt,
    bend right=38,
    "{
      \scalebox{1}{
        \fcolorbox{gray}{white}{
        \cref{GMConnectionOnTwistedCohomologyData}
        }
      }
    }"{description}
  ]
  \ar[
    u,
    decorate,
    shorten >=-1.5pt,
    "{
      \scalebox{.75}{
        \def\arraystretch{.9}
        \color{greenii} \bf
        \hspace{-14pt}
        \begin{tabular}{c}
        \end{tabular}
        \hspace{-14pt}
      }
    }"{xshift=15pt, description}
  ]
\end{tikzcd}
\end{center}

\medskip

\noindent
{\bf Conclusion.}
As a result, what we offer here is a previously missing understanding of topological quantum gates which  is:

\begin{itemize}[leftmargin=.7cm]

\item[\bf (i)] naturally rooted in the modern foundations of homotopical mathematics,

\item[\bf (ii)] fully aware of the physical principles underlying topological quantum materials, and

\item[\bf (iii)] natively implementable in state-of-the-art programming certification languages such as {\tt Agda}.

\end{itemize}

\smallskip
\noindent The first two points mean that the construction is amenable to pure mathematical analysis while at the same
time reflecting the actual physics in question; and the third point makes this accessible to actual programming
languages for verification and simulation.

\medskip

\noindent
{\bf Example application: Certified topological quantum compilation.}
\label{ApplicationToQuantumCompilation}
Once topological quantum hardware becomes available, and generally when {\it simulating} topological quantum circuits,
the key step (e.g. \cite{ZulehnerWille20}) in implementing any quantum algorithm is its {\it compilation} to
a circuit (\hyperlink{FigureQC}{\it Fig. QC}) consisting of those logic gates that the topological hardware actually offers.

While the {\it Solovay-Kitaev Theorem}
\cite{Kitaev97}\cite{Solovay00} guarantees, under mild assumptions,
that such quantum circuits exist for any prescribed accuracy (see \cite{DawsonNielsen06}\cite{Brunekreef14}), it still requires work to
find optimal circuits under given operational constraints --- which is done by brute force (e.g. \cite{BonesteelHormoziZikosSimon05}\cite{HoroziZikosBonesteelSimon07}\cite{HormoziBonesteelSimon09}\cite[\S IV]{JohansenSimula21}).
Our result provides a
certification language (Lit. \ref{VerificationLiterature})
that reflects the analytic detail of the topological quantum compilation, achieving the combination of:

\vspace{1mm}
\begin{itemize}[leftmargin=.5cm]

\item[{\bf (1)}] modern exact real computer analysis (cf. Lit. \ref{ExactRealComputerAnalysis} \& pp. \pageref{ConstructingTheContinuum}) for certifying that one unitary operator approximates another;

\item[{\bf (2)}] novel topological language constructs for certifying that/if such operators arise from physical anyon braiding (Thm. \ref{TheTheorem}).

\end{itemize}

\smallskip

This may open the door to full
formal verification of quantum compilation of realistic topological quantum gates based on their actual physical operation principles; a task arguably constitutive for quantum computation in the long run (cf. Lit. \ref{NeedForTopologicalProtection}).

\noindent

\vspace{5cm}
\noindent
{\bf Outlook.}
\label{Outlook}
Here we are just scratching the surface of the following topics on which type theory might now be brought to bear:

\begin{itemize}[leftmargin=.4cm]
  \item
  The $\suTwo$-monodromy which we encode as a homotopy data type in Def. \ref{DataStructureOfConformalBlocks}, Thm. \ref{TheTheorem} is equivalently
  (one chiral half of) the quantum propagator of $\suTwo$-{\bf Chern-Simons theory} (cf. \cite{CabraRossini97}\cite[\S 5]{Gawedzki00}) on a
  cylinder with ``Wilson line'' insertions.

  \item
  As a type constructor, Def. \ref{DataStructureOfConformalBlocks} clearly works beyond the specific choice of twisting \eqref{SpecifyingTheTwistForKZConnectionCase}: that choice is needed only to identify its categorical semantics as the familiar KZ-monodromy on
  conformal blocks. But more general twistings may still have recognizable semantics. For instance, in \cite[Rem. 2.22]{SS22AnyonicDefectBranes}
  we provided evidence that for fractional levels $\ShiftedLevel \mapsto \ShiftedLevel/q$ we obtain the conformal blocks of {\bf logarithmic CFTs}
  expected for some anyonic quantum states.

  \item The braids (Lit. \ref{LiteratureBraiding}) embodied by Def. \ref{DataStructureOfConformalBlocks} are a special case of embedded
  framed {\bf cobordism}. Indeed, the way we found the construction presented here is
  (following \cite{SS19Conf}\cite{CSS21})
  as a special case of (twisted) cohomology of {\it Cohomotopy moduli spaces}, the latter encoding embedded framed cobordism by
  Pontrjagin's theorem (cf. \cite{SS21MF}\cite{SS19TadpoleCancellation}).
\end{itemize}

\medskip

Last but not least,  a couple of interesting extensions of homotopy type theory (Lit. \ref{LiteratureHomotopyTypeTheory}) seamlessly lend themselves
to handling these and further aspects of topological quantum programming --- we aim to
discuss these elsewhere:

\vspace{1mm}
\begin{itemize}[leftmargin=.5cm]

\item[--] {\bf Linear HoTT and Universal Quantum Programming.}
A natural extension of homotopy data types by (dependent) {\it linear} -- namely: {\it quantum} -- data types exists (\cite{Riley22}, anticipated in \cite{Schreiber14})
providing a universal {\it quantum} programming language
(see \cite{Schreiber22})
reflecting the linearity of quantum data types (such as
the no-cloning property) together with the quantum measurement process and classical control mechanisms.

In fact, beyond ordinary quantum data types, this knows about dependent {\it higher} homotopical linear types, namely about {\it parameterized spectra},
such as those consisting of the parameterized Eilenberg-MacLane types (Lit. \ref{LiteratureEilenbergMacLanSpaces}) used here. This means that
essentially the same homotopy data structure of Def. \ref{DataStructureOfConformalBlocks}
but constructed as a linear homotopy type, integrates topological quantum gates into a full-blown quantum circuit language with quantum measurement gates and classical control; this is discussed in a companion
article \cite{QPinLHOTT}.

\vspace{1mm}
\item[--] {\bf Cohesive HoTT and Solid State Physics of Topological Quantum Gates.}\label{CohesiveHoTT}
The analysis of \cite{SS22AnyonictopologicalOrder} concludes (Lit. \ref{AnyonLiterature}) that realistic topological quantum gates implemented in
topologically ordered quantum materials are described by the holonomy of Gauss-Manin (GM) connections (Lit. \ref{LiteratureGaussManinConnections}),
not just on twisted ordinary cohomology
(as discussed here), but on {\it T}wisted \& {\it E}quivariant \& {\it D}ifferential ({\it TED}) topological K-cohomology
(Lit. \ref{LiteratureKTheoryClassificationOfTopologicalPhases}),  which refines the KZ-monodromy
through a comparison map called the {\it secondary Chern character}.
This predicts physical corrections to the traditional expectation of monodromy braiding gates operating by the holonomy of the
Knizhnik-Zamolodchikov (KZ) connection:
namely, that braid gate operations in the co-kernel of the secondary TED-Chern character are not actually realizable by adiabatic braiding of
defects in real topological quantum materials, while new K-theoretic braiding operations appear in the kernel.

\vspace{1mm}
The
{\it differential}
and
{\it equivariant}  structure required to resolve these corrections from solid state physics is naturally incorporated into homotopy theory by the consideration of {\it cohesive}
(following the terminology of \cite{Lawvere94}\cite{Lawvere07}) systems of modal operators (\cite[\S 3]{SatiSchreiberSasheff12}\cite{Schreiber13}\cite{SS20OrbifoldCohomology}\cite{SS21EPB}) in a way that again lends itself to implementation in type theory, then called {\it cohesive homotopy type theory} (\cite{SchreiberShulman14}\cite{Schreiber15}\cite{Wellen17}\cite{Shulman18}\cite{Myers21}\cite{Myers22}\cite{MyersRiley23}).

\end{itemize}

\smallskip
These {\it linear} and {\it cohesive} refinements of homotopy type theory should naturally combine
to a {\it cohesive linear homotopy type theory} (the corresponding semantics is discussed in \cite[\S 4.1.2]{Schreiber13}\cite{BNV16}\cite{GradySati19}),
which we suggest to be the ultimate platform for certification of classically controlled quantum programming
with realistic topological quantum gates: \footnote{Intriguingly, cohesive linear homotopy theory is also the semantic
context in which to naturally make formal
sense of the key ingredients of high energy physics, specifically of string/M-theory
(cf. \cite{Schreiber14b}\cite[p. 6]{SS20OrbifoldCohomology}). This is in line with a deep relationship between strongly coupled
quantum systems (such as anyonic topological order) and string/M-theory, cf. \cite[Rem. 2.8]{SS22AnyonictopologicalOrder}\cite{Sati23}\cite{Schreiber23}.
}

\vspace{-.35cm}
$$
\hspace{0mm}
  \begin{tikzcd}[
    row sep=5pt,
    column sep=-5pt
  ]
    &
    \overset{
      \mathclap{
        \raisebox{8pt}{
        \scalebox{.8}{
          \color{greenii}
          \def\arraystretch{.8}
          \begin{tabular}{c}
            Certified
            topological quantum gates
            \\
            by adiabatic
            anyonic defect braiding
          \end{tabular}
        }
        }
      }
    }{
    \fcolorbox{black}{lightgray}{
      \color{darkblue}
      \hspace{-10pt}
      \def\arraystretch{.9}
      \begin{tabular}{c}
        Homotopy
        \\
        Type Theory
      \end{tabular}
      \hspace{-10pt}
    }
    }
    \ar[
      dd,
      phantom,
      "{
        \scalebox{.8}{
          discussed in
          \cref{KZConnectionsInHomotopyTypeTheory}
        }
      }"{pos=.05}
    ]
    \ar[
      dl,
      hook'
    ]
    \ar[
      dr,
      hook
    ]
    \\
    \mathllap{
      \scalebox{.8}{
        \color{greenii}
        \def\arraystretch{.9}
        \begin{tabular}{c}
          Certified quantum circuits
          \\
          with full classical control
        \end{tabular}
      }
    }
    \fcolorbox{black}{lightgray}{
      \color{darkblue}
      \hspace{-10pt}
      \def\arraystretch{.9}
      \begin{tabular}{c}
        Linear Homotopy
        \\
        Type Theory
      \end{tabular}
      \hspace{-10pt}
    }
    \ar[
      d,
      phantom,
      shift right=45pt,
      "{
        \scalebox{.8}{
          discussed in
          \cite{QPinLHOTT}
        }
      }"{pos=.1}
    ]
    \ar[
      dr,
      hook'
    ]
    &&
    \fcolorbox{black}{lightgray}{
      \color{darkblue}
      \hspace{-10pt}
      \def\arraystretch{.9}
      \begin{tabular}{c}
        Cohesive Homotopy
        \\
        Type Theory
      \end{tabular}
      \hspace{-10pt}
    }
    \mathrlap{
      \scalebox{.8}{
        \color{greenii}
        \def\arraystretch{.9}
        \begin{tabular}{c}
          Physical effects of symm-
          \\
          etry-protected topologically
          \\
          ordered quantum materials
        \end{tabular}
      }
    }
    \ar[
      d,
      phantom,
      shift left=45pt,
      "{
        \scalebox{.8}{
          following
          \cite{SS22AnyonictopologicalOrder}
        }
      }"{pos=.1}
    ]
    \ar[
      dl,
      hook
    ]
    \\
    {}
    &
    \underset{
      \mathclap{
        \raisebox{-17pt}{
          \color{greenii}
          \scalebox{.8}{
          \def\arraystretch{.9}
          \begin{tabular}{c}
            Certified quantum programming of
            \\
            classically controlled quantum circuits
            \\
            compiled from topological quantum gates
            \\
            in physically realistic quantum materials
            \\
            \color{black}
            to be discussed elsewhere
          \end{tabular}
          }
        }
      }
    }{
    \fcolorbox{black}{lightgray}{
      \color{orangeii}
      \hspace{-8pt}
      \def\arraystretch{.9}
      \begin{tabular}{c}
        Cohesive Linear Homotopy
        \\
        Type Theory
      \end{tabular}
      \hspace{-10pt}
    }
    }
    &
    {}
  \end{tikzcd}
$$

\newpage

\section{Background and perspective}

Since we are connecting areas of physics,  mathematics, and computer science that have not been in close contact
before, we provide here a kind of index for various technical terms, with brief explanations and pointers to the literature.
The reader may or may not want to peruse this section linearly, and can certainly skip ahead to come back here only as the need arises.

\smallskip
\begin{literature}[\bf Quantum computation]
\label{LiteratureQuantumComputation}
The general idea of quantum computation was originally articulated by Yuri Manin \cite{Manin80}\cite{Manin00}, Paul Benioff \cite{Benioff80} and Richard Feynman \cite{Feynman82}, brought into shape by David Deutsch \cite{Deutsch89}, shown to be potentially of dramatic practical relevance by
Peter Shor and others \cite{Shor94}\cite{Simon97}...
{\it if} quantum error correction could be brought under control (Lit. \ref{NeedForTopologicalProtection}), which was shown
by Peter Shor and others \cite{Shor95}
to be at least a theoretical possibility.

Textbook accounts of the general principles of quantum computation and quantum information theory include:
\cite{NC10} \cite{RieffelPolak11}\cite{BenentiCasatiRossini18}\cite{BEZ20}. Impressions of the state of the field may be found in \cite{Preskill22} and in the discussion of NISQ machines (Lit. \ref{LiteratureOnNISQMachines}); elementary
exposition leading up to the following discussion may be found in \cite{Schreiber22Erice}.

\medskip

Here we are concerned with the {\it low hardware-level} formulation of computation, where elementary instructions (``logic gate names'') are mapped to
transformations on a state space (such as on the registers of a computing machine), and where programs consist of paths of such instructions
(``logic circuits'') executed as iterated such state transformations --- we expand on this in \cref{TopologicalQuantumGates}.
From this very basic perspective, the one key aspect of quantum computation to take note of is just that:

Where an operation on a classical state space
is a {\it map of finite sets} (of states), the analogous quantum operation is an {\it invertible}
(Lit. \ref{ReversibleComputingLiterature})
{\it linear map} of finite-dimensional {\it complex vector spaces}.
(In Thms. \ref{GaussManinConnectionInTwistedGeneralizedCohomologyViaMappingSpaces}, \ref{TheTheorem}
these vector spaces arise in the form of {\it complex cohomology groups}, Lit. \ref{LiteratureCohomology}.)

\begin{center}
\hypertarget{TableD}{}
\begin{tabular}{ll}
\def\arraystretch{1.4}
\begin{tabular}{|c|c|}
  \hline
  \bf Classical computation
  &
  \bf Quantum computation
  \\
  \hline
  \hline
  $
    \begin{tikzcd}[decoration=snake]
      \overset{
        \mathclap{
          \raisebox{3pt}{
            \scalebox{.7}{
              \color{darkblue}
              \bf
              \def\arraystretch{.9}
              \begin{tabular}{c}
                finite set
                \\
                of states
              \end{tabular}
            }
          }
        }
      }{
        \{0,1\}^D
      }
      \ar[
        rr,
        "{
          \scalebox{.7}{
            \color{greenii}
            \bf
            computation
          }
        }"
      ]
      &&
      \overset{
        \mathclap{
          \raisebox{3pt}{
            \scalebox{.7}{
              \color{darkblue}
              \bf
              \def\arraystretch{.9}
              \begin{tabular}{c}
                finite set
                \\
                of states
              \end{tabular}
            }
          }
        }
      }{
      \{0,1\}^D
      }
      \\
      t_1
      \ar[
        rr,
        decorate,
        "{
          \scalebox{.7}{
            \color{greenii}
            \bf
            instruction path
          }
        }"{yshift=3pt}
      ]
      &&
      t_2
    \end{tikzcd}
  $
  &
  \;\,
  $
    \begin{tikzcd}[decoration=snake]
      \overset{
        \mathclap{
          \raisebox{3pt}{
            \scalebox{.7}{
              \color{darkblue}
              \bf
              \def\arraystretch{.9}
              \begin{tabular}{c}
                finite dim.
                \\
                vector space
                \\
                of states
              \end{tabular}
            }
          }
        }
      }{
        \ComplexNumbers^{2D}
      }
      \;\;
      \ar[
        rrr,
        "{
          \scalebox{.7}{
            \color{greenii}
            \bf
            computation
          }
        }",
        "{
          \scalebox{.7}{
           (complex linear map)
          }
        }"{swap}
      ]
      &&&
      \;\;
      \overset{
        \mathclap{
          \raisebox{3pt}{
            \scalebox{.7}{
              \color{darkblue}
              \bf
              \def\arraystretch{.9}
              \begin{tabular}{c}
                finite dim.
                \\
                vector space
                \\
                of states
              \end{tabular}
            }
          }
        }
      }{
        \ComplexNumbers^{2D}
      }
      \\
      t_1
      \ar[
        rrr,
        decorate,
        "{
          \scalebox{.7}{
            \color{greenii}
            \bf
            instruction path
          }
        }"{yshift=3pt}
      ]
      &&&
      t_2
    \end{tikzcd}
  $
  \;\,
  \\
  \hline
\end{tabular}
&
\begin{minipage}{6cm}
  \small
  {\bf Table D.}
  Where a classical {\it digital} computation (program) is a function (a map) between finite sets (of bits), a corresponding quantum computation  is (in particular) a {\it linear map} between finite-{\it dimensional} complex vector spaces.

  Compare the discussion in \cref{TopologicalQuantumGates}.
\end{minipage}
\end{tabular}
\end{center}

This {\it complex linearity} of quantum computation processes is to a large extent the source of their richness, since there is now a {\it continuum} (an un-countable multitude) of possible states, hence of data (see Lit. \ref{ExactRealComputerAnalysis}) where a classical computer sees only a finite set. In particular, the {\it verification}/{\it certification} (Lit. \ref{VerificationLiterature}) of a quantum computational process (such as a proof that a given set of quantum gates {\it compiles} to a prescribed linear map, to within specified accuracy, cf. p. \pageref{ApplicationToQuantumCompilation})
is in general a problem in {\it constructive analysis}/{\it exact real computer arithmetic} (Lit. \ref{ExactRealComputerAnalysis}).

Other key aspects of quantum computation, such as unitarity of these linear maps (i.e., their preservation of Hilbert space structure and hence of
the probabilistic interpretation of quantum physics) on the one hand, and non-unitary state collapse (under quantum measurement) on the other,
play a tangential role in the present article but will be brought out fully in the companion article \cite{QPinLHOTT}.
\end{literature}

\begin{literature}[\bf Reversible computing]
  \label{ReversibleComputingLiterature}
  The conceptual development of quantum computation leads to a re-evaluation of basic principles of computing. For instance, at the fundamental
  microscopic level, all physical processes --- and hence also all computational processes --- are {\it reversible}. Executing non-reversible
  classical logic gates (such as $\mathrm{AND} : \Bits^2 \to \Bits$) means for a computer to discard information into --- hence to interact with ---  an arbitrarily
  complex environment: this is known as {\it Landauer's principle} \cite{Bennett03}. Trivial as this may superficially seem, it is exactly such environmental
  interactions that have to be shielded from a microscopic system in order to realize a coherent quantum computation process (Lit. \ref{NeedForTopologicalProtection}) --- which is then reversible
  as predicted by quantum mechanics according to unitary Schr{\"o}dinger evolution. This way, quantum circuits ({\it excluding} quantum {\it measurement} gates,
  which we disregard here but turn to in the companion \cite{QPinLHOTT}) are a form of reversible computation
  (e.g. \cite[\S 2.2]{Kitaev97}\cite[\S 1.4.1, 3.2.5]{NC10}\cite[\S 9]{Aman20}).

  Curiously, exactly this fundamental reversibility of computation is natively reflected --- under the formulation (in \cref{TopologicalQuantumGates})
  of low-level computation as ``path-lifting'' --- by homotopy-typed programming languages (Lit. \ref{LiteratureHomotopyTypeTheory}), where it corresponds
  to the invertibility \eqref{ReversesAreInverses}  of ``parameter paths'' in the guise of identification certificates \eqref{IdentificationType} and thus
  to the denotational semantics of homotopy type theory in spaces/$\infty$-groupoids reviewed in \cref{HoTTIdea}.
\end{literature}

\begin{literature}[\bf NISQ computers]
  \label{LiteratureOnNISQMachines}
  Currently existing quantum computers
  (such as those based on ``superconducting qbits'', see e.g. \cite{ClarkeWilhelm08}\cite{HWFZ20})
  serve as proof-of-principle of the idea of quantum computation
  (Lit.  \ref{LiteratureQuantumComputation})
  but
  offer puny computational resources, as
  they are (very) {\it n}oisy and (at best) of {\it i}ntermediate {\it s}cale:  ``NISQ machines'' \cite{Preskill18}\cite{LeymannBarzen20}.
  What is currently missing are noise-protection mechanisms that would allow to scale up the size and coherence time of quantum memory.
  The foremost such protection mechanism
  arguably (Lit. \ref{NeedForTopologicalProtection}) is {\it topological} protection (Lit. \ref{LiteratureTopologicalQuantumComputation}).
\end{literature}

\begin{literature}[\bf Topological quantum computation]
  \label{LiteratureTopologicalQuantumComputation}
  The idea of topological quantum computation --- specifically (we give an exposition in \cref{TopologicalQuantumGates}) of topological quantum logic gates
  operating by adiabatic (Lit. \ref{TQCAsAdiabaticQuantumComputing}) braiding (Lit. \ref{LiteratureConfigurationSpaces}) of worldlines of anyonic defects
  (Lit. \ref{AnyonLiterature}) in topological quantum materials (Lit. \ref{TopologicalQuantumMaterials}) controlled by Chern-Simons/WZW theory (Lit. \ref{KZConnectionsOnConformalBlocksReferences}) --- is attributed to \cite{Kitaev03}\cite{Freedman98}\cite{FreedmanKitaevLarsenWang03}\cite{FreedmanLarsenWang02} (see also \cite{KauffmanLomonaco04}), further expanded on in \cite{NayakSimonSternFreedmanDasSarma08} (analogous discussion for discrete gauge groups, i.e. Dijkgraaf-Witten theory,
  is in \cite{OgburnPreskill99}\cite{Mochon03}\cite{Mochon04}). Reviews include \cite{Brennen08}\cite{RowellWang18} and textbook accounts include
  \cite{Wang10}\cite{Pachos12}\cite{SternLindner13}\cite{Stanescu20}\cite{Simon21}.

  Interestingly, the key principle of topological quantum computation and also in more generality (as discussed in \cref{TopologicalQuantumGates}) was already clearly articulated in \cite{ZanardiRasetti99} under the name ``Holonomic Quantum Computation''.


  Beware that traditional literature tends to discuss the issue in the abstract, disregarding (cf. \cite{Valera21}) the question of physical realization
  of anyonic defect braiding in actual materials, even theoretically (Lit. \ref{MathematicsOfAnyonsLiterature}). A (previously) popular experimental setup
  via ``Majorana zero modes'' now seems dubious  (Lit. \ref{LiteratureMajoranaZeroModes}). On the other hand, the braiding of band node defects in the
  momentum space (Lit. \ref{TopologicalQuantumMaterials}) of topological semimetals may be promising (Lit. \ref{LiteratureBraidingInMomentumSpace}).
\end{literature}

\begin{literature}[\bf Need for topological quantum protection]
\label{NeedForTopologicalProtection}
While the idea of topological quantum computation (Lit. \ref{LiteratureTopologicalQuantumComputation}) may at times
be presented as just one of several interesting alternative approaches to quantum computation, there are good arguments that any practically useful,
hence scalable, quantum computing architecture, beyond the currently available NISQ machines (Lit. \ref{LiteratureOnNISQMachines}), must be topologically
protected by necessity:
$\,$

\begin{center}
  \begin{minipage}{16cm}
    \cite{Sau17}: ``{small {\rm [NISQ]} machines are unlikely to uncover truly macroscopic quantum phenomena, which have no classical analogs. This will
    likely require a scalable approach to quantum computation {\rm[...]} based on {\rm[...]} topological quantum computation (TQC)  {\rm[...]} The central
    idea of TQC is to encode qubits into states of topological phases of matter. Qubits encoded in such states are expected to be topologically protected,
    or robust, against the 'prying eyes' of the environment, which are believed to be the bane of conventional quantum computation.}''
  \end{minipage}

\end{center}

\begin{center}
  \begin{minipage}{16cm}
    \cite{DasSarma22Problem}:
    ``{The qubit systems we have today are a tremendous scientific achievement, but they take us no closer to having a quantum computer that can solve
    a problem that anybody cares about.} [...] {\it What is missing is the breakthrough} [...] {\it bypassing quantum error correction by using
    far-more-stable qubits, in an approach called topological quantum computing.}''
  \end{minipage}
\end{center}
\end{literature}

\begin{literature}[\bf Adiabatic nature of topological quantum gates]
  \label{TQCAsAdiabaticQuantumComputing}
  The operation on degenerate quantum ground states by braiding (Lit. \ref{LiteratureConfigurationSpaces}) of anyonic (Lit.  \ref{AnyonLiterature})
  defect worldlines (Lit. \ref{LiteratureTopologicalQuantumComputation}) is an instance of unitary transformations given by the {\it Quantum Adiabatic Theorem}
  (e.g. \cite{Nenciu80}\cite{ASY87}\cite{RigolinOrtiz12}\cite{BDF20}). That topological quantum computation is, in this sense, an instance of
  ``adiabatic quantum computation''
  \cite{FGGS00}\cite{AvDKLLR07}
  is at times hard to discern from the literature. This is on the one hand because the term
  tends to be used by default for (non-topological) {\it quantum annealing} processes \cite{GrantHumble20}; and on the other hand, because the experimental
  focus on detecting ``Majorana zero modes'' (Lit. \ref{LiteratureMajoranaZeroModes}) entirely stuck at endpoints of nanowires  seems to have led
  to disregard for the eventual need for braiding, altogether.

  References that do make explicit the adiabatic nature of topological quantum gate processes include (besides \cite{ZanardiRasetti99}):
  \cite[p. 2]{ChildsFarhiPreskill02}\cite[pp. 7]{FreedmanKitaevLarsenWang03}\cite[p. 6]{NayakSimonSternFreedmanDasSarma08} \cite{ChengGalitskiDasSarma11}\cite[p. 1]{RigolinOrtiz12}\cite[p. 50, 52]{Pachos12}\cite{CLBFN15}\cite{MCMC19}\cite[p. 321]{Stanescu20}.

  We amplify this because adiabatic transformations in quantum physics are an instance of {\it path lifting} (Lit. \ref{PathLiftingLiterature}) --- namely of classical {\it parameter paths} to (linear) maps  between quantum state spaces depending on (namely: fibered over) these classical parameters \eqref{TheDrivingTheme}. Concretely: Quantum adiabatic transformations are the parallel transport (Lit. \ref{LiteratureLocalSystems}) of a corresponding connection
  (a flat connection for ``topological'' processes, Lit. \ref{LiteratureLocalSystems})
  on the quantum state bundle
  \cite{Simon83}\cite{WilczekZee84}
  (review in \cite{Berry89}\cite{Vanderbildt18}\cite[\S 2]{Stanescu20})
  whose holonomy/monodromy is known as {\it Berry phases} \cite{Berry84}.

  Curiously, this phenomenon of {\it adiabatic quantum transformations via path lifting} is natively captured by homotopically-typed languages (Lit. \ref{LiteratureHomotopyTypeTheory}), see \cref{TopologicalQuantumGates}.
\end{literature}

\begin{literature}[\bf Topological quantum materials]
\label{TopologicalQuantumMaterials}
A {\it topological phase of matter} (textbook accounts include \cite{Vanderbildt18}\cite[\S II]{Stanescu20}) is a ground state of a crystalline material in
which the vector bundle of occupied Bloch states (of electrons/positrons in the fixed atomic lattice) over the Brillouin torus (the lattice momentum space,
cf. Lit. \ref{LiteratureBraidingInMomentumSpace}) is topologically non-trivial, in that it has a nontrivial class in (twisted, equivariant) topological
K-theory (Lit. \ref{TopologicalQuantumMaterials}).
A ground state is said to exhibit {\it topological order} \cite{Wen91}\cite[\S 6.2]{Stanescu20} if it is degenerate and its adiabatic (Lit. \ref{TQCAsAdiabaticQuantumComputing}) dependency on the position of defects constitutes a braid group representation (Lit. \ref{LiteratureBraidRepresentations}).
Review with comprehensive pointers to the literature is in \cite{SS22AnyonictopologicalOrder}.
\end{literature}

\begin{literature}[\bf Majorana zero modes?]
\label{LiteratureMajoranaZeroModes}
In recent years a lot of attention towards potentially realizing topological quantum computation (Lit. \ref{LiteratureTopologicalQuantumComputation}) has
been focused on the proposal that topological quantum states might be realized in the form of ``Majorana zero modes'' \cite{Kitaev01} localized at the
endpoints of super/semi-conducting nanowires; see reviews in \cite{Marra22}\cite{DasSarma22}. However, a series of prominent claims of experimental
detection of such modes have now been retracted or called into question (see e.g. \cite{DasSarmaPan21}). Even if such modes could be detected, they would,
by design, be immobile and hence impossible as a platform for topological braid quantum gates (Lit. \ref{LiteratureTopologicalQuantumComputation}) in
the original sense based on adiabatic braiding movements (Lit. \ref{TQCAsAdiabaticQuantumComputing}). While it has been argued that some kind of effective
braiding of immobile ``Majorana zero modes'' could be emulated by other means, it seems implausible (certainly unproven) that this could be true in a sense
which would still enjoy the topological protection property of actual adiabatic braiding (Lit. \ref{TQCAsAdiabaticQuantumComputing}), thus defeating the point.

Therefore, it seems to us that even if Majorana zero modes localized in nanowires were real, they would be unlikely to support the original and established
notion of topological quantum gates (Lit. \ref{LiteratureTopologicalQuantumComputation}) with which we are concerned here.
A more promising experimental realization of actual topological quantum gates may be given by braiding of band nodes in momentum space
(Lit. \ref{LiteratureBraidingInMomentumSpace})
\end{literature}

\begin{literature}[\bf Mathematical specification of anyons]
\label{MathematicsOfAnyonsLiterature}
While the basic idea of {\it anyons} (Lit. \ref{AnyonLiterature}) may seem clear, it is not just their experimental realization that has been elusive
(cf. Lit. \ref{LiteratureMajoranaZeroModes}), but arguably already their theoretical derivation and determination within mathematical physics had been
sketchy (cf. \cite{Valera21}). Much of the mathematical/theoretical solid state physics literature takes it for granted that the answer
to ``{\it What is an anyon species?}''
(cf. Lit. \ref{LiteratureAnyonSpecies})
is: ``{\it Any unitary braid group representations!}'', hence: {\it Any unitary R-matrix!}
(e.g. \cite{KauffmanLomonaco04}, cf. Lit. \ref{LiteratureBraidRepresentations}); or more recently and such as to account for different anyon species (Lit. \ref{LiteratureAnyonSpecies}): ``{\it Any unitary braided fusion category!}''
(going back to \cite[\S 8, \S E]{Kitaev06}\cite[pp. 28]{NayakSimonSternFreedmanDasSarma08}\cite[\S 6.3]{Wang10} and repeated, usually without attribution, in numerous reviews,
e.g. \cite[\S 2.4.1]{RowellWang18}\cite[\S 2.2]{Rowell22}).

The basic idea, at least for defect anyons, is that their classical parameter space is the configuration space of points (Lit. \ref{LiteratureConfigurationSpaces}) in an effectively 2-dimensional quantum material (or in its dual Brillouin torus, Lit. \ref{LiteratureBraidingInMomentumSpace}) so that the Berry phase transformation
under adiabatic movement of the anyon positions (Lit. \ref{TQCAsAdiabaticQuantumComputing})
on the material's topologically ordered ground state (Lit. \ref{TopologicalQuantumMaterials})  constitutes a braid representation (Lit. \ref{LiteratureBraidRepresentations}).

While this sounds plausible, the only detailed derivation (as far as we are aware) from first physics principles is our recent argument in \cite[\S 3.3]{SS22AnyonictopologicalOrder};
and the result there is something more specific and also slightly modified: Anyonic wavefunctions are given by certain K-theoretic
(Lit. \ref{LiteratureKTheoryClassificationOfTopologicalPhases}) corrections (cf. p. \pageref{Outlook})
to,
specifically, the $\suTwoAffine{k}$-monodromy braiding rules ($\mathfrak{su}_2$-anyons, cf. \cite[Rem. 3.12]{SS22AnyonictopologicalOrder}).
More concretely --- and this is what matters for the present discussion --- anyon braiding is given by the holonomy of
the $\suTwo$-KZ connection (Lit. \ref{KZConnectionsOnConformalBlocksReferences}) arising as the  Gauss-Manin connections (Lit. \ref{LiteratureGaussManinConnections}) on bundles of twisted generalized cohomology groups (Lit. \ref{LiteratureCohomology}) over configuration spaces
of points (Lit. \ref{LiteratureConfigurationSpaces}).
While this cohomological derivation of anyons overlaps --- up to some corrections and specifications --- with the traditional postulate of braided fusion
categories (see \cite[Rem. 3.12]{SS22AnyonictopologicalOrder}),
it has the striking distinction that it directly lends itself (via the translation in \cref{ViaParameterizedPointSetTopology}) to certification
in homotopically typed programming languages (as such discussed in \cref{KZConnectionsInHomotopyTypeTheory}).
\end{literature}

\begin{literature}[\bf Braiding of band nodes in momentum space]
  \label{LiteratureBraidingInMomentumSpace}
  A moment of reflection reveals that the notion of topological braid quantum gates (Lit. \ref{LiteratureTopologicalQuantumComputation}) relies only on
  properties of the abstract configuration space of the quantum material (its ground states, Lit. \ref{TopologicalQuantumMaterials}), not on the tacit
  assumption that this is identified with physical space (``position space''). In particular, all arguments usually made for would-be anyonic defects
  (Lit. \ref{AnyonLiterature}) in position space immediately apply to the material's {\it momentum space} (the Brillouin torus, Lit. \ref{TopologicalQuantumMaterials})
  if suitable topological defects are present there. While this possibility has received attention only very recently (\cite[Rem. 3.9]{SS22AnyonictopologicalOrder}),
  momentum-space defects in topological quantum materials are well-known, well-established, and well-studied: These are {\it band nodes} in topological
  semi-metals (i.e., Dirac/Weyl points). Moreover, their controlled movement and potential braiding (Lit. \ref{LiteratureBraiding})
  by (adiabatic, Lit. \ref{TQCAsAdiabaticQuantumComputing}) manipulation of external parameters has been established in a variety of (meta-)materials
  and seems a rather generic property.
  References are collected in \cite[Rem. 3.9]{SS22AnyonictopologicalOrder}.
\end{literature}

\begin{literature}[\bf Holographic models for strongly correlated quantum systems]
  \label{LiteratureIntersectingBraneModels}
  Not just the physics of topologically ordered phases of matter (Lit. \ref{LiteratureTopologicalQuantumComputation}),
  but generally that of any
  {\it strongly interacting} quantum system falls outside (e.g. \cite{Strocchi13}) most of traditionally available
  ``perturbative'' analytic methods.
  The popular claim that quantum field theory is the most precise physical theory ever conceived tacitly refers
  to one special case where available
  {\it perturbative} methods happen to work well, namely in quantum electrodynamics (QED). But already quantum chromodynamics (QCD) at room
  temperature -- which is meant to describe nothing less than (the nuclei of) ordinary matter -- is so strongly coupled that there is currently no
  coherent theory for it (just a zoo of partial models and computer lattice simulations). Finding the {\it non-perturbative completion} of QCD which
  would explain in detail how ``constituent quarks'' are ``confined'' within hadronic bound states such as atomic nuclei, is an open
  ``Millennium Problem'' (for good accounts see \cite{RobertsSchmidt20}\cite{Roberts21}\cite{DingRobertsSchmidt22}\cite{Roberts22}).

  \smallskip
  One promising approach to this problem is to focus on the dynamics of the ``flux tubes'' which are thought to connect, and thereby strongly bind,
  any quark to another. Subtle quantum effects make these flux tubes behave
  \cite{Polyakov98}\cite{Polyakov99}\cite{Polyakov02} like ``strings'' propagating in a higher dimensional spacetime
  with only their endpoints (the quarks) constrained to the actual spacetime hypersurface, which now appears as a
  $(3+1)$-dimensional hypersuface
  (called a ``brane'') embedded in a higher dimensional ``bulk'' spacetime. This formulation has come to be known as ``holographic QCD''
  (reviewed in \cite{Erlich15}\cite[\S 4]{RhoZahed16}) a variant of the more famous but less realistic {\it AdS/CFT correspondence} (reviewed in \cite{AGMOO00}).
  The analogous description of
  strongly coupled condensed matter systems as intersecting brane models in string theory is known as {\it holographic quantum matter} (or variant names) \cite{ZLSS15}\cite{HartnollLucasSachdev18}. Here the role of the nuclear force is played by the ``Berry connection''
  (Lit. \ref{TQCAsAdiabaticQuantumComputing})
  and that of quarks by fermionic
  quasi-particles; see the dictionary in \cite[Table 1]{SS22AnyonicDefectBranes}.

  \smallskip
  While the full non-perturbative completion of this {\it string theoretic}-picture of strongly-coupled quantum theory also still remains to be formulated
  (working title: ``M-theory'' \cite{Duff99}), in this case, there is a tight web of hints and consistency checks available (due to insights that came to
  be known as ``the second superstring revolution'' \cite{Schwarz96}). It is by exploring such hints (in \cite{SS22AnyonicDefectBranes}, following
  \cite{SS19Conf}) that the model for topological quantum gates discussed here was discovered in \cite{SS22AnyonictopologicalOrder}. While none of
  this string-theoretic background is needed (nor assumed) for the present article, it may help to put the constructions into their broader perspective.
\end{literature}

\begin{literature}[\bf Topology and Homotopy theory]
  \label{LiteratureTopologyAndHomotopyTheory}
  Strictly speaking, {\it topology} is the study of general spaces --- ``topological spaces'', such as Euclidean space of any dimension, but also mapping
  spaces (e.g. \cite[\S 1]{AguilarGitlerPrieto02}), configuration spaces (Lit. \ref{LiteratureConfigurationSpaces}), parameter spaces, etc. --- up to
  continuous deformations (homeomorphisms); while {\it homotopy theory} is the study of such spaces up to the coarser notion of (weak) homotopy equivalences
  (which is about continuous deformations of {\it continuous maps} between spaces): {\it homotopy types}. However, the term {\it topological} is often
  used for what more precisely would be called {\it homotopical}, so that the terminological distinction is blurred (cf. \cite[p. vii]{Miller19}). This is notably the case for
  ``topological quantum computation'' (Lit. \ref{LiteratureTopologicalQuantumComputation}) which is really concerned with quantum gates parameterized by
  {\it homotopy} classes of continuous (hence: topological) computation paths, where the invariance under deformations (namely: under homotopy) reflects the robustness of the computation against noise, cf. \hyperlink{FigureH}{\it Fig. H} in \cref{TopologicalQuantumGates}.

  This way, the qualifiers ``topological'' in ``topological quantum computation'' and ``homotopy'' in ``homotopy type theory'' are actually synonymous!

  \smallskip
  \noindent
  On the other hand, one distinguishes further between:

  \begin{itemize}[leftmargin=.5cm]

  \item {\bf point-set topology} -- which is topology/homotopy theory of the familiar notion of spaces consisting of sets of points equipped with topological cohesion;

  \item {\bf algebraic topology} -- which studies topological spaces through the homological algebra of their {\it (co)homology groups}
  (Lit. \ref{LiteratureCohomology}); and it is the tendency of these to only depend on the underlying homotopy types which makes this {\it de facto} be a subject of homotopy theory);

  \item {\bf abstract homotopy theory} -- which is concerned with models for homotopy types beyond topological spaces -- notably simplicial sets, cf. around \eqref{InterpretationInTheClassicalModelTopos} below --  and with variants,
  such as local, equivariant, stable, etc.  homotopy types.

  \item {\bf parameterized} versions of all of these, where all spaces and hence all constructions on these are allowed to vary over some parameter space.

  \end{itemize}

  \smallskip

  Textbook introductions to
  basic point-set topology include \cite{Janich84}\cite{Munkres00}
  and introductions to the algebraic topology and homotopy theory based on these topological foundations include \cite{Spanier82}\cite{James84}\cite{Rotman88}\cite{Bredon93}\cite{Hatcher02}\cite{AguilarGitlerPrieto02}\cite{tomDieck08}\cite{Strom11} \cite{Arkowitz11}.
  (For early history of the subject see \cite{Hilton88}.)
  The {\it parameterized} point-set topology/homotopy theory which we need in \cref{ViaParameterizedPointSetTopology} is laid out in \cite{MaySigurdsson06}, following \cite{Booth70}\cite{BoothBrown78}.
  Abstract homotopy theory
  (going back to \cite{Brown65}\cite{Quillen67}\cite{Brown73}\cite{Adams74}) is reviewed for instance in
  \cite{KampsPorter97}\cite{Riehl14}\cite{Richter20}, and concise summaries of facts needed in our context are given in \cite[\S A]{FSS20Character}\cite[\S 3.1]{SS21EPB}).

  \smallskip
  For standard technical reasons, we assume all topological spaces in the following to be (a) compactly generated
  (cf. \cite[(1.2)]{SS21EPB})
  and (b) of the homotopy type of a CW-complex.
\end{literature}

\begin{literature}[\bf Homotopy groups]
  \label{LiteratureHomotopyGroups}
  The basic homotopical invariants of topological spaces $X$ (Lit. \ref{LiteratureIntersectingBraneModels}) are their {\it homotopy groups} $\pi_n(X)$
  (e.g., \cite[\S 3]{AguilarGitlerPrieto02}). Foremost among these is the {\it fundamental group} $\pi_1(X)$ of a connected space, which is the group of
  deformation classes of closed loops in the space, based at any fixed point (e.g. \cite[\S 2.5]{AguilarGitlerPrieto02}). Similarly, the {\it higher
  homotopy groups} $\pi_n(X)$ of a connected space are deformation classes of based maps from the $n$-sphere $S^n$ into $X$ -- these are all abelian for $n \geq 2$.

  Finally, for not necessarily connected spaces $X$ we have their set of path-connected components, denoted $\pi_0(X)$. For instance, the connected
  components of a {\it mapping space} $\Maps{}{Y}{X}$ are the {\it homotopy classes} of maps $Y \to X$.
\end{literature}

\begin{literature}[{\bf Eilenberg-MacLane spaces}]
  \label{LiteratureEilenbergMacLanSpaces}
  For any abelian group $A$ and $n \in \
  \mathbb{N}_{\geq 1}$ there exists a connected topological space
  unique up to homotopy equivalence
  (Lit \ref{LiteratureTopologyAndHomotopyTheory})
  -- called an {\it Eilenberg-MacLane (EM) space} (e.g. \cite[\S 6]{AguilarGitlerPrieto02}) and equivalently denoted $K(A, n)$
  or $B^n A$ -- whose homotopy groups (Lit. \ref{LiteratureHomotopyGroups}) are concentrated on $A$ in degree $n$:
  $$
    \pi_{n'}\big(B^n A\big)
    \;\simeq\;
    \left\{
    \hspace{-4pt}
    \def\arraystretch{1}
    \begin{array}{ll}
      A & \mbox{if} \; n' = n\,,
      \\
      \ast & \mbox{otherwise}\,.
    \end{array}
    \right.
  $$
  If $n = 1$ then $B G$  exists also for non-abelian groups $G$, known as the classifying space for $G$-principal bundles.

  For abelian groups $A$, their $n$th EM-space is naturally equivalent to the based loop space of the $(n+1)$st
  $$
    B^n A
    \;\simeq\;
    \Omega B^{n+1} A
  $$
  which makes them constitute a spectrum $HA$, the Eilenberg-MacLane spectrum for $A$.

  Finally, given a group $G$ (possibly non-abelian) acting by homomorphisms on the abelian group $A$, we have a corresponding
  parameterization of EM-spaces over the classifying space of $G$:
  $$
    \begin{tikzcd}[
      row sep=8pt
    ]
      B^n A
      \ar[r]
      &
      \HomotopyQuotient
        { B^n A }
        { G }
      \ar[d]
      \\
      &
      B G
    \end{tikzcd}
  $$
\end{literature}

\begin{literature}[\bf Cohomology]
  \label{LiteratureCohomology}
  The {\it ordinary cohomology groups} $H^n(X; A)$ of a topological space (Lit. \ref{LiteratureTopologyAndHomotopyTheory})
  are traditionally introduced in terms of cochains, but
  the equivalent reformulation central for our purpose is (e.g. \cite[\S 7.1, Cor. 12.1.20]{AguilarGitlerPrieto02} cf. \cite[Ex. 2.2]{FSS20Character})
  as homotopy classes of maps into an Eilenberg-MacLane space (Lit. \ref{LiteratureHomotopyGroups}):
  $$
    H^n(X; A)
    \;=\;
    \pi_0
    \big(
    \Maps{}
      {X}
      { B^n A }
    \big)
    \,.
  $$
  If  $n = 1$ and $G$ a possibly non-abelian group, then
  $$
    H^1(X; G)
    \;=\;
    \pi_0
    \big(
    \Maps{}
      {X}
      { B G }
    \big)
  $$
  is called the {\it first non-abelian cohomology} of $X$.
  This is the evident special case of {\it generalized non-abelian cohomology} (cf. \cite[\S 2.1]{FSS20Character}) where for {\it any} space $\mathcal{A}$
  we take the cohomology of $X$ with coefficients in $\mathcal{A}$ to be
  \begin{equation}
    \label{GeneralNonabelianCohomology}
    \mathcal{A}(X)
    \;:\defneq\;
    \pi_0
    \big(
    \Maps{}{X}{\mathcal{A}}
    \big)
    \,.
  \end{equation}
  For example, if $\mathcal{A} = S^n$ then this yields the cohomology theory known as {\it Cohomotopy.}

  Often this is considered for the special case when $\mathcal{A}$ is part of a sequence $\big\{E_n\big\}_{n \in \mathbb{N}}$ of pointed spaces
  where each is equipped with a weak homotopy equivalence to the based loop space of the next one (a {\it spectrum} of spaces). In this case
  $$
    E^n
    (X)
    \;=\;
    \pi_0
    \big(
    \Maps
      {\big}
      {X}
      { E_n }
    \big)
  $$
  is the {\it Whitehead generalized cohomology} of $X$ with coefficients in $E$.

  For example, if $E_0 = \mathrm{Fred}^{\mathbb{K}}$ is the space of {\it Fredholm operators} on a separably infinite-dimensional
  $\mathbb{K}$-Hilbert space, then
  $$
    \mathrm{KU}^{-n}(X)
    \;=\;
    \pi_0
    \left(
    \Maps{\big}
      { X }
      {
        \Omega^n
        \mathrm{Fred}^\mathbb{C}
      }
   \!\right)
    \,,
    \;\;\;\;\;\;\;
    \mathrm{KO}^{-n}(X)
    \;=\;
    \pi_0
    \left(
    \Maps{\big}
      { X }
      {
        \Omega^n
        \mathrm{Fred}^\mathbb{R}
      }
   \! \right)
  $$
  is the {\it topological K-cohomology} of $X$ (cf. Lit. \ref{LiteratureKTheoryClassificationOfTopologicalPhases})

\medskip
More generally, if the coefficient space $\mathcal{A}$ is acted on by a group $G$ (which we shall assume to be discrete for ease of exposition), then one may form the $\mathcal{A}$-fiber bundle over the classifying space $B G$ which is {\it associated} to the universal $G$-principal bundle $E G$ (also known as the {\it Borel construction} or {\it homotopy quotient} of $\mathcal{A}$ by $G$). Accordingly generalizing the definition \eqref{GeneralNonabelianCohomology} of cohomology to the ``slice over $B G$'' generalized cohomology to {\it twisted cohomology} (see \cite[\S 2.2]{FSS20Character})
$$
  \overset{
    \mathclap{
      \raisebox{3pt}{
      \scalebox{.7}{
        \color{darkblue}
        \bf
        \def\arraystretch{.9}
        \begin{tabular}{c}
          twisted
          \\
          cohomology
        \end{tabular}
      }
      }
    }
  }{
    \mathcal{A}^\tau
    (X)
  }
  \;\;\;
   :\defneq
  \;\;\;
  \pi_0
  \Big(
    \SliceMaps{\big}{B G}
      {X}
      {
        \HomotopyQuotient{\mathcal{A}}{G}
      }
  \Big)
  \;\;
  =
  \;\;
  \left\{\!\!\!\!\!\!
    \mathclap{
    \phantom{
    \begin{array}{c}
      \vert
      \\
      \vert
      \\
      vert
      \\
      vert
    \end{array}
    }
    }
  \right.
  \adjustbox{raise=16pt}{
  \begin{tikzcd}
    &&
    \overset{
    \mathclap{
      \raisebox{3pt}{
      \scalebox{.7}{
        \color{darkblue}
        \bf
        \def\arraystretch{.9}
        \begin{tabular}{c}
          universal
          \\
          local
          coefficient
          \\
          bundle
        \end{tabular}
      }
      }
    }
  }{
    \HomotopyQuotient{
      \mathcal{A}
    }{ G}
  }
    \ar[d]
    \\
    X
    \ar[
      rr,
      "{ \tau }",
      "{
        \scalebox{.7}{
          \color{greenii}
          \bf
          twist
        }
      }"{swap, sloped}
    ]
    \ar[
      urr,
      dashed,
      "{
        \scalebox{.7}{
          \color{greenii}
          \bf
          cocycle
        }
      }"{sloped}
    ]
    &&
    B G
  \end{tikzcd}
  }
  \left.
    \mathclap{
    \phantom{
    \begin{array}{c}
      \vert
      \\
      \vert
      \\
      vert
      \\
      vert
    \end{array}
    }
    }
  \right\}
  _{\!\!\!\big/\mathrm{hmtp}}
  \;
  =
  \;\;
  \left\{\!\!\!\!\!\!\!
    \mathclap{
    \phantom{
    \begin{array}{c}
      \vert
      \\
      \vert
      \\
      vert
      \\
      vert
    \end{array}
    }
    }
  \right.
  \adjustbox{raise=16pt}{
  \begin{tikzcd}
    &
    \overset{
    \mathclap{
      \raisebox{3pt}{
      \scalebox{.7}{
        \color{darkblue}
        \bf
        \def\arraystretch{.9}
        \begin{tabular}{c}
          local
          coefficient
          \\
          bundle
        \end{tabular}
      }
      }
    }
  }{
    E
  }
    \ar[d]
    \ar[r]
    \ar[dr, phantom, "{\scalebox{.7}{(pb)}}"]
    &
    \HomotopyQuotient{
      \mathcal{A}
    }{ G}
    \ar[d]
    \\
    X
    \ar[r, equals]
    \ar[
      ur,
      dashed,
      "{
        \scalebox{.7}{
          \color{greenii}
          \bf
          cocycle
        }
      }"{sloped}
    ]
    &
    X
    \ar[
      r,
      "{ \tau }",
      "{
        \scalebox{.7}{
          \color{greenii}
          \bf
          twist
        }
      }"{swap, sloped}
    ]
    &
    B G
  \end{tikzcd}
  }
  \left.
    \mathclap{
    \phantom{
    \begin{array}{c}
      \vert
      \\
      \vert
      \\
      vert
      \\
      vert
    \end{array}
    }
    }
\!\!\!\!\!  \right\}
  _{\!\!\!\big/\mathrm{hmtp}}
$$

\hspace{-.8cm}
\begin{tabular}{ll}
\begin{minipage}{10.4cm}
If
$\mathcal{A} \,=\, B^n \ComplexNumbers$ is an $n$-fold delooping of (the discrete abelian group underlying) the ring of complex numbers
and $G \,=\, \ComplexNumbers^\times$ denotes the (discrete!) group of units with its canonical multiplication action on $\ComplexNumbers$, then the local coefficient bundle $E$ on the right is, for $n = 0$, just  a flat line bundle $\mathcal{L}$ (Lit. \ref{LiteratureLocalSystems}) or rather its underlying horizontal covering space (Lit. \ref{PathLiftingLiterature}).

\end{minipage}
&
\hspace{1.3cm}
$
  \begin{tikzcd}[row sep=small, column sep=huge]
    \overset{
      \mathclap{
        \scalebox{.7}{
          \color{darkblue}
          \bf
          \begin{tabular}{c}
            Local system
            \\
            flat complex line bundle
          \end{tabular}
        }
      }
    }
    {
      \mathcal{L}
    }
    \quad
    \ar[r]
    \ar[d]
    \ar[
      dr,
      phantom,
      "{\scalebox{.7}{(pb)}}"
    ]
    &
    \HomotopyQuotient
      {\ComplexNumbers}
      {\ComplexNumbers^\times}
    \ar[d]
    \\
    X
    \ar[r, "\tau"{swap}]
    &
  \quad   \underset{
    \mathclap{
      \scalebox{.7}{
        \color{darkblue}
        \bf
        \def\arraystretch{.9}
        \begin{tabular}{c}
          classifying space of
          \\
          discrete group of complex units
        \end{tabular}
      }
    }
    }{
      B \ComplexNumbers^\times
    }
  \end{tikzcd}
$
\end{tabular}

\noindent
For general $n$ it is the local coefficient bundle for twisted ordinary complex cohomology with
\begin{equation}
 \label{LocalSystemCohomomology}
  H^n(X;\, \mathcal{L})
  \;\;
  =
  \;\;
  H^{n + \tau}
  \big(
    X
    ;\,
    \ComplexNumbers
  \big)
  \;\;
   \simeq
  \;\;
  \pi_0
  \Big(
    \SliceMaps
      {\big}{B \ComplexNumbers^\times}
      { X }
      { \HomotopyQuotient{B^n \ComplexNumbers}{\ComplexNumbers^\times} }
  \Big)
  \,.
\end{equation}
This is the reason why flat vector bundles are often referred to simply as ``local systems'' (namely: of coefficients for twisted ordinary cohomology), see Lit. \ref{LiteratureLocalSystems}. Discussion of the corresponding twisted cohomology goes back to
\cite[\S 2, 6]{Deligne70}, a textbook account is \cite[\S 5.1.1]{Voisin03II}.

In the form of ``cohomology of a local system'' $\mathcal{L}$ on the left of \eqref{LocalSystemCohomomology} twisted cohomology is used abundantly in the discussion of the hypergeometric integral construction (Lit. \ref{HypergeometricIntegralReferences}).
\end{literature}

\begin{literature}[\bf K-Theory classification of topological phases]
  \label{LiteratureKTheoryClassificationOfTopologicalPhases}
  At (very) low temperatures, the electrons in a ``weakly correlated'' crystalline material will incrementally fill up the lowest available $1$-electron
  ground states in the Coulomb background of the atomic nuclei which constitute the crystal lattice (e.g. \cite[\S 2]{Vanderbildt18}). As the wave vectors/momenta of the electrons range
  through the {\it Brillouin torus} of available lattice momenta (e.g. \cite[p. 52]{FreedMoore12}), these {\it valence states} form a complex vector bundle
  \eqref{AVectorBundle}
  over the Brillouin torus. If this vector bundle is ``topologically non-trivial'', one says that the crystalline material is in a {\it topological phase of matter}
  (Lit. \ref{TopologicalQuantumMaterials}).

  In more detail, the full relativistic computation of the electron states shows that their ground state bundle forms a class in the
  {\it topological K-cohomology} (Lit. \ref{LiteratureCohomology}) of the Brillouin torus, constituted by the electron valence bundle and a possible
  admixture of positronic ground states (\cite[Fact 2.3]{SS22AnyonictopologicalOrder}, following a famous original suggestion due to \cite{Kitaev09}).
  Moreover, if the dynamics of the electrons respects crystalline symmetries, then so does their
  valence bundle which is then classified by the corresponding {\it equivariant K-theory} of the Brillouin torus \cite{FreedMoore12}. More generally,
  if the electron dynamics respects also some internal symmetries and/or is subject to Berry phases, their ground states are classified in TED K-theory
  \cite[\S 2.3, 3.1]{SS22AnyonictopologicalOrder}.

  However, if the electrons in the material are {\it strongly correlated} (strongly interacting, cf. Lit. \ref{LiteratureIntersectingBraneModels}) then their ground states must be represented more properly
  by joint $n$-electron states --- for larger $n$ the larger the interaction. By the Pauli exclusion principle (which prevents $n$-electron states to be
  non-trivial if any pair of their momenta coincides) these form a vector bundle over the {\it configuration space of $n$ distinct points}
  (Lit. \ref{LiteratureConfigurationSpaces}) in the Brillouin torus. Accordingly, it is generally the TED K-cohomology of these $n$-configuration
  spaces which classifies such {\it topologically ordered} (strongly correlated) topological phases \cite[\S 3.2]{SS22AnyonictopologicalOrder}.
\end{literature}

\begin{literature}[{\bf Anyons}]
  \label{AnyonLiterature}
   It is tradition
   (a comprehevsive list of literature on this and the following aspects is provided in \cite[\S 3.3]{SS22AnyonictopologicalOrder})
   to motivate the concept of {\it anyons} as hypothetical particle species which are conceptually in
  between the firmly established
  {\it bosons} -- whose joint quantum state picks up no transformation under their pair exchange --, and the firmly
  established {\it fermions} -- whose
  joint quantum state picks up a sign $-1 = \exp(\ImaginaryUnit \pi)  \,\in \, \CircleGroup$ under pair exchange:
  For anyons one imagines {\it any} given
  phase $\exp( \phi \ImaginaryUnit) \,\in\, \CircleGroup$ under pair exchange, whence the name. Under mild assumptions this makes non-trivial sense (only) for quantum
  particles of codimension 2, hence notably for pointlike (quasi-)particle excitations in effectively 2-dimensional materials (e.g. in atomic mono-layer
  crystals similar to graphene). In a superficially evident generalization of this notion, one speaks of {\it non-abelian} anyons if the exchange phases are more
  general unitary operators in some unitary group $\UnitaryGroup(n)$ acting on a higher dimensional Hilbert space of their joint quantum states.

  For a recent impression of the review literature on anyon physics, see for instance \cite{MasakiMizushimaNitta23}.

  \medskip

  However, despite the popularity of this motivation, it misleads about the crucial fact that such anyonic phase factors are supposed to depend on dynamical
  braid paths (Lit. \ref{LiteratureBraiding}) traversed by anyons around each other (thus constituting a braid representation, Lit. \ref{LiteratureBraidRepresentations}), while the boson/fermion phases are kinematical properties of their
  wavefunctions enforced even if no effective motion takes place. Indeed, close inspection of more concrete anyon models discussed in the theoretical physics
  literature reveals first of all two crucially different incarnations of anyons (cf. \cite[Table 5]{SS22AnyonictopologicalOrder}), neither of which quite
  fits the naive popular motivation:

  On the one hand, abelian anyons are thought to be modeled by  fermion-like {\it quasi-particles} which couple via a so-called ``fictitious gauge field''
  such that their {\it quantum propagation} picks up {\it Aharonov-Bohm phases} with respect to each other. On the other hand, non-abelian anyons are
  thought to be modeled as {\it defects} in topologically ordered quantum materials whose {\it adiabatic movement} (Lit.  \ref{TQCAsAdiabaticQuantumComputing})
  by classical parameter evolution makes them pick up {\it Berry phases}. At the same time, a microscopic understanding of how such non-abelian anyons
  would appear in realistic materials had arguably remained elusive (cf. Lit. \ref{LiteratureMajoranaZeroModes}).

  \medskip
  We believe that the first satisfactory
  theoretical model for anyons --- one which (a) unifies these two notions as well as (b) properly embeds their
  theory into that of the ambient topological phases of matter
  (Lit. \ref{TopologicalQuantumMaterials}) which are expected to host them --- is the one laid out in \cite[\S 3.3]{SS22AnyonictopologicalOrder}, where
  the reader may also find extensive referencing of the traditional literature on the matter.
  The upshot of the anyon model in \cite[\S 3.3]{SS22AnyonictopologicalOrder} is that:

  \begin{itemize}[leftmargin=1cm]
  \item[\bf (1)] anyonic quantum ground states are classes in TED K-theory (Lit. \ref{LiteratureKTheoryClassificationOfTopologicalPhases}) of
  configuration spaces of points (Lit. \ref{LiteratureConfigurationSpaces}) in the punctured Brillouin torus of the given quantum material ---
  approximated by their TED {\it Chern-character} which is a class in {\it ordinary} TED cohomology;

  \item[\bf (2)] the anyonic phase factors are the holonomies of the {\it Gauss-Manin connection} (Lit. \ref{LiteratureGaussManinConnections}) on the
  resulting bundles of TED (K-)cohomology groups over the configuration
  spaces of punctures in the Brillouin torus.
  \end{itemize}

  \vspace{1mm}
  \noindent The main result of \cite{SS22AnyonicDefectBranes} is that this prescription --- in the approximation of ordinary TED cohomology as opposed to the
  full TED K-theory --- actually reproduces much of the expected description of anyon braiding, and hence derives it from first principles. The lift
  to full TED K-theory then predicts subtle ``torsion'' corrections to this traditional picture (cf. p. \pageref{Outlook}).
\end{literature}

\vspace{3cm}
\begin{literature}[\bf Configuration spaces of (defect-)points]
 \label{LiteratureConfigurationSpaces}
 The parameters whose motion controls
 topological quantum computation gate operation (Lit. \ref{LiteratureTopologicalQuantumComputation}) by adiabatic
 (Lit. \ref{TQCAsAdiabaticQuantumComputing}) braiding (Lit. \ref{LiteratureBraiding}) are configurations of {\it defects} (Lit. \ref{AnyonLiterature})
 in a quantum material (Lit. \ref{TopologicalQuantumMaterials}), often tacitly assumed to be spatial positions but possibly instead being
 critical momenta given by points in ``momentum space'' (Lit. \ref{LiteratureBraidingInMomentumSpace}). Specifically, for such defects to potentially be
 of anyonic nature they need to have co-dimension = 2 in the ambient quantum material, which for the case of point-like defects means that the quantum
 material must be an effectively 2-dimensional crystal consisting of a single or a very small number of atomic layers (such as now familiar from {\it graphene}).
 This is the case we are focusing on here.

 Key is the assumption that these defects remain separated from each other, hence that the defect points have {\it distinct positions} (possibly meaning:
 distinct positions in momentum space, hence distinct wave vectors!). In physical reality, it may be possible to force defects to merge, but for the
 materials of interest this should correspond to an effectively discontinuous change of the system's ground state, which we disregard by the same rationale
 by which we disregard, say, the process of destructing the crystal structure altogether: while physically possible, this takes us out of the regime of
 quantum materials whose description is of interest here.

 In conclusion then, the relevant parameters of defect positions for topological quantum gates move in a space of configurations of distinct points in a surface. For actual 2-dimensional crystals, this surface is a (Brillouin-){\it torus}, but for simplicity here we will disregard the global topology of tori, which means that we consider configurations of points in the plane $\mathbb{R}^2$, often regarded as the complex plane $\ComplexPlane$ in this context.

 \vspace{-.4cm}
\begin{center}
\hypertarget{FigureD}{}
\begin{tabular}{ll}
\begin{minipage}{10cm}
\footnotesize
{\bf Figure D -- Configuration of (defect-)points in the plane.}

In laboratory realization the plane represents an atomic mono-layer (at most a small multi-layer) of a crystalline quantum material (Lit. \ref{TopologicalQuantumMaterials}),
or its Fourier-dual space of wave-vectors (of excitations in the crystal); and the points represent anyonic defects
(Lit. \ref{AnyonLiterature})
in this structure (such as band nodes of a 2d semi-metal, Lit. \ref{LiteratureBraidingInMomentumSpace}).

Mathematically, the whole configuration is itself one point in the {\it configuration space of points}
\eqref{OrderedConfigurationSpaceOfPoints}
in the plane (cf. Lit. \ref{LiteratureAnyonSpecies}). Jointly moving the defect-positions means to move along a curve in this configuration space (Lit. \ref{LiteratureBraiding}), hence to move the points jointly along a ``braid'' (Lit. \ref{LiteratureBraiding}).

\end{minipage}
&
\hspace{-.4cm}
\adjustbox{scale=1.2, raise=-.8cm}{
\begin{tikzpicture}

\begin{scope}[shift={(2.1,.51)}]

\clip
  (-.2, .08) rectangle (.2,-.365);

\begin{scope}[yscale=.9, shift={(-.32,0)}]
\draw
  (0,0)
    .. controls (.25,0) and (.3,-.4) ..
  (.3,-.6);
\end{scope}
\begin{scope}[
  shift={(+.32,0)},
  xscale=-1, yscale=.9
]
\draw
  (0,0)
    .. controls (.25,0) and (.3,-.4) ..
  (.3,-.6);
\end{scope}

\begin{scope}[shift={(0,-.335)}, scale=.11]
\draw[gray, fill=white]
  (0,0)
    ellipse
      (.32 and .11);
\end{scope}
ffibr

\begin{scope}[shift={(0,-.314)}, scale=.123]
\draw[gray, fill=white]
  (0,0)
    ellipse
      (.32 and .11);
\end{scope}

\begin{scope}[shift={(0,-.29)}, scale=.139]
\draw[gray, fill=white]
  (0,0)
    ellipse
      (.32 and .11);
\end{scope}

\begin{scope}[shift={(0,-.265)}, scale=.156]
\draw[gray, fill=white]
  (0,0)
    ellipse
      (.32 and .11);
\end{scope}

\begin{scope}[shift={(0,-.24)}, scale=.17]
\draw[gray, fill=white]
  (0,0)
    ellipse
      (.32 and .11);
\end{scope}

\begin{scope}[shift={(0,-.21)}, scale=.2]
\draw[gray, fill=white]
  (0,0)
    ellipse
      (.32 and .11);
\end{scope}

\begin{scope}[shift={(0,-.175)}, scale=.24]
\draw[gray, fill=white]
  (0,0)
    ellipse
      (.32 and .11);
\end{scope}

\begin{scope}[scale=.324, shift={(0,-.4)}]
\draw[gray, fill=white]
  (0,0)
    ellipse
      (.32 and .11);
\end{scope}

\begin{scope}[scale=.424, shift={(0,-.19)}]
\draw[gray, fill=white]
  (0,0)
    ellipse
      (.32 and .11);
\end{scope}

\begin{scope}[scale=.63, shift={(0,-.015)}]
\draw[gray, fill=white]
  (0,0)
    ellipse
      (.32 and .11);
\end{scope}

\draw[gray]
 (0,0)  to (0,-.07);
\draw[gray]
 (0,-.086)  to (0,-.12);
\draw[gray]
 (0,-.135)  to (0,-.155);
\draw[gray]
 (0,-.172)  to (0,-.195);
\draw[gray]
 (0,-.21)  to (0,-.225);
\draw[gray]
 (0,-.24)  to (0,-.25);

\end{scope}

\begin{scope}[shift={(2,.17)}]

\clip
  (-.2, .08) rectangle (.2,-.365);

\begin{scope}[yscale=.9, shift={(-.32,0)}]
\draw
  (0,0)
    .. controls (.25,0) and (.3,-.4) ..
  (.3,-.6);
\end{scope}
\begin{scope}[
  shift={(+.32,0)},
  xscale=-1, yscale=.9
]
\draw
  (0,0)
    .. controls (.25,0) and (.3,-.4) ..
  (.3,-.6);
\end{scope}

\begin{scope}[shift={(0,-.335)}, scale=.11]
\draw[gray, fill=white]
  (0,0)
    ellipse
      (.32 and .11);
\end{scope}

\begin{scope}[shift={(0,-.314)}, scale=.123]
\draw[gray, fill=white]
  (0,0)
    ellipse
      (.32 and .11);
\end{scope}

\begin{scope}[shift={(0,-.29)}, scale=.139]
\draw[gray, fill=white]
  (0,0)
    ellipse
      (.32 and .11);
\end{scope}

\begin{scope}[shift={(0,-.265)}, scale=.156]
\draw[gray, fill=white]
  (0,0)
    ellipse
      (.32 and .11);
\end{scope}

\begin{scope}[shift={(0,-.24)}, scale=.17]
\draw[gray, fill=white]
  (0,0)
    ellipse
      (.32 and .11);
\end{scope}

\begin{scope}[shift={(0,-.21)}, scale=.2]
\draw[gray, fill=white]
  (0,0)
    ellipse
      (.32 and .11);
\end{scope}

\begin{scope}[shift={(0,-.175)}, scale=.24]
\draw[gray, fill=white]
  (0,0)
    ellipse
      (.32 and .11);
\end{scope}

\begin{scope}[scale=.324, shift={(0,-.4)}]
\draw[gray, fill=white]
  (0,0)
    ellipse
      (.32 and .11);
\end{scope}

\begin{scope}[scale=.424, shift={(0,-.19)}]
\draw[gray, fill=white]
  (0,0)
    ellipse
      (.32 and .11);
\end{scope}

\begin{scope}[scale=.63, shift={(0,-.015)}]
\draw[gray, fill=white]
  (0,0)
    ellipse
      (.32 and .11);
\end{scope}

\draw[gray]
 (0,0)  to (0,-.07);
\draw[gray]
 (0,-.086)  to (0,-.12);
\draw[gray]
 (0,-.135)  to (0,-.155);
\draw[gray]
 (0,-.172)  to (0,-.195);
\draw[gray]
 (0,-.21)  to (0,-.225);
\draw[gray]
 (0,-.24)  to (0,-.25);

\end{scope}

\begin{scope}[shift={(4.1,.53)}]

\clip
  (-.2, .08) rectangle (.2,-.365);

\begin{scope}[yscale=.9, shift={(-.32,0)}]
\draw
  (0,0)
    .. controls (.25,0) and (.3,-.4) ..
  (.3,-.6);
\end{scope}
\begin{scope}[
  shift={(+.32,0)},
  xscale=-1, yscale=.9
]
\draw
  (0,0)
    .. controls (.25,0) and (.3,-.4) ..
  (.3,-.6);
\end{scope}

\begin{scope}[shift={(0,-.335)}, scale=.11]
\draw[gray, fill=white]
  (0,0)
    ellipse
      (.32 and .11);
\end{scope}

\begin{scope}[shift={(0,-.314)}, scale=.123]
\draw[gray, fill=white]
  (0,0)
    ellipse
      (.32 and .11);
\end{scope}

\begin{scope}[shift={(0,-.29)}, scale=.139]
\draw[gray, fill=white]
  (0,0)
    ellipse
      (.32 and .11);
\end{scope}

\begin{scope}[shift={(0,-.265)}, scale=.156]
\draw[gray, fill=white]
  (0,0)
    ellipse
      (.32 and .11);
\end{scope}

\begin{scope}[shift={(0,-.24)}, scale=.17]
\draw[gray, fill=white]
  (0,0)
    ellipse
      (.32 and .11);
\end{scope}

\begin{scope}[shift={(0,-.21)}, scale=.2]
\draw[gray, fill=white]
  (0,0)
    ellipse
      (.32 and .11);
\end{scope}

\begin{scope}[shift={(0,-.175)}, scale=.24]
\draw[gray, fill=white]
  (0,0)
    ellipse
      (.32 and .11);
\end{scope}

\begin{scope}[scale=.324, shift={(0,-.4)}]
\draw[gray, fill=white]
  (0,0)
    ellipse
      (.32 and .11);
\end{scope}

\begin{scope}[scale=.424, shift={(0,-.19)}]
\draw[gray, fill=white]
  (0,0)
    ellipse
      (.32 and .11);
\end{scope}

\begin{scope}[scale=.63, shift={(0,-.015)}]
\draw[gray, fill=white]
  (0,0)
    ellipse
      (.32 and .11);
\end{scope}

\draw[gray]
 (0,0)  to (0,-.07);
\draw[gray]
 (0,-.086)  to (0,-.12);
\draw[gray]
 (0,-.135)  to (0,-.155);
\draw[gray]
 (0,-.172)  to (0,-.195);
\draw[gray]
 (0,-.21)  to (0,-.225);
\draw[gray]
 (0,-.24)  to (0,-.25);

\end{scope}

\begin{scope}[shift={(2,.17)}]

\clip
  (-.2, .08) rectangle (.2,-.365);

\begin{scope}[yscale=.9, shift={(-.32,0)}]
\draw
  (0,0)
    .. controls (.25,0) and (.3,-.4) ..
  (.3,-.6);
\end{scope}
\begin{scope}[
  shift={(+.32,0)},
  xscale=-1, yscale=.9
]
\draw
  (0,0)
    .. controls (.25,0) and (.3,-.4) ..
  (.3,-.6);
\end{scope}

\begin{scope}[shift={(0,-.335)}, scale=.11]
\draw[gray, fill=white]
  (0,0)
    ellipse
      (.32 and .11);
\end{scope}

\begin{scope}[shift={(0,-.314)}, scale=.123]
\draw[gray, fill=white]
  (0,0)
    ellipse
      (.32 and .11);
\end{scope}

\begin{scope}[shift={(0,-.29)}, scale=.139]
\draw[gray, fill=white]
  (0,0)
    ellipse
      (.32 and .11);
\end{scope}

\begin{scope}[shift={(0,-.265)}, scale=.156]
\draw[gray, fill=white]
  (0,0)
    ellipse
      (.32 and .11);
\end{scope}

\begin{scope}[shift={(0,-.24)}, scale=.17]
\draw[gray, fill=white]
  (0,0)
    ellipse
      (.32 and .11);
\end{scope}

\begin{scope}[shift={(0,-.21)}, scale=.2]
\draw[gray, fill=white]
  (0,0)
    ellipse
      (.32 and .11);
\end{scope}

\begin{scope}[shift={(0,-.175)}, scale=.24]
\draw[gray, fill=white]
  (0,0)
    ellipse
      (.32 and .11);
\end{scope}

\begin{scope}[scale=.324, shift={(0,-.4)}]
\draw[gray, fill=white]
  (0,0)
    ellipse
      (.32 and .11);
\end{scope}

\begin{scope}[scale=.424, shift={(0,-.19)}]
\draw[gray, fill=white]
  (0,0)
    ellipse
      (.32 and .11);
\end{scope}

\begin{scope}[scale=.63, shift={(0,-.015)}]
\draw[gray, fill=white]
  (0,0)
    ellipse
      (.32 and .11);
\end{scope}

\draw[gray]
 (0,0)  to (0,-.07);
\draw[gray]
 (0,-.086)  to (0,-.12);
\draw[gray]
 (0,-.135)  to (0,-.155);
\draw[gray]
 (0,-.172)  to (0,-.195);
\draw[gray]
 (0,-.21)  to (0,-.225);
\draw[gray]
 (0,-.24)  to (0,-.25);

\end{scope}

\begin{scope}[shift={(3,.25)}]

\clip
  (-.2, .08) rectangle (.2,-.365);

\begin{scope}[yscale=.9, shift={(-.32,0)}]
\draw
  (0,0)
    .. controls (.25,0) and (.3,-.4) ..
  (.3,-.6);
\end{scope}
\begin{scope}[
  shift={(+.32,0)},
  xscale=-1, yscale=.9
]
\draw
  (0,0)
    .. controls (.25,0) and (.3,-.4) ..
  (.3,-.6);
\end{scope}

\begin{scope}[shift={(0,-.335)}, scale=.11]
\draw[gray, fill=white]
  (0,0)
    ellipse
      (.32 and .11);
\end{scope}

\begin{scope}[shift={(0,-.314)}, scale=.123]
\draw[gray, fill=white]
  (0,0)
    ellipse
      (.32 and .11);
\end{scope}

\begin{scope}[shift={(0,-.29)}, scale=.139]
\draw[gray, fill=white]
  (0,0)
    ellipse
      (.32 and .11);
\end{scope}

\begin{scope}[shift={(0,-.265)}, scale=.156]
\draw[gray, fill=white]
  (0,0)
    ellipse
      (.32 and .11);
\end{scope}

\begin{scope}[shift={(0,-.24)}, scale=.17]
\draw[gray, fill=white]
  (0,0)
    ellipse
      (.32 and .11);
\end{scope}

\begin{scope}[shift={(0,-.21)}, scale=.2]
\draw[gray, fill=white]
  (0,0)
    ellipse
      (.32 and .11);
\end{scope}

\begin{scope}[shift={(0,-.175)}, scale=.24]
\draw[gray, fill=white]
  (0,0)
    ellipse
      (.32 and .11);
\end{scope}

\begin{scope}[scale=.324, shift={(0,-.4)}]
\draw[gray, fill=white]
  (0,0)
    ellipse
      (.32 and .11);
\end{scope}

\begin{scope}[scale=.424, shift={(0,-.19)}]
\draw[gray, fill=white]
  (0,0)
    ellipse
      (.32 and .11);
\end{scope}

\begin{scope}[scale=.63, shift={(0,-.015)}]
\draw[gray, fill=white]
  (0,0)
    ellipse
      (.32 and .11);
\end{scope}

\draw[gray]
 (0,0)  to (0,-.07);
\draw[gray]
 (0,-.086)  to (0,-.12);
\draw[gray]
 (0,-.135)  to (0,-.155);
\draw[gray]
 (0,-.172)  to (0,-.195);
\draw[gray]
 (0,-.21)  to (0,-.225);
\draw[gray]
 (0,-.24)  to (0,-.25);

\end{scope}

\draw[fill=white, fill opacity=.4]
  (.25,0)
    --
  (3.2,0)
    --
  (5.2,.7)
   --
  (2.25,.7)
    --
  (.25,0);

\draw[line width=1pt]
  (.25,0)
    --
  (3.2,0);

\draw[line width=1pt, dashed]
  (.25-.6,0)
    --
  (.25,0);
\draw[line width=1pt, dashed]
  (3.2+.6,0)
    --
  (3.2,0);

\node[rotate=19]
  at (.47,.24) {
    \scalebox{.6}{
      \color{darkblue}
      \;\;\;\;\;\;\;\;\;\;\;\;\; (Brillouin-)plane
    }
  };

\node
  at (3,.42) {
    \scalebox{.7}{
      \color{orangeii}
      \scalebox{.7}{
        \color{orangeii}
        \def\arraystretch{.85}
        \begin{tabular}{c}
          (defect-)points
        \end{tabular}
      }
    }
  };

\begin{scope}[shift={(4.1,.53)}]

\begin{scope}[scale=.63, shift={(0,-.015)}]
\draw[gray]
  (0,0)
    ellipse
      (.32 and .11);
\end{scope}

\draw[black]
 (0,0)  to (0,-.07);

\end{scope}

\begin{scope}[shift={(2.1,.51)}]

\begin{scope}[scale=.63, shift={(0,-.015)}]
\draw[gray]
  (0,0)
    ellipse
      (.32 and .11);
\end{scope}

\draw[black]
 (0,0)  to (0,-.07);

\end{scope}

\begin{scope}[shift={(2,.17)}]

\begin{scope}[scale=.63, shift={(0,-.015)}]
\draw[gray]
  (0,0)
    ellipse
      (.32 and .11);
\end{scope}

\draw[black]
 (0,0)  to (0,-.07);

\end{scope}

\begin{scope}[shift={(3,.25)}]

\begin{scope}[scale=.63, shift={(0,-.015)}]
\draw[gray]
  (0,0)
    ellipse
      (.32 and .11);
\end{scope}

\draw[black]
 (0,0)  to (0,-.07);

\end{scope}

\end{tikzpicture}
}
\end{tabular}
\end{center}

 Such parameter-{\it spaces of configurations of points} in the plane have long been studied in pure mathematics
 \cite{FadellNeuwirth62};
 a textbook account is \cite{FadellHusseini01},
 gentle introduction includes \cite{Williams20}\cite{Wilson18}, more details may be found in \cite{Cohen09}, and brief review in our context is in \cite{SS19Conf}. Much of their mathematical interest lies in the fact
 \eqref{OrderedConfigurationSpaceIsEMSpaceOfPureBraidGroup}
 that these spaces are Eilenberg-MacLane spaces (Lit. \ref{LiteratureEilenbergMacLanSpaces}) of braid groups \cite{FadellNeuwirth62}\cite{FoxNeuwirth62} (Lit. \ref{LiteratureBraiding}).

 \medskip

 Concretely, the {\it configuration space} of $N+1$ {\it ordered} points (carrying labels $1, 2, ..., N+1$, cf. Lit. \ref{LiteratureAnyonSpecies}) in the plane $\mathbb{R}^2$ is the complement in the product space $(\mathbb{R}^2)^{N+1}$, i.e. in the space of possibly coincident poisitions of $N+1$ points,  away from the subspace where any pair of them coincides:
 \begin{equation}
   \label{OrderedConfigurationSpaceOfPoints}
   \overset{
     \mathclap{
       \raisebox{7pt}{
         \scalebox{.7}{
           \color{darkblue}
           \bf
           \def\arraystretch{.9}
           \begin{tabular}{c}
             configuration space
             \\
             of ordered points
           \end{tabular}
         }
       }
     }
   }{
   \ConfigurationSpace{N+1}
   \big(
     \mathbb{R}^2
   \big)
   }
   \;\;
     :\defneq
   \;\;
   \overset{
     \mathclap{
       \raisebox{6pt}{
         \scalebox{.7}{
           \color{darkblue}
           \bf
           \def\arraystretch{.9}
           \begin{tabular}{c}
           space of unconstrained
           \\
           positions of ordered points
           \end{tabular}
         }
       }
     }
   }{
   \Big(
     \underbrace{
     (\mathbb{R}^2)
     \times
     \cdots
     \times
     (\mathbb{R}^2)
     }_{
       \mathclap{
         \scalebox{.8}{
           $N+1$ factors
         }
       }
     }
   \Big)
   }
   \overset{
     \mathclap{
       \raisebox{8pt}{
         \scalebox{.7}{
           \color{orangeii}
           \bf
           minus
         }
       }
     }
   }{
   \;\;\;\scalebox{1.4}{$\setminus$}\;\;\;
   }
   \overset{
     \mathclap{
       \raisebox{7pt}{
         \scalebox{.7}{
           \color{darkblue}
           \bf
           subspace where any pair of
           points coincide
         }
       }
     }
   }{
   \Big\{
     \;
     z_1, \cdots, z_{N+1} \,\in\,\mathbb{R}^2
    \Big\vert
    \;
    \underset{I, \neq  J}{\exists}
    \;\;\;
    z_I \,=\, z_J
    \;
   \Big\}
   }
 \end{equation}
 The symmetric group $\mathrm{Sym}(N+1)$ evidently acts on such an ordered configuration space by permuting the order of the labels; and the quotient of this action is the {\it un-ordered configuration space}
 \begin{equation}
   \label{UnOrderedConfigurationSpace}
   \overset{
     \mathclap{
       \raisebox{4pt}{
         \scalebox{.7}{
           \color{darkblue}
           \bf
           \def\arraystretch{.9}
           \begin{tabular}{c}
             configuration space
             \\
             of un-ordered points
           \end{tabular}
         }
       }
     }
   }{
     \mathrm{Conf}_{N+1}(\mathbb{R}^2)
   }
   \;\;
   :\defneq
   \;\;
   \overset{
     \mathclap{
       \raisebox{4pt}{
         \scalebox{.7}{
           \color{darkblue}
           \bf
           \def\arraystretch{.9}
           \begin{tabular}{c}
             configuration space
             \\
             of ordered points
           \end{tabular}
         }
       }
     }
   }{
   \ConfigurationSpace{N+1}(\mathbb{R}^2)
   }
   \overset{
     \mathclap{
       \raisebox{4pt}{
         \scalebox{.7}{
           \color{orangeii}
           \bf
           \;\;\;modulo
         }
       }
     }
   }{
   \;\;\;\;
   \scalebox{1.4}{$/$}
   \;\;\;\;
   }
   \overset{
     \mathclap{
       \raisebox{8pt}{
         \scalebox{.7}{
           \color{darkblue}
           \bf
           permutations
         }
       }
     }
   }{
     \mathrm{Sym}(N+1)
   }
   \,.
 \end{equation}

 These configuration spaces
 \eqref{OrderedConfigurationSpaceOfPoints}
 \eqref{UnOrderedConfigurationSpace}
 canonically inherit the structure of topological spaces (Lit. \ref{LiteratureTopologyAndHomotopyTheory} -- in fact of smooth manifolds) and the statement is
 (\cite[p. 118]{FadellNeuwirth62}\cite[\S 2]{FoxNeuwirth62}, review in \cite[\S 2.2]{Wilson18}\cite[pp. 9]{Williams20})
 that the underlying homotopy type (Lit. \ref{LiteratureTopologyAndHomotopyTheory})
 is the delooping (the first Eilenberg-MacLane space, Lit. \ref{LiteratureEilenbergMacLanSpaces}) of the (pure) braid group (Lit. \ref{LiteratureBraiding}):
 \begin{equation}
   \label{OrderedConfigurationSpaceIsEMSpaceOfPureBraidGroup}
   \def\arraystretch{1.6}
   \begin{array}{l}
   \ConfigurationSpace{N+1}(\mathbb{R}^2)
   \;\underset{\mathrm{whe}}{\simeq}\;
   B \, \mathrm{PBr}(N +1)
   \,,
   \;\;\;\;\;\;
   \mbox{i.e.:}
   \;\;\;\;\;
   \pi_k
   \Big(
     \ConfigurationSpace{N+1}(\mathbb{R}^2)
   \Big)
   \;=\;
   \left\{\!\!\!
   \begin{array}{cl}
     \mathrm{PBr}(N + 1) & \mbox{if $k = 1$}
     \\[-3pt]
     \ast & \mbox{otherwise}
   \end{array}
   \right.
   \\
   \mathrm{Conf}_{N+1}(\mathbb{R}^2)
   \;\;
   \underset{\mathrm{whe}}{\simeq}
   \;\;
   B \, \mathrm{Br}(N + 1)
   \,.
   \end{array}
 \end{equation}
 Here the non-trivial statement is that the higher homotopy groups $\pi_{k \geq 2}$ are all trivial: That the configuration space is connected ($\pi_0 = \ast$) is fairly evident; and the (pure) baid group may be understood as being {\it defined}
 \eqref{FibrationOfBraidGroups}
 to be the remaining nontrivial fundamental group $\pi_1$.

 Somewhat less widely studied (away from the subject of hypergeometric KZ-solutions, Lit. \ref{HypergeometricIntegralReferences})
 but key for our analysis is the fact that these {\it ordered} configuration spaces are naturally fibered over each other, locally trivially, by maps that forget the last point in a configuration
 (due to \cite[Thm. 3]{FadellNeuwirth62}, reviewed as \cite[Thm. 1.2]{Birman75}, exposition in \cite[\S 2.1]{Wilson18}):
 \begin{equation}
   \label{FibrationOfOrderedConfigurationSpaces}
   \begin{tikzcd}
     \ConfigurationSpace{N{\color{purple}+2}}(\mathbb{R}^2)
     \ar[
       d,
       ->>,
       "{
        \; \mathrm{pr}^{N+2}_{N+1}
       }"
     ]
     &
     \big(
       z_1,\,  \cdots, z_N, \, z_{N\color{purple}+1} ,\, z_{N \color{purple}+ 2}
     \big)
     \ar[
       d,
       phantom,
       "{\mapsto}"{rotate=-90}
     ]
     \\
     \ConfigurationSpace{N{\color{purple}+1}}(\mathbb{R}^1)
     \ar[
       d,
       ->>,
       "{
        \; \mathrm{pr}^{N+1}_N
       }"
     ]
     &
     \big(
       z_1,\,  \cdots, z_N, \, z_{N\color{purple}+1}
     \big)
     \ar[
       d,
       phantom,
       "{\mapsto}"{rotate=-90}
     ]
     \\
     \ConfigurationSpace{N}(\mathbb{R}^1)
     &
     \big(
       z_1,\,  \cdots, z_N,
     \big)
     \,.
   \end{tikzcd}
 \end{equation}
On homotopy types this gives the fibration of delooped braid groups \eqref{FibrationOfPureBraidGroupsInTermsOfGenerators} which drives the anyon braiding in the final construction \eqref{TypeTheoreticKZFibration}.

\end{literature}

\begin{literature}[\bf Anyon species]
\label{LiteratureAnyonSpecies}
The mathematical literature on configuration spaces (Lit \ref{LiteratureConfigurationSpaces})
tends to consider by default the un-ordered configuration space \eqref{UnOrderedConfigurationSpace} and the corresponding general braid group (Lit. \ref{LiteratureBraiding}). But physically realistic defects (\hyperlink{FigureD}{\it Figure D}) tend to appear in different ``species'', hence carrying different labels, and are hence described by the {\it ordered} configuration space \eqref{OrderedConfigurationSpaceOfPoints} and by the corresponding {\it pure} braid group \eqref{FibrationOfBraidGroups}.

In terms of braid representations (Lit. \ref{LiteratureBraidRepresentations}) this distinction is brought out by the shift of focus in the relevant physics literature to {\it braided tensor/fusion categories} (see references in \cite[Rem. 3.12]{SS22AnyonictopologicalOrder}), which is one way of encoding that each strand of a braid carries a label (here: of an object in the category).

Specifically for the realistic {\it monodromy braid representations} arising as the holonomy/transport of KZ-connections on $\suTwo$-conformal blocks (Lit. \ref{KZConnectionsOnConformalBlocksReferences}) the possible anyon species are indexed by {\it weights} (in CFT: ``highest weights'', see \cite[pp. 8]{SS22AnyonicDefectBranes}), namely by natural numbers $\weight \,\in\, \{0,1, \cdots, k\}$ ranging from the trivial case $0$ (corresponding to: no defect) up to a fixed ``level'' $k = \ShiftedLevel - 2 \,\in\, \NaturalNumbers$ which controls the resulting braiding phases, see \eqref{TheTwistParameters} below.

\medskip

\noindent
For example (see e.g. \cite{JohansenSimula21}\cite{GuHaghighatLiu22} and further references listed in \cite[\S 5.2]{SS22AnyonicDefectBranes}):
\begin{itemize}[leftmargin=.4cm]

\item the popular notion of {\it Majorana anyons} (alternatively: {\it Ising anyons}) correspond to $k = 2$ ($\ShiftedLevel = 4$) and their species of weights $0$,$1$,$2$ are traditionally denoted ``$1$'',  ``$\sigma$'' and ``$\psi$'', respectively.

\item the first class of anyons whose braid gates are  universal for quantum computation are the ``Fibonacci anyons'' corresponding to $k = 3$ ($\ShiftedLevel = 5$)\footnote{In comparing to the physics literature, specifically on Fibonacci anyons, beware that for anyons at odd level it is customary to list only the species of even weight.}.

\item Beyond that, are ``parafermions'' and a whole hierarchy of anyon $\suTwo$ classes at higher levels $k \geq 3$ which do not (yet) carry their own names, but which would all be universal for topological quantum computation (see \cite{KolganovMironovMorozov23}).

\end{itemize}

\end{literature}

\medskip
 \begin{literature}[\bf Braiding]
  \label{LiteratureBraiding}
    Recall (see, e.g., \cite{Miller72}) that by  a {\it group} one means

\vspace{-0cm}
\hspace{-.8cm}
\begin{tabular}{ll}
  \begin{minipage}{11.5cm}
a {\it group of operations}, on some object, which are associatively composable and invertible (cf. pp. \pageref{TypeOfGroups}).
   By a {\it braid group} (due to \cite{Artin25}, monographs include \cite{FoxNeuwirth62}\cite{Birman75},
   exposition in \cite{Williams20}) one means the group of joint continuous movements of a fixed number $N + 1$ of non-coincident points in the plane, from any fixed configuration back to that fixed configuration. The  ``worldlines'' traced out by such points in space-time under such an operation look like a braid with $N + 1$ strands, whence the name.

  As with actual braids, here it is understood that two such operations are identified if they differ only by continuous deformations of the ``strands'' without breaking or intersecting these, hence by an {\it isotopy} in the ambient $\mathbb{R}^3$.
 \end{minipage}

 &
\qquad
\adjustbox{raise=-1.8cm}{
\begin{tikzpicture}

\draw[darkgray]
  (-.2+0,-.245)
  ellipse (.14 and .14*.5);
\draw[darkgray]
  (-.2+.4,-.245)
  ellipse (.14 and .14*.5);
\draw[darkgray]
  (-.2+.8,-.245)
  ellipse (.14 and .14*.5);
\draw[darkgray]
  (-.2+1.2,-.245)
  ellipse (.14 and .14*.5);
\draw[darkgray]
  (-.2+1.6,-.245)
  ellipse (.14 and .14*.5);
\draw[darkgray]
  (-.2+2,-.245)
  ellipse (.14 and .14*.5);

\begin{scope}[scale=.4, yscale=.64, rotate=90]

\begin{scope}[shift={(0,0)}]
\rbraid
\end{scope}

\begin{scope}[shift={(0,-1.5)}]
 \strand
\end{scope}

\begin{scope}[shift={(0,-3)}]
  \lbraid
\end{scope}

\begin{scope}[shift={(0,-4.5)}]
 \strand
\end{scope}

\begin{scope}[shift={(2,.5)}]
  \strand
\end{scope}

\begin{scope}[shift={(2,-1)}]
  \lbraid
\end{scope}

\begin{scope}[shift={(2,-3)}]
  \lbraid
\end{scope}

\begin{scope}[shift={(2,-4.5)}]
 \strand
\end{scope}

\begin{scope}[shift={(4,0)}]
  \rbraid
\end{scope}

\begin{scope}[shift={(4,-2)}]
  \rbraid
\end{scope}

\begin{scope}[shift={(4,-4)}]
  \rbraid
\end{scope}

\begin{scope}[shift={(6,.5)}]
  \strand
\end{scope}

\begin{scope}[shift={(6,-1)}]
  \lbraid
\end{scope}

\begin{scope}[shift={(6,-3)}]
  \lbraid
\end{scope}

\begin{scope}[shift={(6,-4.5)}]
  \strand
\end{scope}

\begin{scope}[shift={(8,0)}]
  \rbraid
\end{scope}

\begin{scope}[shift={(8,-1.5)}]
  \strand
\end{scope}

\begin{scope}[shift={(8,-2.5)}]
  \strand
\end{scope}

\begin{scope}[shift={(8,-4)}]
  \rbraid
\end{scope}

\begin{scope}[shift={(10,0)}]
  \rbraid
\end{scope}

\begin{scope}[shift={(10,-2)}]
  \rbraid
\end{scope}

\begin{scope}[shift={(10,-4)}]
  \rbraid
\end{scope}

\end{scope}

\begin{scope}
\draw[darkgray]
  (-.2+0,-.245)
  ellipse (.14 and .14*.5);
\draw
  (-.2+0,-.245-.24)
  node
  {\scalebox{.7}{\color{darkblue} $1$}};

\draw[darkgray]
  (-.2+.4,-.245)
  ellipse (.14 and .14*.5);
\draw
  (-.2+.4,-.245-.24)
  node
  {\scalebox{.7}{\color{darkblue} $2$}};

\draw[darkgray]
  (-.2+.8,-.245)
  ellipse (.14 and .14*.5);
\draw
  (-.2+.8,-.245-.24)
  node
  {\scalebox{.7}{\color{darkblue} $3$}};

\draw[darkgray]
  (-.2+1.2,-.245)
  ellipse (.14 and .14*.5);
\draw
  (-.2+1.2,-.245-.24)
  node
  {\scalebox{.7}{\color{darkblue} $4$}};

\draw[darkgray]
  (-.2+1.6,-.245)
  ellipse (.14 and .14*.5);
\draw
  (-.2+1.6,-.245-.24)
  node
  {\scalebox{.7}{\color{darkblue} $5$}};
\draw[darkgray]
  (-.2+2,-.245)
  ellipse (.14 and .14*.5);
\draw
  (-.2+2,-.245-.24)
  node
  {\scalebox{.7}{\color{darkblue} $6$}};

\end{scope}

\begin{scope}[shift={(0,3.04)}]

\draw[white, line width=1.2]
  (-.2+0,-.245)
  ellipse (.14 and .14*.5);
\draw[white, line width=1.2]
  (-.2+.4,-.245)
  ellipse (.14 and .14*.5);
\draw[white, line width=1.2]
  (-.2+.8,-.245)
  ellipse (.14 and .14*.5);
\draw[white, line width=1.2]
  (-.2+1.2,-.245)
  ellipse (.14 and .14*.5);
\draw[white, line width=1.2]
  (-.2+1.6,-.245)
  ellipse (.14 and .14*.5);
\draw[white, line width=1.2]
  (-.2+2,-.245)
  ellipse (.14 and .14*.5);

\draw[darkgray]
  (-.2+0,-.245)
  ellipse (.14 and .14*.5);
\draw
  (-.2+0,-.245+.24)
  node
  {\scalebox{.7}{\color{darkblue} $3$}};
\draw[darkgray]
  (-.2+.4,-.245)
  ellipse (.14 and .14*.5);
\draw
  (-.2+.4,-.245+.24)
  node
  {\scalebox{.7}{\color{darkblue} $4$}};
\draw[darkgray]
  (-.2+.8,-.245)
  ellipse (.14 and .14*.5);
\draw
  (-.2+.8,-.245+.24)
  node
  {\scalebox{.7}{\color{darkblue} $6$}};
\draw[darkgray]
  (-.2+1.2,-.245)
  ellipse (.14 and .14*.5);
\draw
  (-.2+1.2,-.245+.24)
  node
  {\scalebox{.7}{\color{darkblue} $2$}};
\draw[darkgray]
  (-.2+1.6,-.245)
  ellipse (.14 and .14*.5);
\draw
  (-.2+1.6,-.245+.24)
  node
  {\scalebox{.7}{\color{darkblue} $1$}};
\draw[darkgray]
  (-.2+2,-.245)
  ellipse (.14 and .14*.5);
\draw
  (-.2+2,-.245+.24)
  node
  {\scalebox{.7}{\color{darkblue} $5$}};

\draw
  (-.2+2.2,-.245+.24)
  node
  {\scalebox{.7}{
    \rlap{
      $= \mathrm{perm}(b)$
    }
  }};

\end{scope}

\node[rotate=90] at (2.2,1.2) {
  \scalebox{.7}{
    \color{greenii}
    \bf
    ----
    braiding {\color{black} $b$}
    $\longrightarrow$
  }
};

\end{tikzpicture}
}

\end{tabular}

  \medskip
  A quick way of saying this with  precision (we consider a more explicit description in a moment) is to observe that
  a braid group is thus the {\it fundamental group} $\pi_1$ (Lit. \ref{LiteratureHomotopyGroups}) of a configuration space of points in the plane (Lit. \ref{LiteratureConfigurationSpaces}). Here it makes a key difference whether one considers the points in a configuration as {\it ordered} (labeled by numbers $1, \cdots, N+1$) in which case one speaks of the {\it pure braid group}, or as indistinguishable (albeit in any case with distinct positions!) in which case one speaks of the {\it braid group} proper: After traveling along a general braid $b$ the order of the given points may come out permuted by a permutation $\mathrm{perm}(b)$, and the braid is {\it pure} precisely if this permutation is trivial:
  \begin{equation}
    \label{FibrationOfBraidGroups}
    \begin{tikzcd}[column sep=huge,
      row sep=8pt
    ]
    \overset{
      \mathclap{
        \raisebox{5pt}{
          \scalebox{.7}{
            \color{darkblue}
            \bf
            \def\arraystretch{.9}
            \begin{tabular}{c}
              pure
              \\
              braid group
            \end{tabular}
          }
        }
      }
    }{
      \mathrm{PBr}(N+1)
    }
    \ar[
      d,
      phantom,
      "{:\defneq}"{rotate=-90}
    ]
    \ar[
      r,
      hook,
      "{ \mathrm{fib}_{\NeutralElement}(\mathrm{perm}) }"
    ]
    &
    \overset{
      \mathclap{
        \raisebox{6pt}{
          \scalebox{.7}{
            \color{darkblue}
            \bf
            \def\arraystretch{.9}
            \begin{tabular}{c}
              braid group
            \end{tabular}
          }
        }
      }
    }{
      \mathrm{Br}(N+1)
    }
    \ar[
      d,
      phantom,
      "{:\defneq}"{rotate=-90}
    ]
    \ar[
      r,
      ->>,
      "{ \mathrm{perm} }"
    ]
    &
    \overset{
      \mathclap{
        \raisebox{6pt}{
          \scalebox{.7}{
            \color{darkblue}
            \bf
            \def\arraystretch{.9}
            \begin{tabular}{c}
              permutation group
              \\
              \color{black}
              \normalfont
              (``symmmetric group'')
            \end{tabular}
          }
        }
      }
    }{
      \mathrm{Sym}(N+1)
    }
    \ar[
      d,
      phantom,
      "{\simeq}"{rotate=-90}
    ]
    \\
    \pi_1
    \Big(
      \underset{
        \mathclap{
          \raisebox{-3pt}{
            \scalebox{.7}{
              \color{darkblue}
              \bf
              \def\arraystretch{.9}
              \begin{tabular}{c}
                configuration space
                \\
                of ordered points
                \\
                in the plane
              \end{tabular}
            }
          }
        }
      }{
        \ConfigurationSpace{N+1}(\mathbb{R}^2)
      }
    \Big)
    \ar[
      r,
      "{
        \scalebox{.7}{
          \color{greenii}
          \bf
          \def\arraystretch{.9}
          \begin{tabular}{c}
            forget
            \\
            ordering
          \end{tabular}
        }
      }"{swap, yshift=-2pt}
    ]
    &
    \pi_1
    \Big(
      \underset{
        \mathclap{
          \raisebox{-11pt}{
            \scalebox{.7}{
              \color{darkblue}
              \bf
              \def\arraystretch{.9}
              \begin{tabular}{c}
                configuration space
                \\
                of un-ordered points
                \\
                in the plane
              \end{tabular}
            }
          }
        }
      }{
      \mathrm{Conf}_{N+1}(\mathbb{R}^2)
      }
    \Big)
    \ar[
      r,
      "{
        \scalebox{.7}{
          \color{greenii}
          \bf
          \def\arraystretch{.9}
          \begin{tabular}{c}
            forget
            \\
            positions
          \end{tabular}
        }
      }"{swap, yshift=-2pt}
    ]
    &
    \pi_1
    \Big(
      \underset{
        \mathclap{
          \raisebox{-11pt}{
            \scalebox{.7}{
              \color{darkblue}
              \bf
              \def\arraystretch{.9}
              \begin{tabular}{c}
                configuration space
                \\
                of un-ordered points
                \\
                in higher dim Eucl. space
              \end{tabular}
            }
          }
        }
      }{
      \mathrm{Conf}_{N+1}(\mathbb{R}^\infty)
      }
    \Big)
    \end{tikzcd}
  \end{equation}

Since these configuration spaces have no other non-trivial homotopy groups \eqref{OrderedConfigurationSpaceIsEMSpaceOfPureBraidGroup},
the vertical identifications mean equivalently that the homotopy type of these configuration spaces  constitute  {\it deloopings} or {\it classifying spaes} or  {\it Eilenberg-MacLane spaces} in degree 1 (Lit. \ref{LiteratureEilenbergMacLanSpaces})
for the braid groups; in particular:
\begin{equation}
  \label{OrderedConfigurationSpacesAsEMSpaceForPureBraidGroup}
  \ConfigurationSpace{N+1}(\mathbb{R}^1)
  \;\;
  \underset{\mathrm{whe}}{\simeq}
  \;\;
  B \, \mathrm{PBr}(N+1)
  \;\;
  \underset{\mathrm{whe}}{\simeq}
  \;\;
  K\big(
    \mathrm{PBr}(N+1)
    ,\,
    1
  \big)
  \,.
\end{equation}

A more explicit way to describe the braid group $\mathrm{Br}(N+1)$ is to observe, first,  that any braid may, clearly, be obtained as a composition of those elementary braids which do nothing but pass a pair of neighbouring points past each other:

\begin{equation}
\label{ArtinGeneratorsGraphically}
\scalebox{.7}{
  \color{darkblue}
  \bf
  \def\arraystretch{.9}
  \begin{tabular}{c}
    $i$th
    \\
    generating
    \\
    braid
  \end{tabular}
}
b_i
\;:\defneq\;
\color{orangeii}
\left[
\color{black}
\raisebox{-28pt}{
\begin{tikzpicture}[yscale=.8]

\begin{scope}[shift={(-1.6,0)}]
\draw[line width=1.4]
  (-.5,1) to (-.5,-1);
\draw[line width=1.2]
  (+.5,1) to (+.5,-1);
\draw
  (-0,-.1) node {$\mathclap{\cdots}$};
\end{scope}

\draw[line width=1.4]
  (-.35,-1)
  .. controls (-.35,0) and (+.35,0)  ..
  (+.35,1);

\draw[line width=4.5, white]
  (+.35,-1)
  .. controls (+.35,0) and (-.35,0)  ..
  (-.35,1);
\draw[line width=1.4]
  (+.35,-1)
  .. controls (+.35,0) and (-.35,0)  ..
  (-.35,1);

\begin{scope}[shift={(+1.6,0)}]
\draw[line width=1.4]
  (-.5,1) to (-.5,-1);
\draw[line width=1.2]
  (+.5,1) to (+.5,-1);
\draw
  (-0,-.1) node {$\mathclap{\cdots}$};
\end{scope}

\draw
  (-2.1, -1.3) node {\color{darkblue}\scalebox{.5}{$1$}};

\draw
  (-1.1, -1.3) node {\color{darkblue}\scalebox{.5}{$i\!-\!1$}};

\draw
  (-.4, -1.3) node {\color{darkblue}\scalebox{.5}{$i$}};

\draw
  (+.35, -1.3) node {\color{darkblue}\scalebox{.5}{$i\!+\!1$}};

\draw
  (+1.1, -1.3) node {\color{darkblue}\scalebox{.5}{$i\!+\!2$}};

\draw
  (+2.1, -1.3) node {\color{darkblue}\scalebox{.5}{$N+1$}};

\end{tikzpicture}
}
\color{orangeii}
\right]
\color{black}
\hspace{1cm}
\scalebox{.7}{
  \color{darkblue}
  \bf
  \def\arraystretch{.9}
  \begin{tabular}{c}
    and its
    \\
    inverse
    \\
    braid
  \end{tabular}
}
b_i^{-1}
\;:\defneq\;
\color{orangeii}
\left[
\color{black}
\raisebox{-28pt}{
\begin{tikzpicture}[yscale=.8]

\begin{scope}[shift={(-1.6,0)}]
\draw[line width=1.4]
  (-.5,1) to (-.5,-1);
\draw[line width=1.2]
  (+.5,1) to (+.5,-1);
\draw
  (-0,-.1) node {$\mathclap{\cdots}$};
\end{scope}

\draw[line width=1.4]
  (+.35,-1)
  .. controls (+.35,0) and (-.35,0)  ..
  (-.35,1);

\draw[line width=4.5, white]
  (-.35,-1)
  .. controls (-.35,0) and (+.35,0)  ..
  (+.35,1);
\draw[line width=1.4]
  (-.35,-1)
  .. controls (-.35,0) and (+.35,0)  ..
  (+.35,1);

\begin{scope}[shift={(+1.6,0)}]
\draw[line width=1.4]
  (-.5,1) to (-.5,-1);
\draw[line width=1.2]
  (+.5,1) to (+.5,-1);
\draw
  (-0,-.1) node {$\mathclap{\cdots}$};
\end{scope}

\draw
  (-2.1, -1.3) node {\color{darkblue}\scalebox{.5}{$1$}};

\draw
  (-1.1, -1.3) node {\color{darkblue}\scalebox{.5}{$i\!-\!1$}};

\draw
  (-.4, -1.3) node {\color{darkblue}\scalebox{.5}{$i$}};

\draw
  (+.35, -1.3) node {\color{darkblue}\scalebox{.5}{$i\!+\!1$}};

\draw
  (+1.1, -1.3) node {\color{darkblue}\scalebox{.5}{$i\!+\!2$}};

\draw
  (+2.1, -1.3) node {\color{darkblue}\scalebox{.5}{$N+1$}};

\end{tikzpicture}
}
\color{orangeii}
\right]
\color{black}
\end{equation}

While any braid may be obtained as a composition of just these generators, not every pair of such compositions yield distinct braids.
For example, if a pair of such elementary braids acts on disjoint strands, then the order in which they are applied does not matter up to the pertinent continuous deformation of braids:

\begin{equation}
\label{FirstArtinRelationGraphically}
\color{orangeii}
\left[
\color{black}
\adjustbox{
  raise=-39pt,
  scale=.9
}{
\begin{tikzpicture}[yscale=.6]

\begin{scope}

\begin{scope}[shift={(-1.6,0)}]
\draw[line width=1.4]
  (-.5,1) to (-.5,-1);
\draw[line width=1.2]
  (+.5,1) to (+.5,-1);
\draw
  (-0,-.1) node {$\mathclap{\cdots}$};
\end{scope}

\draw[line width=1.4]
  (-.35,-1)
  .. controls (-.35,0) and (+.35,0)  ..
  (+.35,1);

\draw[line width=4.5, white]
  (+.35,-1)
  .. controls (+.35,0) and (-.35,0)  ..
  (-.35,1);
\draw[line width=1.4]
  (+.35,-1)
  .. controls (+.35,0) and (-.35,0)  ..
  (-.35,1);

\begin{scope}[shift={(+1.6,0)}]
\draw[line width=1.4]
  (-.5,1) to (-.5,-1);
\draw[line width=1.2]
  (+.5,1) to (+.5,-1);
\draw
  (-0,-.1) node {$\mathclap{\cdots}$};
\end{scope}

\begin{scope}[shift={(3.2,0)}]
\draw[line width=1.4]
  (-.35,-1)
  --
  (-.35,1);

\draw[line width=1.4]
  (+.35,-1)
  --
  (+.35,1);
\end{scope}

\begin{scope}[shift={(+4.8,0)}]
\draw[line width=1.4]
  (-.5,1) to (-.5,-1);
\draw[line width=1.2]
  (+.5,1) to (+.5,-1);
\draw
  (-0,-.1) node {$\mathclap{\cdots}$};
\end{scope}

\draw
  (-2.1, -1.4) node {\color{darkblue}\scalebox{.5}{$1$}};

\draw
  (-1.1, -1.4) node {\color{darkblue}\scalebox{.5}{$i\!-\!1$}};

\draw
  (-.4, -1.4) node {\color{darkblue}\scalebox{.5}{$i$}};

\draw
  (+.35, -1.4) node {\color{darkblue}\scalebox{.5}{$i\!+\!1$}};

\draw
  (+1.1, -1.4) node {\color{darkblue}\scalebox{.5}{$i\!+\!2$}};

\draw
  (+2.1, -1.4) node {\color{darkblue}\scalebox{.5}{$j\!-\!1$}};

\draw
  (+2.8, -1.3) node {\color{darkblue}\scalebox{.5}{$j$}};

\draw
  (+3.5, -1.3) node {\color{darkblue}\scalebox{.5}{$j\!+\!1$}};

\draw
  (+4.3, -1.3) node {\color{darkblue}\scalebox{.5}{$j\!+\!2$}};

\draw
  (+5.3, -1.3) node {\color{darkblue}\scalebox{.5}{$N\!+\!1$}};

\end{scope}

\begin{scope}[shift={(0,2)}]

\begin{scope}[shift={(-1.6,0)}]
\draw[line width=1.4]
  (-.5,1) to (-.5,-1);
\draw[line width=1.2]
  (+.5,1) to (+.5,-1);
\draw
  (-0,-.1) node {$\mathclap{\cdots}$};
\end{scope}

\begin{scope}
\draw[line width=1.4]
  (-.35,-1)
  --
  (-.35,1);

\draw[line width=1.4]
  (+.35,-1)
  --
  (+.35,1);
\end{scope}

\begin{scope}[shift={(+1.6,0)}]
\draw[line width=1.4]
  (-.5,1) to (-.5,-1);
\draw[line width=1.2]
  (+.5,1) to (+.5,-1);
\draw
  (-0,-.1) node {$\mathclap{\cdots}$};
\end{scope}

\begin{scope}[shift={(3.2,0)}]
\draw[line width=1.4]
  (-.35,-1)
  .. controls (-.35,0) and (+.35,0)  ..
  (+.35,1);
\draw[line width=4.5, white]
  (+.35,-1)
  .. controls (+.35,0) and (-.35,0)  ..
  (-.35,1);
\draw[line width=1.4]
  (+.35,-1)
  .. controls (+.35,0) and (-.35,0)  ..
  (-.35,1);
\end{scope}

\begin{scope}[shift={(+4.8,0)}]
\draw[line width=1.4]
  (-.5,1) to (-.5,-1);
\draw[line width=1.2]
  (+.5,1) to (+.5,-1);
\draw
  (-0,-.1) node {$\mathclap{\cdots}$};
\end{scope}

\end{scope}

\end{tikzpicture}
}
\color{orangeii}
\right]
\color{black}
\;\;\;
  =
\;\;\;
\color{orangeii}
\left[
\color{black}
\adjustbox{
  raise=-39pt,
  scale=.9
}{
\begin{tikzpicture}[yscale=.6]

\begin{scope}[shift={(0,2)}]

\begin{scope}

\begin{scope}[shift={(-1.6,0)}]
\draw[line width=1.4]
  (-.5,1) to (-.5,-1);
\draw[line width=1.2]
  (+.5,1) to (+.5,-1);
\draw
  (-0,-.1) node {$\mathclap{\cdots}$};
\end{scope}

\draw[line width=1.4]
  (-.35,-1)
  .. controls (-.35,0) and (+.35,0)  ..
  (+.35,1);

\draw[line width=4.5, white]
  (+.35,-1)
  .. controls (+.35,0) and (-.35,0)  ..
  (-.35,1);
\draw[line width=1.4]
  (+.35,-1)
  .. controls (+.35,0) and (-.35,0)  ..
  (-.35,1);

\begin{scope}[shift={(+1.6,0)}]
\draw[line width=1.4]
  (-.5,1) to (-.5,-1);
\draw[line width=1.2]
  (+.5,1) to (+.5,-1);
\draw
  (-0,-.1) node {$\mathclap{\cdots}$};
\end{scope}

\begin{scope}[shift={(3.2,0)}]
\draw[line width=1.4]
  (-.35,-1)
  --
  (-.35,1);

\draw[line width=1.4]
  (+.35,-1)
  --
  (+.35,1);
\end{scope}

\begin{scope}[shift={(+4.8,0)}]
\draw[line width=1.4]
  (-.5,1) to (-.5,-1);
\draw[line width=1.2]
  (+.5,1) to (+.5,-1);
\draw
  (-0,-.1) node {$\mathclap{\cdots}$};
\end{scope}

\end{scope}

\end{scope}

\begin{scope}[shift={(0,0)}]

\begin{scope}[shift={(-1.6,0)}]
\draw[line width=1.4]
  (-.5,1) to (-.5,-1);
\draw[line width=1.2]
  (+.5,1) to (+.5,-1);
\draw
  (-0,-.1) node {$\mathclap{\cdots}$};
\end{scope}

\begin{scope}
\draw[line width=1.4]
  (-.35,-1)
  --
  (-.35,1);

\draw[line width=1.4]
  (+.35,-1)
  --
  (+.35,1);
\end{scope}

\begin{scope}[shift={(+1.6,0)}]
\draw[line width=1.4]
  (-.5,1) to (-.5,-1);
\draw[line width=1.2]
  (+.5,1) to (+.5,-1);
\draw
  (-0,-.1) node {$\mathclap{\cdots}$};
\end{scope}

\begin{scope}[shift={(3.2,0)}]
\draw[line width=1.4]
  (-.35,-1)
  .. controls (-.35,0) and (+.35,0)  ..
  (+.35,1);
\draw[line width=4.5, white]
  (+.35,-1)
  .. controls (+.35,0) and (-.35,0)  ..
  (-.35,1);
\draw[line width=1.4]
  (+.35,-1)
  .. controls (+.35,0) and (-.35,0)  ..
  (-.35,1);
\end{scope}

\begin{scope}[shift={(+4.8,0)}]
\draw[line width=1.4]
  (-.5,1) to (-.5,-1);
\draw[line width=1.2]
  (+.5,1) to (+.5,-1);
\draw
  (-0,-.1) node {$\mathclap{\cdots}$};
\end{scope}

\draw
  (-2.1, -1.4) node {\color{darkblue}\scalebox{.5}{$1$}};

\draw
  (-1.1, -1.4) node {\color{darkblue}\scalebox{.5}{$i\!-\!1$}};

\draw
  (-.4, -1.4) node {\color{darkblue}\scalebox{.5}{$i$}};

\draw
  (+.35, -1.4) node {\color{darkblue}\scalebox{.5}{$i\!+\!1$}};

\draw
  (+1.1, -1.4) node {\color{darkblue}\scalebox{.5}{$i\!+\!2$}};

\draw
  (+2.1, -1.4) node {\color{darkblue}\scalebox{.5}{$j\!-\!1$}};

\draw
  (+2.8, -1.4) node {\color{darkblue}\scalebox{.5}{$j$}};

\draw
  (+3.5, -1.4) node {\color{darkblue}\scalebox{.5}{$j\!+\!1$}};

\draw
  (+4.3, -1.4) node {\color{darkblue}\scalebox{.5}{$j\!+\!2$}};

\draw
  (+5.3, -1.4) node {\color{darkblue}\scalebox{.5}{$N\!+\!1$}};

\end{scope}

\end{tikzpicture}
}
\color{orangeii}
\right]
\color{black}
\end{equation}

On the other hand, when consecutive triples of elementary braids do act on the same strands, then an evident continuous deformation relates them as follows:

\begin{equation}
\label{SecondArtinRelationGraphically}
\color{orangeii}
\left[
\color{black}
\raisebox{-48pt}{
\begin{tikzpicture}[yscale=.5]

\begin{scope}[shift={(-1.6,0)}]
\draw[line width=1.4]
  (-.5,5) to (-.5,-1);
\draw[line width=1.2]
  (+.5,5) to (+.5,-1);
\draw
  (-0,2.2) node {$\mathclap{\cdots}$};
\end{scope}

\begin{scope}[shift={(0,4)}]
\draw[line width=1.4]
  (-.35,-1)
  .. controls (-.35,0) and (+.35,0)  ..
  (+.35,1);

\draw[line width=4.5, white]
  (+.35,-1)
  .. controls (+.35,0) and (-.35,0)  ..
  (-.35,1);
\draw[line width=1.4]
  (+.35,-1)
  .. controls (+.35,0) and (-.35,0)  ..
  (-.35,1);
\end{scope}

\draw[line width=1.4]
  (-.35,3)
  --
  (-.35,1);

\draw[line width=1.4]
  (+1.05,5)
  --
  (+1.05,3);

\begin{scope}[shift={(.7,2)}]
\draw[line width=1.4]
  (-.35,-1)
  .. controls (-.35,0) and (+.35,0)  ..
  (+.35,1);

\draw[line width=4.5, white]
  (+.35,-1)
  .. controls (+.35,0) and (-.35,0)  ..
  (-.35,1);
\draw[line width=1.4]
  (+.35,-1)
  .. controls (+.35,0) and (-.35,0)  ..
  (-.35,1);
\end{scope}

\begin{scope}
\draw[line width=1.4]
  (-.35,-1)
  .. controls (-.35,0) and (+.35,0)  ..
  (+.35,1);

\draw[line width=4.5, white]
  (+.35,-1)
  .. controls (+.35,0) and (-.35,0)  ..
  (-.35,1);
\draw[line width=1.4]
  (+.35,-1)
  .. controls (+.35,0) and (-.35,0)  ..
  (-.35,1);
\end{scope}

 \draw[line width=1.4]
   (1.05,1) to (1.05,-1);

\begin{scope}[shift={(+2.4,0)}]
\draw[line width=1.4]
  (-.5,5) to (-.5,-1);
\draw[line width=1.2]
  (+.5,5) to (+.5,-1);
\draw
  (-0,2.2) node {$\mathclap{\cdots}$};
\end{scope}

\draw
  (-2.1, -1.4) node {\color{darkblue}\scalebox{.5}{$1$}};

\draw
  (-1.1, -1.4) node {\color{darkblue}\scalebox{.5}{$i\!-\!1$}};

\draw
  (-.4, -1.4) node {\color{darkblue}\scalebox{.5}{$i$}};

\draw
  (+.35, -1.4) node {\color{darkblue}\scalebox{.5}{$i\!+\!1$}};

\draw
  (+1.1, -1.4) node {\color{darkblue}\scalebox{.5}{$i\!+\!2$}};

\draw
  (+1.9, -1.4) node {\color{darkblue}\scalebox{.5}{$i\!+\!3$}};

\draw
  (+2.9, -1.4) node {\color{darkblue}\scalebox{.5}{$N\!+\!1$}};
\end{tikzpicture}
}
\color{orangeii}
\right]
\color{black}
\;\;\;\;\;
=
\;\;\;\;
\color{orangeii}
\left[
\color{black}
\raisebox{-48pt}{
\begin{tikzpicture}[yscale=.5]

\begin{scope}[shift={(-1.6,0)}]
\draw[line width=1.4]
  (-.5,5) to (-.5,-1);
\draw[line width=1.2]
  (+.5,5) to (+.5,-1);
\draw
  (-0,2.2) node {$\mathclap{\cdots}$};
\end{scope}

\begin{scope}[shift={(.7,4)}]
\draw[line width=1.4]
  (-.35,-1)
  .. controls (-.35,0) and (+.35,0)  ..
  (+.35,1);

\draw[line width=4.5, white]
  (+.35,-1)
  .. controls (+.35,0) and (-.35,0)  ..
  (-.35,1);
\draw[line width=1.4]
  (+.35,-1)
  .. controls (+.35,0) and (-.35,0)  ..
  (-.35,1);
\end{scope}

\draw[line width=1.4]
  (1.05,3)
  --
  (1.05,1);

\draw[line width=1.4]
  (-.35,5)
  --
  (-.35,3);

\draw[line width=1.4]
  (-.35,1)
  --
  (-.35,-1);

\begin{scope}[shift={(0,2)}]
\draw[line width=1.4]
  (-.35,-1)
  .. controls (-.35,0) and (+.35,0)  ..
  (+.35,1);

\draw[line width=4.5, white]
  (+.35,-1)
  .. controls (+.35,0) and (-.35,0)  ..
  (-.35,1);
\draw[line width=1.4]
  (+.35,-1)
  .. controls (+.35,0) and (-.35,0)  ..
  (-.35,1);
\end{scope}

\begin{scope}[shift={(.7,0)}]
\draw[line width=1.4]
  (-.35,-1)
  .. controls (-.35,0) and (+.35,0)  ..
  (+.35,1);

\draw[line width=4.5, white]
  (+.35,-1)
  .. controls (+.35,0) and (-.35,0)  ..
  (-.35,1);
\draw[line width=1.4]
  (+.35,-1)
  .. controls (+.35,0) and (-.35,0)  ..
  (-.35,1);
\end{scope}

\begin{scope}[shift={(+2.4,0)}]
\draw[line width=1.4]
  (-.5,5) to (-.5,-1);
\draw[line width=1.2]
  (+.5,5) to (+.5,-1);
\draw
  (-0,2.2) node {$\mathclap{\cdots}$};
\end{scope}

\draw
  (-2.1, -1.4) node {\color{darkblue}\scalebox{.5}{$1$}};

\draw
  (-1.1, -1.4) node {\color{darkblue}\scalebox{.5}{$i\!-\!1$}};

\draw
  (-.4, -1.4) node {\color{darkblue}\scalebox{.5}{$i$}};

\draw
  (+.35, -1.4) node {\color{darkblue}\scalebox{.5}{$i\!+\!1$}};

\draw
  (+1.1, -1.4) node {\color{darkblue}\scalebox{.5}{$i\!+\!2$}};

\draw
  (+1.9, -1.4) node {\color{darkblue}\scalebox{.5}{$i\!+\!3$}};

\draw
  (+2.9, -1.4) node {\color{darkblue}\scalebox{.5}{$N\!+\!1$}};
\end{tikzpicture}
}
\color{orangeii}
\right]
\color{black}
\end{equation}

A classical theorem due to Artin \cite[\S 3]{Artin25} (cf. \cite[\S 7]{FoxNeuwirth62})
says that these are the {\it only} relations between these generators, in that the braid group is {\it presented} by these  {\it generators and relations}, in the general sense of group presentations (e.g. \cite{MKS66}\cite{Johnson90}, cf. pp. \pageref{PresentationsOfGroups}): \footnote{In \eqref{ArtinBraidGroup} we include the neutral element in the set of generators just in order to stick with the convention used in \eqref{TheFreeGroup} below, where it is most natural to regard the free group-construction as an operation on pointed sets.}

\begin{equation}
  \label{ArtinBraidGroup}
  \overset{
    \mathclap{
      \raisebox{4pt}{
        \scalebox{.7}{
          \color{darkblue}
          \bf
          braid group
        }
      }
    }
  }{
    \mathrm{Br}(N+1)
  }
  \;\;
  \simeq
  \;\;
  \mathrm{FreeGrp}
  \big(
    \overset{
      \mathclap{
        \raisebox{4pt}{
          \scalebox{.7}{
            \color{orangeii}
            \bf
            Artin generators
            \color{gray}
            \normalfont \eqref{ArtinGeneratorsGraphically}
          }
        }
      }
    }{
    \{
      \NeutralElement
      ,\,
      b_1,
      \cdots,
      ,\,
      b_n
    \}
    }
  \big)
  \bigg/
  \overset{
    \mathclap{
      \raisebox{4pt}{
        \scalebox{.7}{
          \color{orangeii}
          \bf
          Artin braid relations
          \color{gray}
          \normalfont
          \eqref{FirstArtinRelationGraphically}
          \eqref{SecondArtinRelationGraphically}
        }
      }
    }
  }{
  \left(
  \def\arraystretch{1}
\begin{array}{l}
    b_i \cdot b_j
      \;=\;
   b_j \cdot b_i
   \;\;\mbox{if}\;\;\;
   i + 1 <  j
   \\
   b_i \cdot b_{i+1} \cdot b_i
     \;=\;
   b_{i+1} \cdot b_{i} \cdot b_{i+1}
  \end{array}
  \right)
  }
\end{equation}

Analogously for the pure braid group \eqref{FibrationOfBraidGroups}, it is fairly evident that any pure braid can be obtained by composing ``weaves'' in which one strand lassoes exactly one other strand:

\begin{equation}
 \label{ArtinPureBraidGeneratorsGraphically}
\overset{
  \mathclap{
    \raisebox{4pt}{
      \scalebox{.7}{
        \color{darkblue}
        \bf
        \def\arraystretch{.9}
        %
}
\color{orange}
\right]
\color{black}
\end{equation}
For example, consecutive application of such generators for fixed $i$ and decreasing $j$ yields pure braids of the following form:
\begin{equation}
\label{ProductsOfPureBraidGenerators}
\adjustbox{scale=.85}{
\begin{tikzcd}
\color{orange}
\left[
\color{black}
\raisebox{30pt}{
%
}
\color{orange}
\right]
\color{black}
\end{tikzcd}
}
\end{equation}

As before in \eqref{ArtinBraidGroup}, these pure braid generators \eqref{ArtinPureBraidGeneratorsGraphically} constitute a finite presentation, now of the pure braid group (we show the optimized set of pure braid relations due to \cite[Thm. 1.1, Rem. 3.1]{Lee10}):

\begin{equation}
  \label{ArtinPureBraidGroup}
  \overset{
    \mathclap{
      \raisebox{4pt}{
        \scalebox{.7}{
          \color{darkblue}
          \bf
          \def\arraystretch{.9}
          \begin{tabular}{c}
            pure
            \\
            braid group
          \end{tabular}
        }
      }
    }
  }{
    \mathrm{PBr}(N+1)
  }
  \;\;
  \simeq
  \;\;
  \mathrm{FreeGrp}
  \big(
    \{
      \NeutralElement
    \}
    \sqcup
    \overset{
      \mathclap{
        \raisebox{4pt}{
          \scalebox{.7}{
            \color{orangeii}
            \bf
            pure braid generators
            \color{gray}
            \normalfont \eqref{ArtinPureBraidGeneratorsGraphically}
          }
        }
      }
    }{
    \{
      b_{ij}
    \}_{
      1\leq i < j \leq N + 1
    }
    }
  \big)
  \bigg/
  \overset{
    \mathclap{
      \raisebox{4pt}{
        \scalebox{.7}{
          \color{orangeii}
          \bf
          pure braid relations
          \color{gray}
          \normalfont
          \cite[Thm. 1.1]{Lee10}
        }
      }
    }
  }{
  \left(
  \def\arraystretch{1}
  \begin{array}{l}
    b_{i j} \cdot b_{r s}
    \;=\;
    b_{r s} \cdot b_{i j}
    \;\;\;\;\;
    \mbox{
      if
      $
        r < s < i < j
      $
      or
      $
        i < r < s < j
      $
    }
    \\
    b_{j i} \cdot b_{i r} \cdot b_{r j}
    \;=\;
    b_{i r} \cdot b_{r j } \cdot b_{j i}
    \;=\;
    b_{r j} \cdot b_{j i} \cdot b_{i r}
    \;\;\;\;\;
    \mbox{
      if
      $
        r < i < j
      $
    }
    \\
    b_{r s}
    \cdot
    \big(
      b_{j r}
      \cdot
      b_{j i}
      \cdot
      b_{j s}
    \big)
    \;=\;
    \big(
      b_{j r}
      \cdot
      b_{j i}
      \cdot
      b_{j s}
    \big)
    \cdot
    b_{r s}
    \;\;\;\
    \mbox{
      if
      $
        r < i < s < j
      $
    }
  \end{array}
  \right)
  }
\end{equation}

And as before in \eqref{FirstArtinRelationGraphically}, the first of these relations
\eqref{ArtinPureBraidGroup}
simply say that the order of applying pure braid generators is irrelevant if these act on disjoint intervals of strands:

\begin{equation}
\label{FirstPureArtinRelationGraphically}
\begin{tikzcd}
\color{orange}
\left[
\color{black}
\raisebox{7pt}{
%
}
\color{orange}
\right]
\color{black}
\end{tikzcd}
\end{equation}

The further relations in the presentation \eqref{ArtinPureBraidGroup} of the pure braid group
concern cases where pure braid generators do ``overlap'', specifically with the products
\eqref{ProductsOfPureBraidGenerators}
of other generators:

\medskip

\begin{equation}
\label{ThePureArtinLeeRelations}
\hspace{-.5cm}
\def\tabcolsep{2pt}
%
\end{equation}

\medskip

Notice that all these pure braid relations are {\it commutator} relations \cite[Rem. 3.10]{Lee10}, saying that one pure braid generator commutes with a product
of pure braid generators, such as those in \eqref{ProductsOfPureBraidGenerators}. This implies that group homomorphisms out of a pure braid group into an {\it abelian} group are given by assigning any of the abelian group elements to the pure Artin generators \eqref{ArtinPureBraidGeneratorsGraphically} (used in Lem. \ref{AssigningPhasesToPureArtinGenerators} below).

\end{literature}

\medskip

\begin{literature}[\bf Unitary braid representations]
  \label{LiteratureBraidRepresentations}
  A unitary {\it braid re-presentation}
  (review in \cite{Abad15})
  is a linear representation of a (pure) braid group (Lit. \ref{LiteratureBraiding})
  by unitary operators (on any  Hilbert space), hence is a group homomorphism of the form
  $$
    \rho
    \,:\,
    \overset{
      \mathclap{
        \raisebox{4pt}{
          \scalebox{.7}{
            \color{darkblue}
            \bf
            braid group
          }
        }
      }
    }{
    \mathrm{Br}(N+1)
    }
    \xrightarrow{\quad
      \scalebox{.7}{
        \color{greenii}
        \bf
        homomorphism
      }
    \quad }
    \overset{
      \mathclap{
        \raisebox{4pt}{
          \scalebox{.7}{
            \color{darkblue}
            \bf
            unitary group
          }
        }
      }
    }{
    \mathrm{U}(k)
    }
    \,,
    \hspace{1cm}
    \mbox{or just}
    \hspace{1cm}
    \rho
    \,:\,
    \overset{
      \mathclap{
        \raisebox{4pt}{
          \scalebox{.7}{
            \color{darkblue}
            \bf
            \def\arraystretch{.9}
            \begin{tabular}{c}
              pure
              \\
              braid group
            \end{tabular}
          }
        }
      }
    }{
    \mathrm{PBr}(N+1)
    }
    \xrightarrow{\quad
      \scalebox{.7}{
        \color{greenii}
        \bf
        homomorphism
      }
    \quad}
    \overset{
      \mathclap{
        \raisebox{4pt}{
          \scalebox{.7}{
            \color{darkblue}
            \bf
            unitary group
          }
        }
      }
    }{
    \mathrm{U}(k)
    }
    \,.
  $$
  In the context of topological quantum computation (Lit. \ref{LiteratureTopologicalQuantumComputation})
  one imagines that $\rho(b)$ is the unitary transformation on the ground state of an effectively 2-dimensional topological quantum material
  (Lit. \ref{TopologicalQuantumMaterials})
  which is induced by adiabatially (Lit. \ref{TQCAsAdiabaticQuantumComputing}) moving $N+1$ anyonic (Lit. \ref{AnyonLiterature})
  defect points (Lit. \ref{LiteratureConfigurationSpaces})
  along the given braid $b$ (Lit. \ref{LiteratureBraiding}).

  Some authors (e.g. \cite{KauffmanLomonaco04}) consider representations of the non-pure braid group \eqref{FibrationOfBraidGroups} in the special case that $k = (N + 1)d$, with each strand contributing a $d$-dimensional tensor factor to the full Hilbert space (so that the first Artin braid relation \eqref{FirstArtinRelationGraphically} is satisfied), in which case the representation condition for the remaining second Artin braid relation \eqref{SecondArtinRelationGraphically} is famous as the  {\it Yang-Baxter equation} (e.g. \cite{YangGe91}\cite[\S 5]{Abad15}), much studied in pure algebra.

  Similarly, if the strands (anyons) carry labels (different anyon species, Lit. \ref{LiteratureAnyonSpecies}) so that one is looking at representations of the {\it pure} braid group then many authors consider the corresponding algebraic structure of braided fusion categories, going back to
  \cite[\S 8, \S E]{Kitaev06}\cite[pp. 28]{NayakSimonSternFreedmanDasSarma08}\cite[\S 6.3]{Wang10} and repeated in numerous reviews,
e.g. \cite[\S 2.4.1]{RowellWang18}\cite[\S 2.2]{Rowell22}.

  However, braiding transformations thought to be physically realizable in quantum materials
  arise more specifically as {\it monodromy representations}, being the holonomy/transport of the {\it Knizhnik-Zamolodchikov connection on bundles of conformal blocks}
  (e.g. \cite{TodorovHadjiivanov01}\cite[\S 4]{Abad15}\cite{GuHaghighatLiu22}), see
  Lit. \ref{KZConnectionsOnConformalBlocksReferences} and see \hyperlink{FigureA}{\it Figure A}.
\end{literature}

\begin{literature}[\bf
Flat vector bundles,
Local systems,
Holonomy and Monodromy]
  \label{LiteratureLocalSystems}

  For better or worse, the following terms in pure mathematics are more or less equivalent to each other, their choice of usage depending more on the subject context and author's background than on the underlying phenomenon as such, which is a reflection of the importance of the latter:

  \begin{itemize}[leftmargin=.5cm]
  \item \it flat vector bundles

  \item \it local systems (of coefficients)

  \item \it holonomy (of flat parallel transport)

  \item \it monodromy

  \end{itemize}

\vspace{-.5cm}
\begin{equation}
  \label{AVectorBundle}
\adjustbox{}{
\hspace{-.8cm}
\begin{tabular}{ll}
\begin{minipage}{11.5cm}
To start with, a (topological, Lit. \ref{LiteratureTopologyAndHomotopyTheory}) {\it vector bundle} is a
{\it locally trivial fibration of vector spaces}, namely is a continuous map of topological spaces
$p_{\mathcal{H}} : \mathcal{H} \to X$ such that over an open cover $\{U_i \subset X\}_{i \in I}$,
$\underset{i \in I}{\sqcup} U_i \,=:\, U \twoheadrightarrow X$ it looks like a trivial fibration
$\mathrm{pr}_U :  U \times \mathcal{H}_0 \to U$ of some ``typical fiber'' vector space $\mathcal{H}_0$,
and such that the induced {\it transition functions} over the cover intersections
$U \times _X U \underoverset{\mathrm{pr}_2}{\mathrm{pr}_1}{\rightrightarrows} U$ are
\\[-8pt]
continuous and fiberwise linear.
\end{minipage}
&
\;\;\;
$
  \begin{tikzcd}[column sep=huge]
    U \times
    \overset{
      \mathclap{
        \;\;\;\;\;\;
        \rotatebox{40}{
          \rlap{
            \scalebox{.7}{
              \color{darkblue}
              \bf
              \hspace{-29pt}
              \def\arraystretch{.9}
              \begin{tabular}{r}
                typical
                \\
                fiber
              \end{tabular}
            }
          }
        }
      }
    }{
      \mathcal{H}_0
    }
    \ar[d, "{ \mathrm{pr}_U }"]
    \ar[
      dr,
      phantom,
      "{\scalebox{.7}{(pb)}}"
    ]
    \ar[r]
    &
    \overset{
      \mathclap{
        \raisebox{4pt}{
          \scalebox{.7}{
            \def\arraystretch{.9}
            \color{darkblue}
            \bf
            \begin{tabular}{c}
              vector
              \\
              bundle
            \end{tabular}
          }
        }
      }
    }{
    \mathcal{H}
    }
    \ar[d, "{ p_{\mathcal{H}} }"]
    \ar[d, ->>]
    \\
    \underset{
      \mathclap{
      \raisebox{-3pt}{
      \scalebox{.7}{
        \color{darkblue}
        \bf
        \begin{tabular}{c}
          open cover
        \end{tabular}
      }
      }
      }
    }{
      U
    }
    \ar[r, ->>]
    &
    \underset{
      \mathclap{
      \raisebox{-3pt}{
      \scalebox{.7}{
        \color{darkblue}
        \bf
        base space
      }
      }
      }
    }{
      P
    }
  \end{tikzcd}
  $
\end{tabular}
}
\end{equation}
\vspace{-.3cm}

In our application to quantum physics in \cref{TopologicalQuantumGates}  the base space $P$ is a space of classical {\it parameters} and the {\it fiber} $\mathcal{H}_p$ of $p_{\mathcal{H}}$ over $p$ is (or rather: underlies) a {\it Hilbert space of quantum states} of a quantum systems for these external parameters.

Now, a {\it flat connection} ``on'' this vector bundle (namely: connecting its fibers to each other) is a rule for how to {\it lift paths} (Lit. \ref{PathLiftingLiterature}) from the base space $P$ to the total space $\mathcal{H}$, for every choice of lift of their starting point, such that (1.) these lifts depend only on the homotopy-class of the pass (for fixed endpoints) and  (2.) respect the concatenation of paths and (3.) lift constant paths to constant paths.
Or rather: Such lifting is called the {\it parallel transport} of a flat connection, that connection itself being understood as the prescription for lifting ``infinitesimal paths'' (tangent vectors to paths) which one may make sense of when the topological vector bundle is equipped with differentiable structure.

This parallel transport along {\it closed} paths (representing elements in the fundamental group of the base $P$, Lit. \ref{LiteratureHomotopyGroups}) is called the {\it monodromy} of the flat connection (constituting a linear representation of the fundamental group on the fibers over the base point); and this monodromy equivalently characterizes the flat connection over each connected component of $P$ (\cite[\S I.1]{Deligne70}, cf. e.g. \cite[Prop. 2.5.1]{Dimca04}).

Hence a flat connection on a vector bundle $\mathcal{H} \twoheadrightarrow P$ locally stratifies $\mathcal{H}$ into {\it horizontal subspaces}, making it a covering space $\mathcal{H}^\delta \twoheadrightarrow P$  (Lit. \ref{PathLiftingLiterature}).
Finally (but here we do not further need this): the {\it sheaf of local sections} of this covering space $\mathcal{H}^\delta$ is then a {\it locally constant sheaf of vector spaces} and as such known as a {\it local system}
(terminology going back to \cite{Steenrod43}, textbook accounts include \cite[\S 9.2.1]{Voisin03I}\cite[\S 2.5]{Dimca04}).
This terminology is short for {\it local system of coefficients}, referring to the use of such flat vector bundles as coefficients for {\it twisted cohomology} (Lit. \ref{LiteratureCohomology}).
\end{literature}

\begin{literature}[\bf Gauss-Manin connections]
\label{LiteratureGaussManinConnections}
In the parameterized perspective on algebraic topology (Lit. \ref{LiteratureTopologyAndHomotopyTheory}), one is naturally led to consider fiberwise (co)homology
groups (Lit. \ref{LiteratureCohomology})
over a parameter space. Under suitable conditions, the resulting bundles of (co)homology groups over that base
space carry a canonical flat connection (Lit. \ref{LiteratureLocalSystems}), meaning that there is a canonical way of identifying (co)homology elements along small paths of parameter values.

This notion of {\it Gauss-Manin connections} in a context of algebraic geometry
is due to \cite{Manin} \footnote{
  Yuri Manin hence pioneered both the notion of {\it quantum computation} (Lit. \ref{LiteratureQuantumComputation})
  as well as the notion of {\it Gauss-Manin connections} (Lit \ref{LiteratureGaussManinConnections} -- Gauss' name appears here just as a general tribute to a historical mathematician). Our result shows that these two seemingly unrelated ideas are in fact closely related.
};
see also \cite{Grothendieck}\cite{Katz} \cite{KatzOda}\cite{Griffiths}.
In \cref{ViaParameterizedPointSetTopology} we are concerned with the case of twisted cohomology (Lit. \ref{LiteratureCohomology}) on topological
fiber bundles, following \cite[\S 7.5]{EtingofFrenkelKirillov98}\cite[Def. 9.13]{Voisin03I}\cite[\S 1.5, 2.1]{Kohno02},
where the {\it hyergeometric integral construction} (Lit. \ref{HypergeometricIntegralReferences}) shows that a special case of Gauss-Manin connections are  {\it Knizhnik-Zamolodchikov connections}.
\end{literature}

\begin{literature}[\bf Knizhnik-Zamolodchikov connections on conformal blocks]
\label{KZConnectionsOnConformalBlocksReferences}
The Hilbert spaces of topologically ordered ground states of effectively 2-dimensional topological quantum materials (Lit. \ref{TopologicalQuantumMaterials})
in dependence on the position of anyonic defects (Lit. \ref{LiteratureConfigurationSpaces}) are thought (Lit. \ref{LiteratureTopologicalQuantumComputation})
to be the spaces of states of a Chern-Simons field theory, or rather their ``chiral half'', called the spaces of ``conformal blocks'' of a WZW (aka affine
current algebra) conformal field theory (see \cite[\S C]{FMS97}). A derivation of this (previously unproven, cf. \cite{Valera21}) assumption from first principles is argued
in \cite[\S 3]{SS22AnyonictopologicalOrder}.

As the parameters (the defect configurations) vary, these spaces of conformal blocks form a bundle over the configuration space of points
(Lit. \ref{LiteratureConfigurationSpaces}) which carries a canonical flat connection
(Lit. \ref{LiteratureLocalSystems})
known as the {\it Knizhnik-Zamolodchikov connection}, \cite[\S 15.3.2]{FMS97}\cite[\S 1.5, 2.1]{Kohno02}\cite[\S 4]{Abad15}, cf. Ex. \ref{TheKZConnection}.

The monodromy (Lit. \ref{LiteratureLocalSystems})
of this KZ-connection constitutes a unitary representation of the fundamental group of the configuration space of points, hence a  braid group representation (Lit. \ref{LiteratureBraidRepresentations}),
thought to reflect the operation of topological quantum gates (Lit. \ref{LiteratureTopologicalQuantumComputation}) by anyon braiding (Lit. \ref{MathematicsOfAnyonsLiterature}).
Here we consider all this via its
{\it hypergeometric integral}
representation
(Lit. \ref{HypergeometricIntegralReferences}).
\end{literature}

\begin{literature}[{\bf Hypergeometric integral construction of conformal blocks} {\cite[Prop. 2.15, 2.17]{SS22AnyonicDefectBranes}}]
\label{HypergeometricIntegralReferences}
What is known as the {\it hypergeometric integral construction} in conformal field theory is, in the end, an equivalence between

\begin{itemize}[leftmargin=1cm]
  \item[\bf (1)] the bundle of conformal blocks with its KZ-connection (Lit \ref{KZConnectionsOnConformalBlocksReferences})

  \item[\bf (2)] the bundle of suitably twisted cohomology groups (Lit. \ref{LiteratureCohomology}) of configuration spaces (Lit. \ref{LiteratureConfigurationSpaces}) of $N + \bullet$ points
  \\
  equipped with their Gauss-Manin connection (Lit. \ref{LiteratureGaussManinConnections}, Ex. \ref{TheKZConnection}).
\end{itemize}

The idea originally emerged from the work of several authors in conformal field theory, an early reference is
\cite{DateJimboMatsuoMiwa90}; more accessible exposition is given in \cite[\S 4, \S 7]{EtingofFrenkelKirillov98}.
The proof that for suitable parameters \eqref{TheTwistParameters} the resulting KZ-connection is exactly that on
 $\suTwo$-conformal blocks is due to \cite{FeiginSchechtmanVarchenko94}.\footnote{Beware that these authors equivalently speak in terms of the complexification $\mathfrak{sl}(2,\ComplexNumbers)$ of the real Lie algebra $\suTwo$. }

That this proof is rather technical (but we tried to bring out the key result concisely in \cite[\S 2]{SS22AnyonicDefectBranes}) witnesses the complexity inherited from the traditional notion of $\suTwoAffine{k}$-conformal blocks and only serves to highlight how remarkable it is that the hypergeometric integral construction reveals them as being equivalent to a purely cohomological construction. It is this remarkable fact which allows us, in the next step, to further identify bundles of conformal blocks with a remarakbly simple homotopy data structure (in Theorem \ref{TheTheorem} below).

\end{literature}

\begin{literature}[\bf Software verification, data typing and their mathematical semantics]
 \label{VerificationLiterature}
 A profound confluence of computer science and pure mathematics occurs \cite{MartinLof82} with the consideration of {\it certified} software
 (e.g. \cite{Chlipala07}), where the formal verification of the correct implementation of any program
 (such as by prescribed bounds on numerical rounding errors, cf. Lit. \ref{ExactRealComputerAnalysis})
 happens to coincide with rigorous (and constructive)
 {\it proof}  of a corresponding mathematical theorem -- and vice versa. (Technically, this works due to the {\it BHK correspondence}, see around \eqref{LogicalConnectivesAsTypeFormation} below.)
 Such verification/proof languages (like {\tt Agda}, Lit. \ref{LiteratureAgda})
 are (dependently) {\it typed} in that strictly every piece of data they handle has assigned a precise {\it data type} which provides the strict specification
 that data has to meet in order to qualify as input or output of that type
 (\cite{MartinLof82}\cite{Thompson91}\cite{Streicher93}\cite{Luo94}\cite{Gunter95}\cite{Constable11}\cite{Harper16}).

 The abstract theory of such data typing is known as (dependent-){\it type theory} and the modern flavor relevant here is often called {\it Martin-L{\"o}f type theory} in honor of \cite{MartinLof71}\cite{MartinLof75}\cite{MartinLof84}, for more elaboration and introduction see also \cite{Hofmann97}\cite{UFP13}.

 Once this typing principle is adhered to, the distinction vanishes between writing a program and verifying its correctness.
 Moreover, such a properly typed
 functional program may equivalently be understood as a {\it mathematical} object, namely as a mathematical function \eqref{FunctionDeclaration}
 from the ``space'' of data of
 its input type to that of its output type --- called its {\it denotational semantics} (a seminal idea due to \cite{Scott70}\cite{ScottStrachey71},
 for exposition see \cite[\S 9]{SlonnegerKurtz95}) and more specifically, for dependent type theories: its {\it categorical semantics} in locally Cartesian closed categories \eqref{LCCRules}  \cite{Seely84}\cite[\S 3]{Hofmann97}, reviewed e.g. in \cite{Jacobs98}\cite{Shulman12}. When the univalence axiom
 \eqref{UnivalenceAxiom} is included then this categorical semantics takes place
 \eqref{TypeClassification}
 in categories of ``higher geometric spaces'' (aka: $\infty$-stacks) called {\it model toposes} \cite{Rezk10} ({\it $\infty$-toposes} \cite{Lurie09}, see around \eqref{InterpretationInTheClassicalModelTopos} below), a statement that was first conjectured in \cite{Awodey12} and fully proven in \cite{Shulman19}, exposition is in \cite{Riehl22}.

This tight interrelation between the theories of verified computation of data types and of mathematical spaces has been advertised as the
{\it Computational Trilogy} \eqref{TheTrilogy}, references are provided in \cite[p. 4]{SS22TQC}. It is via  this ``Rosetta stone''
of the Computational Trilogy, generalized to homotopy typing (Lit. \ref{LiteratureHomotopyTypeTheory}), that we translate the algebraic topology
of topological quantum gates (in \cref{ViaParameterizedPointSetTopology}) into their homotopy type theory (in \cref{ViaDependentHomotopyTypeTheory}).
\end{literature}

\begin{literature}[\bf Homotopically typed programming]
  \label{LiteratureHomotopyTypeTheory}
 An operation on data so fundamental and commonplace that it is easily taken for granted is the {\it identification} of a pair of data with each other.
 But taking the idea of program verification by data typing (Lit. \ref{VerificationLiterature})
 seriously leads to consideration also of   {\it certificates of identification} of pairs of data of any given type which thus must  themselves be data
 of ``identification type'' \cite[\S 1.7]{MartinLof75}, see around \eqref{IdentificationType} below.

 Trivial as this may superficially seem,
 something profound emerges with such
 ``thoroughly typed'' programming languages (the technical term is: {\it intensional type theories}, see  \cite[p. 4, 13]{Streicher93}\cite[p. 16]{Hofmann95}),
 in that now given a pair of such identification certificates the same logic applies to themselves and leads to the consideration of
 identifications-of-identifications (first amplified in \cite{HofmannStreicher98}), and so on to higher identifications, {\it ad infinitum}.

 Remarkably, the ``denotational semantics'' (Lit. \ref{VerificationLiterature}) of data types equipped with such towers of identification types, hence the
 corresponding pure mathematics, is  (\cite{AwodeyWarren09}\cite{Awodey12}, exposition in \cite{Shulman12}\cite{Riehl22}) just that of abstract homotopy theory (Lit. \ref{LiteratureTopologyAndHomotopyTheory}) where identification types are interpreted
 \eqref{PathInduction}
 as path spaces and higher-order identifications correspond to
 higher-order homotopies \eqref{ReversesAreInverses}.
 One also expresses this state of affairs, somewhat vaguely, by saying that HoTT has {\it semantics} in homotopy theory, and
 conversely that HoTT is a {\it syntax} for homotopy theory -- we review this dictionary in \cref{HoTTIdea} below.

 Ever since this has been understood, the traditional (``intuitionistic Martin-L{\"o}f''-)type theory of \cite{MartinLof75}\cite{NPS90} has essentially come to be known
 as {\it homotopy type theory} (HoTT) -- specifically so if accompanied by one further ``univalence'' axiom\footnote{
   \label{UnivalenceAttribution}
   The univalence axiom is widely attributed to \cite{Voevodsky10}, but the idea (under a different name) is actually due to
   \cite[\S 5.4]{HofmannStreicher98}, there however formulated with respect to a subtly incorrect type of equivalences (as later shown
   in \cite[Thm. 4.1.3]{UFP13}). The new contribution of \cite[p. 8, 10]{Voevodsky10} was a good definition of the types
   \eqref{DataTypeOfEquivalences}
   of (``weak'') equivalences between types.
 } \eqref{UnivalenceAxiom}
 which enforces that identification of data types themselves coincides with their operational equivalence (exposition in \cite{Aczel11}).

 The standard textbook account for ``informal'' (human-readable) HoTT is \cite{UFP13}, exposition may be found in \cite{BrunerieLicataLumsdaine13},
 gentle introduction in \cite{Rijke18}\cite{Rijke23} (the former more extensive); and see \cref{HoTTIdea} below. Available software that {\it runs} homotopically typed programs
 includes {\tt Agda} (Lit. \ref{LiteratureAgda}) and {\tt Coq}\footnote{\;{\tt Coq} landing page: \href{https://coq.inria.fr}{\tt coq.inria.fr}} (the latter however lacking computational univalence \eqref{UnivalenceAxiom}).
\end{literature}

\begin{literature}[\bf The programming/certification language {\tt Agda}]
\label{LiteratureAgda}
The homotopically typed (Lit. \ref{LiteratureHomotopyTypeTheory}) language {\tt Agda}\footnote{
 {\tt Agda} landing page:
 \href{https://wiki.portal.chalmers.se/agda}{\tt wiki.portal.chalmers.se/agda}
 }
 is due to \cite{Norell09}, its ``cubical'' enhancement with computational univalence \eqref{UnivalenceAxiom} is due to \cite{VezzosiMoertbergAbel19}.
For introductions to
dependent type theory (Lit. \ref{VerificationLiterature}) in {\tt Agda} see \cite{Stump16}
and for introduction to
homotopy type theory (Lit. \ref{LiteratureHomotopyTypeTheory})
in {\tt Agda} see \cite{Escardo19}.
Existing libraries of {\tt Agda}-code relevant for our discussion include:  \cite{1lab}\cite{UniMath}.
\end{literature}

\begin{literature}[\bf Exact real computer arithmetic]
\label{ExactRealComputerAnalysis}
One key application of software certification (Lit. \ref{VerificationLiterature}) is to algorithms dealing with {\it exact real computer arithmetic} (\cite{Vuillemin88}\cite{YapDube95}\cite{PottsEdalat97}), i.e. with operations involving or requiring arbitrary high numerical precision. The basic strategy is to represent a real number by an algorithm which for any prescribed bound on precision produces a rational number that is guaranteed to approximate the intended real number to within that specified bound (cf. Rem. \ref{ComputationalContentOfRealNumbers} below). While traditional {\it floating-point arithmetic} on computers is notoriously prone to rounding errors and hard to verify, exact real computer arithmetic provides algorithms that provably satisfy prescribed accuracy bounds.
The underlying mathematics of exact real computer arithmetic is that of {\it constructive analysis}
\cite{Bishop67}\cite{BishopBridges85}\cite{Bridges99} (which has been fully developed already decades ago, then mainly motivated on philosophical grounds, as an alternative to classical analysis), and naturally implemented in typed programming languages (e.g. \cite{OConnor07}\cite{GNSW07}) such as in {\tt Agda} (Lit. \ref{LiteratureAgda}): \cite{Murray22}\cite{Lundfall15}, cf. pp. \pageref{ConstructingTheContinuum}.

While the application of real analysis to the simulation of classical systems in physics and engineering needs no further emphasis, the simulation of quantum computational systems involves real analysis
yet more fundamentally, in that the vector spaces (Hilbert spaces) of quantum states involve arbitrary-precision data even where the corresponding classical state spaces form a finite set (cf. \hyperlink{TableD}{\it Table D}). Specifically, in quantum computation (Lit. \ref{LiteratureQuantumComputation}) the analog of a classical computation on a finite set of {\it bits} is a (unitary) {\it linear operator} on a finite-dimensional complex vector space (such as spaces of {\it conformal blocks} in the case of anyons, Lit. \ref{AnyonLiterature}), whose certification, as such, already involves arbitrary precision arithmetic --- see the discussion below in \cref{KZConnectionsInHomotopyTypeTheory}.

In fact, the core issue of quantum computation, namely, the {\it compilation} of a set of prescribed {\it quantum logic gates} to a {\it quantum circuit} which evaluates to a prescribed unitary operator is one that must be understood in the sense of exact real arithmetic (constructive analysis) as to be verified for any finite bound on precision (cf. p. \pageref{ApplicationToQuantumCompilation}).
\end{literature}

\begin{literature}[\bf Path lifting and fiber transport]
  \label{PathLiftingLiterature}
  The archetypical phenomenon which leads over from ``point-set topology'' to genuine homotopy theory (Lit. \ref{LiteratureTopologyAndHomotopyTheory})
  is the classification of covering spaces by the ``transport'' operation induced on their fibers via ``path lifting'' (e.g. \cite[\S 3]{tomDieck08}\cite{Moller11}).

\vspace{.1cm}
\hspace{-.8cm}
\begin{tabular}{ll}
  \begin{minipage}{8cm}
  The {\it fundamental theorem of covering spaces} (e.g. \cite[Thm. 3.3.2]{tomDieck08}\cite[7.8]{Moller11}) shows that covering spaces are equivalently
  classified by this transport action of the fundamental group $\pi_1(P)$
  (Lit. \ref{LiteratureHomotopyGroups})
  of their base spaces $P$ given by such lifting of closed paths in the base space.

  A major class of examples of this situation arises from vector bundles $\mathcal{X} \twoheadrightarrow P$ equipped with flat connections (``local systems'', Lit. \ref{LiteratureLocalSystems}): The flat connection locally stratifies $\mathcal{H}$ into ``horizontal subspaces'' making it a covering space $\mathcal{H}^\delta \twoheadrightarrow P$. The corresponding path lifting is the ``parallel transport'' induced by the flat connection.
  \end{minipage}
 &
 \;\;
 \adjustbox{raise=-.3cm}{
\begin{tikzcd}[row sep=30pt]
  C
  \ar[
    dd,
    ->>,
    "{
      \scalebox{.8}{
        \color{darkblue}
        \bf
        covering space
      }
    }"{sloped, rotate=180, yshift=0pt}
  ]
  \\
  \\
  P
\end{tikzcd}
\;
\adjustbox{fbox, raise=-45pt}{
\begin{tikzpicture}[decoration=snake]

\begin{scope}[shift={(3.5,.95)}]
\draw[line width=1.1] (-.8,.6) -- (.8,.6);
\draw[line width=1.1] (-.8,.3) -- (.8,.3);
\draw[line width=1.1] (-.8,0) -- (.8,0);
\draw[line width=1.1] (-.8,-.3) -- (.8,-.3);
\draw[line width=1.1] (-.8,-.6) -- (.8,-.6);
\draw[line width=1.6] (-.8,-1.6) -- (.8,-1.6);

\draw (+.5,-1.9) node {
  \scalebox{1}{$U_j$}
};
\end{scope}

\draw[white, line width=4] (0,-.6) -- (3.6,1.24);
\draw[gray] (0,-.6) -- (3.6,1.24);

\draw[white, line width=4] (0,-.3) -- (3.6,1.24-.9);
\draw[gray] (0,-.3) -- (3.6,1.24-.9);

\draw[white, line width=4] (0,0) -- (3.6,1.24-.6);
\draw[gray] (0,0) -- (3.6,1.24-.6);

\draw[white, line width=4] (0,.3) -- (3.6,1.24-.3);
\draw[gray] (0,.3) -- (3.6,1.24-.3);

\draw[white, line width=4] (0,.6) -- (3.6,1.24+.3);
\draw[gray] (0,.6) -- (3.6,1.24+.3);

\node[rotate=13]
 at (1.45,1.33) {
   \scalebox{.7}{
     \color{greenii}
     \bf
     \def\arraystretch{.9}
     \begin{tabular}{c}
       permutation
       \\
       on fiber sets
     \end{tabular}
   }
};

\begin{scope}
\draw[line width=4, white] (-.8,.6) -- (.8,.6);
\draw[line width=1.1] (-.8,.6) -- (.8,.6);

\draw[line width=4, white] (-.8,.3) -- (.8,.3);
\draw[line width=1.1] (-.8,.3) -- (.8,.3);

\draw[line width=4, white] (-.8,0) -- (.8,0);
\draw[line width=1.1] (-.8,0) -- (.8,0);

\draw[line width=4, white] (-.8,-.3) -- (.8,-.3);
\draw[line width=1.1] (-.8,-.3) -- (.8,-.3);

\draw[line width=4, white] (-.8,-.6) -- (.8,-.6);
\draw[line width=1.1] (-.8,-.6) -- (.8,-.6);

\draw[line width=4, white] (-.8,-1.6) -- (.8,-1.6);
\draw[line width=1.1] (-.8,-1.6) -- (.8,-1.6);
\draw[
  gray,
  decorate,
  bend left=40
] (0,-1.6) to (1.6,-1.4);
\draw[
  ->,
  gray,
  decorate,
  bend left=-40
] (1.6,-1.4) to
  node[yshift=-12pt] {
    \scalebox{.7}{
      \color{greenii}
      \bf
      \hspace{-10pt}
      path in base space
    }
  }
(3.55,-.7);

\draw (+.5,-1.9) node {
  \scalebox{1}{
    \llap{
      \color{darkblue}
      patch
    }
    $U_i$
  }
};

\node[rotate=90]
  at (-1.5,-.5)
  {
    \scalebox{.8}{
      \color{darkblue}
      \bf
      \def\arraystretch{.9}
      \begin{tabular}{c}
        covering space
        \\
        over patch \color{black}$U_i$
      \end{tabular}
    }
  };

\node[rotate=90]
  at (-1.5+6.4,-.5+1)
  {
    \scalebox{.8}{
      \color{darkblue}
      \bf
      \def\arraystretch{.9}
      \begin{tabular}{c}
        covering space
        \\
        over patch \color{black}$U_j$
      \end{tabular}
    }
  };

\end{scope}

\node
  at (1.8,-.8)
  {
    \rotatebox{90}{$\mapsto$}
    \hspace{-9pt}
    \color{orangeii}
    \scalebox{.7}{
    \begin{tabular}{c}
      path
      \\
      lifting
    \end{tabular}
    }
  };

\end{tikzpicture}
}
}
\end{tabular}
\vspace{-.1cm}

  (More generally, one calls a map $p : X \to B$ of topological spaces a {\it Serre fibration} if for every Euclidean family of continuous paths in
  the ``base space'' $B$ and for every lift of the corresponding family of starting points through $p$ to $X$, there is also a compatible lift of
  the entire family of paths to $X$. If one thinks of these lifted paths as lines along which their starting points ``flow'' to their endpoints,
  then such lifts again serve to ``functorially transport'' the fibers of $p$ along paths in $B$ (e.g. \cite[\S 5.6]{tomDieck08}) in a
  homotopy-correct way.

  Noticing here that every (discrete) group $G$ arises as the fundamental group of {\it some} topological space, and specifically of its Eilenberg-MacLane
  spaces $B G = K(G,1)$ (Lit. \ref{LiteratureEilenbergMacLanSpaces}), this means in our context that all systems of {\it reversible} logic gates
  (Lit. \ref{ReversibleComputingLiterature}) acting on the same space of (memory-)states is equivalently embodied as the transport operation
  induced by path lifting in some covering space!
  This perspective we expand on in
  \cref{TopologicalQuantumGates}.

  \smallskip

  Curiously,
  such path lifting/fiber transport is a {\it native} language construct \eqref{TransportRule}
  in homotopy-typed programming languages (Lit. \ref{LiteratureHomotopyTypeTheory}) such as {\tt Agda} (Lit. \ref{LiteratureAgda}).
  This remarkable fact drives our discussion in \cref{ViaDependentHomotopyTypeTheory} and \cref{KZConnectionsInHomotopyTypeTheory} below,
  as highlighted now in
  \cref{TopologicalQuantumGates}.
\end{literature}

\newpage

\section{Topological quantum gates...}
\label{TopologicalQuantumGates}

While the broad idea of topological quantum logic gates is now 25 years old and classical (Lit. \ref{LiteratureTopologicalQuantumComputation}),
its conceptual fine-print has arguably remained somewhat elusive all along, both regarding their physical realization (cf. Lit. \ref{LiteratureMajoranaZeroModes})
as well as their information-theoretic nature (Lit. \ref{MathematicsOfAnyonsLiterature}).
In this section, we provide a quick modernized review and explanation of the basic idea of topological quantum gates,
informed by their cohomological realization (\cite{SS22AnyonictopologicalOrder}, Lit. \ref{LiteratureKTheoryClassificationOfTopologicalPhases})
and
such as to bring out their secret nature \cite{SS22TQC}
as a fundamental construction in parameterized point-set topology (which is developed below in \cref{ViaParameterizedPointSetTopology}) and in
homotopy type theory (developed below in \cref{ViaDependentHomotopyTypeTheory}):

\medskip

\noindent
{\bf The driving theme} of our discussion
is the observation/claim that:

\vspace{-.3cm}
\begin{equation}
\label{TheDrivingTheme}
\adjustbox{}{
\hspace{-3cm}
\fbox{
{\it
\begin{tabular}{l}
Fundamental (quantum) computing processes are
\\
lifts of classical parameter paths, i.e. the programs,
\\
to (linear) maps of state spaces, i.e. the (quantum) gates.
\end{tabular}
}
}
\hspace{.1cm}
\adjustbox{raise=4pt}{
\begin{tikzcd}[
  column sep=84pt
]
  \{0\}
  \ar[
    r,
    "{
      \scalebox{.7}{
        \color{orangeii}
        initial lift = input data
      }
    }"
  ]
  \ar[
    d,
    hook
  ]
  &
  \mathscr{H}
  \mathrlap{
    \raisebox{2pt}{
    \scalebox{.7}{
      \color{darkblue}
      \bf
      quantum state bundle
    }
    }
  }
  \ar[d, ->>]
  \\
  {\;\;\,\;[0,1]}
  \ar[
    r,
    "{
      \scalebox{.7}{
        \color{orangeii}
        parameter
        path =
        program
      }
    }"{swap, yshift=-1pt}
  ]
  \ar[
    ur,
    dashed,
    end anchor={[yshift=-2pt]},
    "{
      \scalebox{.7}{
        \color{orangeii}
        state path
        lift
        =
        execution
      }
    }"{sloped, pos=.4}
  ]
  &
  P
  \;\;
  \mathrlap{
    \raisebox{2pt}{
    \scalebox{.7}{
      \color{darkblue}
      \bf
      parameter space
    }
    }
  }
\end{tikzcd}
}
}
\end{equation}

\medskip

This may sound simple, but we claim it is profound (similar statements are in \cite{ZanardiRasetti99}\cite{NDGD06}\cite{DowlingNielsen08}\cite{LW17}): Namely, it means that natural certification languages for low hardware-level (quantum) computation ought to natively know about {\it path lifting} (Lit. \ref{PathLiftingLiterature}). This is unheard-of in traditional programming languages --- but it is the hallmark \eqref{TransportRule} of homotopically-typed languages! (Lit. \ref{LiteratureHomotopyTypeTheory})  Moreover, under such identification of low-level computation with path-lifting, homotopically-typed languages natively reflect the crucial reversibility \eqref{InversionOfIdentifications} of fundamental quantum computational processes (Lit. \ref{ReversibleComputingLiterature}).

\medskip

\noindent
{\bf The idea of quantum circuits.}
For example, a {\it quantum circuit} (Lit. \ref{LiteratureQuantumComputation}) is {\it de facto}
the compilation of a sequence of basic linear operators $U_{p}$ (the ``gates'') following a sequence of computing instructions $p$. Choosing different classical paths of instructions yields different quantum algorithms in a controlled way, and this dependency of the quantum transport on classical instruction paths is what it means for quantum computations to be programmable (the {\it circuit model} of quantum computation \cite{Deutsch89}, e.g. \cite[\S II.4]{NC10}).

\vspace{.1cm}
\hypertarget{FigureQC}{}
\hspace{-1cm}
\begin{tabular}{cc}
\adjustbox{raise=10pt}{
\begin{tikzpicture}[xscale=1.2, yscale=.6]

\draw
[line width=1.3pt]
  (-4,2) to
  node[yshift=14pt]{
    \scalebox{.8}{
      \color{greenii}
      \bf
      Quantum circuit:
      linear operator composed from
      quantum logic gates
    }
  }
  (4,2);

\draw
[line width=1.3pt]
  (-4,1) to (4,1);

\draw
[line width=1.3pt]
  (-4,0) to (4,0);

\draw
[line width=1.3pt]
  (-4,-1) to (4,-1);

\draw
[line width=1.3pt]
  (-4,-2) to (4,-2);

\draw (-4.3, 2) node
{
  \scalebox{.9}{$\mathscr{H}$}
};
\draw (+4.3, 2) node
{
  \scalebox{.9}{$\mathscr{H}$}
};

\draw (-4.3, 1.5) node
{
  \scalebox{.9}{$\otimes$}
};
\draw (+4.3, 1.5) node
{
  \scalebox{.9}{$\otimes$}
};

\draw (-4.3, 1) node
{
  \scalebox{.9}{$\mathscr{H}$}
};
\draw (+4.3, 1) node
{
  \scalebox{.9}{$\mathscr{H}$}
};

\draw (-4.3, .5) node
{
  \scalebox{.9}{$\otimes$}
};
\draw (+4.3, .5) node
{
  \scalebox{.9}{$\otimes$}
};

\draw (-4.3, 0) node
{
  \scalebox{.9}{$\mathscr{H}$}
};
\draw (+4.3, 0) node
{
  \scalebox{.9}{$\mathscr{H}$}
};

\draw (-4.3, -.5) node
{
  \scalebox{.9}{$\otimes$}
};
\draw (+4.3, -.5) node
{
  \scalebox{.9}{$\otimes$}
};

\draw (-4.3, -1) node
{
  \scalebox{.9}{$\mathscr{H}$}
};
\draw (+4.3, -1) node
{
  \scalebox{.9}{$\mathscr{H}$}
};

\draw (-4.3, -1.5) node
{
  \scalebox{.9}{$\otimes$}
};
\draw (+4.3, -1.5) node
{
  \scalebox{.9}{$\otimes$}
};

\draw (-4.3, -2) node
{
  \scalebox{.9}{$\mathscr{H}$}
};
\draw (+4.3, -2) node
{
  \scalebox{.9}{$\mathscr{H}$}
};

\draw
  (-4.5,-3) node {
    \scalebox{.7}{
      \color{darkblue}
      \bf
      \def\arraystretch{.9}
      \begin{tabular}{c}
      Hilbert space of
      \\
      input
      \\
      quantum states
      \end{tabular}
    }
  };

\draw
  (+4.4,-3) node {
    \scalebox{.7}{
      \color{darkblue}
      \bf
      \def\arraystretch{.9}
      \begin{tabular}{c}
      Hilbert space of
      \\
      output
      \\
      quantum states
      \end{tabular}
    }
  };

\draw (-4.7,0) node
{
  $
  \scalebox{.9}{
    $\mathscr{H}_1$
  }
  \hspace{-3pt}
  \left\{
  \def\arraystretch{3.2}
    \phantom{
      \begin{array}{c}
        \vert
        \\
        \vert
    \end{array}
  }
  \right.
  $
};

\draw (+4.7,0) node
{
  $
  \left.
  \def\arraystretch{3.2}
    \phantom{
      \begin{array}{c}
        \vert
        \\
        \vert
    \end{array}
  }
  \right\}
  \hspace{-3pt}
  \scalebox{.9}{
    $\mathscr{H}_2$
  }
  $
};

\draw
[
  fill=white
]
  (-3.5,2.3)
  rectangle
  node {$U_{p_{12}}$}
  (-2.5,0.7);

\draw
[
  fill=white
]
  (-2,1.3)
  rectangle
  node {$U_{p_{23}}$}
  (-1,-1.3);

\draw (-1.5,-1) node {
  \scalebox{.7}{
    \color{orangeii}
    \bf
    a gate
  }
};

\draw
[
  fill=white
]
  (-.5,1.3)
  rectangle
  node {$U_{p_{34}}$}
  (.5,-0.3);

\draw
[
  fill=white
]
  (1,-0.7)
  rectangle
  node {$U_{p_{45}}$}
  (2,-2.3);

\draw
[
  fill=white
]
  (2.5,2.3)
  rectangle
  node {$U_{p_{56}}$}
  (3.5,-.3);

\begin{scope}
[shift={(-.25,-.5)}]

\draw
[->]
 (-3.5,-3.5) to
 node
   [yshift=-6pt]
 {\scalebox{.9}{$p_{12}$}}
 (-2-.05,-3.5);

\draw
[->]
 (-2+.05,-3.5) to
 node
   [yshift=-6pt]
 {\scalebox{.9}{$p_{23}$}}
 (-.5-.05,-3.5);

\draw
[->]
 (-.5+.05,-3.5) to
 node
   [yshift=-6pt]
 {\scalebox{.9}{$p_{34}$}}
 node[yshift=8pt]{
  \scalebox{.8}{
      \color{greenii}
      \bf
      Path of consecutive quantum logic gate
      names:
      the program
    }
  }
 (+1-.05,-3.5);

\draw
[->]
 (1+.05,-3.5) to
 node
   [yshift=-6pt]
 {\scalebox{.9}{$p_{45}$}}
 (2.5-.05,-3.5);

\draw
[->]
 (2.5+.05,-3.5) to
 node
   [yshift=-6pt]
 {\scalebox{.9}{$p_{56}$}}
 (4-.05,-3.5);

\end{scope}

\end{tikzpicture}
}
&
\hspace{-2mm}
\raisebox{3cm}{
\begin{minipage}{4cm}
  \footnotesize
  {\bf Figure QC.} A quantum logic circuit acting on a compound of quantum systems (each shown as a solid line),
  compiled by composing a sequence of quantum logic gates (shown as boxes).

  Such quantum circuits a naturally understood as compound functions on  {\it linear} types, we expand on this
  in the companion article \cite{QPinLHOTT}. Here we focus on the internal operation of single
  (topological quantum) gates.
\end{minipage}
}
\end{tabular}

\vspace{-1mm}
However, in real laboratory implementations of quantum computers, the idealization of discretized instruction steps is
necessarily approximated by actual physical processes which, however abrupt they may appear, are fundamentally continuous.
This means that all {\it real} quantum computational processes depend on {\it continuous paths} in a classical parameter space.

For example, for many quantum computing architectures, such as for the original {\it spin resonance} principle (see, e.g., \cite{CoryEtAl00}\cite{Equbal22}) as well as for the contemporary {\it superconducting qbits}  (e.g., \cite{ClarkeWilhelm08}\cite{HWFZ20})
it is the case that to operate a quantum gate means to send an electromagnetic pulse of a finite duration through the system
(see, e.g., \cite{GIB12} and \cite{MFL22}, respectively): Therefore, the hardware-level instructions of such quantum computers are continuous parameter paths in the configuration space of the ambient electromagnetic field, quite of the following schematic form:

\vspace{-.4cm}
\begin{center}
\hypertarget{FigureP}{}
\begin{tabular}{ll}
\adjustbox{raise=-1.5cm}{
\begin{tikzpicture}[xscale=1.4, yscale=.7]

\begin{scope}
[shift={(-.25,-.5)}]

\draw
[->]
 (-3.5,-3.5) to
 node
   [yshift=-6pt]
 {\scalebox{.9}{$p_{12}$}}
 (-2-.05,-3.5);

\draw
[->]
 (-2+.05,-3.5) to
 node
   [yshift=-6pt]
 {\scalebox{.9}{$p_{23}$}}
 (-.5-.05,-3.5);

\draw
[->]
 (-.5+.05,-3.5) to
 node
   [yshift=-6pt]
 {\scalebox{.9}{$p_{34}$}}
 node[yshift=8pt]{
  \scalebox{.8}{
      \color{greenii}
      \bf
      Path of consecutive quantum logic gates:
      the program
    }
  }
 (+1-.05,-3.5);

\draw (.2, -6.5) node {
  \scalebox{.8}{
    \color{greenii}
    \bf
      Continuous path of hardware-level instructions: Gate pulse protocol
    }
};

\draw
[->]
 (1+.05,-3.5) to
 node
   [yshift=-6pt]
 {\scalebox{.9}{$p_{45}$}}
 (2.5-.05,-3.5);

\draw
[->]
 (2.5+.05,-3.5) to
 node
   [yshift=-6pt]
 {\scalebox{.9}{$p_{56}$}}
 (4-.05,-3.5);

\end{scope}

\begin{scope}[shift={(-3.8,-5.5)}]
\def\amplitude{.5}
\def\wavelength{.08}
\def\steps{18}
\foreach \i in {0,...,\steps}
\draw
  (0 + \wavelength*\i,0)
    sin
  (1*\wavelength/4 + \wavelength*\i,{\amplitude * exp(-((\i-\steps/2)*4/\steps)^2)})
    cos
  (2*\wavelength/4 + \wavelength*\i,0)
    sin
  (3*\wavelength/4 + \wavelength*\i,-{\amplitude * exp(-((\i-\steps/2)*4/\steps)^2)})
    cos
  ({\wavelength*(\i+1)},0);
\end{scope}

\begin{scope}[shift={(-3.8+1.5,-5.5)}]
\def\amplitude{.7}
\def\wavelength{.11}
\def\steps{13}
\foreach \i in {0,...,\steps}
\draw
  (0 + \wavelength*\i,0)
    sin
  (1*\wavelength/4 + \wavelength*\i,{\amplitude * exp(-((\i-\steps/2)*4/\steps)^2)})
    cos
  (2*\wavelength/4 + \wavelength*\i,0)
    sin
  (3*\wavelength/4 + \wavelength*\i,-{\amplitude * exp(-((\i-\steps/2)*4/\steps)^2)})
    cos
  ({\wavelength*(\i+1)},0);
\end{scope}

\begin{scope}[shift={(-3.8,-5.5)}]
\def\amplitude{.5}
\def\wavelength{.08}
\def\steps{18}
\foreach \i in {0,...,\steps}
\draw
  (0 + \wavelength*\i,0)
    sin
  (1*\wavelength/4 + \wavelength*\i,{\amplitude * exp(-((\i-\steps/2)*4/\steps)^2)})
    cos
  (2*\wavelength/4 + \wavelength*\i,0)
    sin
  (3*\wavelength/4 + \wavelength*\i,-{\amplitude * exp(-((\i-\steps/2)*4/\steps)^2)})
    cos
  ({\wavelength*(\i+1)},0);
\end{scope}

\begin{scope}[shift={(-3.8+3,-5.5)}]
\def\amplitude{.6}
\def\wavelength{.2}
\def\steps{7}
\foreach \i in {0,...,\steps}
\draw
  (0 + \wavelength*\i,0)
    sin
  (1*\wavelength/4 + \wavelength*\i,{\amplitude * exp(-((\i-\steps/2)*4/\steps)^2)})
    cos
  (2*\wavelength/4 + \wavelength*\i,0)
    sin
  (3*\wavelength/4 + \wavelength*\i,-{\amplitude * exp(-((\i-\steps/2)*4/\steps)^2)})
    cos
  ({\wavelength*(\i+1)},0);
\end{scope}

\begin{scope}[shift={(-3.8+4.5,-5.5)}]
\def\amplitude{.8}
\def\wavelength{.04}
\def\steps{37}
\foreach \i in {0,...,\steps}
\draw
  (0 + \wavelength*\i,0)
    sin
  (1*\wavelength/4 + \wavelength*\i,{\amplitude * exp(-((\i-\steps/2)*4/\steps)^2)})
    cos
  (2*\wavelength/4 + \wavelength*\i,0)
    sin
  (3*\wavelength/4 + \wavelength*\i,-{\amplitude * exp(-((\i-\steps/2)*4/\steps)^2)})
    cos
  ({\wavelength*(\i+1)},0);
\end{scope}

\begin{scope}[shift={(-3.8+6,-5.5)}]
\def\amplitude{.4}
\def\wavelength{.09}
\def\steps{17}
\foreach \i in {0,...,\steps}
\draw
  (0 + \wavelength*\i,0)
    sin
  (1*\wavelength/4 + \wavelength*\i,{\amplitude * exp(-((\i-\steps/2)*4/\steps)^2)})
    cos
  (2*\wavelength/4 + \wavelength*\i,0)
    sin
  (3*\wavelength/4 + \wavelength*\i,-{\amplitude * exp(-((\i-\steps/2)*4/\steps)^2)})
    cos
  ({\wavelength*(\i+1)},0);
\end{scope}

\end{tikzpicture}
}
&
\begin{minipage}{6cm}
  \footnotesize
  {\bf Figure P.} The execution of a quantum algorithm on common types of NISQ machines (Lit. \ref{LiteratureOnNISQMachines}), like the currently popular {\it superconducting qbit} architectures, corresponds to subjecting the system to a {\it gate pulse protocol} of radio/microwave modulation
  (schematically shown on the left)
  hence to driving it along a {\it continuous path} in the configuration space of external parameters (here: the electromagnetic cavity field amplitude).
\end{minipage}
\end{tabular}
\end{center}

This is to amplify that {\it all} quantum-computation is fundamentally an evolution of quantum states controlled by continuous classical parameter paths. Now the case of {\it topological} quantum computation corresponds to the special case where these continuous parameter paths are (1.) adiabatic and their effect on the system states is (2.) homotopy-invariant:

\medskip

\noindent
{\bf The idea of adiabatic quantum transport.} A traditional computing process of the form indicated in \hyperlink{FigureP}{\it Figure P} exchanges energy between the quantum system and its control environment. In fact, common NISQ architectures (Lit. \ref{LiteratureOnNISQMachines}) are designed to encode qbit states  as energy eigenstates of anharmonic quantum oscillators, so that passing between their energy levels (the notorious {\it quantum jumps}) is what it means to execute computations on these systems in the first place. At the same time, these energetic interaction channels with their environment is what makes these NISQ machines suffer from noise and decoherence.

In contrast, a topological quantum process (Lit. \ref{LiteratureTopologicalQuantumComputation}) is, first of all, to take place entirely on the (topologically ordered) {\it ground states} of a topological quantum material (Lit. \ref{TopologicalQuantumMaterials}), hence on their lowest energy states, without absorbing any energy from the environment: The notorious {\it energy gap} which measures the fidelity of topological phases of matter (Lit. \ref{TopologicalQuantumMaterials}), separating their topological ground states from their ordinary excited states, is the room within which the control environment may shed energy without disturbing the coherent quantum phase.

\smallskip

This state of affairs is neatly captured by one of the classical theorems of mathematical quantum mechanics: The {\it Quantum Adiabatic Theorem} (Lit. \ref{TQCAsAdiabaticQuantumComputing}) says
(as nicely brought out for quantum computation already in \cite{ZanardiRasetti99})
that in the asymptotic limit of sufficiently gentle (= ``adiabatic'') movement of external classical parameters, the induced quantum system's evolution asymptotically preserves gapped energy eigen-states, hence in particular preserves gapped ground states, and hence acts on the Hilbert space $\mathscr{H}$ of gapped ground states by
unitary operators $U_p$ that vary continuously with the parameter path $p$.

\vspace{-.2cm}
\begin{center}
\label{FigureT}{}
\tabcolsep=10pt
\begin{tabular}{cc}
  \begin{minipage}{4.4cm}
    \footnotesize
    {\bf Figure T.}
    Schematically shown on the left is the
    ``adiabatic'' (Lit. \ref{TQCAsAdiabaticQuantumComputing})
    transport of quantum states along linear maps depending on continuous paths in a classical parameter space.

    The diagram on the right indicates our description of such situations
    by (linear) homotopy type families depending on a base homotopy type, as explained in
    \cref{ViaDependentHomotopyTypeTheory} below (see
    \eqref{TransportRule}
    and
    \eqref{TypeClassification}
    below, noticing that we relegate discussion of {\it linearity} of quantum data types to \cite{QPinLHOTT}).
  \end{minipage}
  &
  \quad
  \begin{tikzcd}[decoration=snake]
    \underset{
      \mathclap{
      \raisebox{-2pt}{
        \tiny
        \color{darkblue}
        \bf
        \def\arraystretch{.8}
        \begin{tabular}{c}
          Hilbert space of
          \\
          quantum states \color{black} at
          \\
          \color{black}
          parameter value $p_1$
        \end{tabular}
        }
      }
    }{
      \mathscr{H}_1
    }
    \ar[
      rr,
      bend left=20,
      "{
        {U_{p_{12}}}_{
      \mathrlap{
       \hspace{-5pt}
       \raisebox{-1pt}{
         \rotatebox{-15}{
            \tiny
            \color{greenii}
            \bf
            adiabatic transport of states
         }
       }
       }
        }
      }"{swap, yshift=-2pt, pos=.45},
      shorten=-3pt
    ]
    &&
    \mathscr{H}_2
    \ar[
      rr,
      bend right=20,
      "{U_{p_{23}}}"{
        swap, yshift=-3pt, pos=.55},
      shorten=-3pt
    ]
    \ar[
      dd,
      phantom,
      "{
        \raisebox{-2pt}{
        \rotatebox{90}{$\mapsto$}
        }
        \scalebox{.7}{
          \color{purple}
          \bf
          lift
        }
      }"
    ]
    &&
    \underset{
      \mathclap{
      \raisebox{-7pt}{
        \tiny
        \color{darkblue}
        \bf
        \def\arraystretch{.8}
        \begin{tabular}{c}
          Hilbert space of
          \\
          quantum states \color{black} at
          \\
          \color{black}
          parameter value $p_3$
        \end{tabular}
        }
      }
    }{
      \mathscr{H}_3
    }
    &[5pt]
    \mathclap{\subset}
    &[-8pt]
    \mathscr{H}
    \ar[
      dd,
      "{
        \mbox{
          \tiny
          \color{orangeii}
          \bf
          \def\arraystretch{.9}
          \begin{tabular}{c}
            fibration of
            \\
            dependent linear types
          \end{tabular}
        }
      }"{description, rotate=-20}
    ]
    \ar[rr]
    \ar[
      ddrr,
      phantom,
      "{\mbox{\tiny\rm(pb)}}"
    ]
    &&
    \widehat{\mathrm{LinType}}
    \ar[
      dd,
      "{
        \mbox{
          \tiny
          \color{orangeii}
          \bf
          \def\arraystretch{.9}
          \begin{tabular}{c}
            univalent
            \\
            fibration of
            \\
            linear types
          \end{tabular}
        }
      }"{description, rotate=-20}
    ]
    \\
    \\
    \underset{
      \mathclap{
      \raisebox{-2pt}{
        \tiny
        \color{darkblue}
        \bf
        \def\arraystretch{.8}
        \begin{tabular}{c}
          external
          \\
          classical
          \\
          parameters
          \\
          \color{black}
          at time $t_1$
        \end{tabular}
      }
      }
    }{
      p_1
    }
    \ar[
      rr,
      decorate,
      bend left=20,
      shorten=-4pt,
      "{
        {p_{12}}_{
      \mathrlap{
       \hspace{-1pt}
       \raisebox{-1pt}{
         \rotatebox{-16}{
            \tiny
            \color{greenii}
            \bf
            path in parameter space
         }
       }
       }
        }
       }"{swap, yshift=-1}
    ]
    &&
    p_2
    \ar[
      rr,
      decorate,
      bend right=20,
      shorten=-3pt,
      "{p_{23}}"{swap, yshift=-2pt, xshift=1pt}
    ]
    &&
    \underset{
      \mathclap{
      \raisebox{-2pt}{
        \tiny
        \color{darkblue}
        \bf
        \def\arraystretch{.8}
        \begin{tabular}{c}
          external
          \\
          classical
          \\
          parameters
          \\
          \color{black}
          at time $t_3$
        \end{tabular}
      }
      }
    }{
      p_3
    }
    &\mathclap{\in}&
    \underset{
      \mathclap{
      \raisebox{-2pt}{
        \tiny
        \color{darkblue}
        \bf
        \def\arraystretch{.8}
        \begin{tabular}{c}
          parameter
          \\
          space
        \end{tabular}
      }
      }
    }{
    P
    }
    \ar[
       rr,
       "{U}",
       "{
         \raisebox{-2pt}{
           \tiny
           \color{greenii}
           \bf
           \def\arraystretch{.8}
           \begin{tabular}{c}
             adiabatic
             \\
             quantum gates
           \end{tabular}
         }
       }"{swap}
    ]
    &&
    \underset{
      \mathclap{
      \raisebox{-2pt}{
        \tiny
        \color{darkblue}
        \bf
        \def\arraystretch{.8}
        \begin{tabular}{c}
          universe of
          \\
          linear types
        \end{tabular}
      }
      }
    }{
      \mathrm{LinType}
    }
  \end{tikzcd}
\end{tabular}
\end{center}

\medskip

\noindent
{\bf The idea of quantum annealing.}
For example, a widely-known implementation of the above {\it Quantum Adiabatic Theorem} in quantum computation is the
paradigm of ``quantum annealing''
\cite{KN98}\cite{FGGS00}
(review in \cite{RSDC22}).
Here one considers a single-parameter path linearly interpolating between two given Hamiltonians, $H_0, H_1$ on a fixed Hilbert
space $\mathscr{H} :\defneq \mathscr{H}_1 = \mathscr{H}_2$:
\vspace{-.2cm}

\hspace{-.8cm}
\begin{tabular}{ll}
\begin{minipage}{10cm}
If one can arrange for both Hamiltonians to have unique gapped ground states and for the ground state of the
first Hamiltonian to be preparable, while that of the second Hamiltonian is unknown but identifiable with
the answer to a given computational problem, then the corresponding adiabatic quantum transport effectively
computes that answer.
\end{minipage}
&
\qquad
\adjustbox{raise=-6pt}{
$
  \begin{tikzcd}[column sep=huge, row sep=1pt, decoration=snake]
    \mathscr{H}
    \ar[
      rr,
      bend left=20,
      "{\color{greenii}
        \mathcal{P}
        \exp
        \big(
          \frac{1}{i \hbar}
          \left(
            (1-t) H_0
            +
            t H_1
          \right)
        \big)
      }"
    ]
    &&
    \mathscr{H}
    \\[-2pt]
    \scalebox{0.7}{$(t=0)$}
    \ar[
      rr,
      decorate,
      shorten=-2pt
    ]
    &&
    \scalebox{0.7}{$(t=1)$}
  \end{tikzcd}
$
}
\end{tabular}
\vspace{.1cm}

Due to its restriction to finding a unique ground state, quantum annealing as such is not a universally programmable form of computation: It is guaranteed to discover the ground state of $H_1$ and does nothing else. However, in this restrictivity annealing does foreshadow a key aspect of topological quantum computation in degenerate form (cf. Lit. \ref{TQCAsAdiabaticQuantumComputing}): Since the ground state of $H_1$ is unique, the annealing process does not actually depend on the exact parameter path chosen to arrive there, it is {\it robust against perturbations of the computational path} (cf. \cite[p. 2]{ChildsFarhiPreskill02}).

\medskip

\noindent
{\bf The idea of topological quantum computation.}
Generally, the profound practical problem with implementing the theoretically straightforward idea of programmable quantum processes (Lit. \ref{LiteratureQuantumComputation})
is
that real quantum machines are not in idealized isolation but are coupled to their environment, which necessarily acts like a ``thermal bath'': Inevitable noise in the environment causes perturbations that tend to de-cohere the machine's quantum state and thus tend to destroy its intended quantum computation (cf. Lit. \ref{NeedForTopologicalProtection}).

\smallskip
Concretely, parameter paths realizable in real laboratories are noisy (cf. Lit. \ref{LiteratureOnNISQMachines}), hence are drawn randomly from an ensemble of small perturbations of the intended path. The result of transporting a pure quantum state along such a noisy ensemble is in general a decohered mixture of pure states which may no longer support the quantum interference effects on which quantum algorithms crucially rely;
unless, that is, one could somehow guarantee that the quantum transport depending on these paths is actually {\it in}dependent of their small perturbations, really depending only on the global properties of these paths.

This is the idea of topological quantum computation (Lit. \ref{LiteratureTopologicalQuantumComputation}):
to ensure that the quantum adiabatic process depends only on the {\it topological homotopy classes} (Lit. \ref{LiteratureTopologyAndHomotopyTheory}, \ref{LiteratureHomotopyGroups})
of the parameter paths (relative to their endpoints).

\smallskip
\hypertarget{FigureH}{}
\begin{tabular}{ccc}
  \begin{tikzcd}[
    decoration=snake,
    column sep=45pt
  ]
    \mathscr{H}_1
    \ar[
      rr,
      bend left=20,
      "{\ }"{swap, name=s},
      "U_{p_{12}}"
    ]
    \ar[
      rr,
      bend right=45,
      "{\ }"{name=t},
      "U_{p'_{12}}"{swap}
    ]
    \ar[
      dd,
      phantom,
      shift left=20pt,
      "{
        \raisebox{-2pt}{
        \rotatebox{90}{$\mapsto$}
        }
        \hspace{-2pt}
        \scalebox{.7}{
          \color{purple}
          \bf
          lift
        }
      }"{pos=.6}
    ]
    &&
    \mathscr{H}_2
    \ar[
      from=s,
      to=t,
      Rightarrow,
      -,
      "{
        \scalebox{.7}{
          \color{greenii}
          \bf
          \def\arraystretch{.8}
          \begin{tabular}{c}
            operationally equal
            \\
            quantum gates
          \end{tabular}
        }
      }"{description, pos=.46}
    ]
    \\[+30pt]
    \\
    p_1
    \ar[
      rr,
      bend left=20,
      shorten=-2pt,
      decorate,
      "{\ }"{swap,name=sbase},
      "p_{12}"{yshift=3pt}
    ]
    \ar[
      rr,
      bend right=45,
      shorten=-2pt,
      decorate,
      "{\ }"{name=tbase},
      "{p'_{12}}"{swap, yshift=-2pt}
    ]
    &&
    p_2
    \ar[
      from=sbase,
      to=tbase,
      Rightarrow,
      "{
        \scalebox{.7}{
          \color{greenii}
          \bf
          \def\arraystretch{.8}
          \begin{tabular}{c}
            homotopy of
            \\
            parameter paths
          \end{tabular}
        }
      }"{description, pos=.46}
    ]
  \end{tikzcd}
  &
  \adjustbox{raise=-12pt}{
  \begin{minipage}{5.8cm}
    \small
    {\bf Figure H.}
    An adiabatic quantum transport
    (\href{FigureT}{Fig. T})
    is
    {\it topological} (or rather: {\it homotopical}, cf. Lit. \ref{LiteratureTopologyAndHomotopyTheory}) if it depends on the parameter path between fixed endpoints only up to small continuous deformations, namely up to homotopy (indicated on the left).

    When this is the case, then quantum transport depends {\it robustly} on ``global'' properties of parameter paths, such as their winding number
    (cf. Lit. \ref{LiteratureHomotopyGroups})
    around ``holes'' in parameter space (schematically indicated on the right) and hence constitutes a form of {\it topological quantum computation} (Lit. \ref{LiteratureTopologicalQuantumComputation}).
  \end{minipage}
  }
  &
  \begin{tikzcd}[
    decoration=snake,
    column sep=45pt
  ]
    \mathscr{H}_1
    \ar[
      rr,
      bend left=20,
      "{\ }"{swap, name=s},
      "U_{p_{12}}"
    ]
    \ar[
      rr,
      bend right=45,
      "{\ }"{name=t},
      "U_{p'_{12}}"{swap}
    ]
    \ar[
      dd,
      phantom,
      shift left=20pt,
      "{
        \raisebox{-2pt}{
        \rotatebox{90}{$\mapsto$}
        }
        \hspace{-2pt}
        \scalebox{.7}{
          \color{purple}
          \bf
          lift
        }
      }"{pos=.6}
    ]
    &&
    \mathscr{H}_2
    \ar[
      from=s,
      to=t,
      phantom,
      "{
        \scalebox{.7}{
          \color{orangeii}
          \bf
          \def\arraystretch{.8}
          \begin{tabular}{c}
            robustly distinct
            \\
            quantum gates
          \end{tabular}
        }
      }"{description, pos=.46}
    ]
    \\[+30pt]
    \\
    p_1
    \ar[
      rr,
      bend left=20,
      shorten=-2pt,
      decorate,
      "{\ }"{swap,name=sbase},
      "p_{12}"{yshift=3pt}
    ]
    \ar[
      rr,
      bend right=45,
      shorten=-2pt,
      decorate,
      "{\ }"{name=tbase},
      "{p'_{12}}"{swap, yshift=-2pt}
    ]
    &&
    p_2
    \ar[
      from=sbase,
      to=tbase,
      phantom,
      "{
        \begin{tikzpicture}
          \draw
          [draw=gray, densely dashed,
          fill=lightgray]
          (0,0)
          node {
            \scalebox{.8}{
              \color{orangeii}
              hole
            }
          }
          circle (.4);
        \end{tikzpicture}
      }"{description, pos=.46}
    ]
  \end{tikzcd}
\end{tabular}

\medskip
The topological quantum computer scientist is thus led to search for topological quantum materials which are
dependent on classical parameter spaces that have a rich structure of ``holes'' in them, namely with a rich {\it fundamental group} (Lit. \ref{LiteratureHomotopyGroups}).
\medskip

\noindent
{\bf The idea of anyon braid grates.}
There could be several possible choices for such topological quantum systems (cf. \cite{ZanardiRasetti99}), but the original proposal by Kitaev (Lit. \ref{LiteratureQuantumComputation}) may be the most promising and has come to often be treated as synonymous with topological quantum computation as such.
Here one imagines that a quantum material's gapped and topologically ordered ground state
(Lit. \ref{TopologicalQuantumMaterials})
depends topologically on tuples
$(z_1, \cdots, z_n)$ of pairwise distinct positions of {\it defect points} (``anyons'', Lit. \ref{AnyonLiterature}) which are effectively
constrained to move inside a surface $\Sigma$ (such as for a crystalline material consisting of a few monolayers of atoms).

For example, much attention has been focused on the idea that such defects might be realized by quantum vortices in the surfaces of quantum fluids, such as certain Bose-Einstein condensates (e.g. \cite{MPSS19}).  The defect parameters could also be more abstract, such as being the critical ``nodal'' values (not of positions but) of {\it momenta} of electrons in topological semimetals (Lit. \ref{LiteratureBraidingInMomentumSpace}). Such nodal momentum values typically vary with fairly easily controllable external parameters such as external strain exerted on the material's crystal structure.

In any case, in such a situation the classical parameter space $P$ is effectively the {\it configuration space of points} (Lit. \ref{LiteratureConfigurationSpaces})
in the surface $\Sigma$
$$
  \ConfigurationSpace{N}
  \big(
   \Sigma
  \big)
  \;=\;
  \Big\{
    (z_1, \cdots, z_N)
    \,\in\,
    \Sigma
    \;\Big\vert\;
    \underset{I \neq J}{\forall}
    \;
    z_I
    \neq z_J
  \Big\}
  \,.
$$
A {\it path} in such a space is an $n$-tuple of ``worldlines'' of defects which may move around each other but never coincide (at any given instant of time), thus forming the appearance of a ``braid'' of $n$ strands in 3d space (Lit. \ref{LiteratureBraiding}).

This implies that even if $\Sigma$ is taken to be topologically trivial (e.g. $\Sigma$ could be the disk through the equator of a Bose-Einstein condensate) there are still plenty of distinct homotopy classes of paths in
$\ConfigurationSpace{N}(\Sigma)$, corresponding to all those braids which cannot be untied. If a quantum material can be found whose degenerate ground states are transported topologically but non-trivially along such braidings of defect points, then this would realize topological quantum computation by {\it anyon braid gates} in the original sense of Freedman and Kitaev (Lit. \ref{LiteratureTopologicalQuantumComputation}).

\hspace{-.9cm}
\begin{tabular}{cc}
\hypertarget{FigureA}{}
\begin{minipage}{8.6cm}
  \small

  {\bf Figure A.}
  If the classical parameter space of a dependent quantum system
  (\href{FigureT}{\it Figure T})
  is a configuration space
  (Lit. \ref{LiteratureConfigurationSpaces})
  of (anyonic) defect points
  (Lit. \ref{AnyonLiterature})
  in a plane,
  then a parameter path
  is a {\it braid} (Lit. \ref{LiteratureBraiding}).

  If the (degenerate) ground state of a topological quantum system
  depends topologically on the defect poisitions (Lit. \ref{TopologicalQuantumMaterials}), then their
  adiabatic transport along such braid paths realizes
  quantum gates that form a linear representation of the braid group.

  In the technologically viable situation of $\mathfrak{su}_2$-anyons,
  this is the {\it monodromy representation} of the canonical
  flat KZ-connection on the
  {\it bundle of conformal block} over configuration (Lit. \ref{KZConnectionsOnConformalBlocksReferences}).

\end{minipage}
&
\hspace{.0cm}
\adjustbox{raise=-50pt}{
\begin{tikzpicture}[scale=.4]

\draw (-1,3.5) node {
  $\mathscr{H}_1$
};
\draw
  [
    ->,
    bend left=20
  ]
  (0,3.5) to
  node[
    yshift=6pt
  ]
  {
    \scalebox{.7}{
      \color{greenii}
      \bf
      anyon braid quantum gate
    }
  }
  (10,3.5);
\draw (11,3.5) node {
  $\mathscr{H}_2$
};

\draw
  (5,1.4)
  node
  {
    \scalebox{.7}{
      \color{greenii}
      \bf
      path in config. space of defect points
    }
  };

\draw (17,3.4) node {
  \scalebox{1.2}{$\mathscr{H}$}
};

\draw (17,-2) node {
  $\ConfigurationSpace{N}(\Sigma)$
};

\draw[->] (17,2.4) to (17,-.9);

\begin{scope}[shift={(0,0)}]
\rbraid
\end{scope}

\begin{scope}[shift={(0,-1.5)}]
 \strand
\end{scope}

\begin{scope}[shift={(0,-3)}]
  \lbraid
\end{scope}

\begin{scope}[shift={(0,-4.5)}]
 \strand
\end{scope}

\begin{scope}[shift={(2,.5)}]
  \strand
\end{scope}

\begin{scope}[shift={(2,-1)}]
  \lbraid
\end{scope}

\begin{scope}[shift={(2,-3)}]
  \lbraid
\end{scope}

\begin{scope}[shift={(2,-4.5)}]
 \strand
\end{scope}

\begin{scope}[shift={(4,0)}]
  \rbraid
\end{scope}

\begin{scope}[shift={(4,-2)}]
  \rbraid
\end{scope}

\begin{scope}[shift={(4,-4)}]
  \rbraid
\end{scope}

\begin{scope}[shift={(6,.5)}]
  \strand
\end{scope}

\begin{scope}[shift={(6,-1)}]
  \lbraid
\end{scope}

\begin{scope}[shift={(6,-3)}]
  \lbraid
\end{scope}

\begin{scope}[shift={(6,-4.5)}]
  \strand
\end{scope}

\begin{scope}[shift={(8,0)}]
  \rbraid
\end{scope}

\begin{scope}[shift={(8,-1.5)}]
  \strand
\end{scope}

\begin{scope}[shift={(8,-2.5)}]
  \strand
\end{scope}

\begin{scope}[shift={(8,-4)}]
  \rbraid
\end{scope}

\begin{scope}[shift={(10,0)}]
  \rbraid
\end{scope}

\begin{scope}[shift={(10,-2)}]
  \rbraid
\end{scope}

\begin{scope}[shift={(10,-4)}]
  \rbraid
\end{scope}

\end{tikzpicture}
}
\end{tabular}

\medskip
\vspace{.1cm}

\noindent
{\bf The idea of certified braid gate data types.}
But none of this intricate internal structure of topological braid quantum gates is visible to existing quantum programming languages; and any traditional implementation of this information (via conformal quantum field theory methods) would be formidable to construct and then inefficient to use.
Yet, detailed verification (Lit. \ref{VerificationLiterature}) of the operation and compilation of these braid gates will arguably be crucial for practical scalable quantum computation (Lit. \ref{NeedForTopologicalProtection}), and will serve for topological quantum simulation already now.

Our claim here is that this problem finds an elegant solution by regarding it through the novel lens of homotopically-typed programming languages (Lit. \ref{LiteratureHomotopyTypeTheory}), where the construction of data types certifying braid quantum gate operation magically turns out to be essentially a native language construct (Thm. \ref{TheTheorem} below).
This we explain now.

\medskip

\newpage

\section{...via parameterized point-set topology}
\label{ViaParameterizedPointSetTopology}

Here we show that the construction of Gauss-Manin connections (Lit. \ref{LiteratureGaussManinConnections}) on bundles of fiberwise cohomology groups (Lit. \ref{LiteratureCohomology})
  has an elementary description in the context of parameterized homotopy theory (Lit. \ref{LiteratureTopologyAndHomotopyTheory}): this is the content of Thm. \ref{GaussManinConnectionInGeneralizedCohomologyViaMappingSpaces} for the untwisted case and Thm. \ref{GaussManinConnectionInTwistedGeneralizedCohomologyViaMappingSpaces} in the broader twisted case.

  \medskip
  This re-formulation is so elementary (in the technical sense) that it becomes effectively a tautology when formulated in the corresponding formal language of
  homotopy type theory (which we turn to in \cref{ViaDependentHomotopyTypeTheory}): The covering space which exhibits the flat Gauss-Manin connection is simply
  the fiberwise 0-truncation of the {\it fiberwise} mapping space (the internal hom-object in the slice) into the corresponding classifying space. This applies
  at once in the generality of twisted generalized and/or non-abelian cohomology theories, such as twisted K-theory and twisted Cohomotopy. Applied to twisted
  complex cohomology groups of configuration spaces of points in the plane, it yields the Knizhnik-Zamolodchikov (KZ) connection on bundles of
  $\suTwoAffine{k}$-conformal blocks (Lit. \ref{KZConnectionsOnConformalBlocksReferences}), and thus the monodromy braid representation characteristic
  of $\suTwo$-anyons (Lit. \ref{LiteratureBraiding}).

\medskip

Here we use (just the most basic aspects of) parameterized ``point-set'' homotopy theory (as laid out in \cite{MaySigurdsson06}, going back to \cite{Booth70}) in order to show that Gauss-Manin connections in (twisted) generalized cohomology groups on fibers of bundles are exhibited by the fiberwise 0th Postnikov stage \eqref{PostnikovTower}
of the
fiberwise mapping space (fiberwise space of sections) into the given classifying space (classifying fibration).

\medskip
With the relevant notions and results from parameterized point-set homotopy theory in hand, the proof is straightforward, so we use the occasion to briefly introduce and review basics of parameterized topology as we go along, in order to make the proof reasonably self-contained also for a general mathematical audience.

\medskip
The construction in itself of the Gauss-Manin connection on fiberwise twisted cohomology groups of locally trivial fiber bundles may be understood without the abstract machinery invoked here; a sketch of such a more low-brow argument is what  \cite[\S 7.5]{EtingofFrenkelKirillov98} offers.
However, it is our abstract re-formulation that provides an elegant handle on Gauss-Manin connections in the language of homotopy type theory, which is discussed in \cref{ViaDependentHomotopyTypeTheory} below.

\medskip
\medskip

Below we make repeated use of the {\it pasting law} (in one direction): In any category with pullbacks,  the pullback along a composite morphism is the pasting of the pullbacks along the two factors (e.g. \cite[Prop. 11.10]{AdamekHerrlichStrecker90}, see also \cite[Prop. 8]{GovzmannPistaloPoncin21}):
\begin{equation}
  \label{PastingLaw}
  \begin{tikzcd}
    (f_2 \circ f_1)^\ast
    X
    \ar[rr]
    \ar[d]
    \ar[
      drr,
      phantom,
      "{\mbox{\tiny\rm(pb)}}"{pos=.4}
    ]
    &&
    X
    \ar[
      d,
      "{p_X}"
    ]
    \\
    B''
    \ar[
      rr,
      "{f_2 \circ f_1}"{swap}
    ]
    &&
    B
  \end{tikzcd}
  \;\;\;\;\;\;
  \simeq
  \;\;\;\;\;\;
  \begin{tikzcd}
    f_2^\ast
    \big(
      f_1^\ast
      X
    \big)
    \ar[r]
    \ar[d]
    \ar[
      dr,
      phantom,
      "\mbox{\tiny\rm(pb)}"{pos=.4}
    ]
    &
    f_1^\ast X
    \ar[r]
    \ar[d]
    \ar[
      dr,
      phantom,
      "\mbox{\tiny\rm(pb)}"{pos=.4}
    ]
    &
    X
    \ar[d]
    \\
    B''
    \ar[r, "{f_2}"{swap}]
    &
    B'
    \ar[r, "{f_1}"{swap}]
    &
    B
  \end{tikzcd}
\end{equation}

\def\AmbientCategory{\kTopologicalSpaces}

\subsection{For ordinary \& generalized cohomology}
\label{ForGeneralizedAndNonAbelianCohomology}

Given a sufficiently nice fibration $p_{\mathrm{X}} \,:\,\mathrm{X} \to \mathrm{B}$, the ordinary complex cohomology groups $H^n(\mathrm{X}_b) \,:\defneq\,  H^n(\mathrm{X}_b; \mathbb{C})$ of its fibers $\mathrm{X}_b$ for $b \in B$, form a bundle of abelian groups over $B$ equipped with a flat connection, known as the {\it Gauss-Manin connection} (Lit. \ref{GaussManinConnectionInTwistedGeneralizedCohomologyViaMappingSpaces}).
This means that a Gauss-Manin connection provides a rule for coherently transporting cohomology classes of spaces $X_b$ as these spaces {\it vary with a parameter} $b \in \mathrm{B}$. Moreover, the flatness of the connection means that the induced transport of cohomology classes depends on parameter paths $\gamma : [0,1] \to \mathrm{B}$ only via  their homotopy classes $[\gamma]$
relative to their endpoints. If $\mathrm{B}$ is connected, this means equivalently that the Gauss-Manin connection is a homomorphism from the fundamental group of $B$ to the automorphism group of $H^n(X_{b_\bullet}; A)$ at any fixed $b_{\bullet}$.
\begin{equation}
  \label{GMConnectionOverview}
  \begin{tikzcd}[sep=small]
    &[+10pt]
    &
    \overset{
      \mathclap{
      \raisebox{2pt}{
        \tiny
        \color{darkblue}
        \bf
        \begin{tabular}{c}
          Fiberwise
          cohomology
        \end{tabular}
     }
      }
    }{
    H^n(\mathrm{X}_{b_2})
    }
    \ar[
      ddr,
      "{\sim}"{sloped},
      shorten=-2pt
    ]
    \\[-27pt]
    \mathllap{
      \raisebox{1pt}{
        \tiny
        \color{darkblue}
        \bf
        \begin{tabular}{c}
          Category
          \\
          of sets
        \end{tabular}
        \hspace{0pt}
      }
    }
    \Sets
    &
    &&&
    \mathrm{X}
    \ar[
      dddd,
      "{
         p_{\mathrm{X}}
     }"{swap},
     "{
       \hspace{-5pt}
       \mbox{
         \tiny
         \color{greenii}
         \bf
         \def\arraystretch{.9}
         \begin{tabular}{c}
           Fiber
           \\
           bundle
         \end{tabular}
       }
     }"
    ]
    \ar[
      white,
      dr,
      "{\color{white}\tau}"{swap, xshift=1pt}
    ]
    \ar[
      rr,
      dashed,
      "{
        \mbox{
          \tiny
          \color{greenii}
          \bf
          Global cocycle
        }
      }"
    ]
    &&
    \overset{
      \mathclap{
      \raisebox{2pt}{
        \tiny
        \color{darkblue}
        \bf
        Classifying space
      }
      }
    }{
      K(n,\mathbb{C})
    }
    \\[-12pt]
    &
    H^n(\mathrm{X}_{b_1})
    \ar[
      uur,
      shorten=-2pt,
      "{
        \sim
      }"{sloped}
    ]
    \ar[
      rr,
      "{\sim}"{swap}
    ]
    &&
    H^n(\mathrm{X}_{b_3})
    &&
    \phantom{B G}
    \ar[
      white,
      dddd,
      "{\color{white}
        p_{B G}
      }"
    ]
    \\
    \\
    &
    &
    \{b_2\}
    \ar[
      ddr,
      "{
        [\gamma_{\,23}]
      }"
    ]
    \\[-20pt]
    \mathllap{
      \raisebox{1pt}{
        \tiny
        \color{darkblue}
        \bf
        \begin{tabular}{c}
          Fundamental
          \\
          path groupoid
        \end{tabular}
        \hspace{0pt}
      }
    }{
    \SingularSimplicialComplex(\mathrm{B})
    }
    \ar[
      uuuu,
      "{
        \mbox{
          \tiny
          \color{greenii}
          \bf
          \def\arrastretch{.9}
          \begin{tabular}{c}
            Gauss-Manin
            \\
            connection
          \end{tabular}
        }
      }"{xshift=2pt}
    ]
    &
    &&&
    \mathrm{B}
    \ar[
      dr,
      shorten=-2pt,
      "{p_{\mathrm{B}}}"{swap}
    ]
    \\[-12pt]
    &
    \{b_1\}
    \ar[
      uur,
      "{
        [\gamma_{\,12}]
      }"
    ]
    \ar[
      rr,
      "{
        [\gamma_{\,23} \, \circ \, \gamma_{\,12}]
      }"{swap}
    ]
    &&
    \{b_3\}
    &
    &
    \ast
  \end{tikzcd}
\end{equation}
In fact, this applies also to {\it twisted} cohomology groups, in which case the Knizhnik-Zamolodchikov connection becomes a special case. We come back to this in a moment (\cref{GaussManinForTwistedGeneralizedCohomology}).

\medskip

Traditionally, Gauss-Manin connections are constructed algebraically.
Here we work entirely homotopy-theoretically and instead make use of the fact that twisted ordinary cohomology of a topological space of CW-type is {\it representable}, in that for $n \in \mathbb{N}$ there exists a topological space $\mathrm{K}(n,\mathbb{C})$ (the $n$th {\it Eilenberg-MacLane space}) such that cohomology is identified with the connected components of the {\it mapping space} into it:
\vspace{-2mm}
\begin{equation}
  \label{OrdinaryCohomologyOfAFiberSpace}
  H^{n}(\TopologicalSpace_b)
  \;\simeq\;
  \pi_0
    \Maps{\big}
      { \mathrm{X}_b }
      { \mathrm{K}(\mathbb{C},n) }
  \,.
\end{equation}

\medskip
\noindent
{\bf Mapping spaces.}
For mapping spaces to work well we may assume without practical restriction that we work in the category
\begin{equation}
  \label{CompactlyGeneratedSpaces}
  \kTopologicalSpaces
  \xhookrightarrow{\;\;}
  \TopologicalSpaces
\end{equation}
of compactly generated space (``k-spaces'', see \cite[Nota. 1.0.15]{SS21EPB} for pointers). Here, for $\mathrm{X}_b \,\in\, \kTopologicalSpaces$, we have an {\it adjunction} of the following form (the cartesian tensor/hom-adjunction, e.g. \cite[\S 7.1, \S 7.2]{BorceuxVol2}\cite[\S VII.2]{Bredon93}):
\vspace{-2mm}
\begin{equation}
  \label{MappingSpaceAdjunction}
  \begin{tikzcd}
    \kTopologicalSpaces
    \ar[
      from=rr,
      shift right=4pt,
      "{
        \mathrm{X} \times (-)
      }"{swap}
    ]
    \ar[
      rr,
      shift right=4pt,
      "{
        \Maps{}{\mathrm{X}}{-}
      }"{swap}
    ]
    \ar[
      rr,
      phantom,
      "{\scalebox{.6}{$\bot$}}"
    ]
    &&
    \kTopologicalSpaces \;,
  \end{tikzcd}
\end{equation}
meaning that there is an {\it exponential law} for k-topological spaces, namely a natural bijection of maps of the following form:
\begin{equation}
  \begin{tikzcd}[sep=0pt]
  \PointsMaps{}
    { \mathrm{X} \times \mathrm{Y} }
    { \mathrm{Z} }
   \ar[rr, "{ \sim }"]
   &&
  \PointsMaps{\big}
    { \mathrm{X} }
    {
      \Maps{}
        { \mathrm{Y} }
        { \mathrm{Z} }
    }
    \\
    \big(
      X \times Y
      \xrightarrow{
        (x,y) \,\mapsto\, f(x,y)
      }
      Z
    \big)
    &\mapsto&
    \Big(
      X
      \xrightarrow{
        x \,\mapsto\,
        \left(
          y \,\mapsto\, f(x,y)
        \right)
      }
      \Maps{\big}{Y}{Z}
   \! \Big).
  \end{tikzcd}
  \end{equation}

Incidentally,  in formal category theory, it is tradition  to denote such adjointness relations by
displaying generic pairs of {\it adjunct maps} separated by a horizontal line:
$$
  \def\arraystretch{1.2}
  \begin{array}{c}
    X \times Y \xrightarrow{\;\;\;\;\; f \;\;\;\;\;} Z
    \\
    \hline
    X \xrightarrow[\;\; \tilde f\;\;]{} \Maps{}{Y}{Z}
    \mathrlap{\,.}
  \end{array}
$$
This in turn alludes to an old tradition in formal logic of denotig {\it natural deduction}-steps this way \cite{Gentzen34} (see \cite{Szabo69}): Here we may read this as saying that ``Given a map $f \,:\, X \times Y \to Z $ we may deduce a map $\tilde f \,:\, X \to \Maps{}{Y}{Z}$, and vice versa.'' This is more than an analogy, it is the first glimpse of the syntaxt-semantics relation between (algebraic) topology and (dependent) type theory, which we invoke in \cite{HoTTIdea}, see around \eqref{FunctionTypesAndMappingSpaces}.

\medskip
\noindent
{\bf Generalized cohomology.}
The following construction of the Gauss-Manin connection over fiber bundles relies only on the existence of such a classifying space, as in \eqref{OrdinaryCohomologyOfAFiberSpace}, but not on its concrete nature. This means that the construction applies also to ``generalized cohomology theories''.

\begin{itemize}
\item If, for instance, $E^n(-)$ is a Whitehead-generalized cohomology theory, such as topological K-theory, elliptic cohomology or cobordism cohomology, then there exists a {\it spectrum} of classifying spaces $\big\{E_n \big\}_{n \in \mathbb{N}}$ such that
\vspace{-1mm}
$$
  E^n(X_b) \;\simeq\;
  \pi_0 \Maps{}
    { \mathrm{X}_b }
    { E_n }
  \,.
$$
\item Regarded the other way around, for {\it any} topological space $\mathrm{A} \,\in\, \kTopologicalSpaces$ we may regard
\vspace{-1mm}
\begin{equation}
  \label{NonAbelianCohomologySet}
  A^0(X_b)
  \;\coloneqq\;
  \pi_0 \Maps{}
   { \mathrm{X}_b }
   { \mathrm{A} }
\end{equation}

\vspace{-1mm}
\noindent
as {\it non-abelian generalized cohomology} with coefficients in $A$.
\item For example, if $\mathrm{A} = B G$ is the classifying space of a discrete or compact Lie group $G$, then
\vspace{-1mm}
$$
  BG^0(X_b) \,\simeq\, H^1(X_b; G)
$$

\vspace{-1mm}
\noindent
is, equivalently, the traditional non-abelian cohomology in degree 1 with coefficients in $G$, which classifies $G$-principal bundles.
\item Or if $\mathrm{A} = S^n \subset \mathbb{R}^{n+1}$ is the topological $n$-sphere, then
\vspace{-1mm}
$$
  (S^n)^0(X_b) \,\simeq\, \pi^n(X_b)
$$

\vspace{-1mm}
\noindent
is unstable {\it Cohomotopy}.
\end{itemize}

\vspace{1mm}
\noindent
{\bf The fiberwise mapping space.}
The key fact now is that in {\it parameterized homotopy theory}
(e.g. \cite{CrabbJames}\cite{Vincent}\cite{MaySigurdsson06})
the mapping space construction
\eqref{MappingSpaceAdjunction}
generalizes to {\it slices} if the base space $\mathrm{B}$ is (compactly generated and) Hausdorff, which we assume from now on:
\vspace{-1mm}
\begin{equation}
  \label{HausdorffBaseSpace}
  B
  \,\in\,
  \kHausdorffSpaces
  \; \xhookrightarrow{\quad} \;
  \kTopologicalSpaces
  \,.
\end{equation}
Here the {\it slice category} $\kTopologicalSpaces_{/\mathrm{B}}$ is the category whose objects $(\mathrm{X},p_{\mathrm{X}})$ are k-topological space $\mathrm{X}$ equipped with a continuous map $p_{\mathrm{X}} : \mathrm{X} \xrightarrow{\;} \mathrm{B}$ and whose morphisms $(\mathrm{X}, p_{\mathrm{X}}) \xrightarrow{\;} (\mathrm{Y}, p_{\mathrm{Y}})$ are compatible maps $\mathrm{X} \xrightarrow{\;} \mathrm{Y}$, hence:
\vspace{-1mm}
\begin{equation}
  \label{SliceHomPullback}
\hspace{3cm}
\begin{tikzcd}
  \mathllap{
  \kTopologicalSpaces_{/\mathrm{B}}
  \big(
    (\mathrm{X},p_{\mathrm{X}})
    ,\,
    (\mathrm{Y},p_{\mathrm{Y}})
  \big)
  \;=\;\,
  }
  \kTopologicalSpaces
  \big(
    \mathrm{X}
    ,\,
    \mathrm{Y}
  \big)
  \underset{
    \kTopologicalSpaces
    (
      \mathrm{X}
      ,\,
      \mathrm{B}
    )
  }{\times}
  \{p_{\mathrm{X}}\}
  \ar[d]
  \ar[rr]
  \ar[drr, phantom, "{\mbox{\tiny\rm(pb)}}"{pos=.4}]
  &&
  \kTopologicalSpaces
  \big(
    \mathrm{X}
    ,\,
    \mathrm{Y}
  \big)
  \ar[
    d,
    "{
      p_{\mathrm{Y}} \circ (-)
    }"
  ]
  \\
  \{
    p_{\mathrm{X}}
  \}
  \ar[rr, hook]
  &&
  \kTopologicalSpaces
  \big(
    \mathrm{X}
    ,\,
    \mathrm{B}
  \big).
\end{tikzcd}
\end{equation}

If we understand $\TopologicalSpace$ as an object in the { slice category} $\kTopologicalSpaces_{/\mathrm{B}}$ over $\mathrm{B}$ via $p_{\TopologicalSpace}$ \eqref{GMConnectionOverview} and if we denote by $p_\mathrm{B}^\ast \mathrm{A}$ the trivial fiber bundle over $\mathrm{B}$ with fiber $\mathrm{A}$ regarded in the slice category, then their {\it fiberwise mapping space} is a topological space which is itself fibered over $\mathrm{B}$,
such that the fibers of the fiberwise mapping space are the ordinary mapping spaces \eqref{MappingSpaceAdjunction} on the fibers:
\vspace{-2mm}
\begin{equation}
  \label{FiberOfFiberwiseMappingSpace}
  \begin{tikzcd}[column sep=50pt]
    \overset{
      \mathclap{
      \raisebox{2pt}{
        \tiny
        \color{darkblue}
        \bf
        Ordinary mapping space
        on fiber
      }
      }
    }{
    \Maps{\big}
      { \mathrm{X}_b }
      { \mathrm{A} }
    }
    \ar[r]
    \ar[d]
    \ar[
      dr,
      phantom,
      "{
        \mbox{\tiny\rm(pb)}
      }"{pos=.3}
    ]
    &
    \overset{
      \mathclap{
      \raisebox{2pt}{
        \tiny
        \begin{tabular}{c}
          {\color{darkblue}
          \bf
          Fiberwise mapping space}
          \color{black}
          (itself a topological space over $\mathrm{B}$)
        \end{tabular}
      }
      }
    }{
    \Maps{\big}
      { (\mathrm{X}, p_{\mathrm{X}}) }
      { p_{\mathrm{B}}^\ast \mathrm{A} }
    }.
    \ar[d]
    \\
    \ast
    \ar[r, hook, "{b}"]
    &
    \underset{
      \mathclap{
      \raisebox{-2pt}{
        \tiny
        \color{darkblue}
        \bf
        Parameter space
      }
      }
    }{
    \mathrm{B}
       }
  \end{tikzcd}
\end{equation}

\vspace{-2mm}
\noindent The right choice of topology on the total fiberwise mapping space is subtle\footnote{
  ``{\it The point-set topology of parametrized spaces is surprisingly subtle. Parametrized mapping spaces are especially delicate.}'' \cite[p. 15]{MaySigurdsson06}
} and the result that such a topology exists
(see \cite[\S 1.3.7-\S1.3.9]{MaySigurdsson06}, following \cite[Thm. 3.5]{BoothBrown78})
may be regarded as the engine which powers our slick re-construction of Gauss-Manin connections from the point of view of point-set topology.
Here the right topology is that which ensures the sliced analog of the adjunction \eqref{MappingSpaceAdjunction}
\begin{equation}
  \label{FiberwiseMappingSpaceAdjunction}
  \begin{tikzcd}
    \kTopologicalSpaces_{/\mathrm{B}}
    \ar[
      from=rrr,
      shift right=5pt,
      "{
        (\mathrm{X}, p_{\mathrm{X}}) \times (-)
      }"{swap}
    ]
    \ar[
      rrr,
      shift right=5pt,
      "{
      \Maps{}
        { (\mathrm{X}, p_{\mathrm{X}}) }
        { - }
      }"{swap}
    ]
    \ar[
      rrr,
      phantom,
      "{\scalebox{.7}{$\bot$}}"
    ]
    &&&
    \kTopologicalSpaces_{/\mathrm{B}}
    \,,
  \end{tikzcd}
\end{equation}
hence, equivalently,
the {\it exponential law} in the slice
\cite[Thm. 3.5]{BoothBrown78}\cite[(1.3.9)]{MaySigurdsson06}:
a natural bijection of the form
\begin{equation}
  \label{ExponentialLawForFiberwiseMappingSpace}
  \kTopologicalSpaces_{/B}
  \Big(
    \overset{
      \mathclap{
      \raisebox{2pt}{
        \tiny
        \color{darkblue}
        \bf
        Fiberwise product space
      }
      }
    }{
    (\mathrm{X},p_{\mathrm{X}})
    \times
    (\mathrm{Y},p_{\mathrm{Y}})
    }
    ,
      { (\mathrm{Z}, p_{\mathrm{Z}}) }
  \Big)
  \;\;
  \simeq
  \;\;
  \kTopologicalSpaces_{/B}
  \Big(
    (\mathrm{X},p_{\mathrm{X}})
    ,
    \overset{
      \mathclap{
      \raisebox{2pt}{
        \tiny
        \color{darkblue}
        \bf
        Fiberwise mapping space
      }
      }
    }{
    \Maps{\big}
      { (\mathrm{Y},p_{\mathrm{Y}}) }
      { (\mathrm{Z}, p_{\mathrm{Z}}) }
    }
  \Big)
  \,,
\end{equation}
where now the product on the right is that in the slice, hence is the {\it fiber product} of k-topological spaces:
$$
  (\mathrm{X},p_{\mathrm{X}})
  \times
  (\mathrm{Y},p_{\mathrm{Y}})
  \;\simeq\;
  \big(
    \mathrm{X} \times_{\mathrm{B}} \mathrm{Y}
    ,
    p_{\mathrm{X}}\circ \mathrm{pr}_{\mathrm{X}}
    =
    p_{\mathrm{Y}}\circ \mathrm{pr}_{\mathrm{Y}}
  \big)
  \,.
$$
Of course, the unit for this product is the identity map on the base space:
\begin{equation}
  \label{FiberProductUnit}
  (\mathrm{X}, p_{\mathrm{X}})
  \times
  (\mathrm{B}, \mathrm{id}_{\mathrm{B}})
  \;\simeq\;
  (\mathrm{X}, p_{\mathrm{X}})
  \,.
\end{equation}

This exponential law in slices implies a wealth of useful structure:

\begin{proposition}[\bf Base change adjoint triple]
For any map between base spaces $f \,:\, \mathrm{B} \xrightarrow{\;}\mathrm{B}'$ \eqref{HausdorffBaseSpace}, there is a ``{\it base change}'' adjoint triple
\vspace{-2mm}
\begin{equation}
  \label{BaseChangeAdjunction}
  \begin{tikzcd}
    \kTopologicalSpaces_{/\mathrm{B}}
    \ar[
      rr,
      shift left=13pt,
      "{
        f_!
      }"
    ]
    \ar[
      from=rr,
      "{
        f^\ast
      }"{description}
    ]
    \ar[
      rr,
      shift right=12pt,
      "{
        f_\ast
      }"{swap}
    ]
    \ar[
      rr,
      phantom,
      shift left=8pt,
      "{
        \scalebox{.6}{$\bot$}
      }"
    ]
    \ar[
      rr,
      phantom,
      shift right=7pt,
      "{
        \scalebox{.6}{$\bot$}
      }"
    ]
    &&
    \kTopologicalSpaces_{/\mathrm{B}'}
    \,,
  \end{tikzcd}
\end{equation}
where $f^\ast$ denotes the pullback operation (formed in
\noindent$\kTopologicalSpaces$),
its the left adjoint $f_!$ is given by postcomposition with $f$,
and the right adjoint $f_\ast$ is given by the following pullback construction:
\vspace{-2mm}
\begin{equation}
  \label{RightBaseChangeAsAPullback}
  \begin{tikzcd}[row sep=small]
    f_\ast (\mathrm{X},p_{\mathrm{X}})
    \ar[rr]
    \ar[d]
    \ar[
      drr,
      phantom,
      "{\mbox{\tiny\rm(pb)}}"
    ]
    &&
    \Maps{\big}
      { (\mathrm{B}, f) }
      { (\mathrm{X}, f \circ p_{\mathrm{X}}) }
    \ar[d]
    \\
    (\mathrm{B}',\mathrm{id}_{\mathrm{B}'})
    \ar[
      rr,
      "{
        \widetilde{\mathrm{id}}
      }"{swap}
    ]
    &&
    \Maps{\big}
      { (\mathrm{B},f) }
      { (\mathrm{B},f) } \,.
  \end{tikzcd}
\end{equation}
\end{proposition}
\begin{proof}
  For the left adjoint $(f_! \dashv f^\ast)$ the required hom-isomorphism is immediate from the universal property of the pullback:
 \vspace{-2mm}
  $$
    \begin{tikzcd}[column sep=large, row sep=small]
      \mathrm{X}
      \ar[
        ddr,
        bend right=10
      ]
      \ar[
        rrd,
        shift left=2pt,
        bend left=10
      ]
      \ar[dr, dashed, shorten >=-5]
      &[-10pt]
      \\[-12pt]
      &
      f^\ast \mathrm{Y}
      \ar[r]
      \ar[d]
      \ar[
        dr,
        phantom,
        "{\mbox{\tiny\rm(pb)}}"
      ]
      &
      \mathrm{Y}
      \ar[d, "{p_{\mathrm{y}}}"]
      \\
      &
      \mathrm{B}
      \ar[r, "{f}"{swap}]
      &
      \mathrm{B}' \;.
    \end{tikzcd}
  $$

  \vspace{-1mm}
  \noindent
  For the right adjoint, the required hom-isomorphism is obtained as the following sequence of natural isomorphisms:
  \def\AmbientCategory{\kTopologicalSpaces_{/\mathrm{B}'}}
  \vspace{-2mm}
$$
\hspace{-3mm}
  \def\arraystretch{1.8}
  \begin{array}{llll}
    \PointsMaps{\big}
      { (U,p_{\mathrm{U}}) }
      { f_\ast (\mathrm{X}, p_{\mathrm{X}}) }
  &   \!\!\!\!\!\!\!\! \simeq
    \PointsMaps{\bigg}
      {\!\! (U,p_{\mathrm{U}}) }
      {
        \Maps{\Big}
          { (\mathrm{B}, f) }
          { (\mathrm{X}, f \circ p_{\mathrm{X}}) }
        \underset{
          \Maps{\big}
            { (\mathrm{B}, f) }
            { (\mathrm{B},f) }
        }{\times}
         \!\! \{\widetilde {\mathrm{id}}\}
      \!\!}
    &&
  \!\!\!\!\!\!\!\!  \proofstep{by \eqref{RightBaseChangeAsAPullback}}
    \\
&  \!\!\!\!\!\!\!\!  \simeq
    \PointsMaps{\Big}
      { (U,p_{\mathrm{U}}) }
      {
        \Maps{\big}
          { (\mathrm{B}, f) }
          { (\mathrm{X}, f \circ p_{\mathrm{X}}) }
      }
    \underset{
      \Maps{\big}
        { (U,p_{\mathrm{U}}) }
        {
            \Maps{\big}
              { (\mathrm{B}, f) }
              { (\mathrm{B},f) }
        }
    }
    {\times}
    \!\! \{\widetilde {\mathrm{id}}\}
     &&
    \!\!\!\!\!\! \proofstep{by \eqref{SliceHomPullback}}
    \\
&     \!\!\!\!\!\!\!\!  \simeq
    \PointsMaps{\Big}
      { (U,p_{\mathrm{U}}) \times (\mathrm{B}, f) }
      { (\mathrm{X}, f \circ p_{\mathrm{X}}) }
    \underset{
      \PointsMaps{\Big}
        {
          (U,p_{\mathrm{U}})
          \times
          (\mathrm{B}, f)
        }
        { (\mathrm{B},f) }
    }
    {\times}
   \!\!\! \{\widetilde {\mathrm{id}}\}
    &&
    \!\!\!\!\!\!\! \proofstep{by \eqref{ExponentialLawForFiberwiseMappingSpace}}
    \\
 &   \!\!\!\!\!\!\!\!  \simeq
    \def\AmbientCategory{\kTopologicalSpaces}
    \bigg(
      \PointsMaps{}
        { f^\ast \mathrm{U} }
        { \mathrm{X} }
        \underset{
          \PointsMaps{}
            { f^\ast \mathrm{U} }
            { \mathrm{B}' }
        }{\times}
        \ast
    \bigg)
    \underset{
      \Big(
        \PointsMaps{}
          { f^\ast \mathrm{U} }
          { \mathrm{B} }
          \underset{
            \PointsMaps{}
             { f^\ast \mathrm{U} }
             { \mathrm{B}' }
          }{\times}
         \, \ast
      \Big)
    }
    {\times}
     \{\widetilde {\mathrm{id}}\}
     &&
    \!\!\!\!\!\! \proofstep{by \eqref{SliceHomPullback}}
    \\
 &    \!\!\!\!\!\!\!\!  \simeq
    \def\AmbientCategory{\kTopologicalSpaces}
    \Maps{\big}
      { f^\ast \mathrm{U} }
      { \mathrm{X} }
    \underset{
      \Maps{}
        { f^\ast\mathrm{U} }
        { \mathrm{B} }
    }{\times}
    \ast
    &&
  \!\!\!\!\!\!  \proofstep{by \eqref{DiagramOfPullbackDiagrams}}
    \\
    \def\AmbienCategory{\kTopologicalSpaces_{/\mathrm{B}}}
&  \!\!\!\!\!\!  \simeq\;
    \PointsMaps{\big}
      { f^\ast (\mathrm{U}, p_{\mathrm{U}}) }
      { (\mathrm{X}, p_{\mathrm{X}}) }
     &&
   \!\!\!\!\!  \proofstep{by \eqref{SliceHomPullback}}
     \,.
  \end{array}
$$
Here the penultimate step is observing that the fiber products (limits) may be interchanged: Instead of computing the horizontal fiber product of the vertical fiber product in the following diagram, we may first compute the horizontal fiber products (shown on the right, again by \eqref{SliceHomPullback}):
\begin{equation}
  \label{DiagramOfPullbackDiagrams}
  \begin{tikzcd}[column sep=large]
    \PointsMaps{\big}
      { f^\ast \mathrm{U} }
      { \mathrm{X} }
    \ar[
      r,
      "{
        p_{\mathrm{X}} \circ (-)
      }"
    ]
    \ar[
      d,
      "{
        f \circ p_{\mathrm{X}} \circ (-)
      }"{swap}
    ]
    &
    \PointsMaps{\big}
      { f^\ast \mathrm{U} }
      { \mathrm{B} }
    \ar[from=r]
    \ar[
      d,
      "{
        f \circ (-)
      }"
    ]
    &
    \ast
    \ar[d]
    &[+6pt]
    \kTopologicalSpaces_{/\mathrm{B}}
    \big(
      f\ast (\mathrm{U}, p_{\mathrm{U}})
      ,\,
      (\mathrm{X}, p_{\mathrm{X}})
    \big)
    \ar[d]
    \\
    \PointsMaps{\big}
      { f^\ast \mathrm{U} }
      { \mathrm{B}' }
    \ar[
      r,
      "{
        \mathrm{id}
      }"
    ]
    &
    \PointsMaps{\big}
      { f^\ast \mathrm{U} }
      { \mathrm{B}' }
    \ar[from=r]
    &
    \ast
    &
    \ast
    \\
    \ast
    \ar[u]
    \ar[r]
    &
    \ast
    \ar[u]
    \ar[from=r]
    &
    \ast
    \ar[u]
    &
    \ast
    \ar[u]
    \\[+6pt]
  \end{tikzcd}
\end{equation}
\vspace{-1cm}

\noindent
Therefore the evident remaining vertical fiber product is as claimed.
\end{proof}

\begin{example}[\bf Space of sections]
  \label{SpaceOfSectionsAsRightBaseChange}
  When $\mathrm{B}' \,=\, \ast$ in \eqref{BaseChangeAdjunction}, the right base change
  \eqref{RightBaseChangeAsAPullback}
  constructs {\it spaces of sections}:
  \begin{equation}
    \label{SpaceOfSections}
    \Gamma_{\mathrm{X}_b}
    (
      -
    )
    \;
    \simeq
    \;
    (
      p_{\mathrm{X}_b}
    )_\ast
    \;:\;\;
    \kTopologicalSpaces_{/\mathrm{X}_b}
    \xrightarrow{\qquad}\,
    \kTopologicalSpaces
    \mathrlap{\,.}
  \end{equation}
  Generally, one may understand the right base change as forming {\it fiberwise spaces of sections}.
\end{example}

\begin{proposition}[\bf Cartesian Frobenius reciprocity]
  For $f \,:\, \mathrm{B} \xrightarrow{\;} \mathrm{B}'$
   a map of base spaces \eqref{HausdorffBaseSpace},
  we have a natural isomorphism
  \begin{equation}
    \label{CartesianFrobeniusReciprocityIsomorphism}
    f_!
    \big(
      (\mathrm{X}, p_{\mathrm{X}})
      \times
      f^\ast
      (\mathrm{Y},p_{\mathrm{Y}})
    \big)
    \;\simeq\;
    \big(
      f_!
      (\mathrm{X}, p_{\mathrm{X}})
    \big)
    \times
    (\mathrm{Y}, p_{\mathrm{Y}})
    \,,
  \end{equation}
  where $(f_! \dashv f^\ast)$ is the left base change adjunction \eqref{BaseChangeAdjunction}.
\end{proposition}
\begin{proof}
  This follows by the pasting law \eqref{PastingLaw}, which here says that the following pullback squares in $\kTopologicalSpaces$ agree:
  $$
    \begin{tikzcd}
      \mathrm{X} \times_{\mathrm{B}} f^\ast \mathrm{Y}
      \ar[d]
      \ar[r]
      \ar[
        dr,
        phantom,
        "{
          \mbox{\tiny\rm(pb)}
        }"{pos=.4}
      ]
      &
      f^\ast \mathrm{Y}
      \ar[d]
      \ar[r]
      \ar[
        dr,
        phantom,
        "{
          \mbox{\tiny\rm(pb)}
        }"{pos=.45}
      ]
      &
      \mathrm{Y}
      \ar[d]
      \ar[
        d,
        "{p_{\mathrm{Y}}}"
      ]
      \\
      \mathrm{X}
      \ar[
        r,
        "{p_{\mathrm{X}}}"{swap}
      ]
      &
      \mathrm{B}
      \ar[
        r,
        "{f}"{swap}
      ]
      &
      \mathrm{B}'
    \end{tikzcd}
    \quad
    \simeq
    \quad
    \begin{tikzcd}
      \mathrm{X}
      \times_{\mathrm{B}'}
      \mathrm{Y}
      \ar[rr]
      \ar[d]
      \ar[
        drr,
        phantom,
        "{\mbox{\tiny\rm(pb)}}"{pos=.4}
      ]
      &&
      \mathrm{Y}
      \ar[d]
      \\
      \mathrm{X}
      \ar[
        rr,
        "{ f \,\circ\, p_{\mathrm{X}} }"{swap}
      ]
      &&
      \mathrm{B}'.
    \end{tikzcd}
  $$

  \vspace{-7mm}
\end{proof}

In generalization of \eqref{FiberOfFiberwiseMappingSpace}, we have:
\begin{proposition}[\bf Base change is closed functor]
The pullback (base change) of a fiberwise mapping space along any continuous map of base spaces \eqref{HausdorffBaseSpace}
$
  f \,:\, B' \xrightarrow{\phantom{-}} B
$
is the fiberwise mapping space of the pullbacks of the arguments:
\begin{equation}
  \label{PullbackIsClosedFunctorOnFiberwiseMappingSpaces}
  f^\ast
  \Maps{\big}
    { (\mathrm{X},p_{\mathrm{X}}) }
    { (\mathrm{Y}, p_{\mathrm{Y}}) }
  \;\;
  \simeq
  \;\;
  \Maps{\big}
    { f^\ast (\mathrm{X},p_{\mathrm{X}}) }
    { f^\ast (\mathrm{Y}, p_{\mathrm{Y}}) }\,.
\end{equation}
\end{proposition}
\begin{proof}
Apply the Yoneda Lemma over $\kTopologicalSpaces_{/B}^{\mathrm{op}}$ to the following sequence of natural isomorphisms:
\vspace{-3mm}
$$
\hspace{-3mm}
  \def\arraystretch{1.8}
  \begin{array}{lll}
    \kTopologicalSpaces_{/\mathrm{B}'}
    \Big(
      (\mathrm{U},p_{\mathrm{U}})
      ,\,
      f^\ast
      \Maps{\big}
        { (\mathrm{X}, p_{\mathrm{X}}) }
        { (\mathrm{Y}, p_{\mathrm{Y}}) }
    \Big)
    &    \simeq\;
    \kTopologicalSpaces_{/\mathrm{B}}
    \Big(
      f_!(\mathrm{U},p_{\mathrm{U}})
      ,\,
      \Maps{\big}
        { (\mathrm{X}, p_{\mathrm{X}}) }
        { (\mathrm{Y}, p_{\mathrm{Y}}) }
    \Big)
    &
    \proofstep{by \eqref{BaseChangeAdjunction}}
    \\
  &  \simeq\;
    \kTopologicalSpaces_{/\mathrm{B}}
    \Big(
      \big((f_!(\mathrm{U},p_{\mathrm{U}})\big)
      \times
      { (\mathrm{X}, p_{\mathrm{X}}) }
      ,\,
      { (\mathrm{Y}, p_{\mathrm{Y}}) }
    \Big)
    &
    \proofstep{by \eqref{ExponentialLawForFiberwiseMappingSpace}}
    \\
  &  \simeq\;
    \kTopologicalSpaces_{/\mathrm{B}}
    \Big(
      f_!
      \big(
      (\mathrm{U},p_{\mathrm{U}})
      \times
      f^\ast
      (\mathrm{X}, p_{\mathrm{X}})
      \big)
      ,\,
      { (\mathrm{Y}, p_{\mathrm{Y}}) }
    \Big)
    &
    \proofstep{by \eqref{CartesianFrobeniusReciprocityIsomorphism}
    }
    \\
 &  \simeq\;
    \kTopologicalSpaces_{/\mathrm{B}'}
    \Big(
      (\mathrm{U},p_{\mathrm{U}})
      \times
      f^\ast
      (\mathrm{X}, p_{\mathrm{X}})
      ,\,
      { f^\ast (\mathrm{Y}, p_{\mathrm{Y}}) }
    \Big)
    &
    \proofstep{by \eqref{BaseChangeAdjunction}}
    \\
  &  \;\simeq\;
    \kTopologicalSpaces_{/\mathrm{B}'}
    \Big(
      (\mathrm{U},p_{\mathrm{U}})
      ,\,
      \Maps{\big}
        {
          f^\ast (\mathrm{X}, p_{\mathrm{X}})
        }
        { f^\ast (\mathrm{Y}, p_{\mathrm{Y}}) }
    \Big)
    &
    \proofstep{by \eqref{ExponentialLawForFiberwiseMappingSpace} }.
\end{array}
$$

\vspace{-6mm}
\end{proof}

For the case at hand, this has the following consequence:
\begin{proposition}[\bf Fiberwise mapping space out of fiber bundle is fiber bundle]
\label{FiberwiseMappingSpaceOutOfFiberBundleIsFiberBundle}
Let $(\mathrm{X},p_{\mathrm{X}}) \,\in\, \kTopologicalSpaces_{/\mathrm{B}}$ be a fiber bundle with local trivialization
\begin{equation}
  \label{AssumingXIsFiberBundle}
 \phi \,:\, \mathrm{U} \underset{\mathrm{opn}}{\relbar\joinrel\twoheadrightarrow}
 \mathrm{B}
 \,,
 \qquad
 \mbox{such that}
 \qquad
 \phi^\ast (\mathrm{X},p_{\mathrm{X}}) \,\simeq\, p_{\mathrm{U}}^\ast \mathrm{X}_0
 \;:\defneq\;
 \mathrm{U} \times \mathrm{X}_0
 \,.
\end{equation}
 Then the fiberwise mapping space is a fiber bundle that is locally trivial with respect to the same open cover:
 \vspace{-1mm}
\begin{equation}
  \label{FiberwiseMappingSpaceOfFiberBundleIsFiberBundle}
  \phi^\ast
  \Maps{\big}
    { (\mathrm{X}, p_{\mathrm{X}}) }
    { p_{\mathrm{B}}^\ast A }
      \simeq\;
  p_{\mathrm{U}}^\ast
  \Maps{\big}
    { \mathrm{X}_0 }
    { \mathrm{A} }.
\end{equation}
\end{proposition}

\begin{proof}
$\,$

\vspace{-6mm}
    \begin{equation*}
  \label{FiberwiseMappingSpaceOfFiberBundleIsFiberBundle}
  \def\arraystretch{1.6}
  \begin{array}{lll}
  \phi^\ast
  \Maps{\big}
    { (\mathrm{X}, p_{\mathrm{X}}) }
    { p_{\mathrm{B}}^\ast A }
     & \simeq\;
  \Maps{\big}
    { \phi^\ast(\mathrm{X}, p_{\mathrm{X}}) }
    { \phi^\ast p_{\mathrm{B}}^\ast \mathrm{A} }
  &
  \proofstep{\rm by \eqref{PullbackIsClosedFunctorOnFiberwiseMappingSpaces}}
  \\
  & \simeq\;
  \Maps{\big}
    { p_{\mathrm{U}}^\ast \mathrm{X}_0 }
    { p_{\mathrm{U}}^\ast \mathrm{A} }
  &
  \proofstep{\rm by \eqref{AssumingXIsFiberBundle}}
  \\
  & \simeq\;
  p_{\mathrm{U}}^\ast
  \Maps{\big}
    { \mathrm{X}_0 }
    { \mathrm{A} }
  &
  \proofstep{\rm by \eqref{PullbackIsClosedFunctorOnFiberwiseMappingSpaces}}.
  \end{array}
\end{equation*}

\vspace{-5mm}
\end{proof}

\medskip

\noindent
{\bf Fiberwise homotopy theory.}
Moreover, the topology on the fiberwise mapping space \eqref{FiberOfFiberwiseMappingSpace} is also ``homotopy correct'' in that its map to the base $\mathrm{B}$ is an h-fibration as soon as $p_{\mathrm{X}}$ is an h-fibration (by \cite[\S 6.1]{Booth70}, see also \cite[Prop. 1.3.11]{MaySigurdsson06}), which is the case for $\mathrm{B}$ a metrizable space and $p_{\mathrm{X}}$ a fiber bundle.
This implies that the fibers of the fiberwise mapping space are in fact homotopy fibers, and that
forming path $\infty$-groupoids (singular simplicial complexes, cf. \cite[Pro. 3.3.43]{SS21EPB}),
\vspace{-2mm}
$$
  \begin{tikzcd}
    \kTopologicalSpaces
    \ar[
      from=rr,
      shift right=4pt,
      "{
        \vert-\vert
      }"{swap}
    ]
    \ar[
      rr,
      shift right=4pt,
      "{
        \SingularSimplicialComplex
      }"{swap}
    ]
    \ar[
      rr,
      phantom,
      "{\scalebox{.6}{$\bot$}}"
    ]
    &&
    \SimplicialSets
  \end{tikzcd}
$$

\vspace{-2mm}
\noindent respects this property:
\begin{equation}
  \label{sSetVersioOfFiberwiseMappingSpaceFibration}
  \begin{tikzcd}
    \SingularSimplicialComplex
    \,
    \Maps{\big}
      { \mathrm{X}_b }
      { \mathrm{A} }
    \ar[rr]
    \ar[d]
    \ar[
      drr,
      phantom,
      "{
        \mbox{\tiny\rm(pb)}
      }"
    ]
    &&
    \SingularSimplicialComplex
    \,
    \Maps{\big}
      { (\mathrm{X}, p_{\mathrm{X}}) }
      { p_{\mathrm{B}}^\ast \mathrm{A}  }.
    \ar[
      d,
      "{
        \in \KanFibrations
      }"
    ]
    \\
    \ast
    \ar[rr, "{b}"{swap}]
    &&
    \SingularSimplicialComplex(\mathrm{B})
  \end{tikzcd}
\end{equation}

\begin{lemma}[\bf Fiberwise truncation is preserved by base change]
  \label{FiberwiseTruncationPreservedByBaseChange}
  For any map of simplicial sets
  $f \,\colon\,B' \xrightarrow{\;} B$,
  the operation of base change
  (pullack)
  $f^\ast$ of Kan fibrations
  preserves fiberwise Postnikov truncation \eqref{PostnikovTower}:
  $$
    f^\ast \circ \pi_{0/B}
    \;\simeq\;
    \pi_{0/B'} \circ f^\ast
    \,.
  $$
\end{lemma}
\begin{proof}
  It is useful to understand this as a special case of a general phenomenon of {\it $0$-truncation} in slice homotopy theories \cite[\S 5.5.6]{Lurie09}.
  Every morphism $p : X \xrightarrow{\;} S$ factors uniquely through $\pi_{0/S}(E)$ as a ``$0$-connected'' map followed by a ``$0$-truncated map'',
  and both these classes are preserved by homotopy pullback (\cite[Ex. 5.2.8.16 \& Lem. 6.5.1.16(6)]{Lurie09}). Therefore,
  the claim follows by the pasting law \eqref{PastingLaw}, which also holds for homotopy pullbacks (\cite[Lem. 4.4.2.1]{Lurie09}\cite[Prop. 8]{GovzmannPistaloPoncin21}):
  \vspace{-1mm}
  \begin{equation}
    \label{ConnectedTruncatedPullback}
    \begin{tikzcd}
      f^\ast X
      \ar[
        d,
        "{
          \mbox{
            \tiny
            0-cnctd
          }
        }"{description, left}
      ]
      \ar[
        dr,
        phantom,
        "{\mbox{\tiny\rm(hpb)}}"{pos=.55}
      ]
      \ar[r]
      &
      X
      \ar[
        d,
        "{
          \mbox{
            \tiny
            0-cnctd
          }
        }"{description, right}
      ]
      \\
      \pi_{0/B'}
      \big(
        f^\ast X
      \big)
      \ar[r]
      \ar[
        d,
        "{
          \mbox{
            \tiny
            0-cnctd
          }
        }"{description, left}
      ]
      \ar[
        dr,
        phantom,
        "{\mbox{\tiny\rm(hpb)}}"{pos=.35}
      ]
      &
      \pi_{0/B}(X)
      \ar[
        d,
        "{
          \mbox{
            \tiny
            0-trnctd
          }
        }"{description, right}
      ]
      \\
      B'
      \ar[r, "{f}"]
      &
      B
      \mathrlap{\,.}
    \end{tikzcd}
  \end{equation}
  \vspace{-.7cm}

\end{proof}

Therefore, applying Lemma \ref{FiberwiseTruncationPreservedByBaseChange} to the diagram
\eqref{sSetVersioOfFiberwiseMappingSpaceFibration}, we obtain a homotopy pullback diagram as shown on the left here:
\vspace{-2mm}
\begin{equation}
 \label{TheGMFibrationForUntwistedCase}
\hspace{1.5cm}
  \begin{tikzcd}[column sep=25pt]
    \mathllap{
      \overset{
        \mathclap{
        \raisebox{2pt}{
          \tiny
          \color{darkblue}
          \bf
          \def\arraystretch{.9}
          \begin{tabular}{c}
            $A$-cohomology
            \\
            of fiber
          \end{tabular}
        }
        }
      }{
        A^0(\mathrm{X}_b)
      }
      \;\simeq\;\,
    }
    \pi_0
    \big(
    \SingularSimplicialComplex
    \,
    \Maps{}
       { \mathrm{X}_b }
       { \mathrm{A} }
    \!\big)
    \ar[rr]
    \ar[d]
    \ar[drr, phantom, "{\mbox{\tiny\rm(hpb)}}"{pos=.4}]
    &&
    \overset{
      \mathclap{
      \raisebox{2pt}{
        \tiny
        \def\arraystretch{.9}
        \begin{tabular}{c}
          {\color{darkblue}
          \bf
          Fiberwise 0-truncation of}
          {(path $\infty$-groipoid of)}
          {\color{darkblue}
          \bf
          fiberwise mapping space}
        \end{tabular}
      }
      }
    }{
    \pi_{0/\SingularSimplicialComplex(\mathrm{B})}
    \Big(
    \SingularSimplicialComplex
    \,
    \Maps{\big}
      { (\mathrm{X}, p_{\mathrm{X}}) }
      { p_{\mathrm{B}}^\ast \mathrm{A} }
    \!\Big)
    }
    \ar[rr]
    \ar[
      d
    ]
    \ar[drr, phantom, "{\mbox{\tiny\rm(hpb)}}"{pos=.4}]
    &&
    \Sets^{\ast/}
    \ar[
      d,
      "{
        \mbox{
          \tiny
          \color{greenii}
          \bf
          \begin{tabular}{c}
            Covering space
            \\
            classifier
          \end{tabular}
        }
      }"{xshift=-4pt}
    ]
    \\
    \{b\}
    \ar[rr]
    &&
    \underset{
      \mathclap{
      \raisebox{-2pt}{
        \tiny
        \color{darkblue}
        \bf
        \def\arraystretch{.9}
        \begin{tabular}{c}
          Fundamental
                    groupoid
                    \\of
                    base space
        \end{tabular}
      }
      }
    }{
      \SingularSimplicialComplex(\mathrm{B})
    }
    \ar[
      rr,
      "{
        \nabla^{\mathrm{GM}}_{X, A}
      }"{description}
      ,
      "{
        \mbox{
          \tiny
          \color{greenii}
          \bf
          \begin{tabular}{c}
            Gauss-Manin connection
          \end{tabular}
        }
      }"{swap, yshift=-3pt}
    ]
    &&
    \Sets
    \,.
  \end{tikzcd}
\end{equation}

\vspace{-2mm}
\noindent The left square shows that the fiberwise 0-truncation of the fiberwise mapping space is a fibration over the fundamental groupoid of $\mathrm{B}$, whose (homotopy) fibers are the generalized cohomology sets \eqref{NonAbelianCohomologySet} of the fiber space $\mathrm{X}_b$. The homotopy pullback shown on the right follows by:

\begin{lemma}[\bf Univalent universe of sets]
\label{UnivalentUniversesOfInfinityGroupoids}
Any  homotopy fibration of sets, as in the middle of  \eqref{TheGMFibrationForUntwistedCase},
is classified by -- i.e., is the homotopy pullback along -- an essentially unique map $\nabla^{\mathrm{GM}}_{X,A}$ to the covering space classifier, as shown in the square on the right of
\eqref{TheGMFibrationForUntwistedCase}.
\end{lemma}
\begin{proof}
This may be understood as a simple special case of the general fact that $\infty$-groupoids form an $\infty$-topos in which there exists a ``small fibration classifier'' $\InfinityGroupoids^{\ast/} \xrightarrow{\quad} \InfinityGroupoids$ (\cite[Prop. 3.3.2.7]{Lurie09}\cite[\S 5.2]{Cisinski19}\cite{KapulkinLumsdained21}). \end{proof}

\begin{remark}[\bf Flat connections as functors on the fundamental groupoid]
\label{FlatConnection}
Noticing that $\SingularSimplicialComplex(\mathrm{B})$ is equivalently the disjoint union over connected components $[b] \in \pi_0(\mathrm{B})$ of delooping groupoids $\mathbf{B} \pi_1(\mathrm{B}, b)$, this map $\nabla^{\mathrm{GM}}_{X,A}$ \eqref{TheGMFibrationForUntwistedCase}
is over each connected component equivalently a group homomorphism
\vspace{-1mm}
$$
  \Omega_b
  \big(\nabla^{\mathrm{GM}_{X,A}}\big)
  \,:\,
  \pi_1(\mathrm{B},b)
  \xrightarrow{\quad}
  \mathrm{Aut}
  \big(
    A^0(\mathrm{X}_b)
  \big)
  \,.
$$

\vspace{-1mm}
\noindent This is a traditional incarnation of flat connections on a space $\mathrm{B}$ (e.g. \cite[\S I.1]{Deligne70}\cite[Prop. 2.5.1]{Dimca04}).
\end{remark}

Moreover, from Prop. \ref{FiberwiseMappingSpaceOutOfFiberBundleIsFiberBundle} it follows that this local system of sets trivializes over any cover over which $p_{\mathrm{X}}$ trivializes, so that it corresponds to a {\it covering space} which we denote as follows:
\vspace{-2mm}
\begin{equation}
  \label{FiberOfFiberwiseComponentsOfFiberwiseMappingSpace}
  \underset{b \in \mathrm{B}}{\forall}
  \hspace{2.6cm}
  \begin{tikzcd}[column sep=huge]
    \mathllap{
      \overset{
        \mathclap{
        \raisebox{2pt}{
          \tiny
          \color{darkblue}
          \bf
          \def\arraystretch{.9}
          \begin{tabular}{c}
            A-cohomology
            \\
            of fiber
          \end{tabular}
        }
        }
      }{
        A^0(\mathrm{X}_b)
      }
      \;=\quad
    }
    \overset{
      \mathclap{
      \raisebox{2pt}{
        \tiny
        \color{darkblue}
        \bf
        \begin{tabular}{c}
        {\color{greenii} Connected components of}
        \\
        ordinary mapping space
        on fiber
        \end{tabular}
      }
      }
    }{
    \pi_0
    \Maps{\big}
      { \mathrm{X}_b }
      { \mathrm{A} }
    }
    \ar[r]
    \ar[d]
    \ar[
      dr,
      phantom,
      "{
        \mbox{\tiny\rm(pb)}
      }"{pos=.3}
    ]
    &
    \overset{
      \mathclap{
      \raisebox{2pt}{
        \tiny
        \begin{tabular}{c}
          \bf
          {\color{greenii}
            Parameterized connected components of
          }
          \\
          {\color{darkblue} \bf
          fiberwise mapping space}
          \color{black} \bf
          (a covering space over $\mathrm{B}$)
        \end{tabular}
      }
      }
    }{
    \pi_{0/\mathrm{B}}
    \Maps{\big}
      { (\mathrm{X}, p_{\mathrm{X}}) }
      { p_{\mathrm{B}}^\ast \mathrm{A} }
    }\mathrlap{\,.}
    \ar[d]
    \\
    \ast
    \ar[r, hook, "{b}"]
    &
    \underset{
      \mathclap{
      \raisebox{-2pt}{
        \tiny
        \color{darkblue}
        \bf
        Parameter space
      }
      }
    }{
    \mathrm{B}
        }
  \end{tikzcd}
\end{equation}

\vspace{-2mm}
\noindent Using this, comparison with \cite[Def. 9.13]{Voisin03I} readily shows that the flat connection
$\nabla^{\mathrm{GM}}_{X,A}$
in
\eqref{TheGMFibrationForUntwistedCase}
is indeed the Gauss-Manin connection.
In conclusion, we have shown so far:
\begin{theorem}[\bf Gauss-Manin connection in generalized cohomology over fiber bundles via fiberwise mapping spaces]
\label{GaussManinConnectionInGeneralizedCohomologyViaMappingSpaces}
  Let $\mathrm{B} \,\in\, \mathrm{kHaus}$ be a
  metrizable Hausdorff
  space and $(\mathrm{X}, p_{\mathrm{X}}) \,\in\, \kTopologicalSpaces_{/\mathrm{B}}$ be a locally trivial fiber bundle whose typical fiber admits the structure of a CW-complex. Then for any $\mathrm{A} \,\in\, \kTopologicalSpaces$ the Gauss-Manin-connection on the $A$-cohomology sets \eqref{NonAbelianCohomologySet} of the fibers $\mathrm{X}_b$ is exhibited
  (under Lem. \ref{UnivalentUniversesOfInfinityGroupoids})
  by the fiberwise 0-truncation of the fiberwise mapping space
  \eqref{FiberwiseMappingSpaceAdjunction}
  from $\mathrm{X}$ into $\mathrm{A}$:
  $$
    \underset{
      \mathclap{
      \raisebox{-7pt}{
        \tiny
        \color{darkblue}
        \bf
        \def\arraystretch{.9}
        \begin{tabular}{c}
          Gauss-Manin
          \\
          connection on
          \\
          $\mathrm{A}$-cohomology
        \end{tabular}
      }
      }
    }{
    \nabla^{\mathrm{GM}}_{X,A}
    }
    \qquad
    \underset{
      \mbox{
        \tiny
        \rm
        Lem. \ref{UnivalentUniversesOfInfinityGroupoids}
      }
    }{
    \longleftrightarrow
    }
    \quad
    \underset{
      \mathclap{
      \raisebox{-2pt}{
        \tiny
        \color{darkblue}
        \bf
        \begin{tabular}{c}
          Fiberwise 0-truncation of
          fiberwise mapping space into $\mathrm{A}$
        \end{tabular}
      }
      }
    }{
    \pi_{
      0/\SingularSimplicialComplex(\mathrm{B})
    }
    \Big(
      \SingularSimplicialComplex
      \,
      \Maps{\big}
       { (\mathrm{X},p_{\mathrm{X}}) }
       { p_{\mathrm{B}}^\ast \mathrm{A} }
  \!  \Big)
    }
    \,.
  $$
\end{theorem}

\subsection{For twisted generalized cohomology}
\label{GaussManinForTwistedGeneralizedCohomology}

We generalize the above discussion to the case of fiberwise {\it twisted} cohomology (Thm. \ref{GaussManinConnectionInTwistedGeneralizedCohomologyViaMappingSpaces})
and bring out the motivating example of the $\suTwoAffine{\Level}$-Knizhnik-Zamolodchikov equation (Ex. \ref{TheKZConnection}).

\medskip

In the existing literature, this is discussed for the special case when
the total space $\mathrm{X}$ is equipped with a flat complex line bundle $\mathcal{L}$ classified by a map $\tau : \mathrm{X} \to B \mathrm{U}(1)$.
In this case one may consider the $\tau_{b}$-twisted complex ordinary  cohomology of the fibers, namely the cohomology with coefficient in the local system of parallel local sections of $\mathcal{L}$ (e.g. \cite[\S 5.1.1]{Voisin03II}):
\vspace{-1mm}
$$
  H^{n + \tau_b}(\mathrm{X}_b; \mathbb{C})
  \,=\,
  H^n(\mathrm{X}_b; \mathcal{L})
  \,.
$$

\vspace{-2mm}
\noindent
At least when $p_{\mathrm{X}}$ is a fiber bundle, these twisted cohomology groups again carry a flat Gauss-Manin connection.
In the example where $\mathrm{X} \,=\, \ConfigurationSpace{n+N}(\mathbb{R}^{2})$ is a configuration space of points, and $p_X$ the map that forgets the first $n$ of $n + N$ points, then a {\it hyprgeometric integral construction} identifies this Gauss-Manin-connection on fiberwise twisted complex cohomology with a Knizhnik-Zamolodchikov connection (\cite[\S 7.5]{EtingofFrenkelKirillov98}).

\medskip
The above is the main example of interest for us. However, in \cite{SS22AnyonicDefectBranes}\cite{SS22AnyonictopologicalOrder}we explained that it is useful to regard this twisted ordinary cohomology as the home of the twisted Chern characters of twisted {\it K-theory} groups. For this reason we are interested in considering, more generally, Gauss-Manin connections on bundles of twisted {\it generalized} cohomology groups.
\vspace{-3mm}
$$
\hspace{2mm}
  \begin{tikzcd}[row sep=7pt, column sep=32pt]
    &
    &
    \overset{
      \mathclap{
      \raisebox{2pt}{
        \tiny
        \color{darkblue}
        \bf
        \def\arraystretch{.9}
        \begin{tabular}{c}
          Fiberwise
          \\
          twisted cohomology
        \end{tabular}
      }
      }
    }{
      A^{\tau_{b_2}}(\mathrm{X}_{b_2})
    }
    \ar[
      ddr,
      "{\sim}"{sloped},
      shorten=-2pt
    ]
    \\[-27pt]
    \mathllap{
      \mbox{
        \tiny
        \color{darkblue}
        \bf
        \def\arraystretch{.9}
        \begin{tabular}{c}
          Category
          \\
          of sets
        \end{tabular}
      }
    }
    \Sets
    &
    &&&
    \mathrm{X}
    \ar[
      dddd,
      "{p_{\mathrm{X}}}"{swap, pos=.65},
      "{
        \mbox{
          \tiny
          \color{greenii}
          \bf
          \def\arraystretch{.9}
          \begin{tabular}{c}
            fiber
            \\
            bundle
          \end{tabular}
        }
      }"{xshift=-6pt, pos=.65}
    ]
    \ar[
      dr,
      shorten <=-1pt,
      shorten >=-4pt,
      "{\tau}"{yshift=-1.5pt, pos=.7},
      "{
        \mbox{
          \tiny
          \color{greenii}
          \bf
          global twist
        }
      }"{swap, sloped, pos=.5}
    ]
    \ar[
      rr,
      dashed
    ]
    &&
    \overset{
      \mathclap{
      \raisebox{2pt}{
        \tiny
        \color{darkblue}
        \bf
        \begin{tabular}{c}
          Classifying fibration
        \end{tabular}
      }
      }
    }{
      \mathrm{A} \times_{\mathrm{G}} E {\mathrm{G}}
    }
    \ar[
      dl,
      shorten=-2pt,
      "{
        p_{\mathrm{A} \times_G E G}
      }"{pos=.9}
    ]
    \\[-12pt]
    &
    A^{\tau_{b_1}}(\mathrm{X}_{b_1})
    \ar[
      uur,
      shorten=-2pt,
      "{
        \sim
      }"{sloped}
    ]
    \ar[
      rr,
      "{\sim}"{swap}
    ]
    &&
    A^{\tau_{b_3}}(\mathrm{X}_{b_3})
    &&
    B \mathrm{G}
    \ar[
      dddd,
      "{
        p_{B G}
      }"
    ]
    \\
    \\
    &
    &
    \{b_2\}
    \ar[
      ddr,
      "{
        [\gamma_{\,23}]
      }"
    ]
    \\[-27pt]
    \mathllap{
      \mbox{
        \tiny
        \color{darkblue}
        \bf
        \def\arraystretch{.9}
        \begin{tabular}{c}
          Fundamental
          \\
          path groupoid
        \end{tabular}
      }
    }
    \SingularSimplicialComplex
    (\mathrm{B})
    \ar[
      uuuu,
      "{
        \mbox{
          \tiny
          \color{greenii}
          \bf
          \def\arraystretch{.9}
          \begin{tabular}{c}
            Gauss-Manin
            \\
            connection
          \end{tabular}
        }
      }"{xshift=2pt}
    ]
    &
    &&&
    \mathrm{B}
    \ar[
      dr,
      shorten=-1pt,
      "{p_{\mathrm{B}}}"{swap}
    ]
    \\[-12pt]
    &
    \{b_1\}
    \ar[
      uur,
      "{
        [\gamma_{\,12}]
      }"
    ]
    \ar[
      rr,
      "{
        [\gamma_{\,23}\, \circ \, \gamma_{\,12}]
      }"{swap}
    ]
    &&
    \{b_3\}
    &
    &
    \ast
  \end{tikzcd}
$$

\vspace{3cm}
\noindent
{\bf Twisted generalized non-abelian cohomology.}
In generalization of \eqref{NonAbelianCohomologySet}, we have:
\begin{definition}[{\bf Twisted generalized non-abelian cohomology} {\cite[\S 2.2]{FSS20Character}\cite[Rem. 2.94]{SS20OrbifoldCohomology}}]
\label{TwistedGeneralizedCohomology}
For
\begin{itemize}[leftmargin=.4cm]
\setlength\itemsep{3pt}
\item
$\mathrm{G} \,\in\, \Groups(\kTopologicalSpaces)$
a topological group;
\item
$\mathrm{G} \acts \, \mathrm{A} \,\in\, \Actions{\mathrm{G}}\big(\kTopologicalSpaces\big)$
a topological $\mathrm{G}$-space,
\item with $\big(\mathrm{A} \times_{\mathrm{G}} E \mathrm{G}, p_{\mathrm{A} \times_{\mathrm{G}} E \mathrm{G}} \big) \,\in\, \kTopologicalSpaces_{/B \mathrm{B}G}$ its Borel construction;
\item
$\tau_b \,:\, \mathrm{X}_b \xrightarrow{\;} B \mathrm{G}$
a continuous map;
\end{itemize}
we say that
\vspace{-2mm}
\begin{equation}
  \label{TwistedNonAbelianCohomologySets}
  \def\arraystretch{1.5}
  \begin{array}{lll}
  A^{\tau_b}(\mathrm{X}_b)
  & \coloneqq\;
  H^{\tau_b}
  \big(
    \mathrm{X}_b
    ;\,
    \mathrm{A}
  \big)
  \\
  &
  \;\coloneqq\;
  \pi_0
  \,
  \Gamma_{\mathrm{X}}
  \big(
    (\tau_b)^\ast
    (\mathrm{A} \times_{\mathrm{G}} E \mathrm{G})
  \big)
  \\
  &
  \;\underset{\eqref{SectionsAsRightBaseChangeOfMappingSpace}}{\simeq}\;
  (p_{B\mathrm{G}})_\ast
  \Maps{\Big}
    {
      \underset{
        \mathclap{
        (\tau_b)_!(\mathrm{X}_b, \mathrm{id}_{\mathrm{X}_b})
        }
      }{
      \underbrace{
      (\mathrm{X}_b, \tau_b)
      }
      }
    }
    { (\mathrm{A} \times_{\mathrm{G}} E\mathrm{G},
      p_{\mathrm{A} \times_{\mathrm{G}} E\mathrm{G}}) }
    &
  \end{array}
\end{equation}

\vspace{0mm}
\noindent is the
{\it $\tau_b$-twisted $A$-cohomology} of $\mathrm{X}_b$.
\end{definition}

To see the identification
shown in  \eqref{TwistedNonAbelianCohomologySets},
of the space of sections
with a right base change of the fiberwise mapping space over the classifying space of twists, apply the Yoneda Lemma to the following sequence of natural bijections:
\vspace{-2mm}
\begin{equation}
  \label{SectionsAsRightBaseChangeOfMappingSpace}
  \def\arraystretch{1.8}
  \begin{array}{lll}
    \def\AmbientCategory{\kTopologicalSpaces}
    \PointsMaps{\Big}
      {
        \mathrm{U}
      }
      {
        (p_{B\mathrm{G}})_\ast
        \Maps{\big}
          { (\tau_b)_!(\mathrm{X}_b, \mathrm{id}_{\mathrm{X}_b}) }
          { (\mathrm{A} \times_{\mathrm{G}} E\mathrm{G} , p_{\mathrm{A} \times_{\mathrm{G}}E \mathrm{G} }) }
      }
    \\
    \;\simeq\;
    \def\AmbientCategory{\kTopologicalSpaces_{/B\mathrm{G}}}
    \PointsMaps{\Big}
      {
        (p_{B\mathrm{G}})^\ast \mathrm{U}
      }
      {
        \Maps{\big}
          { (\tau_b)_!(\mathrm{X}_b,\mathrm{id}_{\mathrm{X}_b}) }
          { (\mathrm{A} \times_{\mathrm{G}} E\mathrm{G} , p_{\mathrm{A} \times_{\mathrm{G}}E \mathrm{G} }) }
      }
    &&
    \proofstep{by \eqref{BaseChangeAdjunction}
    }
    \\
    \;\simeq\;
    \def\AmbientCategory{\kTopologicalSpaces_{/B\mathrm{G}}}
    \PointsMaps{\Big}
      {
        \big(
        (p_{B\mathrm{G}})^\ast \mathrm{U}
        \big)
        \times
        { (\tau_b)_!(\mathrm{X}_b, \mathrm{id}_{\mathrm{X}_b}) }
      }
      {
          { (\mathrm{A} \times_{\mathrm{G}} E\mathrm{G} , p_{\mathrm{A} \times_{\mathrm{G}}E \mathrm{G} }) }
      }
    &&
    \proofstep{by \eqref{ExponentialLawForFiberwiseMappingSpace}}
    \\
    \;\simeq\;
    \def\AmbientCategory{\kTopologicalSpaces_{/B\mathrm{G}}}
    \PointsMaps{\Big}
      {
        (\tau_b)_!
        \big(
        (p_{\mathrm{X}_b})^\ast \mathrm{U}
        \times
        (\mathrm{X}_b, \mathrm{id}_{\mathrm{X}_b})
        \big)
      }
      {
          { (\mathrm{A} \times_{\mathrm{G}} E\mathrm{G} , p_{\mathrm{A} \times_{\mathrm{G}}E \mathrm{G} }) }
      }
    &&
    \proofstep{by \eqref{CartesianFrobeniusReciprocityIsomorphism}}
    \\
    \;\simeq\;
    \def\AmbientCategory{\kTopologicalSpaces_{/\mathrm{X}_b}}
    \PointsMaps{\Big}
      {
        (p_{\mathrm{X}_b})^\ast \mathrm{U}
        \times
        (\mathrm{X}_b, \mathrm{id}_{\mathrm{X}_b})
      }
      {
        (\tau_b)^\ast
          (\mathrm{A} \times_{\mathrm{G}} E\mathrm{G} , p_{\mathrm{A} \times_{\mathrm{G}}E \mathrm{G} })
      }
    &&
    \proofstep{by \eqref{BaseChangeAdjunction}
    }
    \\
    \;\simeq\;
    \def\AmbientCategory{\kTopologicalSpaces_{/\mathrm{X}_b}}
    \PointsMaps{\Big}
      {
        (p_{\mathrm{X}_b})^\ast \mathrm{U}
      }
      {
          (\tau_b)^\ast
          (\mathrm{A} \times_{\mathrm{G}} E\mathrm{G} , p_{\mathrm{A} \times_{\mathrm{G}}E \mathrm{G} })
      }
    &&
    \proofstep{by \eqref{FiberProductUnit}}
    \\
    \;\simeq\;
    \def\AmbientCategory{\kTopologicalSpaces}
    \PointsMaps{\Big}
      {
        \mathrm{U}
      }
      {
        (p_{\mathrm{X}_b})_\ast
        \big(
          (\tau_b)^\ast
          (\mathrm{A} \times_{\mathrm{G}} E\mathrm{G} , p_{\mathrm{A} \times_{\mathrm{G}}E \mathrm{G} })
        \big)
      }
    &&
    \proofstep{by \eqref{BaseChangeAdjunction}}
    \\
    \;\simeq\;
    \def\AmbientCategory{\kTopologicalSpaces}
    \PointsMaps{\Big}
      {
        \mathrm{U}
      }
      {
        \Gamma_{\mathrm{X}_b}
        \big(
          (\tau_b)^\ast
          (\mathrm{A} \times_{\mathrm{G}} E\mathrm{G} , p_{\mathrm{A} \times_{\mathrm{G}}E \mathrm{G} })
        \big)
      }
   &&
   \proofstep{by \eqref{SpaceOfSections}}
   \mathrlap{\,.}
\end{array}
\end{equation}

\begin{example}[\bf Twisted ordinary complex cohomology with coefficients in a local system]
  \label{TwistedOrdinaryCohomology}
  For each $n \in \mathbb{N}$, the canonical multiplication action of $\CircleGroup \subset \ComplexNumbers^\times$ on $\ComplexNumbers$ induces an action on the $n$th Eilenberg-MacLane space $\CircleGroup \acts \; \mathrm{K}(\ComplexNumbers, n)$.
  For $\tau_{b} : \SingularSimplicialComplex(\mathrm{X}_b) \xrightarrow{\;} \mathbf{B}\CircleGroup$ the classifying map of a flat connection (Rem. \ref{FlatConnection}) on a complex line bundle (i.e.,  on the connected component $[b]$ a group homomorphism $\pi_1(\mathrm{B},b) \xrightarrow{\;} \CircleGroup$), the corresponding twisted cohomology according to Def. \ref{TwistedGeneralizedCohomology} is the traditional cohomology with coefficients in the local system $\mathcal{L}(\tau)$ of parallel sections of this flat connection (e.g. \cite[\S 5.1.1]{Voisin03I}):
  \vspace{-1mm}
  $$
    \mathrm{K}(\ComplexNumbers,n)^\tau(\mathrm{X}_b)
    \;=\;
    H^{n + \tau}(\mathrm{X}_b;\, \ComplexNumbers)
    \;\simeq\;
    H^n\big(
      \mathrm{X}_b
      ;\,
      \mathcal{L}(\tau)
    \big)
    \,.
  $$
\end{example}

\medskip

Now given $\mathrm{G} \acts \, \mathrm{A}$,
as in Def. \ref{TwistedGeneralizedCohomology},
and a fibration $(\mathrm{X},p_{\mathrm{X}}) \,\in\, \kTopologicalSpaces_{/\mathrm{B}}$, as before in \cref{ForGeneralizedAndNonAbelianCohomology}, consider a choice of  {\it global twist}, namely a continuous map
$$
  \tau
  \,:\,
  \mathrm{X}
  \xrightarrow{\quad}
  B \mathrm{G}
  \,.
$$
Via this twist, we may regard $\mathrm{X}$ as fibered over the product space $\mathrm{B} \times B\mathrm{G}$, whence its fibers are also still fibered over $B\mathrm{G}$
$$
  \big(
    \mathrm{X}
    ,\,
    (p_{\mathrm{X}}, \tau)
  \big)
  \;\;
  \in
  \;
  \kTopologicalSpaces_{/\mathrm{B} \times B\mathrm{G}}
  \,,
  \hspace{.8cm}
  \Rightarrow
  \hspace{.8cm}
  \underset{b \in \mathrm{B}}{\forall}
  \;\;\;\;
  (i_b \times \mathrm{id}_{B}\mathrm{G})^\ast
  \big(
    \mathrm{X},
    (p_{\mathrm{X}},\tau)
  \big)
  \;=\;
  (\mathrm{X}_b, \tau_b)
  \;=\;
  (\tau_b)_! \mathrm{X}_b
  \;\;\;
  \in
  \;
  \kTopologicalSpaces_{/B \mathrm{G}}
  \,.
$$
Forming the fiberwise mapping space \eqref{FiberwiseMappingSpaceAdjunction} in this sense, we obtain the following twisted generalization of \eqref{FiberOfFiberwiseMappingSpace}:
\vspace{-2mm}
\begin{equation}
  \label{FiberOfFiberwiseTwistedMappingSpace}
  \begin{tikzcd}[column sep=large]
    \overset{
      \mathclap{
      \raisebox{2pt}{
        \tiny
        \color{darkblue}
        \bf
        Fiberwise mapping space
        over classifying space of twists
      }
      }
    }{
    \Maps{\big}
      { (\mathrm{X}_b, \tau_b) }
      { (A \times_{\mathrm{G}} E \mathrm{G},
      p_{A \times_{\mathrm{G}} E \mathrm{G}}) }
    }
    \ar[r]
    \ar[d]
    \ar[
      dr,
      phantom,
      "{
        \mbox{\tiny\rm(pb)}
      }"
    ]
    &
    \overset{
      \mathclap{
      \raisebox{2pt}{
        \tiny
        \begin{tabular}{c}
          {\color{darkblue}
          \bf
          Fiberwise mapping space}
          \color{black}
          (itself a topological space over $\mathrm{B} \times B \mathrm{G}$)
        \end{tabular}
      }
      }
    }{
    \Maps{\big}
      {
        \big(
          \mathrm{X},
          (p_{\mathrm{X}}, \tau)
        \big)
      }
      {
        p_{\mathrm{B}}^\ast
        (\mathrm{A} \times_{\mathrm{G}} E \mathrm{G},
        p_{\mathrm{A} \times_{\mathrm{G}} E \mathrm{G}})
      }
    }.
    \ar[d]
    \\
    \{b\}
    \times
    \underset{
      \mathclap{
      \raisebox{-2pt}{
        \tiny
        \color{darkblue} \bf
        \def\arraystretch{.9}
        \begin{tabular}{c}
          Classifying space
          \\
          of twists
        \end{tabular}
      }
      }
    }{
    B \mathrm{G}
    }\;\;\;\;
    \ar[
      r,
      hook,
      "{
        i_b
        \times
        \mathrm{id}_{B \mathrm{G}}
      }"{swap}
    ]
    &
    \underset{
      \mathclap{
      \raisebox{-2pt}{
        \tiny
        \color{darkblue}
        \bf
        \def\arraystretch{.9}
        \begin{tabular}{c}
          Space of
          \\
          parameters
          and twists
        \end{tabular}
      }
      }
    }{
    \mathrm{B}
      \times
    B \mathrm{G}
    }
  \end{tikzcd}
\end{equation}

In order to turn this into a pullback diagram over just $\mathrm{B}$ we need the {\it Beck-Chevalley relation} (see \cite[\S 1]{Pavlovic91}\cite[\S 7.5]{Balmer15} and \cite[Def. 5.5]{Schreiber14}\cite[\S 2.4.1]{GaitsgoryLurie17}):

\begin{proposition}[\bf Cartesian Beck-Chevalley property]
Given a fiber product diagram in $\kTopologicalSpaces$ as shown on the left below, the possible composite base changes \eqref{BaseChangeAdjunction}
through the diagram are naturally isomorphic as shown on the right:
\begin{equation}
  \label{CartesianBeckChevalleySituation}
  \begin{tikzcd}[row sep=small]
    &
    \mathrm{X}
      \times_{\mathrm{B}}
    \mathrm{Y}
    \ar[
      dl,
      "~~~~~{\mathrm{pr}_{\mathrm{X}}}"{swap}
    ]
    \ar[
      dr,
      "{\mathrm{pr}_{\mathrm{Y}}}"
    ]
    \ar[dd, phantom, "{\mbox{\tiny\rm(pb)}}"]
    \\
    \mathrm{X}
    \ar[dr, "{p_{\mathrm{X}}}"{swap}]
    &&
    \mathrm{Y}
    \ar[dl, "{p_{\mathrm{Y}}}"]
    \\
    &
    \mathrm{B}
  \end{tikzcd}
  \hspace{1cm}
  \Rightarrow
  \hspace{1cm}
  \left\{\!\!\!\!\!
  \def\arraystretch{1.5}
  \begin{array}{c}
    (p_{\mathrm{Y}})^\ast
    \circ
    (p_{\mathrm{X}})_!
    \;\simeq\;
    (\mathrm{pr}_{\mathrm{Y}})_!
    \circ
    (\mathrm{pr}_{\mathrm{X}})^\ast
    \,,
    \\
    (p_{\mathrm{X}})^\ast
    \circ
    (p_{\mathrm{Y}})_\ast
    \;\simeq\;
    (\mathrm{pr}_{\mathrm{X}})_\ast
    \circ
    (\mathrm{pr}_{\mathrm{Y}})^\ast
    \,.
  \end{array}
  \right.
\end{equation}
\end{proposition}
\begin{proof}
The first isomorphism in \eqref{CartesianBeckChevalleySituation} follows immediately from the pasting law \eqref{PastingLaw}, which for $(\mathrm{E}, p_{\mathrm{E}}) \,\in\, \kTopologicalSpaces_{/\mathrm{X}}$ gives the following natural identification:
\vspace{-5mm}
$$
  \begin{tikzcd}
    (p_{\mathrm{Y}})^\ast
    (p_{\mathrm{X}})_!
    \mathrm{E}
    \ar[rr]
    \ar[d]
    \ar[drr,phantom,"{\mbox{\tiny\rm(pb)}}"{pos=.4}]
    &&
    \mathrm{Y}
    \ar[d, "{p_{\mathrm{Y}}}"]
    \\
    \mathrm{E}
    \ar[
      rr,
      "{ p_{\mathrm{X}} \circ p_{\mathrm{E}} }"{swap}
    ]
    &&
    \mathrm{B}
  \end{tikzcd}
  \hspace{1cm}
  \simeq
  \hspace{1cm}
  \begin{tikzcd}[column sep=huge]
    (\mathrm{pr}_{\mathrm{X}})^\ast E
    \ar[r]
    \ar[d]
    \ar[dr,phantom,"{\mbox{\tiny\rm(pb)}}"{pos=.4}]
    \ar[
      rr,
      rounded corners,
      to path={
           ([yshift=+0pt]\tikztostart.north)
       --  ([yshift=+4pt]\tikztostart.north)
       --  node[above]{\scalebox{.8}{$
             p_{
               (\mathrm{pr}_{\mathrm{Y}})_!
               (\mathrm{pr}_{\mathrm{X}})^\ast
               \mathrm{E}
             }
           $}}
           ([yshift=+6pt]\tikztotarget.north)
       --  ([yshift=+0pt]\tikztotarget.north)
      }
    ]
    &
    \mathrm{X}
      \times_{\mathrm{B}}
    \mathrm{Y}
    \ar[r, "{\mathrm{pr}_{\mathrm{Y}}}"]
    \ar[d, "{\mathrm{pr}_{\mathrm{X}}}"{description}]
    \ar[dr,phantom,"{\mbox{\tiny\rm(pb)}}"{pos=.4}]
    &
    \mathrm{Y}
    \ar[d, "{p_{\mathrm{Y}}}"]
    \\
    E
    \ar[r, "{p_{\mathrm{E}}}"{swap}]
    &
    \mathrm{X}
    \ar[r, "{p_{\mathrm{X}}}"{swap}]
    &
    \mathrm{B}
    \mathrlap{\,.}
  \end{tikzcd}
$$

\vspace{-2mm}
\noindent This implies the second natural isomorphism by adjointness \eqref{BaseChangeAdjunction} and the Yoneda Lemma:
\vspace{-1mm}
$$
  \begin{aligned}
  \def\AmbientCategory{\kTopologicalSpaces_{/\mathrm{X}}}
  \PointsMaps{\big}
    { (\mathrm{U},p_{\mathrm{U}}) }
    {
      (p_{\mathrm{X}})^\ast
      (p_{\mathrm{Y}})_\ast (\mathrm{E}, p_{\mathrm{E}})
    }
  &
  \;\simeq\;
  \def\AmbientCategory{\kTopologicalSpaces_{/\mathrm{B}}}
  \PointsMaps{\big}
    {
      (p_{\mathrm{X}})_!
      (\mathrm{U},p_{\mathrm{U}})
    }
    {
      (p_{\mathrm{Y}})_\ast (\mathrm{E}, p_{\mathrm{E}})
    }
  \\
  &
  \;\simeq\;
  \def\AmbientCategory{\kTopologicalSpaces_{/\mathrm{Y}}}
  \PointsMaps{\big}
    {
      (p_{\mathrm{Y}})^\ast
      (p_{\mathrm{X}})_!
      (\mathrm{U},p_{\mathrm{U}})
    }
    {
       (\mathrm{E}, p_{\mathrm{E}})
           }\,,
  \end{aligned}
$$
and similarly for the other side of the isomorphism.
\end{proof}

Now consider the Beck-Chevalley relation \eqref{CartesianBeckChevalleySituation} for the following special case
\vspace{-2mm}
\begin{equation}
  \label{BeckChevalleyProperty}
  \hspace{-5mm}
  \begin{tikzcd}[
row sep=tiny,
    column sep=huge
  ]
    &
    B \mathrm{G}
    \ar[
      dr,
      "{
        i_B
        \times
        \mathrm{id}_{B \mathrm{G}}
      }"
    ]
    \ar[
      dl,
      "{
        \mathrm{p}_{B \mathrm{G}}
      }"{swap}
    ]
    \ar[
      dd,
      phantom,
      "{
        \mbox{
          \tiny
          \rm
          (pb)
        }
      }"
    ]
    \\
    \ast
    \ar[
      dr,
      "{
        i_{b}
      }"{swap}
    ]
    &&
    \mathrm{B}
    \times
    B \mathrm{G}\;,
    \ar[
      dl,
      "{
        \mathrm{id}_{\mathrm{B}}
        \times
        p_{B \mathrm{G}}
      }"
    ]
    \\
    &
    \mathrm{B}
  \end{tikzcd}
  \hspace{1.2cm}
  \big(
    i_b
  \big)^\ast
    \circ
  \big(
    \mathrm{id}_{\mathrm{B}}
    \times
    p_{B \mathrm{G}}
  \big)_\ast
  \;\;
  \simeq
  \;\;
  \big(
    p_{B G}
  \big)_\ast
  \circ
  \big(
    i_b \times \mathrm{id}_{B \mathrm{G}}
  \big)^\ast
  \,.
\end{equation}

\vspace{-2mm}
\noindent This implies, from \eqref{FiberOfFiberwiseTwistedMappingSpace}, the following pullback diagram:
\vspace{-2mm}
\begin{equation}
  \label{DependentProductOfFiberOfFiberwiseTwistedMappingSpace}
  \begin{tikzcd}
    \overset{
      \mathclap{
      \raisebox{2pt}{
        \tiny
        \color{darkblue}
        \bf
        Space of sections over fiber
      }
      }
    }{
    (p_{B \mathrm{G}})_\ast
    \Maps{\big}
      { (\mathrm{X}_b, \tau_b) }
      { (A \times_G E G, P_{A \times_G E G}) }
    }
    \ar[r]
    \ar[d]
    \ar[
      dr,
      phantom,
      "{
        \mbox{\tiny\rm(pb)}
      }"
    ]
    &
    \overset{
      \mathclap{
      \raisebox{2pt}{
        \tiny
        \begin{tabular}{c}
          \bf
          {\color{darkblue}
          Fiberwise space of sections}
          \color{black}
          (itself a topological space over $\mathrm{B}$)
        \end{tabular}
      }
      }
    }{
    (\mathrm{id}_{\mathrm{B}} \times p_{B \mathrm{G}})_\ast
    \Maps{\big}
      { (\mathrm{X}, p_{\mathrm{X}}) }
      { p_{\mathrm{B}}^\ast
        \mathrm{A} \times_{\mathrm{G}} E \mathrm{G}
      }
    }.
    \ar[d]
    \\
    \{b\}
    \ar[
      r,
      hook,
      "{i_b}"{swap}
    ]
    &
    \underset{
      \mathclap{
      \raisebox{-2pt}{
        \tiny
        \color{darkblue}
        \bf
        \def\arraystretch{.9}
        \begin{tabular}{c}
          Space of
          \\
          parameters
          \\
          and twists
        \end{tabular}
      }
      }
    }{
    \mathrm{B}
    }
  \end{tikzcd}
\end{equation}

\vspace{-2mm}
\noindent
Since the classifying space $B \mathrm{G}$ -- in its construction due to Milgram: $B \mathrm{G} \,=\, \vert N(\mathrm{G} \rightrightarrows \ast) \vert$ (recalled, e.g., in \cite[(2.64)]{SS21EPB}) -- is a CW-complex and hence Serre-cofibrant, the map on the right is still a Serre fibration, so that passing to parameterized connected components works
as before in \eqref{TheGMFibrationForUntwistedCase} to yield a covering space, generalizing  \eqref{FiberOfFiberwiseComponentsOfFiberwiseMappingSpace}, whose fiber over $b \in \mathrm{B}$ is the $\tau_b$-twisted $A$-cohomology (Def. \ref{TwistedGeneralizedCohomology}) of the fiber $\mathrm{X}_b$:
\vspace{-2mm}
\begin{equation}
  \label{FiberwiseConnectedComponentOfDependentProductOfFiberOfFiberwiseTwistedMappingSpace}
 \hspace{1cm}
  \begin{tikzcd}[column sep=10pt]
    \mathllap{
      \overset{
        \mathclap{
        \raisebox{2pt}{
          \tiny
          \color{darkblue}
          \bf
          \def\arraystretch{.9}
          \begin{tabular}{c}
            $\tau$-twisted
            \\
            $A$-cohomology
            \\
            of fiber
          \end{tabular}
        }
        }
      }{
        A^\tau(X_b)
      }
      \;\simeq\;
    }
    \overset{
      \mathclap{
      \raisebox{0pt}{
        \tiny
        \color{darkblue}
        \bf
        {\color{greenii} Connected components of}
        space of sections over fiber
      }
      }
    }{
    \pi_0
    \Big(
    (p_{B \mathrm{G}})_\ast
    \Maps{\big}
      { (\mathrm{X}_b, \tau_b) }
      { (A \times_G E G, P_{A \times_G E G}) }
    \Big)
    }
    \ar[r]
    \ar[d]
    \ar[
      dr,
      phantom,
      "{
        \mbox{\tiny\rm(pb)}
      }"{pos=.3}
    ]
    &
    \overset{
      \mathclap{
      \raisebox{0pt}{
        \tiny
        \begin{tabular}{c}
          \bf
          {\color{greenii} Fiberwise connected components of}
          \\
          {\color{darkblue}
          fiberwise space of sections}
          \color{black}
          (itself a topological space over $\mathrm{B}$)
        \end{tabular}
      }
      }
    }{
    \pi_{0/\mathrm{B}}
    \Big(
    (\mathrm{id}_B \times p_{B \mathrm{}})_\ast
    \Maps{\big}
      {
        \big(
          \mathrm{X},
          (p_{\mathrm{X}}, \tau)
        \big)
      }
      {
        p_{\mathrm{B}}^\ast
        \mathrm{A}
        \times_{\mathrm{G}}
        E\mathrm{G}
      }
    \Big)
    }.
    \ar[d]
    \\
    \ast
    \ar[
      r,
      hook,
      "{i_b}"{swap}
    ]
    &
    \underset{
      \mathclap{
      \raisebox{-2pt}{
        \tiny
        \color{darkblue}
        \bf
        \def\arraystretch{.9}
        \begin{tabular}{c}
          Space of
          \\
          parameters
          \\
          and twists
        \end{tabular}
      }
      }
    }{
    \mathrm{B}
  }
  \end{tikzcd}
\end{equation}

\vspace{-2mm}
\noindent
As before in the untwisted case, Prop. \ref{FiberwiseMappingSpaceOutOfFiberBundleIsFiberBundle} again implies that this covering space trivializes compatibly with any local trivialization of $(\mathrm{X}, p_{\mathrm{X}})$, thus exhibiting its corresponding classifying map $\nabla^{\mathrm{GM}}_{\scalebox{.65}{$X,G\acts A$}}$ (via Lem. \ref{UnivalentUniversesOfInfinityGroupoids}) as the Gauss-Manin connection (cf. the description in \cite[\S 7.5]{EtingofFrenkelKirillov98}).

\medskip

In conclusion, we have now shown the following generalization of Thm. \ref{GaussManinConnectionInGeneralizedCohomologyViaMappingSpaces} to twisted cohomology:

\begin{theorem}[\bf Gauss-Manin connection in twisted generalized cohomology over fiber bundles via fiberwise mapping spaces]
\label{GaussManinConnectionInTwistedGeneralizedCohomologyViaMappingSpaces}
  Let $\mathrm{B} \,\in\, \mathrm{kHaus}$ be a
  metrizable
  space and $(\mathrm{X}, p_{\mathrm{X}}) \,\in\, \kTopologicalSpaces_{/\mathrm{B}}$ be a locally trivial fiber bundle whose typical fiber admits the structure of a CW-complex.

  Then, for any discrete or compact Lie
  $\mathrm{G} \in \Groups(\kTopologicalSpaces)$
  and local coefficients
  $G \acts \, \mathrm{A} \,\in\, \Actions{\mathrm{G}}(\kTopologicalSpaces)$, the Gauss-Manin-connection on the twisted $A$-cohomology sets \eqref{TwistedNonAbelianCohomologySets} of the fibers $\mathrm{X}_b$ is exhibited by the fiberwise 0-truncation of the right base change along $B \mathrm{G}$
  \eqref{DependentProductOfFiberOfFiberwiseTwistedMappingSpace}
  of the
  fiberwise mapping space
  \eqref{FiberOfFiberwiseTwistedMappingSpace}
  from $\mathrm{X}$ into the Borel construction $\mathrm{A} \times_{\mathrm{G}} E \mathrm{G}$:
  \vspace{-2mm}
  \begin{equation}
    \label{TheHomotopyTheoreticFormulationOfGaussManinConnection}
    \underset{
      \mathclap{
      \raisebox{-5pt}{
        \scalebox{.7}{
        \color{darkblue}
        \bf
        \def\arraystretch{.9}
        \begin{tabular}{c}
          Gauss-Manin
          \\
          connection on
          \\
          twisted $A$-cohomology
        \end{tabular}
        }
      }
      }
    }{
    \nabla^{\mathrm{GM}}_{
      \scalebox{.7}{$X,\, G \acts \, A,\, \tau$}
    }
    }
    \qquad
    \underset{
      \raisebox{-2pt}{
        \scalebox{.7}{
        \rm
        Lem. \ref{UnivalentUniversesOfInfinityGroupoids}
      }
      }
    }{
      \longleftrightarrow
    }
    \qquad
    \underset{
      \mathclap{
      \raisebox{-1pt}{
       \scalebox{.7}{
       \color{darkblue}
       \bf
       \begin{tabular}{c}
       Fiberwise 0-truncation of right base change along $B\mathrm{G}$
       \\
       of fiberwise mapping space into Borel construction
       on $\mathrm{A}$
       \end{tabular}
      }
      }
      }
    }{
    \pi_{
      0/\SingularSimplicialComplex(\mathrm{B})
    }
    \bigg(
      \SingularSimplicialComplex
      \,
      (
        \mathrm{id}_{\mathrm{B}}
          \times
        p_{B\mathrm{G}}
      )_\ast
      \Maps{\Big}
       { \big(\mathrm{X},(p_{\mathrm{X}}, \tau) \big) }
       { p_{\mathrm{B}}^\ast \mathrm{A} \times_{\mathrm{G}} E\mathrm{G} }
    \!\!\bigg)
    }
    \,.
  \end{equation}
\end{theorem}

\medskip

In the next section \cref{ViaDependentHomotopyTypeTheory} we explain how the homotopy-theoretic construction on the right of \eqref{TheHomotopyTheoreticFormulationOfGaussManinConnection} has a slick implementation in homotopy-typed programming languages such as {\tt Agda}: This is Def. \ref{TypeTheoreticGaussManinConnection} below.

\medskip

But first, we now make explicit the special case of these general considerations that is of interest for quantum computation:

\newpage

\begin{example}[\bf The Knizhnik-Zamolodchikov connection of $\suTwoAffine{\Level}$-conformal blocks via parameterized homotopy theory]
\label{TheKZConnection}
Consider the  specialization of the above setup to the following choice of domain fibration, local coefficients and twist,
parameterized by (cf. Lit. \ref{LiteratureAnyonSpecies}) :

\begin{equation}
\label{TheTwistParameters}
\adjustbox{}{
\def\arraystretch{1.2}
\begin{tabular}{l|l}
  \hline
\rowcolor{lightgray} $N \;\;\;\in\;\;\; \mathbb{N}_{\geq 1}$
    &
    Number of ``defect'' points
    \\
    $n \;\;\;\in\;\;\; \mathbb{N}$
    &
    Number of ``probe'' points
  \\
 \rowcolor{lightgray}   $\ShiftedLevel \;\;\;\in\;\;\; \mathbb{N}_{\geq 2}$
    & The ``shifted level''
        \\
    $
    (\weight_I)_{I =1}^N
    \,\in\,
    \big\{
      0, 1, \cdots, \ShiftedLevel-2
    \big\}^{ N }
    $
    &
    The ``weights'' carried by the defects
  \\
  \hline
\end{tabular}
}
\end{equation}

and then chosen as follows:

\medskip

\begin{enumerate}
  \item[{\bf (i)}] The {\bf domain fibration}

  \begin{equation}
    \label{DomainFibrationForKZConnectionCase}
    \big(
      \mathrm{X}
        \xrightarrow{p_X} \mathrm{B}
    \big)
    \,:\defneq\,
    \Big(
    \ConfigurationSpace{N + n}(\ComplexPlane)
    \xrightarrow{
      \mathrm{pr}^{N+n}_N
    }
    \ConfigurationSpace{N}(\ComplexPlane)
    \Big)
    \;
    \simeq
    \;
    \Big(
      B \, \mathrm{PBr}(N + n)
      \xrightarrow{\;\;}
      B \, \mathrm{PBr}(N)
    \Big)
  \end{equation}
  is the fibration
  \eqref{FibrationOfOrderedConfigurationSpaces}
  of configuration spaces of ordered points in the plane (Lit. \ref{LiteratureConfigurationSpaces})
  which forgets the last $n$ of $N + n$ points;
  to be regarded as the equivalent fibration of delooped pure braid groups (Lit. \ref{LiteratureBraiding}) as shown on the right \eqref{FibrationOfPureBraidGroupsInTermsOfGenerators}.

  \medskip

  \item[{\bf (ii)}]  The {\bf local coefficient space}

  \begin{equation}
    \label{LocalCoefficientsForKZConnectionCase}
    G \acts \, A
    \;\;\; :\defneq \;\;\;
    \ComplexNumbers^\times
    \acts
    \;
    K(\mathbb{C}, n)
  \end{equation}

  is the complex Eilenberg-MacLane space
  from Ex. \ref{TwistedOrdinaryCohomology},
  in degree $n$ and equipped
  with its canonical $\mathbb{C}^\times$-action;

  \medskip

\item[{\bf (iii)}]  The {\bf twist}

  \begin{equation}
  \label{TwistForKZConnectionCase}
  \tau_{
    \scalebox{.85}{$($}
      \ShiftedLevel,\, (\weight_I)_{I=1}^N
    \scalebox{.85}{$)$}
  }
  \,:\,
  \ConfigurationSpace{N+n}(\ComplexNumbers)
  \;\simeq\;
  B \, \mathrm{PBr}(N + n)
  \xrightarrow{\qquad}
  B \,\ComplexNumbers^\times
  \;,
  \end{equation}

is the delooping of the following group homomorphism:

\begin{equation}
  \label{SpecifyingTheTwistForKZConnectionCase}
  \begin{tikzcd}[
    sep=0pt
  ]
    \mathrm{PBr}(N + n)
    \ar[
      rr,
      "{
        \Omega \tau
      }"
    ]
    &&
    \ComplexNumbers^\times
    \\
  \scalebox{0.75}{$  b_{I ,\, i}$}
      &\longmapsto&
     \scalebox{0.75}{$  \exp\big(
      2\pi\ImaginaryUnit
      \,
      \frac{
        {\weight_I}
      }
      {
        \ShiftedLevel\
      }
    \big)
    $}
    &
    \\
   \scalebox{0.75}{$    b_{i,\,j}$}
      &\longmapsto&
   \scalebox{0.75}{$    \exp\big(
      2 \pi \ImaginaryUnit
      \,
      \frac{
        2
      }{
        \ShiftedLevel
      }
    \big)
    $}
    \\
    \scalebox{0.75}{$   b_{I,\, J}$}
    &\longmapsto&
      \scalebox{0.75}{$ \exp\big(
      2 \pi \ImaginaryUnit
      \,
      \frac{
        \weight_I \weight_J
      }
      {
        2\ShiftedLevel
      }
    \big)
    $}
  \end{tikzcd}
  \;\;\;\;
  \adjustbox{raise=-6pt}{
  \mbox{\small for}
  \;\;
  $
  \def\arraystretch{1.4}
  \begin{array}{l}
 \scalebox{0.8}{$ 1 \leq I \leq N$}
    \\
\scalebox{0.8}{$ N < i \leq N + n$}
  \end{array}
  $
  }
\end{equation}

Here $b_{i j}$ denote the pure braid generators \eqref{ArtinPureBraidGeneratorsGraphically}. Notice that any such assignment respects
the pure braid relations \eqref{ArtinPureBraidGroup} because these are all group commutator relations which are all trivially
satisfied in an abelian group such as $\ComplexNumbers^\times$.

\end{enumerate}

\medskip

\noindent
With these choices, the above Gauss-Manin connection specializes to the
situation discussed in \cite[\S 7.5]{EtingofFrenkelKirillov98}
and reviewed in our context in \cite{SS22AnyonicDefectBranes}\cite{SS22AnyonictopologicalOrder}\cite{SS22TQC},
yielding the Gauss-Manin connection on the bundle of fiberwise twisted ordinary cohomology groups (Ex. \ref{TwistedOrdinaryCohomology})
\vspace{-1mm}
$$
  \big(
  \{z_I\}_{I =1}^N
  \;\in\;
  \ConfigurationSpace{N}(\ComplexPlane)
  \big)
  \;\;\;\;
  \longmapsto
  \;\;\;\;
  H^{n}
  \Big(
    \ConfigurationSpace{n}
    \big(
      \ComplexPlane
      \setminus
      \{z_I\}_{I = 1}^N
    \big)
    ;\,
    \mathcal{L}\big(
      \tau_{
        \scalebox{1}{$($}
          \ShiftedLevel
          ,\,
          (\weight_I)_{I=1}^N
        \scalebox{1}{$)$}
      }
    \big)
  \Big)
$$

By \cite[\S 7.5]{EtingofFrenkelKirillov98}
this is the Knizhnik-Zamolodchikov connection (Lit. \ref{KZConnectionsOnConformalBlocksReferences}) on  --- by \cite{FeiginSchechtmanVarchenko94} ---
the spaces of $\suTwoAffine{\ShiftedLevel-2}$ conformal blocks for $N + 1$-point correlators on the Riemann sphere with weights $\weight_1, \cdots, \weight_N$ as specified and $\weight_{N+1} = n + \sum_{I = 1}^N \weight_I $.
This is the result of the {\it hypergeometric integral construction} (Lit. \ref{HypergeometricIntegralReferences}) of KZ-solutions further reviewed (and referenced) in \cite{SS22AnyonicDefectBranes}\cite{SS22AnyonictopologicalOrder}.

(Notice that the  choice of phase for $b_{I J}$ in the last line of  \eqref{SpecifyingTheTwistForKZConnectionCase},
follows \cite[(7.14)]{EtingofFrenkelKirillov98}\cite[(6.1)]{Kohno12}. Making a different choice here only results in tensoring the resulting flat connection by a flat line bundle, changing the resulting monodromy by these global phases.)
\end{example}

In conclusion, in the case of this Example \ref{TheKZConnection},
Theorem \ref{GaussManinConnectionInTwistedGeneralizedCohomologyViaMappingSpaces} with Example \ref{TwistedOrdinaryCohomology}
says that th KZ-connection on the space of $\suTwoAffine{\ShiftedLevel-2}$-conformal blocks is realized as equivalently reflected in the fiberwise 0-truncation of  (a right base change of) a fiberwise mapping space:
\begin{equation}
  \label{KZConnectionByFiberwiseMapping}
  \def\arraystretch{2}
  \begin{array}{c}
  \overset{
     \mathclap{
    \raisebox{+10pt}{
      \scalebox{.7}{
      \color{darkblue}
      \bf
      \def\arraystretch{.9}
      \begin{tabular}{c}
        KZ-connection on
        $\suTwoAffine{\ShiftedLevel-2}$-conformal
        blocks
        on the sphere
        \\
        with $N + 1$ insertions
        of
        weights
        $(\weight_I)_{I=1}^{N}$
        ,
        $\weight_{N+1}
        =
        n + \sum_{I} \weight_I$
        \\
      \end{tabular}
      }
    }
     }
   }{
    \nabla^{\mathrm{KZ}}
      \big(
        \ShiftedLevel
        ,\,
        (\weight_I)_{I=1}^N
        ,\,
        n
      \big)
  }
  \\[5pt]
    \rotatebox{90}{
      \clap{$\longleftrightarrow$}
    }
    \mathrlap{ \color{greenii}
      \raisebox{+3pt}{
        \scalebox{.7}{
        Lem.
          \ref{UnivalentUniversesOfInfinityGroupoids}
        }
      }
     }
   \\[7pt]
  \underset{
    \raisebox{-4pt}{
      \scalebox{.7}{
      \color{darkblue}
      \bf
      Fiberwise 0-truncation
      of right base change along
      $B \ComplexNumbers^\times$
      of fiberwise mapping space
      from configuration space
      into Eilenberg-MacLane fiber bundle
      }
    }
  }{
  \pi_{
    \scalebox{.7}{$
    0/
    \SingularSimplicialComplex
    \!\!
    \ConfigurationSpace{N}(\ComplexPlane)
    $}
  }
  \left(
    \adjustbox{raise=0pt}{$
    \SingularSimplicialComplex
    \;
    \big(
      \mathrm{id}_{\!\!\!{}_{
        \scalebox{.7}{$
        \ConfigurationSpace{N}(\ComplexPlane)
        $}
        }
      }
      \times
      p_{{}_{B\CyclicGroup{\ShiftedLevel}}}
    \big)_\ast
    \,
  \mathrm{Map}\!
  \left(
  {
        \adjustbox{raise=+6pt}{
        \begin{tikzcd}
          \ConfigurationSpace{N + n}
          (
            \ComplexPlane
          )
          \ar[
            d,
            shorten=-2pt,
            "{
              (
              \mathrm{pr}^{N+n}_N
              ,\,
              \tau_{(\ShiftedLevel, \weight_\bullet)}
              )
            }"
            {swap}
          ]
          \\
          \ConfigurationSpace{N}(\ComplexPlane)
          \times
          B \ComplexNumbers^\times
       \end{tikzcd}
       }
      }
      ,\,
      {
       \quad
        p_{
          \!\!\!\!
          \scalebox{.7}{
            $\ConfigurationSpace{N}$
          }
        }^\ast
        \adjustbox{raise=6pt}{
        \begin{tikzcd}
          \mathrm{K}(\ComplexNumbers,n)
          \times_{
            {}_{
              \ComplexNumbers^\times
          }}
          E\ComplexNumbers^\times
          \ar[d]
          \\
          B \ComplexNumbers^\times
        \end{tikzcd}
        }
      }
      \right)
  $}
   \right)
  }
  \end{array}
\end{equation}

\medskip

Next we turn to the task of encoding the construction on the right in homotopically typed programming languages such as {\tt Agda}: This is achieved in \cref{KZConnectionsInHomotopyTypeTheory} below (Thm. \ref{TheTheorem}). In preparation, we now first discuss homotopy data type theory and its encoding of Gauss-Manin connections in generality (\cref{ViaDependentHomotopyTypeTheory}).

\newpage

\section{...via dependent homotopy type theory}
\label{ViaDependentHomotopyTypeTheory}

\def\AmbientCategory{\mathbf{H}}

Here we recast the topological constructions of
\cref{ViaParameterizedPointSetTopology} into the programming language of \emph{homotopy type theory} (HoTT, Lit. \ref{LiteratureHomotopyTypeTheory}), which we motivate and survey in \cref{HoTTIdea}: In \cref{GMConnectionOnTwistedCohomologyData}, we translate the previous
Thm. \ref{GaussManinConnectionInTwistedGeneralizedCohomologyViaMappingSpaces}
to a construction of data transport (Lit. \ref{PathLiftingLiterature}) through dependent data type families whose ``semantics'' is given by the monodromy of Gauss-Manin connections.
Specializing this construction to the context of Ex. \ref{TheKZConnection} provides -- further below in \cref{KZConnectionsInHomotopyTypeTheory} -- the promised type-theoretic encoding of monodromy of Knizhnik-Zamolodchikov connections and hence of anyon braid quantum gates.

\subsection{From data types to homotopy types}
\label{HoTTIdea}

We give an informal but detailed exposition of the general principles of programming languages with homotopy data types to naturally motivate --- from just the idea of declaring for strictly all data a corresponding {\it data type specification} --- the homotopy theoretic language structures which in \cref{KZConnectionsInHomotopyTypeTheory} serve to naturally specify anyon braid quantum gates.
While there is by now a fair supply of literature on the subject of this {\it Homotopy Type Theory} (HoTT, Lit. \ref{LiteratureHomotopyTypeTheory}), we feel that an exposition along the following lines has been missing; in any case it serves to coherently record and reference the {\it Rosetta stone}-like dictionary (the homotopical ``computational trilogy'', see pointers in \cite[p. 4]{SS22TQC}, Lit. \ref{VerificationLiterature})

\vspace{-.2cm}
\begin{equation}
\label{TheTrilogy}
\adjustbox{raise=-50pt, scale=.8}{
\small
\begin{tikzpicture}
  \draw (90+70:1.3)
    node[rectangle,fill=white]
      {\color{darkblue}
        \begin{tabular}{c}
          \circled{1}
          \\
          \fbox{\bf Computation} \;\;\;\;\;\;\;\;\;
          \end{tabular}
      };
  \draw (90-70:1.3)
    node[rectangle, fill=white]
      {\color{darkblue}
        \begin{tabular}{c}
          \circled{2}
          \\
          \;\;\;\;\;\;
          \fbox{\bf Type Theory}
        \end{tabular}
      };
  \draw (90-180:1.5)
    node[rectangle, fill=white]
      {\color{darkblue}
        \begin{tabular}{c}
          \circled{3}
          \\
        \fbox{\bf Alg. Topology}
        \end{tabular}
      };
  \draw[<->]
    (90-40:1.3)
      arc
      (90-40:90+40:1.3);
  \draw[<->]
    (-45-20:1.3)
      arc
      (-45-20:-45+40:1.3);
  \draw[<->] (225-40:1.3) arc (225-40:225+20:1.3);
\end{tikzpicture}
}
\end{equation}
\vspace{-.2cm}

\noindent
which translates our algebro-topological theorems in \cref{ViaParameterizedPointSetTopology} to the programming language constructions in
\cref{GMConnectionOnTwistedCohomologyData} and
\cref{KZConnectionsInHomotopyTypeTheory}:

\medskip

\medskip

\noindent
{\bf The idea of typed programming languages...}
It is a time-honored idea that every piece of data handled by a programming language be assigned a {\it data type} which specifies its intended nature (e.g. \cite{MartinLof82}\cite{Thompson91}\cite{Streicher93}\cite{Luo94}\cite{Gunter95}\cite{Constable11}\cite{Harper16}, see also p. \pageref{DataStructures} below).

For example, declaring a numerical datum $d$ to be of the type of natural numbers --- to be denoted ``$d : \mathbb{N}$'' \eqref{NaturalNumberstype} --- instead of, say, the type $\mathbb{Z}$ of integers or the type $\RationalNumbers$ of rational or the type $\RealNumbers$ of real or the type $\ComplexNumbers$ of complex numbers (all discussed in  \cref{KZConnectionsInHomotopyTypeTheory}),
conveys information about how subsequent parts of the program may or may not operate with this datum. For instance, if a program computes a number which can be certified to be of type $\NaturalNumbers$, as opposed to the less constrained type $\Integers$, this guarantees the non-negativity of that number, which may be important for the soundness of subsequent computations.

\medskip

With this understood, the {\bf purpose of typed programming languages is}:

To declare {\it programs} $f$ which, {\it given data $d$ of input type $D$},  are guaranteed to {\it produce data $f(d)$ of output type $A$}.

\smallskip

\noindent
Following tradition in formal logic, such algorithms (``assertions'' \cite[p. xviii]{RussellWhitehead10}, ``proofs'' \cite[\S 5]{Church40}, ``validations'' \cite[p. 223]{Kochen61}) are denoted \cite[\S 2.1.2]{Hofmann95}\footnote{
As is usual, in \eqref{FunctionDeclaration} we are including denotation of a generic context of ``constants'' $c$ of any type $\Gamma$ --  on which the execution of $f$ may well depend, but which one may not want to regard as external parameters but as implicit arguments passed to the program (e.g. ``flags''). As also usual, often we will notationally suppress this generic background context.}  by the symbol ``$\,\vdash\,$'' \cite[\S 2]{Frege1879} for {\it judgements}, cf.  \cite[p. 2]{MartinLof84}\cite[\S 1]{MartinLof96}:

\begin{equation}
\label{FunctionDeclaration}
\adjustbox{scale=1}{
\def\arraystretch{1}
\def\arraystretch{1.3}
\begin{tabular}{|c||c|}
\hline
\begin{minipage}{6cm}
{\bf Programming language} (syntax)
\end{minipage}
&
\begin{minipage}{6cm}
{\bf Mathematical denotation} (semantics)
\color{gray}
\end{minipage}
\\
\hline
\hline
&
\\[-8pt]
$
  \arraycolsep=8pt
  \def\arraystretch{1.3}
  \begin{array}{cccc}
    \overset{
      \mathclap{
      \raisebox{4pt}{
        \scalebox{.8}{
          \color{orangeii}
          \bf
          Given...
        }
        }
      }
    }{
    \underset{
      \mathclap{
        \raisebox{-2pt}{
          \rotatebox{-30}{
            \scalebox{.75}{
              \rlap{
                \color{darkblue}
                \bf
                \hspace{-25pt}
                \def\arraystretch{.7}
                \begin{tabular}{c}
                  any
                  \\
                  data context
                \end{tabular}
              }
            }
          }
        }
      }
    }{
      \Gamma
      \mathrlap{,}
    }
    }
    \phantom{--}
    \;\;\;\;\;\;\;\;
    \overset{
      \mathclap{
      \raisebox{4pt}{
        \scalebox{.8}{
          \color{orangeii}
          \bf
          ...and moreover...
        }
        }
      }
    }{
    \underset{
      \mathclap{
        \raisebox{-2pt}{
          \rotatebox{-30}{
            \scalebox{.75}{
              \rlap{
                \color{darkblue}
                \bf
                \hspace{-25pt}
                \def\arraystretch{.7}
                \begin{tabular}{c}
                  \;\;\;data $d$
                  \\
                  of type $D$
                \end{tabular}
              }
            }
          }
        }
      }
    }{
      d : D
    }
    }
    &\;\;\;\;\; \vdash \;\;\;&
    \overset{
      \mathclap{
      \raisebox{4pt}{
        \scalebox{.8}{
          \color{orangeii}
          \bf
          ...construct
        }
        }
      }
    }{
    \underset{
      \mathclap{
        \raisebox{-2pt}{
          \rotatebox{-30}{
            \scalebox{.75}{
              \rlap{
                \color{darkblue}
                \bf
                \hspace{-40pt}
                \def\arraystretch{.7}
                \begin{tabular}{c}
                  \;\;\;\;\;\;\;\;data $f_c(d)$
                  \\
                  of type $A$
                \end{tabular}
              }
            }
          }
        }
      }
    }{
      f_c(x) : A
    }
    }
  \end{array}
$
&
$
  \begin{tikzcd}
    \underset{
      \mathclap{
      \raisebox{-4pt}{
        \scalebox{.7}{
          \color{darkblue}
          \bf
          \def\arraystretch{.7}
          \begin{tabular}{c}
            space of
            \\
            input data
          \end{tabular}
        }
      }
      }
    }{
      \Gamma \times D
    }
    \;\; \ar[
      rrr,
      dashed,
      "f",
      "{
        \scalebox{.8}{
          \color{greenii}
          \bf
          function
          {\color{black}/}
          map
        }
      }"{swap}
    ]
    &&& \quad
    \underset{
      \mathclap{
      \raisebox{-4pt}{
        \scalebox{.7}{
          \color{darkblue}
          \bf
          \def\arraystretch{.7}
          \begin{tabular}{c}
            space of
            \\
            output data
          \end{tabular}
        }
      }
      }
    }{
      A
    }
  \end{tikzcd}
$
\\
&
\\
\hline
\end{tabular}
}
\end{equation}
\vspace{.0cm}

\noindent
{\bf ...and their mathematical meaning.} As shown on the right, making the data typing explicit allows one to recognize evident ``mathematical meaning'' or ``denotational semantics'' of programs (a seminal idea due to \cite{Scott70}\cite{ScottStrachey71}, for exposition see \cite[\S 9]{SlonnegerKurtz95}): A program $f$ that (is guaranteed to eventually halt and then) outputs data of type $A$ when run on input of type $D$ is evidently a {\it function} or {\it map} from the ``space'' of all data of type $D$ to that of all data of type $A$.
This tautologous-sounding statement evolves into a remarkably powerful relation between programming languages and algebraic topology as we now progress through the full logical consequences of the principle of typed programming.

\newpage

\noindent
{\bf The idea of the data type of data types.}
Taking the data-typing principle seriously, one realizes that data types are clearly a kind of data themselves --- and hence ought to be assigned a data type:
We write ``$\Types$'' for the (``large'') data  type of all
       data types --- often called the {\it type universe} and denoted ``$\mathcal{U}$'' or similar
    \cite[\S 1.10]{MartinLof75}\cite[pp. 47]{MartinLof84}\cite[\S 2.3.5]{Hofmann95}\cite{Palmgren98}.

\vspace{-.48cm}
\begin{equation}
    \label{TypeOfSmallTypes}
  \hspace{-.5cm}
  \begin{tabular}{ll}
  \begin{minipage}{9.7cm}
The semantics of type universes is given by {\it Grothendieck universes} in set theory (e.g. \cite[\S 3.2]{Schubert72}, references in \cite[\S 6.4.4]{Kromer07}),
or rather
-- via the univalence axiom \eqref{UnivalenceAxiom} -- their generalization to {\it object classifiers} ``$\Objects$'' \cite[\S 6.1.6]{Lurie09}, see below around \eqref{TypeClassification}.
  \end{minipage}
  &
  \hspace{3mm}
  \def\arraystretch{1.2}
  \tabcolsep=5pt
  \begin{tabular}{|c||c|}
  \hline
  {\bf Type theory}
  &
  {\bf Homotopy theory}
  \\
  \hline
  \hline
  $
  \Gamma
  \;\;
  \vdash
  \;\;
  D \,:\, \Types
  $
  &
  $
    \begin{tikzcd}
      \Gamma
      \ar[r, "{ D }"]
      &
      \Objects
    \end{tikzcd}
  $
   \\
   \hline
  \end{tabular}
 \end{tabular}
\end{equation}

\noindent
{\bf The idea of data type formation.}
Given such data types, there are fairly self-evident rules (nicely motivated in \cite{MartinLof84}\cite{MartinLof96}, following ``constructive logic'' \eqref{LogicalConnectivesAsTypeFormation}) for forming/constructing further data types out of them.
For example, given a pair of data types $X_1$ and $X_2$, one will consider the {\it product type} $X_1 \times X_2$ of {\it pairs} of data $(d_1, d_2)$ with $d_1 \,:\, X_1$, $d_2 \,:\, X_2$. As the notation suggests, the denotational semantics of the syntactic rules for product types are the characteristic properties of {\it product spaces}:

\vspace{-.1cm}
\begin{equation}
\label{PairTypes}
\adjustbox{scale=.95}{
\def\arraystretch{1}
\def\arraystretch{1.2}
\begin{tabular}{|c||c|c|}
\hline
  &
\begin{minipage}{7cm}
{\bf Programming language} (syntax)
\end{minipage}
&
\begin{minipage}{6cm}
{\bf Mathematical denotation} (semantics)
\end{minipage}
\\[-5pt]
&
\begin{minipage}{7cm}
for {\bf Product types}
\end{minipage}
&
\begin{minipage}{6cm}
of {\bf Product spaces}
\end{minipage}
\\
\hline
\hline
  &
&
\\[-9pt]
  \rotatebox{90}{
  \clap{
  \bf Pair type formation rule
  }
  }
&
$
  \arraycolsep=8pt
  \def\arraystretch{1.3}
  \begin{array}{cccc}

    \def\arraystretch{1}
    \begin{array}{ccccc}
      \vdash &
               \overset{
               \mathclap{
               \raisebox{3pt}{
               \scalebox{.8}{
               \color{orangeii}
               \bf
               \begin{tabular}{c}
                 Given
                 \color{darkblue}
                 one data type...
               \end{tabular}
               }
               }
               }
               }{
               X_1 : \Types
               }
      &\phantom{-}&
                    \vdash &
                             \overset{
                             \mathclap{
                             \raisebox{3pt}{
                             \scalebox{.8}{
                             \color{darkblue}
                             \bf
                             \begin{tabular}{c}
                               ...and another
                               one...
                             \end{tabular}
                             }
                             }
                             }
                             }{
                             X_2 : \Types
                             }
    \end{array}

    \\
    \hline
    \begin{array}{cc}
      \vdash
      &
        \underset{
        \mathclap{
        \raisebox{-7pt}{
        \scalebox{.8}{
        \color{orangeii}
        \bf
        \def\arraystretch{.9}
        \begin{tabular}{c}
          ...we infer
          their
          \\
          \color{darkblue}
          data type of pairs.
        \end{tabular}
        }
        }
        }
        }{
        X_1 \times X_2 : \Types
        }
    \end{array}

\end{array}
$
&
  \begin{tikzpicture}[baseline={(0,0)}]
    \node[label={above:\descriptionboxdefault{
        Given {\color{darkblue}one space...}
      }}](X1) at (0,1) {$X_1$};
    \node[label={below:\descriptionboxdefault{
        {\color{orangeii}... and \color{darkblue}  another...}
      }}](X2) at (0,-1){$X_2$};
    \node[label={[label distance = -10pt]left:\descriptionboxdefault{
          ...we infer their
          \\
          {\color{darkblue} product space.}
      }}] (prod) at (0,0) {$X_1 \times X_2$};
  \end{tikzpicture}
  \\
  \hline
  &
  &
  \\[-9pt]
  \rotatebox{90}{
  \clap{
  \bf Pair introduction rule
  }
  }
  &
    \(

  \def\arraystretch{1.4}
  \begin{array}{c}
    \arraycolsep=4pt
    \def\arraystretch{1}
    \begin{array}{ccccccc}
      \Gamma
      &
      \overset{
        \mathclap{
        \raisebox{8pt}{
          \scalebox{.8}{
            \bf
            \def\arraystretch{.8}
            \begin{tabular}{c}
              \color{orangeii}
              Given
              \color{darkblue}
              a program which
              \\
              \color{darkblue}
              computes data of type $X_1$...
            \end{tabular}
          }
        }
        }
      }{
        \vdash
      }
      &
        x_1 : X_1
      &
      \phantom{--}
      &
      \Gamma
      &
      \overset{
        \mathclap{
        \raisebox{8pt}{
          \scalebox{.8}{
            \bf
            \def\arraystretch{.8}
            \begin{tabular}{c}
              \color{orangeii}
              Given
              \\
              \color{darkblue}
              $i : I$-indexed $X_i$-data...
            \end{tabular}
          }
        }
        }
      }{
        \vdash
      }
      & x_2 : X_2
    \end{array}
    \\
    \hline
    \begin{array}{cccc}
      \Gamma
      &
      \underset{
        \mathclap{
        \;\;\;\;\;\;\;\;\;\;\;\;\;\;\;\;\;\;\;
        \raisebox{-10pt}{
          \scalebox{.8}{
            \bf
            \def\arraystretch{.8}
            \begin{tabular}{c}
              \color{orangeii}
              ... we infer
              \color{darkblue}
              a program which
              \\
              \color{darkblue}
              computes data of type $X_1 \times X_2$.
            \end{tabular}
          }
        }
        }
      }{
        \vdash
      }
      &
      (x_1,x_2) : X_1 \times X_2
    \end{array}
  \end{array}
    \)
&
  \begin{tikzcd}[
   column sep=40pt,
   row sep=30pt
  ]
    &&
    X_1
    \\
    \Gamma
    \ar[
      urr,
      bend left=25,
      "{
        \overset{
          \mathclap{
          \raisebox{3pt}{
            \scalebox{.7}{
              \bf
              \color{orangeii}
              Given
              \color{darkblue}
              a map to one space
            }
            }
          }
        }{
          x_1
        }
      }"
    ]
    \ar[
      drr,
      bend right=25,
      "{
        \underset{
          \mathclap{
          \raisebox{3pt}{
            \scalebox{.7}{
              \bf
              \color{orangeii}
              and
              \color{darkblue}
              a map to another space
            }
            }
          }
        }{
          x_2
        }
      }"{swap}
    ]
    \ar[
      rr,
      dashed,
      "{
          (x_1, x_2)
      }",
      "{
          \raisebox{3pt}{
            \scalebox{.7}{
              \bf
              \color{orangeii}
              ...we infer
              \color{darkblue}
              a map to the product space
            }
            }
      }"{swap, pos=.83, yshift=-3pt}
    ]
    &&
    X_1 \times X_2
    \\
    &&
    X_2
  \end{tikzcd}
  \\
  \hline
  & &
  \\
  [-9pt]
  \rotatebox{90}{
  \clap{
  \bf  Pair elimination rule
  }
  }
  &
    \(
  \def\arraystretch{1.4}
  \begin{array}{c}
    \arraycolsep=4pt
    \def\arraystretch{1}
    \begin{array}{ccc}
      \mathclap{\phantom{\vert_{\vert_{\vert}}}}
      \Gamma
      &
      \overset{
        \mathclap{
        \;\;\;\;\;\;\;\;\;\;\;
        \hspace{25pt}
        \raisebox{8pt}{
          \scalebox{.8}{
            \bf
            \def\arraystretch{.8}
            \begin{tabular}{c}
              \color{orangeii}
              Given
              \color{darkblue}
              a program which
              \\
              \color{darkblue}
              computes data of
              pair type...
            \end{tabular}
          }
        }
        }
      }{
        \vdash
      }
      &
      f : X_1 \times X_2
    \end{array}
    \\
    \hline
    \begin{array}{ccccccc}
      \Gamma
      &
      \underset{
        \mathclap{
        \;\;\;\;\;\;\;\;\;\;\;\;\;\;\;\;\;\;\;
        \raisebox{-10pt}{
          \scalebox{.8}{
            \bf
            \def\arraystretch{.8}
            \begin{tabular}{c}
              \color{orangeii}
              ... we infer
              \color{darkblue}
              programs which
              \\
              \color{darkblue}
              compute the component data.
            \end{tabular}
          }
        }
        }
      }{
        \vdash
      }
      &
      \mathrm{pr}_i(f) \,:\, X_i
    \end{array}
  \end{array}
    \)
&
  \begin{tikzcd}[
   column sep=40pt,
   row sep=30pt
  ]
    &&
    X_1
    \\
    \Gamma
    \ar[
      urr,
      dashed,
      bend left=25,
      "{
        \overset{
          \mathclap{
          \raisebox{3pt}{
            \scalebox{.7}{
              \bf
              \color{orangeii}
            }
            }
          }
        }{
          \mathrm{pr}_1 \circ f
        }
      }"
    ]
    \ar[
      drr,
      dashed,
      bend right=25,
      "{
        \underset{
          \mathclap{
          \raisebox{3pt}{
            \scalebox{.7}{
              \bf
              \color{orangeii}
              ...we infer
              \color{darkblue}
              its component maps
            }
            }
          }
        }{
          \mathrm{pr}_2 \circ f
        }
      }"{swap}
    ]
    \ar[
      rr,
      "{
          f
      }",
      "{
          \raisebox{3pt}{
            \scalebox{.7}{
              \bf
              \def\arraystretch{.8}
              \begin{tabular}{l}
              \color{orangeii}
              Given
              \color{darkblue}
              a map
              \\
              \color{darkblue}
              to the product space...
              \end{tabular}
            }
            }
      }"{swap, pos=.65, yshift=-3pt}
    ]
    &&
    X_1 \times X_2
    \ar[
      u,
      "{ \mathrm{pr}_1 }"{swap}
    ]
    \ar[
      d,
      "{ \mathrm{pr}_2 }"{}
    ]
    \\
    &&
    X_2
  \end{tikzcd}
  \\
  \hline
  & & \\
  [-9pt]
  \rotatebox{90}{
  \clap{
  \bf  Pair computation rules
  }
  }
  &
    \(
  \def\arraystretch{1.4}
  \begin{array}{c}
    \arraycolsep=4pt
    \def\arraystretch{1}
    \begin{array}{ccccccc}
      \Gamma
      &
      {
        \vdash
      }
      &
        x_1 : X_1
      &
      \phantom{--}
      &
      \Gamma
      &
      \vdash
      & x_2 : X_2
    \end{array}
    \\
    \hline
    \begin{array}{cccc}
      \Gamma
      &
      \vdash
      &
      \underset{
        \mathclap{
        \;\;\;\;\;
        \raisebox{-9pt}{
          \scalebox{.8}{
            \bf
            \def\arraystretch{.8}
            \begin{tabular}{c}
              \color{orangeii}
              Such that
              \color{darkblue}
              feeding the pairing program
              into
              \\
              \color{darkblue}
              the projector program
              recovers the input data...
            \end{tabular}
          }
        }
        }
      }{
        \mathrm{pr}_{i}(x_1,x_2)
        \defneq
        x_i
      }
      \;\;:\;\;
      X_i
    \end{array} 
    \\

    \arraycolsep=4pt
    \def\arraystretch{1}
    \begin{array}{ccc}
      \Gamma
      &
      \overset{
        \phantom{
        \mathclap{
        \raisebox{8pt}{
          \scalebox{.8}{
            \color{orangeii}
            \bf
            \def\arraystretch{.8}
            \begin{tabular}{c}
              Given a program which
              \\
              computes data of type $X$...
            \end{tabular}
          }
        }
        }
        }
      }{
        \vdash
      }
      &
        p : X_1 \times X_2
    \end{array}
    \\
    \hline
    \begin{array}{cccc}
      \Gamma
      &
      \vdash
      &
      \underset{
        \mathclap{
        \;\;\;\;\;\;\;\;\;\;
        \raisebox{-5pt}{
          \scalebox{.8}{
            \bf
            \def\arraystretch{.8}
            \begin{tabular}{c}
              \color{orangeii}
              ...
              and
              \color{darkblue}
              any datum of pair type
            \\
            \color{darkblue} is the pair of its projections.
            \end{tabular}
          }
        }
        }
      }{
       p \defneq (\mathrm{pr}_1 p,\, \mathrm{pr}_2 p)
      }
      \;\;:\;\;
      X_i
    \end{array}
  \end{array}
    \)
&
  \begin{tikzcd}[
   column sep=40pt,
   row sep=30pt
  ]
    &&
    X_1
    \\
    \Gamma
    \ar[
      urr,
      bend left=25,
      "{
          x_1
      }",
      "{
        \scalebox{.7}{
          \bf
          \color{orangeii}
          Such that
          \color{darkblue}
          this diagram commutes...
          \hspace{-2cm}
        }
      }"{yshift=8pt, pos=.8}
    ]
    \ar[
      drr,
      bend right=25,
      "{
        \underset{
          \mathclap{
          \raisebox{3pt}{
            \scalebox{.7}{
              \bf
            }
            }
          }
        }{
          x_2
        }
      }"{swap}
    ]
    \ar[
      rr,
      "{
          (x_1, x_2)
      }",
      "{
          \raisebox{3pt}{
            \scalebox{.7}{
              \bf
              \color{orangeii}
              ...
              for exactly
              \color{darkblue}
              this map.
            }
            }
      }"{swap, pos=.6, yshift=-3pt}
    ]
    &&
    X_1 \times X_2
    \ar[u, "{\mathrm{pr}_1}"{swap}]
    \ar[d, "{\mathrm{pr}_2}"{swap}]
    \\
    &&
    X_2
  \end{tikzcd}

  \\\hline
\end{tabular}
}
\end{equation}

Here and henceforth, we use the following
traditional notation for {\it syntactic rules} of the programming language:
\begin{itemize}
\item A horizontal line denotes a {\it natural deduction} rule
(\cite{Gentzen34}, see \cite{Szabo69})
for passing from the {\it judgement} \eqref{FunctionDeclaration}
above to that below the line. By such rules, valid typed programs are incrementally formed (e.g. \cite{DownenAriola18}).

\item The symbol ``$\defneq$'' relates terms which are regarded as {\it syntactically equal} (``{\it definitional equality}'', e.g. \cite[\S 10.1]{Chlipala07}\cite[\S 5.2.1]{Thompson91}\cite[p. 19]{UFP13}), to be distinguished from ``{\it identification}'' \eqref{IdentificationType}
and ``{\it propositional equality}'' \eqref{PropositionalEquality}.

\end{itemize}

\vfill

\newpage

\medskip

\noindent
{\bf The idea of dependently typed programming languages.}
In general, not only the data itself but also its type may {\it depend} on the given data context (see \cite{Hofmann97}\cite[\S 1.2.2]{Chlipala13}).
For example, depending on a natural
number $d : \mathbb{N}$ previously computed,
a (quantum state-)vector $\psi_d$ might be specifically declared to be of the type $\mathbb{C}^d$ of elements of the $d$-dimensional
complex Hilbert space. Here these finite-dimensional Hilbert spaces jointly form a {\it dependent type}, namely depending on data of
the type of natural numbers (this being their dimension).

\smallskip
Trivial as this may superficially seem at this syntactic level (it is not, though),
a programming language that computes such $\psi_d$, while adhering to the principle of its dependent typing, has a conceptually
interesting semantics, namely the function
\eqref{FunctionDeclaration}
that it computes is now a {\it section of a fibration}:

\begin{center}
\def\arraystretch{1.3}
\begin{tabular}{|c||c|}
\hline
\begin{minipage}{7cm}
{\bf Programming language} (syntax)
\end{minipage}
&
\begin{minipage}{7cm}
{\bf Mathematical denotation} (semantics)
\end{minipage}
\\
\hline
\hline
$
  \overset{
    \mathclap{
    \raisebox{4pt}{
      \scalebox{.8}{
        \color{orangeii}
        \bf
        Given...
      }
    }
    }
  }{
    \underset{
      \mathclap{
        \raisebox{-2pt}{
          \rotatebox{-40}{
            \scalebox{.75}{
              \rlap{
                \color{darkblue}
                \bf
                \hspace{-15pt}
                anything
              }
            }
          }
        }
      }
    }{
      \Gamma
    }
    ,\;\;
    \underset{
      \mathclap{
        \raisebox{-2pt}{
          \rotatebox{-40}{
            \scalebox{.75}{
              \rlap{
                \color{darkblue}
                \bf
                \hspace{-15pt}
                a nat. number
              }
            }
          }
        }
      }
    }{
      d : \mathbb{N}
    }
  }
  \qquad \vdash\qquad
  \overset{
    \mathclap{
    \raisebox{3pt}{
      \scalebox{.8}{
      \color{orangeii}
      \bf
      ...construct
      }
    }
    }
  }{
    \underset{
      \mathclap{
        \hspace{11pt}
        \raisebox{-2pt}{
          \rotatebox{-40}{
            \scalebox{.75}{
              \rlap{
                \color{greenii}
                \bf
                \hspace{-15pt}
                a vector
              }
            }
          }
        }
      }
    }{
      \psi_d
    }
    \underset{
      \mathclap{
        \hspace{11pt}
        \raisebox{-2pt}{
          \rotatebox{-40}{
            \scalebox{.75}{
              \rlap{
                \color{darkblue}
                \bf
                \hspace{-12pt}
                in
              }
            }
          }
        }
      }
    }{
      \,:\,
    }
    \underset{
      \mathclap{
        \hspace{11pt}
        \raisebox{-2pt}{
          \rotatebox{-40}{
            \scalebox{.75}{
              \rlap{
                \color{darkblue}
                \bf
                \hspace{-15pt}
                  $d$-dim space
              }
            }
          }
        }
      }
    }{
      \mathbb{C}^d
    }
  }
$
&
\begin{tikzcd}[column sep=huge]
  &
  \mathcal{V}
  \mathrlap{
    \;
    =
    \Gamma
      \times
    \underset{
      \mathclap{
        d \in \mathbb{N}
      }
    }{\coprod}
    \,
    \mathbb{C}^d
  }
  \ar[
    d,
    ->>,
    shorten <= -8pt,
    "{
      \scalebox{.8}{
        \color{darkblue}
        \bf
        \;\;\;\;\;\;fibration
      }
    }"{description, pos=.1}
  ]
  \\
  \Gamma \times \mathbb{N}
  \ar[r, Rightarrow, -]
  \ar[
    ur,
    dashed,
    "{\psi_{(-)}}",
    "{
      \scalebox{.75}{
        \color{greenii}
        \bf
        section
      }
    }"{swap, sloped}
  ]
  &
  \Gamma \times \mathbb{N}
  \mathrlap{
    \scalebox{.8}{
      \color{darkblue}
      \bf
      \; base
    }
  }
\end{tikzcd}
\\
&
\\
\hline
\end{tabular}
\end{center}

\noindent
Therefore the denotational semantics of dependent data types is that of fibrations (``bundles'', often called ``display maps''
in this context, see e.g. \cite[\S 10]{Jacobs98}\cite{Joyal17}), such as known from homotopy theory (e.g. \cite[\S 2, \S 3]{Shulman15}).

\vspace{-1mm}
\begin{equation}
\label{DependentData}
\adjustbox{}{
\def\arraystretch{1}
\def\arraystretch{1.3}
\begin{tabular}{|c||c|}
\hline
{\bf Dependent data types and dependent data}
&
{\bf Iterated fibrations and their relative sections}
\\
\hline
\hline
&
\\[-14pt]
$
  \def\arraystretch{1.4}
 \begin{array}{cccccc}
   \Gamma\mathrlap{,}
   &
   \overset{
     \clap{
       \raisebox{4pt}{
       \scalebox{.8}{
         \bf
         context type
       }
       }
     }
   }{
     x_c : X_c
   }
   &\qquad \vdash& \quad
   \overset{
     \clap{
     \raisebox{3pt}{
       \scalebox{.8}{
         \color{darkblue}
         \bf
         dependent type
       }
     }
     }
   }{
     E_c(x_c) : \Types
   }
   \\
   \Gamma\mathrlap{,}
   &
   x_c : X_c
   &\qquad \vdash& \quad
   \underset{
       \clap{
       {
       \scalebox{.8}{
         \color{greenii}
         \bf
         \begin{tabular}{c}
           dependent term
         \end{tabular}
       }
       }
     }
   }{
     \sigma_c(x_c) : E_c(x_c)
   }
 \end{array}
$
&
\adjustbox{raise=6pt}{
\begin{tikzcd}[
  row sep=25pt,
  column sep=18pt
]
  &&
  \overset{
    \clap{
    \raisebox{3pt}{
      \scalebox{.7}{
        \color{darkblue}
        \bf
        fibration
      }
      }
    }
  }{
  E
  }
  \ar[
    d,
    ->>
  ]
  \\
  X
  \ar[rr, Rightarrow, -]
  \ar[
    urr,
    dashed,
    "\sigma",
    "{
      \scalebox{.8}{
        \color{greenii}
        \bf
        section
      }
    }"{swap, sloped}
  ]
  \ar[
    dr,
    ->>
  ]
  &&
  X
  \ar[
    dl,
    ->>
  ]
  \\[-20pt]
  &
  \Gamma
\end{tikzcd}
}
\\[-13pt]
&
\\
\hline
&
\\[-14pt]
$
  \begin{array}{cccccc}
  \Gamma\mathrlap{,}
  &
  x : X
  &
 \;\; \vdash
  & \;\;
  \sigma_c(x_c)
  :
    \underset{
      \mathclap{
        \hspace{22pt}
        \raisebox{-1pt}{
          \rotatebox{-30}{
            \scalebox{.75}{
              \rlap{
                \color{darkblue}
                \bf
                \hspace{-36pt}
                \def\arraystretch{.7}
                \begin{tabular}{r}
                  \color{black}
                  variable
                  \\
                  \color{orangeii}
                  substitution
                \end{tabular}
              }
            }
          }
        }
      }
    }{
      E_c\big(f_c(x_c)\big)
    }
  \end{array}
$
&
\begin{tikzcd}[
  row sep=34pt,
  column sep=15pt
]
  &&
  E
  \ar[
    d,
    ->>
  ]
  \\
  X
  \ar[rr, "f"{description}]
  \ar[
    dr,
    ->>,
    shorten=-1pt
  ]
  \ar[
    urr,
    dashed,
    "\sigma",
    "{
      \scalebox{.7}{
        \bf
        \color{orangeii}
        relative
        \color{greenii}
        section
      }
    }"{swap, sloped, pos=.55}
  ]
  &&
  Y
  \ar[
    dl,
    shorten =-1pt
  ]
  \\[-20pt]
  &
  \Gamma
\end{tikzcd}
$\;\;\;\Leftrightarrow$
\adjustbox{raise=7pt}{
\begin{tikzcd}[
  row sep=34pt,
  column sep=17pt
]
  &
  \overset{
    \raisebox{4pt}{
      \clap{
      \scalebox{.7}{
        \color{darkblue}
        \bf
        \def\arraystretch{.9}
        \begin{tabular}{c}
          \color{orangeii}
          pullback
          \\
          \color{black}
          fibration
        \end{tabular}
      }
      }
    }
  }{
    f^\ast E
  }
  \ar[r]
  \ar[
    d,
    ->>,
    shorten <=-2pt
  ]
  \ar[
    dr,
    phantom,
    "\scalebox{.7}{(pb)}"{xshift=4pt}
  ]
  &
  E
  \ar[
    d,
    ->>
  ]
  \\
  X
  \ar[dr]
  \ar[r, Rightarrow, -]
  \ar[
    ur,
    dashed,
    "\sigma",
    "{
      \scalebox{.7}{
        \color{greenii}
        \bf
        section
      }
    }"{swap, sloped, pos=.55}
  ]
  &
  X
  \ar[d]
  \ar[r, "f"]
  &
  Y
  \ar[dl]
  \\[-18pt]
  &
  \Gamma
\end{tikzcd}
}
\\[-8pt]
&
\\
\hline
\end{tabular}
}
\end{equation}

\medskip

\noindent
{\bf The idea of aggregating dependent data.}
Given such a dependent data type
$d \,:\, D \;\;\vdash\;\; C_d \,\colon\, \Types$
\eqref{DependentData},
there are two natural ways to ``aggregate'' all the types $C_d$ into a single independent data type:

\begin{itemize}[leftmargin=.85cm]

\item [\eqref{PiInferenceRules}] by forming the {\it dependent product} ${\prod}_{d \colon D} \, C_d$ of all $C_d$ --- such that to give data of this aggregated type is to give data of type $C_d$ {\it for each} $d \,:\, D$, hence to give a {\it dependent function} of the form $(d : D) \to C_d$  (a section of the fibration);

\item[\eqref{DependentSumInference}] by forming the {\it dependent co-product} ${\coprod}_{d \colon D} \, C_d$ of all $C_d$ --- such that to give data of this aggregated type is to give data of type $C_d$ {\it for one} $d \,\colon\, D$, hence to give a {\it pair} of the form $(d : D) \times C_d$ (a point in the fibration's total space):
\end{itemize}

\vspace{-.3cm}
\begin{equation}
\label{DependentFunctionNotation}
\adjustbox{}{
\def\arraystretch{2}
\tabcolsep=3pt
\begin{tabular}{|c|cccccc|}
  \hline
  Given data types:
  &
  the type former:
  &&&&
  gives type of:
  &
  which:
  \\
  \hline
  \hline
  &&&&&&
  \\[-20pt]
  \hspace{-9pt}
  \multirow{2}{*}{
$
\def\arraystretch{1.4}
\arraycolsep=3pt
\begin{array}{rcl}
  \\[-5pt]
  &\vdash& D\; \,:\, \Types
  \\
  d : D &\vdash&  C_d \,:\, \Types
\end{array}
$
  }
  \hspace{-8pt}
  &
  {\bf
  \def\arraystretch{1}
  \begin{tabular}{c}
    dependent
    \\
    product
  \end{tabular}
  }
  &
  $
  \underset{
    d \,:\, D
  }{\prod}
  \,
  C_d
  $
  &
  $\;=\;$
  &
  $
    (d : D)
    \to
    C_d
  $
  &
  \bf
  \def\arraystretch{1}
  \begin{tabular}{c}
    dependent
    \\
    functions
  \end{tabular}
  &
  \def\arraystretch{1}
  \begin{tabular}{c}
    {\it map} data $d$ of type $D$
    \\
    {\it to} data of type $C_d$
  \end{tabular}
  \\[-18pt]
  &&&&&&
  \\
  \cline{2-7}
  &&&&&&
  \\[-20pt]
  &
  \bf
  \def\arraystretch{1}
  \begin{tabular}{c}
    dependent
    \\
    co-product\footnotemark
  \end{tabular}
  &
  $
  \underset{
    d \,:\, D
  }{\coprod}
  C_d
  $
  &
  $=$
  &
  $
    (d : D)
    \;\times\;
    C_d
  $
  &
  \bf
  \def\arraystretch{1}
  \begin{tabular}{c}
    dependent
    \\
    pairs
  \end{tabular}
  &
  \def\arraystretch{1}
  \begin{tabular}{c}
    {\it pair} data $d$ of type $D$
    \\
    {\it with} data of type $C_d$
  \end{tabular}
  \\[-17pt]
  &&&&&&
  \\
  \hline
  \hline
\end{tabular}
}
\end{equation}

\footnotetext{
  The traditional notation for the dependent pair type is some variant of ``$\sum_{d \,\colon\, D} C_d$'', pronounced the ``dependent sum''. While widely used, this is clearly a misnomer --- and it becomes a fatal misnomer once we generalize dependent type theory to dependent linear type theory \cite{QPinLHOTT}, where an actual dependent {\it sum} (in the sense of linear algebra) does appear beside the dependent co-product. Luckily, the alternative pair-type notation $(d : D) \times C_d$ (cf. \cite[p. 1]{Warn22}) not only circumvents this clash but is arguably also more convenient and more suggestive in general --- as seen for instance in \eqref{LCCRules}
  and on p. \pageref{DataStructures} below).
}

\medskip

Concretely, the inference rules for dependent pair types are as follows, in evident dependent generalization of \eqref{PairTypes}:

\vspace{-.3cm}
\begin{equation}
\label{DependentSumInference}
\adjustbox{}{
\hspace{-.6cm}
\hypertarget{SigmaInferenceRules}{}
\noindent
\tabcolsep=1.8pt
\begin{tabular}{|c|}
\hline
\hspace{-3pt}
\adjustbox{scale=1.0}{
$
  \arraycolsep=1pt
  \begin{array}{ll}
  \def\arraystretch{1.7}
  \begin{array}{c}
    \\[-6pt]
    \def\arraystretch{1}
    \arraycolsep=3pt
    \begin{array}{cccccccc}
      \Gamma
      &
      \vdash &
      \overset{
        \mathclap{
        \raisebox{3.3pt}{
          \hspace{-6pt}
          \scalebox{.8}{
            \bf
            \begin{tabular}{c}
              \color{orangeii}
              Given
              \color{darkblue}
              one data type...
            \end{tabular}
          }
        }
        }
      }{
        I : \Types
      }
      &\phantom{-}&
       \Gamma\mathrlap{,}
       &
       i : I
      &
      \overset{
        \mathclap{
        \;\;\;\;\;\;\;\;\;\;
        \raisebox{3pt}{
          \scalebox{.8}{
            \color{darkblue}
            \bf
            \begin{tabular}{c}
              indexing another one...
            \end{tabular}
          }
        }
        }
      }{
        \vdash
      }
      &
      X_i : \Types
    \end{array}
    \\
    \hline
    \arraycolsep=4pt
    \begin{array}{ccc}
      \Gamma
      &
      \vdash
      &
      \underset{
        \mathclap{
        \raisebox{-7pt}{
          \scalebox{.8}{
            \color{orangeii}
            \bf
            \def\arraystretch{.9}
            \begin{tabular}{c}
              \color{orangeii}
              ...we infer
              \color{darkblue}
              the type of
              \\
              \color{darkblue}
              $X_i$ data
              paired with its index $i : I$.
            \end{tabular}
          }
        }
        }
      }{
        \,\dsum{i : I} X_i
        \;:\;
        \Types
      }
    \end{array}
  \end{array}
  \hspace{-5pt}
  \raisebox{0pt}{
  \scalebox{.8}{
    \color{gray}
    \bf
    \def\arraystretch{.8}
    \begin{tabular}{c}
      Dependent pair
      \\
      formation rule
    \end{tabular}
  }
  }
  &
  \hspace{.5cm}
  \def\arraystretch{1.4}
  \begin{array}{c}
    \arraycolsep=4pt
    \def\arraystretch{1}
    \begin{array}{cccccccc}
      \Gamma
      &
        \overset{
        \mathclap{
        \hspace{.3cm}
        \raisebox{8pt}{
        \scalebox{.8}{
        \color{orangeii}
        \bf
        \def\arraystretch{.8}
        \begin{tabular}{c}
          Given \color{darkblue}
          a program
          \\
          \color{darkblue}
          which computes $I$-data...
        \end{tabular}
        }
        }
        }
        }{
        \vdash
        }
      &
        i : I
      &
        \phantom{---}
      &
      \Gamma
      &
      \overset{
        \mathclap{
        \hspace{.7cm}
        \raisebox{8pt}{
          \scalebox{.8}{
            \color{orangeii}
            \bf
            \def\arraystretch{.8}
            \begin{tabular}{c}
              \color{darkblue}
              ... and a program which
              \\
              \color{darkblue}
              computes $X_i$-data...
            \end{tabular}
          }
        }
        }
      }{
        \vdash
      }
      &
      x_i : X_i
    \end{array}
    \\
    \hline
    \arraycolsep=5pt
    \begin{array}{cccc}
      \Gamma
      &
      \underset{
        \mathclap{
        \;\;\;\;\;\;\;\;\;\;\;\;\;\;\;\;\;\;\;
        \raisebox{-13pt}{
          \scalebox{.8}{
            \color{orangeii}
            \bf
            \def\arraystretch{.8}
            \begin{tabular}{c}
              \color{orangeii}
              ... we infer
              \color{darkblue}
              a program which
              \\
              \color{darkblue}
              computes
              the pairs $(i,\, x_i)$.
            \end{tabular}
          }
        }
        }
      }{
        \vdash
      }
      &
        (i, x)
      :
        \dsum{i : I} X_i
    \end{array}
  \end{array}
  \hspace{-7pt}
  \raisebox{3pt}{
  \scalebox{.8}{
    \color{gray}
    \bf
    \def\arraystretch{.9}
    \begin{tabular}{c}
      Dependent pair
      \\
      introduction rule
    \end{tabular}
  }
  }
  \\[+50pt]
  \def\arraystretch{1.4}
  \begin{array}{c}
    \arraycolsep=4pt
    \def\arraystretch{1}
    \begin{array}{ccc}
      \Gamma
      &
      \overset{
        \mathclap{
        \;\;\;\;\;\;\;\;\;\;\;
        \hspace{25pt}
        \raisebox{8pt}{
          \scalebox{.8}{
            \color{orangeii}
            \bf
            \def\arraystretch{.8}
            \begin{tabular}{c}
              Given {\color{darkblue}a program which}
              \\
              \color{darkblue}computes pairs...
            \end{tabular}
          }
        }
        }
      }{
        \vdash
      }
      &
      p
      :
        \dsum{i : I} X_i
    \end{array}
    \\
    \hline
    \arraycolsep=4pt
    \begin{array}{ccccccc}
      \Gamma
      &
      {
        \vdash
      }
      &
      \mathrm{pr}_I(p)
      :
      I
    \end{array}
    \\
    \arraycolsep=4pt
    \begin{array}{cccccc}
      \hspace{23pt}
      \Gamma
      &
        \underset{
        \mathclap{
        \;\;\;\;\;\;\;\;\;\;\;\;\;\;\;\;\;\;\;
        \raisebox{-10pt}{
        \scalebox{.8}{
        \bf
        \def\arraystretch{.8}
        \begin{tabular}{c}
          \color{orangeii}
          ... we infer
          \color{darkblue}
          programs which
          \\
          \color{darkblue}
          compute the components...
        \end{tabular}
        }
        }
        }
        }{
        \vdash
        }
      &
        \mathrm{pr}_X(p)
        : X_{\mathrm{pr}_I(p)}
    \end{array}
  \end{array}
  \hspace{-7pt}
  \raisebox{10pt}{
  \scalebox{.8}{
    \color{gray}
    \bf
    \def\arraystretch{.9}
    \begin{tabular}{c}
      Dependent pair
      \\
      elimination rule
    \end{tabular}
  }
  }
  &
  \hspace{-.2cm}
  \def\arraystretch{1.4}
  \begin{array}{c}
    \arraycolsep=4pt
    \def\arraystretch{1}
    \begin{array}{cccccccc}
      \Gamma
      &

      {
        \vdash
      }
      &
        i : I
      &
        \overset{
        \mathclap{
        \raisebox{10pt}{
        \scalebox{.8}{
        \color{orangeii}
        \bf
        \def\arraystretch{.8}
        \begin{tabular}{c}
          \color{orangeii}
          ...such that
          \color{darkblue}
          these are indeed the components
          of the pair...
        \end{tabular}
        }
        }
        }
        }{\quad\quad}
      &
      \Gamma
      &
      \vdash
      & x : X_i
    \end{array}
    \\
    \hline
    \arraycolsep=4pt
    \begin{array}{cccc}
      \Gamma
      &
      \vdash
      &
      {
        \mathrm{pr}_I(i, x)
        \defneq
        i
      }
      \;\;:\;\;
        I
    \end{array}
    \\

    \arraycolsep=4pt
    \begin{array}{cccc}
      \Gamma
      &
        \vdash
      &
        {
        \mathrm{pr}_X(i, x)
        \defneq
       x
        }
        \;\;:\;\;
       X_i
    \end{array}
    \\
    \\
    \arraycolsep=4pt
    \def\arraystretch{1}
    \begin{array}{ccc}
      \Gamma
      &
      \vdash
      & p : \dsum{i : I} X_i
    \end{array}
    \\

    \hline
    \arraycolsep=4pt
    \begin{array}{cccc}
      \Gamma
      &
      \vdash
      &
      \underset{
        \mathclap{
        \;\;\;\;\;\;\;\;\;\;\;\;\;\;\;\;\;\;\;
        \raisebox{-9pt}{
          \scalebox{.8}{
            \bf
            \def\arraystretch{.8}
            \begin{tabular}{c}
              \color{darkblue}
              ...and any pair is the pair of its components.
            \end{tabular}
          }
        }
        }
      }{
        p
          \defneq
        \big(
          \mathrm{pr}_I(p)
          ,\,
          \mathrm{pr}_X(p)
        \big)
      }
      \;\;:\;\;
        \dsum{i : I} X_i
    \end{array}
  \end{array}
  \hspace{-6pt}
  \color{gray}
  \raisebox{40pt}{
  \scalebox{.8}{
    \bf
    \def\arraystretch{.8}
    \begin{tabular}{c}
      \clap{Dependent pair}
      \\
      computation rule
    \end{tabular}
  }
  }
    \raisebox{-30pt}{
    \hspace{-75pt}
    \scalebox{.8}{
    \bf
    \def\arraystretch{.8}
    \begin{tabular}{c}
      \clap{Dependent pair}
      \\
      uniqueness rule
    \end{tabular}
    }
    }
  \end{array}
$
}
\hspace{-1pt}
\\[-8pt]
\\
\hline
\end{tabular}
}
\end{equation}

\vspace{-.3cm}
\begin{equation}
\label{FunctionTypesAndMappingSpaces}
\hspace{-10.5mm}
\adjustbox{}{
\begin{tabular}{ll}
\begin{minipage}{7.3cm}
The adjoint inference rules for dependent function types are shown in \eqref{PiInferenceRules}.
Notice
\eqref{FunctionTypesAndMappingSpaces}
how the term introduction rule \eqref{PiInferenceRules}, when specialized to the case that the various dependencies are trivial, expresses but the subtle yet crucial distinction between a function as such \eqref{FunctionDeclaration} and that function regarded as data {\it of function type} --- the denotational semantics of which is given by {\it mapping space adjunctions} \eqref{MappingSpaceAdjunction}.
\end{minipage}
&
\quad
\def\arraystretch{2}
\begin{tabular}{|l|l|}
\hline
{\bf
 \def\arraystretch{.9}
 \begin{tabular}{c}
   Term introduction
   \\
   of function type
   $\mathclap{\phantom{\vert_{\vert_{\vert}}}}$
 \end{tabular}
}
&
{\bf
 \def\arraystretch{.9}
 \begin{tabular}{c}
   Internal hom
   \\
   adjunction
   $\mathclap{\phantom{\vert_{\vert_{\vert}}}}$
 \end{tabular}
}
\\
\hline
\hline
$
  \def\arraystretch{1.2}
  \begin{array}{c}
    \overset{
      \mathclap{
        \raisebox{4pt}{
        \scalebox{.7}{
          \color{darkblue}
          \bf
          function
        }
        }
      }
    }{
    \Gamma ,\;
    d \,:\, D
    \;\;\;
    \vdash
    \;\;\;
    c_d \,:\, C
    }
    \\
    \hline
    \underset{
      \mathclap{
        \raisebox{-4pt}{
          \scalebox{.7}{
            \color{darkblue}
            \bf
            data of function type
          }
        }
      }
    }{
    \Gamma
    \;\;\; \vdash \;\;\;
    (d \mapsto c_d)
    \,:\,
    D \to C
    }
  \end{array}
$
&
$
\def\arraystretch{1.2}
  \begin{array}{c}
    \overset{
      \mathclap{
        \raisebox{0pt}{
        \scalebox{.7}{
          \color{darkblue}
          \bf
           $\mathclap{\phantom{\vert^{\vert^{\vert}}}}$
          map
        }
        }
      }
    }{
     \Gamma \times D \xrightarrow{\;\;f\;\;} C
    }
    \\
    \hline
    \underset{
      \mathclap{
        \raisebox{-4pt}{
          \scalebox{.7}{
            \color{darkblue}
            \bf
            element of mapping space
            $\mathclap{\phantom{\vert_{\vert_{\vert}}}}$
          }
        }
      }
    }{
      \Gamma \xrightarrow{\;
        \mathclap{\phantom{\vert^{\vert}}}
        \tilde f\;}
        \Maps{}{D}{C}
    }
  \end{array}
$
\\
\hline
\end{tabular}
\end{tabular}
}
\end{equation}

\vspace{.2cm}
\hspace{-.8cm}
\begin{tabular}{ll}
\begin{minipage}{7.4cm}
The semantics of
{\it general} dependent function types $(\delta : \Delta_\gamma) \to (-)$ is
that of forming spaces of relative sections
via the {\it slice mapping space} adjunction \eqref{ExponentialLawForFiberwiseMappingSpace},  hence is
(e.g. \cite[Ex. 2.9]{Shulman15})
the right base change
operation $(p_D)_\ast$ from base $\Delta$ to base $\Gamma$  \eqref{BaseChangeAdjunction} .
\end{minipage}
&
\hspace{1cm}
\adjustbox{fbox}{
\hspace{-6pt}
$
  \begin{tikzcd}[
    column sep=60pt
  ]
  \Types_{\Delta}
  \ar[
    rr,
    shift left=10pt,
    "{
      \raisebox{-.5pt}{\scalebox{1.3}{$($}}
        (\gamma \,:\, \Gamma)
        \;\vdash\;
        (\delta_\gamma
          \,:\,
          \Delta_\gamma)
        \times (-)
      \raisebox{-.5pt}{\scalebox{1.3}{$)$}}
    }"{description}
  ]
  \ar[
    from=rr
  ]
  \ar[
    rr,
    shift right=10pt,
    "{
      \raisebox{-.5pt}{\scalebox{1.3}{$($}}
        (\gamma \,:\, \Gamma)
        \;\vdash\;
        (\delta_\gamma
          \,:\,
          \Delta_\gamma)
        \to (-)
      \raisebox{-.5pt}{\scalebox{1.3}{$)$}}
    }"{description}
  ]
  &&
  \Types_{\Gamma}
  \end{tikzcd}
$
\hspace{-6pt}
}
\end{tabular}

\vspace{0cm}
\begin{equation}
\label{LCCRules}
\hspace{-4mm}
\adjustbox{}{
\def\arraystretch{1.6}
\begin{tabular}{|c|c|}
\hline
{\bf Aggregation of dependent data types}
&
{\bf Base change via Local Cartesian Closure}
\\
\hline
\hline
$
  \begin{array}{lcc}
    {}
    \\[-8pt]
    (\gamma : \Gamma)
    \times
    (\delta : \Delta_\gamma)
    ,\;\;
    (d_{\gamma} : D_{\gamma})
    &\vdash&
    c_{(\delta,\, d_\gamma)}
    :
    C_{\delta}
    \\
    \hline
    (\gamma : \Gamma)
    ,\;\;
    (d_\gamma : D_\gamma)
    &\vdash&
    \adjustbox{raise=-6pt}{$
    \def\arraystretch{1.1}
    \def\arraycolsep{2pt}
    \begin{array}{ccc}
      (\delta : \Delta_\gamma)
      &\to&
      C_{\delta}
      \\
      \delta
      &\mapsto&
      c_{(\Delta_\gamma,\,d_\gamma)}
    \end{array}
    $}
  \end{array}
$
&
$
\begin{array}{c}
  \hspace{-18pt}
  \begin{tikzcd}[
    column sep=25pt,
    row sep=4pt
  ]
    \Delta \times_\Gamma D
    \ar[
      dr,
      shorten=-2pt,
      "{
        p_{\Delta}
          \times
        p_D
      }"{swap, sloped, pos=.4}
    ]
    \ar[
      rr,
      dashed
    ]
    &&
    C
    \ar[
      dl,
      shorten=-2pt,
      "{ p_C }"
    ]
    \\
    &
    \Delta
  \end{tikzcd}
  \\
  \hline
  \begin{tikzcd}[
    column sep=22pt,
    row sep=4pt
  ]
    D
    \ar[
      rr,
      dashed
    ]
    \ar[
      dr,
      shorten=-2pt,
      "{ p_D }"{swap}
    ]
      &&
    \Maps{\big}{\Delta}{p_C}
    \ar[
      dl,
      shorten <=-7pt,
      shorten >=-1pt,
      "{
        (p_{\Delta})_\ast p_C
      }"{swap, sloped, pos=.25}
    ]
    \\
    &
    \Gamma
  \end{tikzcd}
\end{array}
$
\\
\hline
$
  \begin{array}{rcl}
    (\gamma : \Gamma)
    \times
    (\delta : \Delta_\gamma)
    ,\;\;
    (d_\gamma : D_\gamma)
    &\vdash&
    (c_\gamma : C_\Gamma)
    \\
    \hline
    (\gamma : \Gamma)
    ,\;\;
    (\delta : \Delta_\gamma)
    \times
    (d_\gamma : D_\gamma)
    &\vdash&
    (c_\gamma : C_\Gamma)
  \end{array}
$
&
$
\begin{array}{c}
  \begin{tikzcd}[
    row sep=6pt, column sep=large
  ]
    D
    \ar[
      rr,
      shorten=-2pt,
      dashed
    ]
    \ar[
      dr,
      shorten=-2pt,
      "{ p_D }"{swap}
    ]
      &&
    \Delta \times_\Gamma C
    \ar[
      dl,
      shorten=-2pt,
      "{
        p_{\widehat{\Gamma}}
        \times
        p_C
      }"{sloped, pos=.6}
    ]
    \ar[rr]
    \ar[
      dr,
      phantom,
    ]
    &[-40pt]
    &[-0pt]
    C
    \ar[
      dl,
      shorten=-2pt,
      "{ p_C }"{}
    ]
    \\
    &
    \Delta
    \ar[
      rr,
      "{ p_{\Delta} }"{swap}
    ]
    &{}&
    \Gamma
  \end{tikzcd}
  \\
  \hline
  \begin{tikzcd}[
    row sep=6pt
  ]
    D
    \ar[
      rrrr,
      dashed
    ]
    \ar[
      dr,
      shorten=-2pt,
      "{ p_D }"{swap}
    ]
    \ar[
      drrr,
      "{
        (p_{\Delta})_! p_D
      }"{sloped, description},
      end anchor={[yshift=3pt, xshift=2pt]}
    ]
      &&
    \phantom{\Delta \times_\Gamma C}
    \ar[
      dr,
      phantom,
    ]
    &[-40pt]
    &[-0pt]
    C
    \ar[
      dl,
      shorten=-2pt,
      "{ p_C }"{}
    ]
    \\
    &
    \Delta
    \ar[
      rr,
      "{ p_{\Delta} }"{swap}
    ]
    &{}&
    \Gamma
  \end{tikzcd}
\end{array}
$
\\
\hline
\end{tabular}
}
\end{equation}

\newpage
This means (as originally understood by \cite{Seely84}, see also \cite{CurienGarnerHofmann14} and \cite[p. 11]{Shulman15})
that dependent data type systems with dependent function type formation \eqref{PiInferenceRules}
have denotational semantics in  categories of spaces
such as $\kHausdorffSpaces$ \eqref{HausdorffBaseSpace}
where each slice category has a mapping space adjunction, called {\it locally cartesian closed categories} (LCCC), cf. \eqref{InterpretationInTheClassicalModelTopos} below.

\begin{equation}
\label{PiInferenceRules}
\hspace{-6mm}
\adjustbox{}{
\tabcolsep=2.3pt
\begin{tabular}{|c||c|}
\hline
\hspace{-3pt}
\adjustbox{scale=.9}{
$
  \arraycolsep=1pt
  \begin{array}{ll}
  \def\arraystretch{1.7}
  \begin{array}{c}
    \\[-6pt]
    \def\arraystretch{1}
    \arraycolsep=3pt
    \begin{array}{cccccccc}
      \Gamma
      &
      \vdash &
      \overset{
        \mathclap{
        \raisebox{5.5pt}{
          \hspace{-7pt}
          \scalebox{.8}{
            \bf
            \begin{tabular}{c}
              \color{orangeii}
              Given
              \color{darkblue}
              one data type...
            \end{tabular}
          }
        }
        }
      }{
        I : \Types
      }
      &\phantom{-}&
       \Gamma\mathrlap{,}
       &
       i : I
      &
      \overset{
        \mathclap{
        \;\;\;\;\;\;\;\;\;\;
        \raisebox{5pt}{
          \scalebox{.8}{
            \color{darkblue}
            \bf
            \begin{tabular}{c}
              indexing another one...
            \end{tabular}
          }
        }
        }
      }{
        \vdash
      }
      &
      X_i : \Types
    \end{array}
    \\
    \hline
    \\[-22.5pt]
    \arraycolsep=4pt
    \begin{array}{ccc}
      \Gamma
      &
      \vdash
      &
      \underset{
        \mathclap{
        \raisebox{-6pt}{
          \hspace{-11pt}
          \scalebox{.8}{
            \color{orangeii}
            \bf
            \def\arraystretch{.9}
            \begin{tabular}{c}
              \color{orangeii}
              ...we infer
              \color{darkblue}
              the type of
              functions from
              $i : I$ to $X_i$-data.
            \end{tabular}
          }
        }
        }
      }{
        (i : I)
        \to
        X_i
        \;:\;
        \Types
      }
    \end{array}
  \end{array}
  \hspace{-6pt}
  \raisebox{-3pt}{
  \scalebox{.8}{
    \color{gray}
    \bf
    \def\arraystretch{.8}
    \begin{tabular}{c}
      Dependent function
      \\
      type formation rule
    \end{tabular}
  }
  }
  &
  \;\;\;\;
  \def\arraystretch{1.4}
  \begin{array}{c}
    \arraycolsep=4pt
    \def\arraystretch{1}
    \begin{array}{ccccc}
      \Gamma\mathrlap{,}
      &
      i : I
      &
      \overset{
        \mathclap{
        \hspace{-.1cm}
        \raisebox{5pt}{
          \scalebox{.8}{
            \color{orangeii}
            \bf
            \def\arraystretch{.8}
            \begin{tabular}{c}
              \color{orangeii}
              Given
              \color{darkblue}
              $i: I$-indexed $X_i$-data...
            \end{tabular}
          }
        }
        }
      }{
        \vdash
      }
      &
      x_i : X_i
    \end{array}
    \\
    \hline
    \arraycolsep=5pt
    \begin{array}{cccc}
      \Gamma
      &
      \underset{
        \mathclap{
        \hspace{3cm}
        \raisebox{-9pt}{
          \scalebox{.8}{
            \color{orangeii}
            \bf
            \def\arraystretch{.8}
            \begin{tabular}{c}
              \color{orangeii}
              ...we infer
              \color{darkblue}
              a function
              from
              $i : I$ to $X_i$-data.
            \end{tabular}
          }
        }
        }
      }{
        \vdash
      }
      &
      (i \mapsto x_i)
      :
      \dprod{i : I} X_i
    \end{array}
  \end{array}
  \hspace{-6pt}
  \raisebox{4pt}{
  \scalebox{.8}{
    \color{gray}
    \bf
    \def\arraystretch{.9}
    \begin{tabular}{c}
      Dependent function
      \\
      term introduction rule
    \end{tabular}
  }
  }
  \\[+50pt]
  \def\arraystretch{1.4}
  \begin{array}{c}
    \arraycolsep=4pt
    \def\arraystretch{1}
    \begin{array}{cccccccc}
      \Gamma
      &
      \overset{
        \mathllap{
        \hspace{29pt}
        \hspace{25pt}
        \raisebox{8pt}{
          \scalebox{.8}{
            \color{orangeii}
            \bf
            \def\arraystretch{.8}
            \begin{tabular}{l}
              \color{orangeii}
              Given
              \color{darkblue}
              $I$-data...
            \end{tabular}
          }
        }
        }
      }{
        \vdash
      }
      &
      \iota : I
      &\phantom{--}&
      \Gamma
      &
      \overset{
        \mathclap{
        \;\;\;\;\;\;\;\;\;\;\;
        \hspace{25pt}
        \raisebox{8pt}{
          \scalebox{.8}{
            \color{orangeii}
            \bf
            \def\arraystretch{.8}
            \begin{tabular}{c}
              \color{orangeii}
              and
              \color{darkblue}
              a function
              from $i:I$ to
              $X_i$-data...
            \end{tabular}
          }
        }
        }
      }{
        \vdash
      }
      &
      f
      :
      (i : I) \to X_i
    \end{array}
    \\
    \hline
    \arraycolsep=4pt
    \begin{array}{ccccccc}
      \Gamma
      &
      \underset{
        \mathclap{
        \;\;\;\;\;\;\;\;\;\;\;\;\;\;\;\;\;\;\;
        \raisebox{-10pt}{
          \scalebox{.8}{
            \bf
            \def\arraystretch{.8}
            \begin{tabular}{c}
              \color{orangeii}
              ... we infer
              \color{darkblue}
              the function's value data
            \end{tabular}
          }
        }
        }
      }{
        \vdash
      }
      &
      f(\iota)
      : X_\iota
    \end{array}
  \end{array}
  \hspace{-6pt}
  \raisebox{2pt}{
  \scalebox{.8}{
    \color{gray}
    \bf
    \def\arraystretch{.9}
    \begin{tabular}{c}
      Dependent function
      \\
      term elimination rule
    \end{tabular}
  }
  }
  &
  \def\arraystretch{1.4}
  \begin{array}{c}
    \arraycolsep=4pt
    \def\arraystretch{1}
    \begin{array}{ccccccccc}
      \Gamma
      &
      \overset{
        \phantom{
        \mathclap{
        \raisebox{8pt}{
          \scalebox{.8}{
            \color{orangeii}
            \bf
            \def\arraystretch{.8}
            \begin{tabular}{c}
              Given a program which
              \\
              computes data of type $X$...
            \end{tabular}
          }
        }
        }
        }
      }{
        \vdash
      }
      &
        i : I
      &\overset{
        \mathclap{
        \;\;\;\;\;\;\;\;\;\;\;\;\;\;\;\;\;\;\;
        \raisebox{9pt}{
        \scalebox{.8}{
        \bf
        \def\arraystretch{.8}
        \begin{tabular}{c}
          \color{orangeii}
          ...such that
          \color{darkblue}
          functions of indexed data
          evaluate to that data
        \end{tabular}
        }
        }
        }
        }
      {\phantom{-}}
      &
      \Gamma\mathrlap{,}
      &
      i : I
      &
      \vdash
      & x_i : X_i
    \end{array}
    \\
    \hline
    \arraycolsep=4pt
    \begin{array}{cccc}
      \Gamma
      &
      \;\vdash\;
      &
      {
        (
          i \mapsto x_i
        )(\iota)
        \defneq
        x_\iota
      }
      \;\;:\;\;
      X_\iota
    \end{array}
    \\[+16pt]
    \begin{array}{ccc}
      \Gamma
      &
      \;\;\;\;\vdash\;\;\;\;
      &
      f \,:\, (i : I) \to X_i
    \end{array}
    \\
    \hline
    \arraycolsep=4pt
    \begin{array}{cccc}
      \Gamma
      &
      \vdash
      &
      \underset{
        \mathclap{
        \;\;\;\;\;\;\;\;\;\;\;\;\;\;\;\;\;\;\;
        \raisebox{-9pt}{
          \scalebox{.8}{
            \bf
            \def\arraystretch{.8}
            \begin{tabular}{c}
              \color{orangeii}
              ... and
              \color{darkblue}
              are determined by these values.
            \end{tabular}
          }
        }
        }
      }{
        f
          \,\defneq\,
        \raisebox{-.5pt}{\scalebox{1.2}{$($}}
          i \mapsto f(i)
        \raisebox{-.5pt}{\scalebox{1.2}{$)$}}
      }
      \;\;:\;\;
        \dprod{i : I} X_i
    \end{array}
  \end{array}
  \hspace{-4pt}
  \color{gray}
  \raisebox{26pt}{
  \scalebox{.8}{
    \bf
    \def\arraystretch{.8}
    \begin{tabular}{c}
      \clap{Dependent function}
      \\
      computation rule
    \end{tabular}
  }
  }
    \raisebox{-25pt}{
    \hspace{-72pt}
    \scalebox{.8}{
    \bf
    \def\arraystretch{.8}
    \begin{tabular}{c}
      \clap{Dependent function}
      \\
      uniqueness rule
    \end{tabular}
    }
    }\phantom{-}
  \end{array}
$
}
\hspace{-10pt}
&
\rotatebox{-90}{
  \clap{
  \bf Inference rules for dependent function types
  }
}
\\[-8pt]
&
\\[+4pt]
\hline
\hline
&
\multirow{4}{*}{
\rotatebox{-90}{
  \rlap{
  \hspace{-24pt}
  \bf
  Universal properties of rel. mapping spaces
  }
}
}
\\[-8pt]
$
  \def\arraystretch{1}
  \begin{array}{c}
  \begin{tikzcd}[sep=34]
    &&
    \overset{
      \mathclap{
      \raisebox{4pt}{
        \scalebox{.8}{
          \color{darkblue}
          \bf
          fibration
        }
      }
      }
    }{
    X
    }
    \ar[
      d,
      ->>
    ]
    \\
    I
    \ar[dr]
    \ar[rr, Rightarrow, -]
    \ar[
      urr,
      "{
        \sigma
        \;:\;
        i
        \;\mapsto\;
        x_i
      }"{sloped},
      "{
        \scalebox{.75}{
          \color{greenii}
          \bf
          section
        }
      }"{swap, sloped}
    ]
    &&
    I
    \ar[dl]
    \\[-24pt]
    &
    \Gamma
  \end{tikzcd}
  \\
  \\[-6pt]
  \hline
  \\[-6pt]
  \begin{tikzcd}
    &&
    \underset{c \in \Gamma}{\coprod}
    \;
    \overset{
      \mathclap{
      \raisebox{6pt}{
        \scalebox{.75}{
          \color{darkblue}
          \bf
          \def\arraystretch{.9}
          \begin{tabular}{c}
            slice mapping space
            \\
            = space of sections
          \end{tabular}
        }
      }
      }
    }{
      \mathrm{Map}(I_c,X_c)_{/I_c}
    }
    \ar[d]
    \\
    \Gamma
    \ar[
      urr,
      "{
        c
        \;\mapsto\;
        \sigma_c
      }"{sloped}
    ]
    \ar[rr, Rightarrow, -]
    &&
    \Gamma
  \end{tikzcd}
  \end{array}
  \hspace{-8pt}
  \raisebox{-10pt}{
  \scalebox{.8}{
    \color{gray}
    \bf
    \def\arraystretch{.8}
    \begin{tabular}{c}
      right base change
      \\
      adjunction
    \end{tabular}
  }
  }
$
\hspace{.7cm}
$
  \begin{tikzcd}
    \underset{
      \Gamma
    }{\coprod}
    \;
    I_c
      \times
    \mathrm{Map}(I_c, X_c)_{/I_c}
    \ar[
      rr,
      "{
        \underset{\Gamma}{\coprod}
        \;
        \mathrm{ev}
      }",
      "{
        \scalebox{.7}{
          \color{greenii}
          \bf
          adjunction unit
        }
      }"{swap}
    ]
    \ar[dr]
    &&
    \underset{
      \Gamma
    }{\coprod}
    \;
    I_c
    \ar[dl]
    \\[-5pt]
    &
    \Gamma
  \end{tikzcd}
$
&
\\[-6pt]
&
\\
\hline
\end{tabular}
}
\end{equation}

\medskip

\noindent
{\bf The data type of dependent data types.}
For example, the application of the function data introduction rule \eqref{FunctionTypesAndMappingSpaces} to the type universe \eqref{TypeOfSmallTypes} shows that $D$-dependent data types \eqref{DependentData} are the same as functions taking $D$-data to data types \eqref{TypeOfSmallTypes}:
\begin{equation}
  \label{DependentDataTypesAsFunctions}
  \begin{array}{c}
    \overset{
      \mathclap{
        \raisebox{4pt}{
        \scalebox{.7}{
          \color{darkblue}
          \bf
          $D$-dependent data type
        }
        }
      }
    }{
    d \,:\, D
    \;\;\;
    \vdash
    \;\;\;
    C_d \,:\, \Types
    }
    \\
    \hline
    \underset{
      \mathclap{
        \raisebox{-4pt}{
          \scalebox{.7}{
            \color{darkblue}
            \bf
            function data into type universe
          }
        }
      }
    }{
    \;\;\; \vdash \;\;\;
    (d \mapsto C_d)
    \,:\,
    D \to \Types
    \mathclap{\phantom{\vert^{\vert^{\vert}}}}
    }
  \end{array}
\end{equation}
Therefore it makes sense to define the (large) {\it type of $D$-dependent types } as
\begin{equation}
  \label{TypeOfSmallDependentTypes}
  \Types_D
  \;\;
    :\defneq
  \;\;
  \big(
    D \to \Types
  \big)
  \,.
\end{equation}

\newpage

\noindent
{\bf The idea of homotopy-typed programming languages.}
Taking seriously the idea that all data should be (dependently) typed suggests that the same should be true for {\it data identifications}:
A certificate $p$ which verifies that two data structures
$d_1, d_2 \,:\, D$
are identifiable -- namely that they are computationally equivalent given the defining nature of their type $D$ --  should itself be data of the {\it type of identifications} of $D$-data;
and every datum should be certified to be identified with itself:
\begin{equation}
\label{IdentificationType}
\adjustbox{scale=1}{
\def\arraystretch{1.3}
\begin{tabular}{|c||c|}
\hline
{\bf Type of identifications}
&
{\bf Path space fibration}
\\
\hline
\hline
\;\; $
  \overset{
    \raisebox{3pt}{
      \scalebox{.8}{
        \color{orangeii}
        \bf
        Given...
      }
    }
  }{
    \underset{
      \mathclap{
        \raisebox{-2pt}{
          \rotatebox{-30}{
            \scalebox{.75}{
              \rlap{
                \color{darkblue}
                \bf
                \hspace{-20pt}
                \def\arraystretch{.8}
                \begin{tabular}{c}
                  any data
                \end{tabular}
              }
            }
          }
        }
      }
    }{
      \Gamma,
    }\;\;\;\;
   }
    \underset{
      \mathclap{
        \raisebox{-2pt}{
          \rotatebox{-30}{
            \scalebox{.75}{
              \rlap{
                \color{darkblue}
                \bf
                \hspace{-22pt}
                \def\arraystretch{.8}
                \begin{tabular}{r}
                  a data type
                  \\
                  in this context
                \end{tabular}
              }
            }
          }
        }
      }
    }{
     X : \Types,
    }
  \;\;\;\;\;
    \underset{
      \mathclap{
        \raisebox{-2pt}{
          \rotatebox{-30}{
            \scalebox{.75}{
              \rlap{
                \color{darkblue}
                \bf
                \hspace{-25pt}
                \def\arraystretch{.8}
                \begin{tabular}{r}
                  a pair of\;\;\;\;\;\;\;
                  \\
                  data of this type
                \end{tabular}
              }
            }
          }
        }
      }
    }{
    (x_1,  x_2)
    :
    X \times X
  }
  \;\;
  \;\;\;\;
  \vdash
  \;\;\;\;
  \overset{
    \mathclap{
    \raisebox{3pt}{
      \scalebox{.8}{
      \color{orangeii}
      \bf
       ...we have
      }
    }
    }
  }{
    \underset{
      \mathclap{
        \raisebox{-2pt}{
          \rotatebox{-30}{
            \scalebox{.75}{
              \rlap{
                \color{darkblue}
                \bf
                \hspace{-25pt}
                \def\arraystretch{.8}
                \begin{tabular}{c}
                  the data type
                  \\
                  of identifications
                  \\
                  of these terms
                \end{tabular}
              }
            }
          }
        }
      }
    }{
  \mathrm{Id}_X(x_1, x_2)
  :
  \Types
  }
  }
$
\;\;\;\;\;\;
&
\adjustbox{raise=-3pt}{
\begin{tikzcd}
  \overset{
    \mathclap{
    \raisebox{3pt}{
      \scalebox{.8}{
        \def\arraystretch{.8}
        \begin{tabular}{c}
          \color{darkblue}
          \bf
          path space
          \\
          (mapping space
          \\
          out of interval)
        \end{tabular}
      }
      }
    }
  }{
    X^I :\defneq \mathrm{Map}(I,X)
  }
  \ar[
    d,
    ->>,
    "{(\mathrm{ev}_0, \mathrm{ev}_1)}"{swap}
  ]
  &[-20pt]
  \overset{
    \mathclap{
    \raisebox{3pt}{
      \scalebox{.8}{
        \color{darkblue}
        \bf
        \def\arraystretch{.8}
        \begin{tabular}{c}
          interval
        \end{tabular}
      }
      }
    }
  }{
    I
  }
  \\
  X \times X
  &
  \{0,1\}
  \ar[u, hook]
\end{tikzcd}
}
\;\;\;\;
\\[-5pt]
&
\\
\hline
\end{tabular}
}
\end{equation}

\noindent
{\bf (I)}
{\it All data is reflexively identified with itself.}
\;\;\;
\vspace{-2mm}
\begin{equation}
\label{CanonicalIdentification}
\hspace{-4mm}
\adjustbox{scale=1}{\
\def\arraystretch{1.3}
\begin{tabular}{|c||c|}
\hline
\;{\bf Reflexivity certificates}\;
&
{\bf
Constant paths
sectioning the path fibration}
\\
\hline
\hline
&
\\[-7pt]
\;
$
  \Gamma,
  \;\,
  \overset{
    \mathrlap{
      \raisebox{4pt}{
        \scalebox{.8}{
          \bf
          \color{orangeii}
          Given...
        }
      }
    }
  }{
    \underset{
      \mathrlap{
        \raisebox{-4pt}{
          \scalebox{.75}{
            \color{darkblue}
            \bf
            some data
          }
        }
      }
    }{
      x : X
    }
  }
  \;\;\;\;
  \vdash
  \;\;\;\;
  \overset{
    \mathclap{
      \raisebox{4pt}{
        \scalebox{.8}{
          \bf
          \color{orangeii}
          ...we obtain
        }
      }
    }
  }{
  \underset{
    \mathclap{
      \raisebox{-6pt}{
        \scalebox{.75}{
          \bf
          \color{darkblue}
          \def\arraystretch{.8}
          \begin{tabular}{c}
            a certificate of its trivial
            \\
            self-identification.
          \end{tabular}
        }
      }
    }
  }{
    \mathrm{id}_X(x)
    :
    \mathrm{Id}_X(x,x)
  }
  }
  \!\!\!\!\!\!
$
\;\;\;
&
\!\!\!
$
  \begin{tikzcd}
    &
    \mathrm{diag}^\ast X^I
    \ar[rr]
    \ar[d]
    \ar[
      drr,
      phantom,
      "{\scalebox{.7}{(pb)}}"
    ]
    &&
    \overset{
      \mathclap{
        \raisebox{4pt}{
          \scalebox{.8}{
            \color{darkblue}
            \bf
            path fibration
          }
        }
      }
    }{
      X^I
    }
    \ar[d, "\in \Fibrations"]
    \\
    X
    \ar[r, Rightarrow, -]
    \ar[
      ur,
      dashed,
      "\mathrm{id}_X"{sloped}
      ]
    &
    X
    \ar[
      rr,
      shorten=-2pt,
      "{
        \mathrm{diag}
        \,:\,
        x \,\mapsto\,
        (x,x)
      }"{swap, yshift=-1pt},
      "{
        \scalebox{.75}{
          \color{greenii}
          \bf
          diagonal map
        }
      }"{yshift=-1pt}
    ]
    &&
    X \times X
  \end{tikzcd}
  \hspace{2pt}
  \Leftrightarrow
  \hspace{-2pt}
  \begin{tikzcd}[column sep=large]
    &&
    \overset{
      \mathclap{
      \raisebox{4pt}{
        \scalebox{.8}{
          \color{darkblue}
          \bf
          path fibration
        }
      }
      }
    }{
      X^I
    }
    \ar[
      d,
      shorten <=-3pt,
      "{\in \Fibrations}"
    ]
    \\
    X
    \ar[
      rr,
      "{
        \mathrm{diag}
        \,:\,
        x \,\mapsto\, (x,x)
      }"{swap}
    ]
    \ar[
      urr,
      shorten >= -2pt,
      dashed,
      "{
        x \,\mapsto\,
        \mathrm{const}_x
      }"{yshift=-1pt, pos=.6, swap, sloped},
      "{
        \scalebox{.75}{
          \color{greenii}
          \bf
          \def\arraystretch{.8}
          \begin{tabular}{c}
            assign
            \\
            constant paths
          \end{tabular}
        }
      }"{sloped}
    ]
    &&
    X \times X
  \end{tikzcd}
$
\;
\\
\hline
\end{tabular}
}
\end{equation}

\noindent
As indicated on the right, the foundational insight of  {\it homotopy type theory} is that the semantics of such $\mathrm{Id}$-types is that of {\it path space fibrations} in algebraic topology (\cite{AwodeyWarren09}). One may understand this in two equivalent ways:

\medskip
\noindent
{\bf Inductive notion of identification.} Naively, reflexive self-identifications \eqref{CanonicalIdentification} should freely generate the identification types \eqref{IdentificationType}. Expressed operationally this should mean that a (dependent) function out of an identification type is given as soon as it is specified on these reflexivity certificates. This {\it inductive} definition (cf. pp. \pageref{InductiveTypes} below) of identifications is the profound insight of \cite[\S 1.7 and p. 94]{MartinLof75}, who labeled this inference rule ``J'' under which name it became widely known:

\smallskip

\noindent
{\bf (J)} \;$\Leftrightarrow$\footnote{
In type theory literature, the $\mathrm{Id}$-induction J-rule \eqref{PathInduction} is traditionally postulated directly (going back to \cite[\S 1.7 and p. 94]{MartinLof75}\cite[\S 8.1]{NPS90} in the general context of inductive data types, cf. pp. \pageref{InductiveTypes}).
Its equivalence to the combination of
``transport'' \eqref{TransportRule} with ''reversal'' \eqref{IdReversal} was expressed in \cite{Coquand11} and further amplified in \cite{LadymanPresnell15}; a detailed proof is spelled out in \cite[\S 4]{Goetz18}. The implication (J) $\Rightarrow$ (II) has an evident meaning: The J-rule is the application of the transport rule (IIa) to just those identifications of identifications given by the uniqueness rule (IIb) }\; (IIa) and (IIb)
\;\;\;
{
\color{gray}
\scalebox{.9}{
(Martin-L{\"o}f, Coquand)
}
}

\vspace{-.5cm}

\begin{equation}
\label{PathInduction}
\hspace{-.2cm}
\adjustbox{scale=1}{
\def\arraystretch{1.4}
\begin{tabular}{|c||c|}
\hline
{\bf Induction principle for identification certificates}
&
\;{\bf Constant path map lifts against fibrations}\;
\\
\hline
\hline
&
\\[-10pt]
$
  \arraycolsep=6pt
    \def\arraystretch{1.5}
    \begin{array}{cccl}
    \overset{
      \rlap{
        \raisebox{3pt}{
          \scalebox{.8}{
          \bf
          \color{orangeii}
            Given...
          }
        }
      }
    }{
      x_1 : X,\, x_2 : X,
    }
    &
    \overset{
      \clap{
        \raisebox{3pt}{
          \scalebox{.8}{
          \bf
          \color{darkblue}
          ...an identification-dependent...
          }
        }
      }
    }{
      p_{12} :
      \mathrm{Id}_X(x_1, x_2)
    }
    &
    \vdash
    &
    \overset{
      \clap{
        \raisebox{3pt}{
          \scalebox{.8}{
          \bf
          \color{darkblue}
          ...data type...
          }
        }
      }
    }{
      E(x_1,x_2,p_{12})
      :
      \Types
    }
    \\
    \;\;\,
    \underset{
      \mathclap{
      \raisebox{-4pt}{
        \scalebox{.8}{
          \color{orangeii}
          \bf
          ...and...
        }
        }
      }
    }{
      x : X
    }
    &
    &
    \vdash
    &
    \underset{
      \mathclap{
      \raisebox{-2pt}{
        \scalebox{.8}{
          \color{darkblue}
          \bf
          \def\arraystretch{.8}
          \begin{tabular}{c}
            ...data for all
            \\
            reflexive identifications...
          \end{tabular}
        }
        }
      }
    }{
      \sigma :
      E\big(x,x,\mathrm{id}_X(x)\big)
    }
    \\
    \hline
    \\[-10pt]
    \underset{
      \mathclap{
      \raisebox{-3pt}{
        \scalebox{.8}{
          \color{orangeii}
          \bf
          ...obtain...
        }
        }
      }
    }{
      x_1 : X,\,x_2 : X,
    }
    &
    \underset{
      \mathclap{
      \raisebox{-3pt}{
        \scalebox{.8}{
          \color{darkblue}
          \bf
          ...for any identification...
        }
        }
      }
    }{
      p_{12} :
      \mathrm{Id}_X(x_1, x_2)
    }
    &
    \vdash
    &
    \underset{
      \mathclap{
      \raisebox{-3pt}{
        \scalebox{.8}{
          \color{darkblue}
          \bf
          ...compatible data.
        }
        }
      }
    }{
      \widehat{\sigma} : E(x_1,x_2,p_{12})
    }
    \\[-3pt]
    &&&
    \mathclap{
    \scalebox{.8}{
      \llap{such that}
    $
      \hspace{8pt}
      \widehat{\sigma}
      \big(
        \mathrm{id}_X(x)
      \big)
      \defneq
      {\sigma}
      \big(
        \mathrm{id}_X(x)
      \big)
    $}
    }
    \end{array}
$
&
\hspace{-.3cm}
$
  \begin{tikzcd}[row sep=34pt, column sep=30pt]
    X
    \ar[d, "\mathrm{id}_X"{description}]
    \ar[r, "\sigma"]
    &
    E
    \ar[r, Rightarrow, -, shorten=-1pt]
    \ar[d, ->>]
    \ar[drr, phantom, "\scalebox{.7}{(pb)}"{xshift=3pt}]
    &[-30pt]
    f^\ast E'
    \ar[r]
    &[-8pt]
    E'
    \ar[d, ->>]
    \\
    X^I
    \ar[r, Rightarrow, -]
    \ar[
      ur,
      dashed,
      "\widehat{\sigma}",
      "{
        \scalebox{.8}{
          \color{greenii}
          \bf
          lift
        }
      }"{sloped, swap}
    ]
    &
    X^I
    \ar[rr, "f"]
    &&
    B
  \end{tikzcd}
$
\hspace{-.2cm}
\\
\hline
\end{tabular}
}
\end{equation}

\medskip

Remarkably, as indicated on the right of \eqref{PathInduction}, the semantics of the $\mathrm{Id}$-induction rule  is (\cite[\S 3.4]{AwodeyWarren09}, review in \cite[3, pp. 26]{Shulman12})
the {\it lifting property} that characterizes `very good path space objects'' in abstract homotopy theory (\cite[\S 4.12]{DwyerSpalinski95}):

Notice that naive interpretation of the reflexive self-identifications in $\mathrm{Id}_X(x_1,\, x_2)$ would be the diagonal inclusion $\mathrm{diag} : X \to X \times X$ of the pairs of equal elements $(x, x)$; but this map is not in general a fibration; and in homotopy theory demands that it be {\it resolved} (up to weak equivalence) by a path space fibration $X^I \twoheadrightarrow X$, which also makes the resulting inclusion of reflexive self-identifications $\mathrm{id}_X : X \to X^I$ an ``acyclic cofibration'', thus implying the lifting property on the right above.



\newpage
\noindent
{\bf Identification as transport.}
Much older than this inductive understanding of identification is the characterization of identifications as those processes which {\it preserve all properties} (``salva veritate'', Leibniz $\sim 1700$, cf. \cite[p. 373]{Lewis1918}). Understood as: {\it preserve all dependent data} this is the following type-theoretic rule --
which is in fact implied (cf. \cite[\S 2.3]{UFP13})
by $\mathrm{Id}$-induction \eqref{PathInduction}
and which has the striking semantic interpretation of
{\it path lifting}
and
{\it fiber transport}
(Lit. \ref{PathLiftingLiterature}):

\vspace{.2cm}

\noindent
{\bf (IIa)}
{\it Substitution of identifications preserves computations.}

\vspace{-.3cm}

\begin{equation}
\label{TransportRule}
\adjustbox{scale=1}{
\def\arraystretch{1.4}
\begin{tabular}{|c||c|}
\hline
{\bf Transport of data along identifications of variables}
&
{\bf Fiber transport in fibrations }
\\
\hline
\hline
&
\\[-12pt]
$
  \;\;
  \def\arraystretch{1}
  \begin{array}{c}
  \arraycolsep=4pt
  \def\arraystretch{1}
  \begin{array}{ccccccccc}
    \Gamma &\vdash&
    X : \Types
    &\phantom{-}&
    \Gamma\mathrlap{,}
    &
    x : X
    &
    \overset{
      \mathclap{
        \raisebox{4pt}{
          \scalebox{.8}{
            \bf
            \color{orangeii}
            Given \color{orangeii} an
            \color{black}$X$-\color{darkblue}
            dependent data type...
          }
        }
      }
    }{
      \vdash
    }
    &
    E_x : \Types
  \end{array}
  \\[+2pt]
  \hline
  \\[-8pt]
  \arraycolsep=4pt
  \begin{array}{ccccccccc}
    \Gamma\mathrlap{,}
    &
    \underset{
      \mathclap{
        \hspace{40pt}
        \raisebox{-6pt}{
          \scalebox{.8}{
            \bf
            \color{orangeii}...and
            \color{darkblue}
            an identification of
            $X$-data...
          }
        }
      }
    }{
      x_1, x_2 : X
    }
    \mathrlap{,}
    &
    p_{12}
      :
    \mathrm{Id}_X(x_1, x_2)
    &\vdash&
    \underset{
      \mathclap{
        \raisebox{-8pt}{
          \scalebox{.8}{
            \bf
            \color{orangeii}
            \def\arraystretch{.8}
            \begin{tabular}{c}
              ...obtain
              \color{darkblue} transformation
              \\
              \color{darkblue}
              of all dependent data.
            \end{tabular}
          }
        }
      }
    }{
      (p_{12})_\ast
      :
      E_{x_1} \to E_{x_2}
    }
    \\[+2pt]
    &&&&
    \mathclap{
      \scalebox{.8}{
      \llap{
        such that
      }
      $
      \mathrm{id}_X(x)_\ast
      :
      e \,\mapsto\,e
      $
      }
    }
  \end{array}
  \end{array}
  \;\;
$
&
$
  \adjustbox{raise=-5pt}{
  \begin{tikzcd}[decoration=snake]
    E_{x_1}
    \ar[
      rr,
      shorten=-2pt,
      bend left=20,
      "{
        (p_{12})_\ast
      }"
    ]
    &&
    E_{x_2}
    &[-20pt]
    \subset
    &[-20pt]
    E
    \ar[
      d,
      ->>
    ]
    \\
    x_1
    \ar[
      rr,
      shorten=-3pt,
      decorate,
      bend left=20,
      "{ p_{12} }"{yshift=2pt}
    ]
    &&
    x_2
    &\in&
    X
  \end{tikzcd}
  }
$
\\[-10pt]
&
\\
\hline
\end{tabular}
}
\end{equation}

\noindent
{\bf (IIa')}
{\it Substitution of identifications preserves identifications.}

\vspace{.1cm}

\noindent
\def\arraystretch{1}
\begin{tabular}{|c||c|}
\hline
&
\\[-2pt]
\; $
  \def\arraystretch{1}
  \begin{array}{c}
  \arraycolsep=5pt
  \begin{array}{ccccccccccccccc}
    \overset{
      \mathrlap{
      \raisebox{3pt}{
        \scalebox{.8}{
          \color{orangeii}
          \bf
          Given...
        }
      }
      }
    }
    {
      \Gamma
    }
    &\vdash&
    \overset{
      \mathrlap{
      \raisebox{3pt}{
        \scalebox{.8}{
          \color{darkblue}
          \bf
          a parameter type
          \rlap{...}
        }
      }
      }
    }
    {
      X : \Types
    }
    &\phantom{-}&
    \Gamma\mathrlap{,}\;\;
    &
    x : X
    &
    \vdash
    &
    \overset{
      \mathllap{
      \raisebox{3pt}{
        \scalebox{.8}{
          \color{darkblue}
          \bf
          and a type depending on it, ...
        }
      }
      }
    }
    {
    E_x : \Types
    }
    &&
  \end{array}
  \\
  \hline
  \\[-10pt]
  \hspace{-10pt}
  \begin{array}{ccccccccc}
    \Gamma,
    &
    \underset{
      \mathclap{
      \raisebox{-8pt}{
        \scalebox{.8}{
          \color{darkblue}
          \bf
          \def\arraystretch{.8}
          \begin{tabular}{c}
            for any parameter-
            \\
            identification...
          \end{tabular}
        }
      }\;\;\;\;\;
      }
    }
    {
      p_{12} :
    }
    \mathrm{Id}_X(x_1, x_2),
    &
    \underset{
      \mathclap{
      \raisebox{-9pt}{
        \hspace{-10pt}
        \scalebox{.75}{
          \color{darkblue}
          \bf
          \def\arraystretch{.8}
          \begin{tabular}{c}
            and data at the ini-
            \\
            tial parameter value,
          \end{tabular}
        }
      }\;\;\;
      }
    }
    {
      e_1
    }
    : E_{x_1}
    &
    \underset{
      \mathclap{
        \raisebox{-13pt}{
        \scalebox{.76}{
          \color{orangeii}
          \bf
          \hspace{10pt}
          obtain\;\;
        }
        }
      }
    }{
      \vdash
    }
    &
    \underset{
      \mathclap{
      \raisebox{-4.5pt}{
        \scalebox{.8}{
          \color{darkblue}
          \bf
          \;\;\;
          \def\arraystretch{.7}
          \begin{tabular}{c}
          an identification
          with the transported data.
          \end{tabular}
        }
      }
      }
    }
    {
    \widehat{p_{12}}
    :
    \mathrm{Id}_{
      \underset{x : X}{\coprod}
      \, E_x
    }
    \Big(
      (x_1, e_1)
      ,\,
      \big(x_2, (p_{12})_\ast(e_1)\big)
    \Big)
    \hspace{-10pt}
    }
  \end{array}
  \end{array}
$
&
\adjustbox{raise=-20pt}{
$
  \hspace{-2pt}
  \begin{tikzcd}[decoration=snake]
    \overset{
      \mathclap{
        \rotatebox{40}{
          \rlap{
            \hspace{-13pt}
            \scalebox{.65}{
              \color{darkblue}
              \bf
              \def\arraystretch{.8}
              \begin{tabular}{c}
                lifted
                \\
                start point
              \end{tabular}
            }
          }
        }
      }
    }{
      e_1
    }
    \ar[
      rr,
      shorten <=-2pt,
      shorten >=-5pt,
      bend left=20,
      dashed,
      decorate,
      "{\;
        \overset{
          \mathclap{
            \raisebox{3pt}{
              \scalebox{.7}{
                \color{greenii}
                \bf
                lifted path
              }
            }
          }
        }{
          \widehat{p_{12}}
        }
      }"{yshift=2.5pt, pos=.67}
    ]
    &&
    \overset{
      \mathclap{
        \rotatebox{40}{
          \rlap{
            \hspace{16pt}
            \hspace{-39pt}
            \scalebox{.65}{
              \color{darkblue}
              \bf
              \def\arraystretch{.8}
              \begin{tabular}{c}
                transported
                \\
                \;\;\;\;\;\;\;\;\;initial data
              \end{tabular}
            }
          }
        }
      }
    }{
    (p_{12})_\ast(e_1)
    }
    &[-28pt]
    \in
    &[-25pt]
    \overset{
      \mathclap{
        \raisebox{3pt}{
          \scalebox{.8}{
            \color{darkblue}
            \bf
            fibration
          }
        }
      }
    }{
      E
    }
    \ar[
      d,
      "\in \Fibrations"
    ]
    \\
    x_1
    \ar[
      rr,
      shorten=-2pt,
      bend left=20,
      decorate,
      "p_{12}"{yshift=2.5pt},
      "{
        \scalebox{.7}{
          \color{greenii}
          \bf
          parameter path
        }
      }"{swap,yshift=-5pt}
    ]
    &&
    x_2
    &\in&
    \underset{
      \mathclap{
        \raisebox{-4pt}{
          \scalebox{.8}{
            \color{darkblue}
            \bf
            \def\arraystretch{.8}
            \begin{tabular}{c}
              parameter
              \\
              space
            \end{tabular}
          }
        }
      }
    }{
      X
    }
  \end{tikzcd}
$
  \hspace{-3pt}
}
\\[-7pt]
&
\\
\hline
\end{tabular}
\vspace{.1cm}

\noindent
When combined with the following rule \eqref{IdReversal}, the transport/lifting rule
\eqref{TransportRule}
is {\it equivalent} to $\Id$-induction (cf. prev. footnote):

\medskip

\noindent
{\bf (IIb)} {\it Identifications are preserved by composition with identities.}

\vspace{-.5cm}

\begin{equation}
\label{IdReversal}
\adjustbox{scale=1}{
\def\arraystretch{1.4}
\begin{tabular}{|c||c|}
\hline
{\bf Essential uniqueness of identification certificates}
&
{\bf Contraction of based path space}
\\
\hline
\hline
&
\\[-6pt]
\hspace{-.2cm}
$
  \begin{tikzcd}
    \arraycolsep=2pt
    \begin{array}{ccccccccc}
      \Gamma,
      &
      x, x' : X\mathrlap{,}
      &
      \overset{
        \mathclap{
          \raisebox{4pt}{
            \scalebox{.8}{
              \bf
              \color{orangeii}
              Given...
            }
          }
        }
      }{
    \underset{
      \mathclap{
          \hspace{12pt}
          \rotatebox{-30}{
            \hspace{-24pt}
            \scalebox{.75}{
              \rlap{
                \color{darkblue}
                \bf
                \def\arraystretch{.8}
                \begin{tabular}{c}
                  \hspace{-11pt}an
                  \\
                  identification
                \end{tabular}
              }
          }
        }
      }
    }{
        p
        }
      }
    :
    \mathrm{Id}_X(x,x')
      &\vdash&
      \overset{
        \mathclap{
          \raisebox{4pt}{
            \scalebox{.8}{
              \bf
              \color{orangeii}
              ...we obtain
            }
          }
        }
      }{
    \underset{
      \mathclap{
          \hspace{5pt}
          \rotatebox{-30}{
            \hspace{-15pt}
            \scalebox{.75}{
              \rlap{
                \color{darkblue}
                \bf
                \def\arraystretch{.8}
                \begin{tabular}{c}
                  a certificate
                \end{tabular}
              }
          }
        }
      }
    }{
      p_{\ast}
    }
      \,:\,
    \underset{
      \mathclap{
          \hspace{20pt}
          \rotatebox{-30}{
            \hspace{-30pt}
            \scalebox{.75}{
              \rlap{
                \color{darkblue}
                \bf
                \def\arraystretch{.8}
                \begin{tabular}{c}
                  \;\;\;\;\;\;\;of identification-
                  \\
                  of-identifications
                \end{tabular}
              }
          }
        }
      }
    }{
      \mathrm{Id}_{
        (x' : X)
        \times
        \mathrm{Id}_X(x,x')
      }
    }
      }
      \Big(
        (x, \mathrm{id}_x)
    \underset{
      \mathclap{
          \hspace{-10pt}
          \rotatebox{-30}{
            \hspace{-25pt}
            \scalebox{.75}{
              \rlap{
                \color{darkblue}
                \bf
                \def\arraystretch{.8}
                \begin{tabular}{c}
                  with the
                  \\
                  reflexive identif.
                \end{tabular}
              }
          }
        }
      }
    }{
      ,\,
    }
    \underset{
      \mathclap{
          \hspace{1pt}
          \rotatebox{-30}{
            \hspace{-24pt}
            \scalebox{.75}{
              \rlap{
                \color{darkblue}
                \bf
              }
          }
        }
      }
    }{
        (
          x', p
        )
    }
      \Big)
    \end{array}
  \end{tikzcd}
$
&
\hspace{-.2cm}
$
  \begin{tikzcd}[column sep=large,decoration=snake]
    &
    x
    \ar[
      dl,
      shorten=-2pt,
      decorate,
      bend right=24,
      "{ \mathrm{id}_x }"{swap, xshift=-2pt, pos=.85},
      "{
        \scalebox{.7}{
          \color{greenii}
          \bf
          \def\arraystretch{.8}
          \begin{tabular}{c}
            constant
            \\
            path
          \end{tabular}
        }
      }"{yshift=2pt, sloped},
      "{  }"{name=s, pos=.7}
    ]
    \ar[
      dr,
      shorten <=-2pt,
      shorten >=-4pt,
      decorate,
      bend left=24,
      "{ p }"{pos=.81},
      "{
        \scalebox{.71}{
          \color{greenii}
          \bf
          \def\arraystretch{.8}
          \begin{tabular}{c}
            path
          \end{tabular}
        }
      }"{yshift=1pt, pos=.4, sloped},
      "{ }"{swap, name=t, pos=.61}
    ]
    \\
    x
    \ar[
      rr,
      shorten=-2pt,
      decorate,
      shift right=2pt,
      "{
        p
        \;
        \scalebox{.7}{
          \color{greenii}
          \bf
          path
        }
      }"{yshift=-2pt,swap}
    ]
    &&
    x'
    \ar[
      from=s,
      to=t,
      shorten=-2pt,
      Rightarrow,
      decorate,
      "{
        \scalebox{.7}{
          \color{orangeii}
          \bf
          path-of-paths
        }
      }"{yshift=-1pt, swap, pos=-.05}
    ]
  \end{tikzcd}
$
\hspace{-.4cm}
\\[-4pt]
&
\\
\hline
\end{tabular}
}
\end{equation}
It follows that {\bf data types} of a dependently typed language implementing these inference rules for identification types {\bf are} (fibrations of) {\bf  homotopy types} \cite{Awodey12}, in that they behave like topological spaces up to (weak) homotopy equivalences:

\medskip

\noindent
{\bf The idea of homotopy data type structure.}
Informally, path induction \eqref{PathInduction} says that to
define data $\hat{\sigma}$ of type $E(p_{12})$ dependent on an identification
$p_{12} : \mathrm{Id}_D(d_1, d_2)$ (for \emph{free variables} $d_1$ and $d_2$),
it suffices to assume that $d_2$ is $d_1$ and $p_{12}$ is $\mathrm{id}_D(d_1)$ and
to define (just) the corresponding data $\sigma : E\big(\mathrm{id}_X(x_1)\big)$ in this special case.
Using this one finds, for instance:

\vspace{1mm}
\begin{itemize}[leftmargin=.4cm]

\item
{\it Inversion of identifications} by declaring that that reflexive self-identification is its own inverse (cf. \cite[Lem. 2.1.1]{UFP13}):

\vspace{-.4cm}
\begin{equation}
  \label{InversionOfIdentifications}
  \adjustbox{}{
  \def\arraystretch{1}
  \begin{tabular}{|c|c|}
  \hline
  \begin{minipage}[center]{11.5cm}
  $$
  \def\arraystretch{1.1}
  \begin{array}{rccc}
    \mathrm{inv}_D
    \;:\;
    \dprod{d_1,\, d_2 \,:\, D}
    &
  \Big(
    \mathrm{Id}_D(d_1,\, d_2)
    &\to&
    \mathrm{Id}_D(d_2,\, d_1)
    \mathrlap{
      \!\!
      \Big)^{\mathclap{
        \phantom{
          \vert^{\vert^{\vert}}
        }
      }}
    }
    \\[1pt]
    &
    \mathrm{id}_{d_1}
      &\mapsto&
    \mathrm{id}_{d_1}
    \mathclap{\phantom{
      \vert_{\vert_{\vert_{\vert}}}
    }}
  \end{array}
  $$
  \end{minipage}
  &
  \begin{minipage}{3.5cm}
  $
  \begin{tikzcd}[
    column sep=25pt,
    decoration=snake
  ]
    d_1
    \ar[
      rr,
      decorate,
      shift left=5pt,
      "{
        p_{12}
      }"{yshift=+2pt}
    ]
    \ar[
      from=rr,
      decorate,
      shift left=5pt,
      "{
        \mathrm{inv}(p_{12})
      }"{yshift=-2pt}
    ]
    &&
    d_2
  \end{tikzcd}
  $
  \end{minipage}
  \\
  \hline
  \end{tabular}
  }
\end{equation}

\item
{\it Concatenation of identifications} by declaring them to be trivial
for reflexive identificiations
(cf. \cite[Lem. 2.1.2]{UFP13}):

\vspace{-.4cm}
\begin{equation}
  \label{ConcatenationOfIdentifications}
  \adjustbox{}{
  \begin{tabular}{|c|c|}
  \hline
  \begin{minipage}{11.5cm}
  $$
  \def\arraystretch{1.3}
  \begin{array}{rccc}
    \mathrm{conc}_D
    \;:\;
    \dprod{
      d_1,\, d_2 \,:\, D
    }
    &
    \Big(
    \mathrm{Id}_D(d_1,\, d_2)
    &\to&
    \big(
      \mathrm{Id}_D(d_2,\, d_3)
      \to
      \mathrm{Id}_D(d_1,\, d_3)
    \big)
    \mathrlap{
      \!
      \Big)
    }^{\mathclap{
        \phantom{
          \vert^{\vert^{\vert}}
        }
      }}
        \\
    &
    \mathrm{id}_{d_1}
      &\mapsto&
    (
      p
      \;\mapsto\;
      p
    )
    \\
    {}
  \end{array}
  $$
  \end{minipage}
  &
  \hspace{-4pt}
  \begin{minipage}{3.5cm}
  $$
  \begin{tikzcd}[
    column sep=12pt,
    row sep=-4pt,
    decoration=snake
  ]
    &&
    d_2
    \ar[
      drr,
      decorate,
      "{ p_{23} }"{yshift=2pt}
    ]
    \\
    d_1
    \ar[
      urr,
      decorate,
      "{ p_{12} }"{yshift=2pt}
    ]
    \ar[
      rrrr,
      shift right=2pt,
      decorate,
      "{
        \mathrm{conc}(p_{12},\, p_{23})
      }"{swap, yshift=-2pt}
    ]
    &&&&
    d_3
  \end{tikzcd}
  $$
  \end{minipage}
  \\
  \hline
  \end{tabular}
  }
\end{equation}

\end{itemize}

\newpage
\label{TheMagicOfHomotopyTypeTheory}
It is  {\bf the magic of homotopy type theory} that tautologous-seeming constructions like
\eqref{InversionOfIdentifications} and \eqref{ConcatenationOfIdentifications} provide --- all through the induction principle \eqref{PathInduction} --- just the right information to know everything there is to know about homotopy types --- a subject that is far from trivial (Lit. \ref{LiteratureTopologyAndHomotopyTheory}).
Notably, it follows that:

\begin{itemize}[leftmargin=.4cm]

\item {\it Reverse identifications \eqref{InversionOfIdentifications} are inverses} under concatenation \eqref{ConcatenationOfIdentifications}
(up to higher identification) simply by checking that this holds trivially for reflexive self-identifications  (cf. \cite[Lem. 2.1.4 (ii)]{UFP13}):

\vspace{-.5cm}
\begin{equation}
  \label{ReversesAreInverses}
  \hspace{-.6cm}
  \adjustbox{}{
  \begin{tabular}{|c|c|}
  \hline
  \begin{minipage}{12.5cm}
  $
  \def\arraystretch{1.3}
  \!\!\!
  \begin{array}{rccc}
    \mathrm{ivt}_D
    \,:\,
    \dprod{
      d_1,\,d_2 \,:\, D
    }
    &
  \!\!\!\!\!\!  \bigg(\!\!\!
        \big(
      p : \mathrm{Id}_{D}(d_1,\, d_1)
    \big)
    &\to&
    \mathrm{Id}_{
      \left(
        \mathrm{Id}_{D}(d_1,\,d_1)
      \right)
    }
    \Big(
      \mathrm{conc}\big(
        p
        ,\,
        \mathrm{inv}(p)
      \big)
      ,\,
      \mathrm{id}_{d_1}
    \Big)
    \mathrlap{
      \!\!
      \bigg)
    }
    \\
    &
    \mathrm{id}_{d_1}
    &\mapsto&
    \mathrm{id}_{(\mathrm{id}_{d_1})}
  \end{array}
  $
  \end{minipage}
  &
\hspace{-10pt}
\begin{minipage}{3.4cm}
\hspace{-6pt}
$
    \begin{tikzcd}[decoration=snake]
      &
      d_2
      \ar[
        dr,
        decorate,
        "{
          \mathrm{inv}(p_{12})
        }"{sloped, yshift=2pt}
      ]
      \ar[
        d,
        Rightarrow,
        decorate, color=orangeii
      ]
      \\
      d_1
      \ar[
        ur,
        decorate,
        "{
          p_{12}
        }"{sloped, yshift=2pt}
      ]
      \ar[
        rr,
        decorate,
        bend right=6,
        "{
          \mathrm{id}_{d_1}
        }"{swap, yshift=-1pt}
      ]
      &
      {}
      &
      d_3
    \end{tikzcd}
    $
    \end{minipage}
    \\
    \hline
    \end{tabular}
  }
\end{equation}

\item
{\it Concatenation of identifications is associative} (up to higher identification), simply by checking
that this holds trivially for reflexive self-identifications:
\end{itemize}

\vspace{-.5cm}
\begin{equation}
  \label{ConcatenationIsAssociative}
  \hspace{-2pt}
  \adjustbox{fbox}{$
  \def\arraystretch{1.4}
  \arraycolsep=2pt
  \begin{array}{rccc}
    \mathrm{ast}_D
    :
    \dprod{
      d_1,d_2,d_3,d_4 \,:\, D
    }
    \Bigg(\!\!\!
    &
   \!\!\!\! \left(
      \adjustbox{scale=.8}{$
      \def\arraystretch{.9}
      \begin{array}{l}
        p_{12}
        :
        \mathrm{Id}_{D}(d_1,\, d_2)
        \\
        p_{23}
        :
        \mathrm{Id}_{D}(d_2,\, d_3)
        \\
        p_{34}
        :
        \mathrm{Id}_{D}(d_3,\, d_4)
      \end{array}
      $}
    \right)
    &\to&
    \mathrm{Id}_{(\mathrm{Id}_{D})}
    \Big(
      \mathrm{conc}\big(
        p_{12}
        ,\,
        \mathrm{conc}(
          p_{23}
          ,\,
          p_{34}
        )
      \big)
      ,\,
      \mathrm{conc}\big(
        \mathrm{conc}(
          p_{12}
          ,\,
          p_{23}
        )
        ,\,
        p_{34}
      \big)
    \Big)
    \mathrlap{
      \!\!\!
    \Bigg)
    }
        \\
    &
    \big(
      \mathrm{id}_{d_1}
      ,\,
      \mathrm{id}_{d_1}
      ,\,
      \mathrm{id}_{d_1}
    \big)
    &\mapsto&
    \mathrm{id}_{(\mathrm{id}_{d_1})}
  \end{array}
  $}
\end{equation}

\begin{equation}
\hspace{-.215cm}
\adjustbox{}{
\begin{tabular}{ll}

\hspace{-4mm}
\begin{minipage}{8.9cm}\label{GroupoidsInTypeTheory}
As first observed in \cite{HofmannStreicher98}, translating to algebraic topology this kind of structure (namely ``arrows'' $d_1 \rightsquigarrow d_2$ between ``objects'' $d_1, d_2$, equipped with invertible and associative composition) is known as constituting a {\it groupoid} (jargon for: ``akin to a group but possibly with several objects'', see e.g. \cite{Weinstein96}), specifically the {\it fundamental path groupoid} (e.g. \cite{Santini11}) of the corresponding topological space $D$.

Better yet, since invertibility \eqref{ReversesAreInverses} and associativity \eqref{ConcatenationOfIdentifications} appear themselves as identifications-of-identifications (2-arrows), which by repeated use of the $\mathrm{Id}$-induction rule \eqref{PathInduction}
are found to come with appropriate higher-dimensional analogs of composition, inverses and associativity, the type theory reflects ``higher dimensional'' groupoid structure \cite{vdBergGarner11}, as known from the ``fundamental 2-groupoids'' of topological spaces \cite{HardieKampsKieboom01} and more generally from their ``fundamental $\infty$-groupoids''  typically modeled (e.g. \cite[\S 1.1.2]{Lurie09}) by Kan fibrant singular simplicial sets \cite{KLV12}
or more general fibrant models \cite{AwodeyWarren09}.

\end{minipage}
&
\hspace{-.1cm}
\adjustbox{fbox}{
\begin{minipage}{6.7cm}
$$
  \begin{tikzcd}[
    column sep=30,
    row sep=60,
    decoration=snake
  ]
    &
    d_2
    \ar[
      rr,
      decorate,
      "{ p_{23} }"{yshift=2pt}
    ]
    \ar[
      drrr,
      decorate,
      bend left=00,
      "{ \mathrm{comp}(p_{23}, p_{34}) }"{description, sloped, pos=.35}
    ]
    &&
    d_3
    \ar[
      dr,
      decorate,
      "{ p_{34} }"
    ]
    \\
    d_1
    \ar[
      ur,
      decorate,
      "{ p_{12} }"
    ]
    \ar[
      dr,
      decorate,
      "{ p_{12} }"{swap}
    ]
    \ar[
      drrr,
      decorate,
      bend right=0,
      "{ \mathrm{comp}(p_{12}, p_{23}) }"{description, sloped, pos=.65}
    ]
    \ar[
      rrrr,
      start anchor={[xshift=8pt]},
      end anchor={[xshift=-10pt]},
      decorate,
      bend left=19,
      "{
        \mathrm{comp}
        \big(
          p_{12}
          ,\,
          \mathrm{com}
          ( p_{23},\, p_{34} )
        \big)
      }"{description},
      "{\ }"{name=s, swap}
    ]
    \ar[
      rrrr,
      decorate,
      bend right=19,
      start anchor={[xshift=8pt]},
      end anchor={[xshift=-10pt]},
      "{
        \mathrm{comp}
        \big(
          \mathrm{com}
          ( p_{12},\, p_{23} )
          ,\,
          p_{34}
        \big)
      }"{description},
      "{\ }"{name=t}
    ]
    \ar[
      from=s,
      to=t,
      Rightarrow,
      decorate, color=orangeii
    ]
    &&&&
    d_4
    \\
    &
    d_2
    \ar[
      rr,
      decorate,
      "{ p_{23} }"{swap, yshift=-2pt}
    ]
    &&
    d_3
    \ar[
      ur,
      decorate,
      "{ p_{34} }"{swap}
    ]
  \end{tikzcd}
$$
\end{minipage}
}
\end{tabular}
}\end{equation}

\bigskip
With data types thus revealed as secretly {\it being} $\infty$-groupoids, one discovers similarly that functions (programs!)
from one data type to another secretly {\it are $\infty$-functors}, in that they respect this $\infty$-groupoid structure, again all by
$\mathrm{Id}$-induction \eqref{PathInduction}
(cf. \cite[Lem. 2.2.1]{UFP13}):

\vspace{-.5cm}
\begin{equation}
  \label{FunctionApplicationToIdentifications}
  \hspace{-.4cm}
  \adjustbox{}{
  \def\arraystretch{1}
  \begin{tabular}{|c|c|}
  \hline
  \begin{minipage}[center]{11.9cm}
  $$
  \hspace{-.2cm}
  \def\arraystretch{1.4}
  \begin{array}{rccc}
    f \,:\, D \to D'
    \;\;\;\;\;\;
    \vdash
    \;\;\;\;\;\;
    \mathrm{ap}_f
    \;:
    \;
    \dprod{d_1,\, d_2 \,:\, D}
    &
    \Big(
    \hspace{-1pt}
    \mathrm{Id}_D\big(d_1,\, d_2\big)
    &\to&
    \mathrm{Id}_{D'}\big(
      f(d_1)
      ,\,
      f(d_2)
    \big)
    \mathrlap{
      \!\!
      \Big)
    }
    \\
    &
    \mathrm{id}_{d_1}
      &\mapsto&
    \mathrm{id}_{f(d_1)}
    \mathclap{\phantom{
      \vert_{\vert_{\vert_{\vert}}}
    }}
  \end{array}
  $$
  \end{minipage}
  &
  \begin{minipage}{3.65cm}
  $
  \begin{tikzcd}[
    column sep=16pt,
    row sep=33pt,
    decoration=snake
  ]
    D
    \ar[
      rr,
      "{
        f^{\mathclap{
          \phantom{\vert^{\vert^{\vert}}}
        }}
      }"
    ]
      &&
    D'
    \\[-30pt]
    d_1
    \ar[
      d,
      decorate,
      "{ p_{12} }"{description}
    ]
    &
    {}
    \ar[
      d,
      phantom,
      "{ \mapsto }"
    ]
    &
    f(d_1)
    \ar[
      d,
      decorate,
      end anchor={[yshift=-2pt]},
      "{ \mathrm{ap}_f(p_{12}) }"{description}
    ]
    \\
    d_2
    &{}&
    f(d_2)_{\mathclap{
      \phantom{\vert_{\vert}}
    }}
  \end{tikzcd}
  $
  \end{minipage}
  \\
  \hline
  \end{tabular}
  }
\end{equation}

\vspace{-.3cm}
\begin{equation}
  \label{FunctionApplicationToIdentificationsRespectsComposition}
  \hspace{-.45cm}
  \adjustbox{}{
  \def\arraystretch{1}
  \begin{tabular}{|c|c|}
  \hline
  \begin{minipage}[center]{11.9cm}
  $$
  \hspace{-.2cm}
  \def\arraystretch{1.6}
  \arraycolsep=2pt
  \begin{array}{rccc}
    \mathrm{cmpstr}_{f}
    \;:
    \!\!\!
    \dprod{
      { d_1,\, d_2,}
      \atop
      {d_3 :\, D}
    }
    \Bigg(
    \!\!\!\!\!\!\!
    &
    \left(
    \def\arraystretch{1}
    \begin{array}{l}
      p_{12} \,:\, \mathrm{Id}_{D}
      (d_1,\,d_2)
      ,
      \\
      p_{23} \,:\, \mathrm{Id}_{D}(d_2,\,d_3)
    \end{array}
    \right)
    &\to&
    \mathrm{Id}_{
      \mathrm{Id}_D
    }\bigg(
      \def\arraystretch{.93}
      \begin{array}{l}
      \mathrm{ap}_f\big(
        \mathrm{conc}(p_{12},\, p_{23})
      \big),\,
      \\
      \mathrm{conc}\big(
        \mathrm{ap}_f(p_{12})
        ,\,
        \mathrm{ap}_f(p_{23})
      \big)
      \end{array}
      \!\!
    \bigg)
    \mathrlap{
     \!\! \Bigg)
    }
    \\
    &
    \big(
      \mathrm{id}_{d_1}
      ,\,
      \mathrm{id}_{d_1}
    \big)
      &\mapsto&
    \mathrm{id}_{\mathrm{id}_{f(d_1)}}
    \mathclap{\phantom{
      \vert_{\vert_{\vert_{\vert}}}
    }}
  \end{array}
  $$
  \end{minipage}
  &
  \hspace{-10pt}
  \begin{minipage}{3.9cm}
  $
  \begin{tikzcd}[
    column sep=12pt,
    row sep=38pt,
    decoration=snake
  ]
    &
    f(d_2)^{\mathclap{
      \phantom{\vert^{\vert^{\vert}}}
    }}
    \ar[
      dr,
      decorate,
      end anchor={[xshift=3pt]},
      "{
        \mathrm{ap}_f(p_{23})
      }"{yshift=2pt, sloped}
    ]
    \ar[
      d,
      decorate,
      Rightarrow, color=orangeii,
      start anchor={[yshift=-10pt]},
      end anchor={[yshift=+7pt]},
    ]
    \\
    f(d_1)
    \ar[
      rr,
      decorate,
      "{
        \mathrm{ap}_f
        \left(
          \mathrm{conc}(p_{12},\, p_{23})
        \right)_{\mathclap{
          \phantom{\vert_{\vert_{\vert}}}
        }}
      }"{swap, yshift=-2pt}
    ]
    \ar[
      ur,
      decorate,
      end anchor={[xshift=-2pt]},
      "{
        \mathrm{ap}_f(p_{12})
      }"{yshift=2pt, sloped}
    ]
    &{}&
    f(d_3)
  \end{tikzcd}
  $
  \end{minipage}
  \\
  \hline
  \end{tabular}
  }
\end{equation}

\newpage

Transport \eqref{TransportRule} is also functorial in identifications, which is provable just as easily by appealing to path induction \eqref{PathInduction}.

\vspace{-3mm}

\begin{equation}
\label{FunctorialityOfTransport}
\hspace{-3mm}
\adjustbox{scale=.87}{
\def\arraystretch{1.0}
\begin{tabular}{|c|c|}
\hline
&
\\[-12pt]
\def\arraystretch{0.7}
\(
\!\!\!\!\begin{array}{ccc}
  p_{12} : \mathrm{Id}_X(x_1, x_2),\, p_{23} : \mathrm{Id}_X(x_2, x_3)   &  \vdash & \mathrm{func} : \mathrm{Id}_{}((p_{23})_{\ast} \circ (p_{12})_{\ast},\,\mathrm{conc}( p_{12}, p_{23} )_{\ast})\\ && \\
   \mathrm{id}_{x_1},\;\;\; p_{23}& \longmapsto & \mathrm{id}_{(p_{23})_{\ast}}
\end{array}
\)
&
$
  \adjustbox{raise=-5pt}{
  \begin{tikzcd}[row sep=small, decoration=snake]
    && E_{x_2}  \ar[
      drr,
      shorten=-2pt,
      bend left=15,
      "{
        (p_{23})_\ast
      }"
    ] \ar[d,
    Rightarrow,
    color=orangeii, decorate,
    shorten=2pt,
    ]
    &&
    \\
    E_{x_1}
    \ar[
      urr,
      shorten=-2pt,
      bend left=15,
      "{
        (p_{12})_\ast
      }"
    ] \ar[
      rrrr,
      shorten=-3pt,
      "{\mathrm{conc}( p_{12}, p_{23} )_{\ast} }"{name=t, swap, yshift=-2pt}
    ]
    &&
    {}
    &&
    E_{x_3}
    \\[1pt]
    && x_2 \ar[
      drr,
      shorten=-3pt,
      decorate,
      bend left=15,
      "{ p_{23} }"{yshift=2pt}
    ] &&
    \\
    x_1
    \ar[
      urr,
      shorten=-3pt,
      decorate,
      bend left=15,
      "{ p_{12} }"{yshift=2pt}
    ] \ar[
      rrrr,
      shorten=-3pt,
      decorate,
      "{\mathrm{conc}( p_{12}, p_{23} ) }"{name=l, swap, yshift=-2pt}
    ]
    &&
    &&
    x_3
  \end{tikzcd}
  }
$
\\[-10pt]
&
\\
\hline
\end{tabular}
}
\end{equation}

\vspace{1mm}
\noindent
{\bf Basic notions of homotopy theory among data types.} In such a homotopically typed programming language, basic constructions on data types correspond to fundamental construction in homotopy theory (Lit. \ref{LiteratureTopologyAndHomotopyTheory}). For instance:

\begin{center}
\hspace{-20pt}
\def\arraystretch{1.1}
\tabcolsep=4pt
\begin{tabular}{|c|c|c|}
\hline
 \begin{minipage}{3.6cm}
 \;{\bf Colloquial}
\end{minipage}
&
 \begin{minipage}{4.5cm}
  \; {\bf Homotopy type theory}
\end{minipage}
&
 \begin{minipage}{6.5cm}
 \;{\bf Homotopy theory}
 \end{minipage}
\\
\hline
\end{tabular}
\end{center}
\begin{equation}
\label{Cospan}
\adjustbox{}{
\def\arraystretch{1}
\tabcolsep=4pt
\begin{tabular}{|c|c|c|}
\hline
&&
\\[-0pt]
 \begin{minipage}{3.6cm}
 \;{\bf Cospan}
 \end{minipage}
&
 \begin{minipage}{4.5cm}
 \;
 $
   \def\arraystretch{1.2}
   \begin{array}{r}
    \vdash f \,:\, Y' \to Y
     \\
    \vdash p_X \,:\, X \to Y
   \end{array}
 $
\end{minipage}
&
 \begin{minipage}{6.5cm}
 \hspace{10pt}
 $
  \begin{tikzcd}[column sep=large]
    &&
    X
    \ar[
      d,
      "{ p_X }"
    ]
    \\
    Y'
    \ar[
      rr,
      "{ f }"
    ]
    &&
    Y
  \end{tikzcd}
$
\end{minipage}
\\[-10pt]
&&
\\
\hline
\end{tabular}
}
\end{equation}
\begin{equation}
\label{FibrantResolution}
\adjustbox{}{
\def\arraystretch{1.3}
\tabcolsep=4pt
\begin{tabular}{|c|c|c|}
\hline
&&
\\[-0pt]
 \begin{minipage}{3.6cm}
 \;{\bf Fibrant resolution}
 \end{minipage}
&
 \begin{minipage}{4.5cm}
 \;
 $
    y : Y
    ,\,
    x : X
    \;\;\;
    \vdash
    \;\;\;
      \Id_{Y}
      \big(
        y
        ,\,
        p_X(x)
      \big)
$
\end{minipage}
&
 \begin{minipage}{6.5cm}
 \;
 $
  \begin{tikzcd}[column sep=large]
    \widehat{X}
    \ar[
      dd,
      rounded corners,
      to path ={
            ([xshift=-00pt]\tikztostart.west)
        --  ([xshift=-10pt]\tikztostart.west)
        --  node[xshift=-7pt]{
              \scalebox{.8}{$
                \widehat{p_X}
              $}
            }
            ([xshift=-10pt]\tikztotarget.west)
        --  ([xshift=-00pt]\tikztotarget.west)
      }
    ]
    \ar[
      dr,
      phantom,
      "{
        \lrcorner
      }"{pos=.15}
    ]
    \ar[r]
    \ar[
      d,
    ]
    &
    Y^I
    \ar[
      d,
      "{
        (\mathrm{ev}_0, \mathrm{ev}_1)
      }"
    ]
    \\
    Y \times X
    \ar[
      d,
      "{
        \mathrm{pr}_Y
      }"
    ]
    \ar[
      r,
      "{
        \mathrm{id}_Y
        \times
        p_X
      }"{swap}
    ]
    &
    Y \times Y
    \\
    Y
    \ar[
      rr,
      phantom,
      shift left=3,
      "{
        \scalebox{.7}{
          \color{gray}
          \def\arraystretch{.9}
          \begin{tabular}{l}
            Brown's factorization lemma
            \\
            \cite[p. 421]{Brown73}
          \end{tabular}
        }
      }"{pos=.8}
    ]
    &&
    {}
  \end{tikzcd}
$
\end{minipage}
\\[-10pt]
&&
\\
\hline
\end{tabular}
}
\end{equation}
\begin{equation}
\label{HomotopyPullback}
\adjustbox{}{
\def\arraystretch{1.3}
\tabcolsep=4pt
\begin{tabular}{|c|c|c|}
\hline
&&
\\[-0pt]
 \begin{minipage}{3.6cm}
 \;{\bf Homotopy pullback}
 \end{minipage}
&
 \begin{minipage}{4.5cm}
 $
   \def\arraystretch{1.2}
   \begin{array}{l}
   \vdash \big(
     (y',\,x)
     : Y' \times X
   \big)
   \\
     \,\times\,
      \Id_{Y}
      \big(
        f(y')
        ,\,
        p_X(x)
      \big)
  \end{array}
$
\end{minipage}
&
 \begin{minipage}{6.5cm}
 \;
 $
  \begin{tikzcd}
    Y' \times^h_Y X
    \ar[r]
    \ar[d]
    \ar[
      dr,
      phantom,
      "{
        \lrcorner
      }"{pos=.15}
    ]
    &
    \widehat{X}
    \ar[
      dr,
      phantom,
      "{
        \lrcorner
      }"{pos=.15}
    ]
    \ar[r]
    \ar[
      d,
    ]
    &
    Y^I
    \ar[
      d,
      "{
        (\mathrm{ev}_0, \mathrm{ev}_1)
      }"
    ]
    \\
    Y' \times X
    \ar[
      r,
      "{
        f \times \mathrm{id}_X
      }"{swap, yshift=-1pt}
    ]
    &
    Y \times X
    \ar[
      r,
      "{
        \mathrm{id}_Y
        \times
        p_X
      }"{swap}
    ]
    &
    Y \times Y
  \end{tikzcd}
$
\end{minipage}
\\[-10pt]
&&
\\
\hline
\end{tabular}
}
\end{equation}
\begin{equation}
\label{HomotopyFiber}
\adjustbox{}{
\def\arraystretch{1.3}
\tabcolsep=4pt
\begin{tabular}{|c|c|c|}
\hline
&&
\\[-0pt]
 \begin{minipage}{3.6cm}
 \;{\bf Homotopy fiber}
 \end{minipage}
&
 \begin{minipage}{4.5cm}
 \;\;
 $
    \vdash \big(
        x : X
    \big)
    \,\times\,
      \Id_Y
      \big(
        y
        ,\,
        p_X(x)
      \big)
$
\end{minipage}
&
 \begin{minipage}{6.5cm}
 \;
 $
  \begin{tikzcd}[column sep=large]
    \fiber{p_X}{y}
    \ar[r]
    \ar[d]
    \ar[
      dr,
      phantom,
      "{
        \lrcorner
      }"{pos=.15}
    ]
    &
    Y^I
    \ar[
      d,
      "{
        (\mathrm{ev}_0, \mathrm{ev}_1)
      }"
    ]
    \\
    X
    \ar[
      r,
      "{
       x \,\mapsto\,
       \scalebox{1.2}{$($}
         y
         ,\,
         p_X(x)
       \scalebox{1.2}{$)$}
      }"{swap, yshift=-1pt}
    ]
    &
    Y \times Y
  \end{tikzcd}
$
\end{minipage}
\\[-10pt]
&&
\\
\hline
\end{tabular}
}
\end{equation}
\begin{equation}
\label{LoopSpace}
\adjustbox{}{
\def\arraystretch{1.3}
\tabcolsep=4pt
\begin{tabular}{|c|c|c|}
\hline
&&
\\[-0pt]
 \begin{minipage}{3.6cm}
 \;{\bf Loop space}
 \end{minipage}
&
 \begin{minipage}{4.5cm}
 \;\;
 $
   \vdash \Id_Y
      \big(
        y
        ,\,
        y
      \big)
$
\end{minipage}
&
 \begin{minipage}{6.5cm}
 \hspace{12pt}
 $
  \begin{tikzcd}[
    column sep=50pt
  ]
    \Omega_y Y
    \ar[r]
    \ar[d]
    \ar[
      dr,
      phantom,
      "{
        \lrcorner
      }"{pos=.15}
    ]
    &
    Y^I
    \ar[
      d,
      "{
        (\mathrm{ev}_0, \mathrm{ev}_1)
      }"
    ]
    \\
    \ast
    \ar[
      r,
      "{
       \scalebox{1.2}{$($}
         y
         ,\,
         y
       \scalebox{1.2}{$)$}
      }"{swap, yshift=-1pt}
    ]
    &
    Y \times Y
  \end{tikzcd}
$
\end{minipage}
\\[-10pt]
&&
\\
\hline
\end{tabular}
}
\end{equation}
\begin{equation}
\label{Contraction}
\adjustbox{}{
\def\arraystretch{1.3}
\tabcolsep=4pt
\begin{tabular}{|c|c|c|}
\hline
&&
\\[-0pt]
 \begin{minipage}{3.6cm}
 \;{\bf Contraction}
 \end{minipage}
 &
 \begin{minipage}{4.5cm}
 \;
 $
  \vdash \mathrm{contr}_{x_0}
   \,:\,
   \big( x : X \big)
    \to
      \Id_X
      \big(
        x_0
        ,\,
        x
      \big)
  $
\end{minipage}
&
 \begin{minipage}{6.5cm}
 \;
 $
  \begin{tikzcd}[column sep=large]
    &
    P_{x_0} X
    \ar[r]
    \ar[d]
    \ar[
      dr,
      phantom,
      "{
        \lrcorner
      }"{pos=.15}
    ]
    &
    X^I
    \ar[
      d,
      "{
        (\mathrm{ev}_0, \mathrm{ev}_1)
      }"
    ]
    \\
    X
    \ar[r, equals]
    \ar[
      ur,
      dashed,
      bend left=10pt,
      "{
        \mathrm{contr}_{x_0}
      }"
    ]
    &
    X
    \ar[
      r,
      "{
       x \,\mapsto\,
      (
         x_0
         ,\,
         x
      )
      }"{swap, yshift=-1pt}
    ]
    &
    X \times X
  \end{tikzcd}
$
\end{minipage}
\\[-10pt]
&&
\\
\hline
\end{tabular}
}
\end{equation}

\newpage

\noindent
{\bf The idea of propositions as data types.}
{\it Unique existence} is a crucial logical notion --- many theorems in mathematics can be expressed as the statement that a concept uniquely defines a given object. In the homotopical interpretation of type theory, the statement that there is a unique element of a type $X$ becomes the statement that $X$ is {\it contractible}. To say that a type $X$ has a unique element, we must give an element of it and a way to identify every other element with that element.
Such certificates are evidently of this type:

\vspace{-2mm}
\begin{equation}
  \label{IsContractible}
  \overset{
    \mathrlap{
      \raisebox{7pt}{
        \scalebox{.7}{
          \color{darkblue}
          \bf
          \def\arraystretch{.9}
          \begin{tabular}{c}
            Certificates that there exists
            \\
            unique $X$-data
          \end{tabular}
        }
      }
    }
  }{
  \exists!
  \;
    X
  }
  \;:\defneq\;
  \mathrm{isContractible}(X)
  \;:\defneq
  \hspace{-3pt}
  \overset{
    \raisebox{4pt}{
      \scalebox{.7}{
        \bf
        \color{orangeii}
        \def\arraystretch{.8}
        \begin{tabular}{c}
          There is
          \\
          a datum
        \end{tabular}
      }
    }
  }{
    (x_0 : X)
    \times
  }
  \Big(
  \overset{
    \raisebox{4pt}{
      \scalebox{.7}{
        \bf
        \color{orangeii}
        \def\arraystretch{.8}
        \begin{tabular}{c}
          onto which all
          \\
          $X$-data contracts.
        \end{tabular}
      }
    }
  }{
  (x : X)
  \to
  \Id_X
  (
    x_0
    ,\,
    x
  )
  }
  \Big)
  \,.
\end{equation}
In particular, if all the {\it identification types}  $\Id_P(p,p')$ \eqref{IdentificationType} of a given data type $P$ are certified
to be contractible \eqref{IsContractible}, this means that all pairs of data $p, p' : P$ are identified -- {\it if} there are any.
There is at most one element of such types $P$, which is then canonically understood
to be a {\it certificate of the truth} -- hence a {\it proof} --  of the {\it proposition} that ``$P$-data exists'' at all:
\begin{equation}
  \label{isProposition}
  \mathrm{isProposition}(P)
  \;:\defneq\;
  \dprod{p, p' : P}
  \exists!
  \;
  \Id_P\big(p, p'\big)
  \;.
\end{equation}
As as example, for any type $X$ the type $\exists ! X$ \eqref{IsContractible} which asserts that $X$ has a unique element is a proposition; there is at most one way to give a unique element of a type. To give an element of $\exists ! X$ is simply to prove that $X$ has a unique element.

This way, propositions and their (first order) {\it propositional logic} {\it emerge} inside (dependent) data type theory -- a profound
statement famous as the ``Curry-Howard isomorphism'' (e.g. \cite{SorensenUrzyczyn06}) or as the slogan ``propositions as types'' (e.g. \cite{AwodeyBauer04}\cite{Wadler15}, for review in our homotopical context see \cite[\S 1.11]{UFP13}).
Moreover, under this identification, the logical connectives on propositions are nothing but the various type formation rules, an observation known as the {\it BHK correspondence} \footnote{The BHK correspondence \eqref{LogicalConnectivesAsTypeFormation} originates in the school of mathematical {\it intuitionism}
\cite[p. 59]{Kolmogorov32}\cite{Troelstra69}
and eventually lead to the formulation of dependent type theory in \cite{MartinLof75} (exposition in \cite[Lec. 3]{MartinLof96}), which is the historical reason that one also speaks of ``intuitionistic type theory''. } \cite[\S 2]{Troelstra77}\cite[\S 3.1]{TroelstraVanDalen88}\cite[p. 96]{Bridges99}:

\vspace{-.3cm}
\begin{equation}
\label{LogicalConnectivesAsTypeFormation}
\hspace{-.6cm}
\adjustbox{}{
\begin{tabular}{ll}
\begin{minipage}{7.6cm}

\begin{itemize}[leftmargin=.4cm]
\item
For $P, P' : \Propositions$,
the  product type $P \times P'$ \eqref{PairTypes} reflects the proposition that $P$ {\it and} $P'$ hold.

\item
For $P, P' : \Propositions$,
the  coproduct type $P \sqcup P'$ \eqref{InferenceRulesForHomotopyPushout} reflects the proposition that $P$ {\it or} $P'$ hold.

\item
For $P, P' : \Propositions$,
the  function type $P \to P'$ reflects the proposition that $P$ {\it implies} $P'$.

\item
For $D : \Sets,\; d : D \,\vdash\, P(d) : \Propositions$

\begin{itemize}[leftmargin=.3cm]

\item the dependent function type $(d : D) \to P(d)$ \eqref{PiInferenceRules}
reflects the proposition $\forall_{d : D}\, P(d)$ that $P(d)$ holds {\it for all} $d$.

\item
  the (-1)-truncated \eqref{MeaningOfPropositionalTruncation}
  dependent pair type
  $\exists \big( (d : D) \times P(d) \big)$
  \eqref{DependentSumInference}
  reflects the proposition $\exists_{d : D}\, P(d)$ that {\it there exists} $d$ for which $P(d)$ holds.
\end{itemize}

\end{itemize}
\end{minipage}
&
\hspace{3mm}
{\small
\def\arraystretch{1.7}
\begin{tabular}{|p{2.2cm}|c|c|c|}
  \hline
  \multicolumn{2}{|c|}{
    \bf Type theory
  }
  &
  {\bf Logic}
  \\
  \hline
  \hline
  \begin{tabular}{l}
    (-1)-type
  \end{tabular}
  &
  $d \colon D \;\vdash\;P(d) : \Propositions$
  &
  Proposition
  \\
  \hline
  \begin{tabular}{l}
    product type
  \end{tabular}
  &
  $P \times P'$
  &
  $P \;\mathrm{and}\; P'$
  \\
  \hline
  \begin{tabular}{l}
  Coproduct type
  \end{tabular}
  &
  $P \sqcup P'$
  &
  $P \;\mathrm{or}\; P'$
  \\
  \hline
  \begin{tabular}{l}
    Function type
  \end{tabular}
  &
  $P \to P'$
  &
  $P \;\mbox{implies}\; P'$
  \\
  \hline
  \def\arraystretch{.9}
  \begin{tabular}{l}
    Dependent
    \\
    function type
  \end{tabular}
  &
  $(d : D) \to P(d)$
  &
  $\forall_{d : D} \; P(d)$
  \\
  \hline
  \def\arraystretch{.9}
  \begin{tabular}{l}
    (-1)-truncated
    \\
    dependent
    \\
    pair type
  \end{tabular}
  &
  $\exists \big( (d : D) \times P(d) \big)$
  &
  $\exists_{d : D} \; P(d)$
  \\
  \hline
\end{tabular}
}
\end{tabular}
}
\end{equation}

It is this intimate relation between proofs in logic on the one hand and data of data types on the other which makes programs written in typed languages automatically come with proofs of their correctness (Lit. \ref{VerificationLiterature})

\smallskip
Notice that this correspondence is contentful (only) because we consider {\it dependent} data types \eqref{DependentData}, so that
these propositional types $d : D \;\;\vdash\;\; P(d) : \mathrm{Prop}$ are propositions {\it about} data of the type $D$ that they depend on.
For instance, the traditional way to build sets of data satisfying a given proposition is given by the (untruncated) dependent pair type
\eqref{DependentSumInference} of such dependent propositions:
\begin{equation}
 \label{DependentSumOfProps}
  \def\arraystretch{1.6}
  \begin{array}{c}
  \overset{
    \mathclap{
      \raisebox{4pt}{
        \scalebox{.7}{
          \color{darkblue}
          \bf
          proposition about $D$-data
        }
      }
    }
  }{
  d \,:\, D
  \;\;\vdash\;\;
  P(d) \,:\, \Propositions
  }
  \\
  \hline
  \underset{
    \mathclap{
      \raisebox{-1pt}{
        \scalebox{.7}{
          \color{darkblue}
          \bf
          $D$-data verifying this proposition
        }
      }
    }
  }{
  \big\{
    d : D \mid P(d)
  \big\}
  \;:\defneq\;
  \big(
    d : D
  \big)
  \times
  P(d)
  \,:\,
  \Types
  }
  \end{array}
  \,.
\end{equation}
Here to give an element of $\{d : D \mid P(d)\}$ is  precisely to give data $d$ equipped with a certificate that $P(d)$ holds. For instance, we may define the proposition that a number $n : \mathbb{N}$ is even as the type of numbers which divide it evenly in half:
\begin{equation}
  \label{isEven}
  \mathrm{isEven}(n)
  \;:\defneq\;
  \dsum{k : \mathbb{N}}
  \Id_{\mathbb{N}}\big(2k, n\big)
  \;.
\end{equation}
We can then define the type of even numbers as $\{n : \mathbb{N} \mid \mathrm{isEven}(n)\}$. While this type technically has as elements triples $(n, k, p)$ where $p : \Id_{\mathbb{N}}(2 \cdot k, n)$ is an identification of $2k$ with $n$, we note that $k$ is uniquely determined by $n$ and the property that $2k$ equals $n$, so it is harmless to identify the elements of this type with the even natural numbers.

As another example, since being a proposition is itself a proposition, we may form the type of all propositions among all types \eqref{TypeOfSmallTypes} is:
\begin{equation}
  \label{Propositions}
  \overset{
    \mathclap{
    \raisebox{5pt}{
      \scalebox{.7}{
        \color{darkblue}
        \bf
        Type of all propositional types
      }
      }
    }
  }{
    \Propositions
    \;\;:\defneq\;\;
    \dsum{
      P : \Types
    }
    \mathrm{isProposition}(P)
  }
\end{equation}

Examples of propositional data types in this sense \eqref{Propositions} include: $\mathrm{isContractible}(-)$ \eqref{IsContractible}, $\mathrm{isProposition}(-)$ \eqref{Propositions}, and  $\mathrm{isEquivalence}(-)$ \eqref{HomotopyEquivalenceOfTypes}.
\medskip

{\bf }

\medskip

\noindent
{\bf The idea of data sets.} How should one define data {\it sets} in a homotopy typed programming language? Here our usual intuition for equality helps us: The equality of elements in sets should be a {\it proposition}. Conversely, a data {\it set} should be a data type $X$ whose identification types $\Id_X(x,\, y)$ \eqref{IdentificationType} are all propositional \eqref{isProposition} --- namely, $\Id_X(x, y)$ should reflect the proposition: ``$x$ equals $y$'':

\begin{equation}
  \label{Sets}
  \def\arraystretch{1.6}
  \begin{array}{l}
  \mathrm{isSet}(X)
  \;:\defneq\;
  \isTruncatedType{0}(X)
  \;:\defneq\;
  \dprod{x, y : X} \mathrm{isProposition}\big(
    \Id_X(x, y)
  \big)
  \,,
  \\
  \Sets \;:\defneq\;
  \dsum{
    S : \Types
  }
  \mathrm{isSet}(S)
  \,.
  \end{array}
\end{equation}
It is the data of these propositional identification types
which deserves\footnote{
  It is famously popular in the homotopy type theory literature to use the notation ``$x =_{{}_X} y$'' for what we denote $\Id_X(x,\,y)$, even when $X$ is a higher homotopy type. We humbly suggest that mathematical intuition is served and much debate is avoided by using the equality sign only for the actual notion of (propositional) equality -- which is that of identifications of elements of sets.
}
to be denoted by the traditional notation for equality (whence: ``propositional equality''):
\begin{equation}
  \label{PropositionalEquality}
  \overset{
    \mathclap{
      \raisebox{3pt}{
        \scalebox{.7}{
          \bf
          \begin{tabular}{l}
          \color{orangeii}
          Given
          \\
          \color{darkblue}
          a data set
          with a pair of
          identifiable elements
          \end{tabular}
        }
      }
    }
  }{
  X : \Sets
  ,\;\;\;\;\;\;
  x, y : X
  ,\;\;\;\;\;\;
  \Id_X(x,\,y)
  }
  \;\;\;\;\;\;
  \vdash
  \;\;\;\;\;\;
  \;\;
  \overset{
    \mathclap{
      \raisebox{3pt}{
        \scalebox{.7}{
          \bf
          \begin{tabular}{l}
            \color{orangeii}
            obtain
            \\
            \color{darkblue}
            equality as the certificate
            of their identification.
          \end{tabular}
        }
      }
    }
  }{
  x = y
  \;\;\,\;
  :\defneq
  \;\;\,\;
  x \rightsquigarrow y
  \;\;:\;\;
  \Id_X(x,\,y)
  }
  \,.
\end{equation}
With concatenation ``$\mathrm{conc}$'' \eqref{ConcatenationOfIdentifications} of identifications denoted just by their juxtaposition, this allows to obtain classical-looking proof certificates of equality in homotopy data sets, such as:
\begin{equation}
  \label{ConcatenationOfEqualities}
  X \,:\, \Sets
  ,\;\;\;
  x,y,z \,:\, X
  ,\;\;\;
  (x = y) : \Id_X(x\, y)
  ,\;\;\;
  (y = z) : \Id_X(y\, z)
  \;\;\;\;\;
    \vdash
  \;\;\;\;\;
  x = y = z \;:\; \Id_X(x,\, z)
  \,.
\end{equation}

These homotopy data sets \eqref{Sets}
encode inside homotopy-type languages the ordinary kind of data which is available in non-homotopic programming languages (cf. \cite{RjkeSpitters}), such as the type of bits \eqref{BitsType}, or of natural numbers \eqref{NaturalNumberstype}:
$
  \mathrm{Bit}, \; \mathbb{N}
  \;:\;
  \Sets
  \,.
$

\medskip

\noindent
{\bf The idea of higher homotopy data types.}
Continuing the pattern by which the notion of propositional data types \eqref{Propositions} leads to that of data sets \eqref{Sets}
yields the notion of higher homotopy data types:

The next stage in the hierarchy corresponds (as first observed by \cite{HofmannStreicher98}) to what in homotopy  theory are called {\it groupoids} or {\it homotopy 1-types} (exposition and introduction in \cite{Weinstein96}\cite{Santini11}\cite[\S 1.2]{Richter20}), where for a pair of data of such type there may be not just one (or none), but a whole set of different ways of identifying them:
\begin{equation}
  \label{TypeOfGroupoids}
  \def\arraystretch{1.6}
  \begin{array}{l}
  \mathrm{isGroupoid}(X)
  \;\;:\defneq\;\;
  \isTruncatedType{1}(X)
  \;\;:\defneq\;\;
  \dprod{x, y: X}
  \mathrm{isSet}\big(\Id_X(x, y)\big)
  \,,
  \\
  \Groupoids \;:\defneq\;
  \TruncatedTypes{1}
  \;\;
  :\defneq
  \;\;
  \dsum{
    \mathcal{G} : \Types
  } \mathrm{isGroupoid}(\mathcal{G})
  \,.
  \end{array}
\end{equation}

Historically, the passage from understanding such homotopy 1-types to understanding their generalization to higher homotopy $n$-types was a long and convoluted one (cf. \cite{Hilton88}); but in homotopy-typed language this is now immediate (the key insight of \cite{Voevodsky10}):
\begin{equation}
  \label{HigherHomotopyType}
  \def\arraystretch{1.6}
  \begin{array}{l}
  \isTruncatedType{n+1}(X)
  \;\;:\defneq\;\;
  \dprod{x, y: X}
  \isTruncatedType{n}\big(\Id_X(x, y)\big)
  \,,
  \\
  \TruncatedTypes{n+1}
  \;\;
    :\defneq
  \;\;
  \dsum{
    X : \Types
  }
  \isTruncatedType{n}(X)
  \,.
  \end{array}
\end{equation}
The mathematical semantics of these homotopical data $n$-types is by what in homotopy theory has long been known (long before any relation to data type theory was even thought of) as {\it homotopy $n$-types} and generally as {\it homotopy types} (e.g. \cite[p. 25 \& \S 8]{Spanier82}\cite{Baues95}).

\medskip

The basic examples of higher homotopy $n$-types are the {\it higher spheres} \eqref{HigherHomotopyBits}; see also the comments on higher structures on pp. \pageref{TheIdeaOfHigherStructures}. Often one is interested in {\it truncating} the homotopy level of a type, see p. \ref{TheIdeaOfNTruncation} below.

\newpage

\noindent
{\bf The idea of equivalence/isomorphy of data types.}
Two data types $D, C \colon \Types$ may appear nominally different, but if one can transform data $f : D \to C$ such that this transformation may be {\it inverted up to identification}, then $D$-data is {\it equivalent}  (``isomorphic'' \cite{DiCosmo95}\cite[\S 5.4]{HofmannStreicher98}\cite[Def. 3.1.1]{KapulkinLumsdained21})
to $C$-data: Whatever program operates on $D$-data may then be transformed into a program operating on $C$-data, and vice versa (e.g. \cite{BarthePons01}\cite{RPYLG21}).

Alternatively, one may ask \cite[p. 8, 10]{Voevodsky10} that $f : D \to C$ be a {\it bijection} (or {\it weak equivalence}, cf. below \eqref{TheTwoNotionsOfEquivalencesCoincide}), in that for $c : C$ we have a unique \eqref{IsContractible} inverse image $d : \mathrm{fib}_c(f)$ \eqref{HomotopyFiber}:

\vspace{-.3cm}
\hspace{-.9cm}
\begin{equation}
  \label{HomotopyEquivalenceOfTypes}
  \hspace{-.4cm}
\def\arraystretch{1.4}
\begin{tabular}{|c|c|}
\hline
{\bf Data type equivalence}
&
{\bf Homotopy equivalence}
\\
\hline
\hline
\begin{minipage}{9.2cm}
$
  \def\arraystretch{1}
  \arraycolsep=1pt
  \begin{array}{lcl}
  \mathclap{\phantom{\vert^{\vert^{\vert^{\vert}}}}}
  f \,:\, D \to C
  \;\;\;\;\;\;
  \vdash
  \\[+1pt]
  \scalebox{.7}{
    \color{orangeii}
    \bf
    \def\arraystretch{.9}
    \begin{tabular}{c}
      a function is
      \\
      an {\it equivalence}
      \\
      or {\it isomorphism}
    \end{tabular}
    \color{black}
  }
  &
  \raisebox{3pt}{
  \scalebox{.7}{
    iff
  }}
  \\[-22pt]
  \mathrm{isEquivalence}(f)
  &:\defneq&
  \overset{
    \raisebox{0pt}{
      \scalebox{.7}{
        \color{darkblue}
        \bf
        we have a reverse function
        which is left inverse
      }
    }
  }
  {
  \underset{
    \raisebox{-0pt}{
      \scalebox{.7}{
        \color{darkblue}
        \bf
        and a reverse function which is right inverse
       $\mathclap{\phantom{\vert_{\vert_{\vert_{\vert}}}}}$
      }
    }
  }{
  \def\arraystretch{1.4}
  \left\{
  \arraycolsep=0pt
  \begin{array}{ll}
  &
  \dsum{
    \overline{f}_l
    \,:\,
    C \to D
  }
  \underset{ (D \to D) }
  {\mathrm{Id}}
  \big(
   \, \overline{f}_l \circ f
    ,\,
    (d \,\mapsto\, d)
  \big)
  \\
  \times
  &
  \dsum{
    \overline{f}_r
    \,:\,
    C \to D
  }
  \underset{ (C \to C) }
  {\mathrm{Id}}
  \big(
    f \circ \overline{f}_r
    ,\,
    (c \,\mapsto\, c)
  \big)
  \end{array}
  \right.
  }
  }
  \end{array}
$
\end{minipage}
&
$
  \begin{tikzcd}[
    decoration=snake,
    column sep=40pt,
    row sep=16pt
  ]
    {}&{}&{}&{}
    \\
    C
    \ar[
      r,
      "{
        \overline{f}_r
      }"{description}
    ]
    \ar[
      rr,
      bend right=50,
      "{ \mathrm{id}_B }"{description}
    ]
    &
    D
    \ar[
      r,
      "{ f }"{description}
    ]
    \ar[
      rr,
      bend left=50,
      "{
        \mathrm{id}_A
      }"{description}
    ]
    \ar[
      d,
      Rightarrow,
      decorate
    ]
    &
    C
    \ar[
      u,
      Rightarrow,
      decorate
    ]
    \ar[
      r,
      "{
        \overline{f}_l
      }"{description}
    ]
    &
    D
    \\
    {}&{}&{}&{}
  \end{tikzcd}
$
\\
\hline
\begin{minipage}{9.2cm}
 $
   \def\arraystretch{1}
   \begin{array}{lcl}
     \mathclap{\phantom{\vert^{\vert^{\vert}}}}
     f : D \to C
     \;\;\vdash
     \\
     \scalebox{.7}{
       \color{orangeii}
       \bf
       \def\arraystretch{1}
       \begin{tabular}{c}
         a function is
         \\
         a {\it bijection}
       \end{tabular}
     }
     &
     \scalebox{.7}{
       iff
     }
     &
     \hspace{-6pt}
     \scalebox{.7}{
       \color{darkblue}
       \bf
       \begin{tabular}{c}
         its pre-images
         are ess. unique
       \end{tabular}
     }
     \\
     \mathrm{isBijection}(f)
     &:\defneq&
     (c : C) \to \exists ! \, \mathrm{fib}_c(f)
     \mathclap{\phantom{\vert_{\vert_{\vert}}}}
   \end{array}
 $
\end{minipage}
&
$
 \mbox{
   ``$\underset{c \in C}{\forall}$''
 }
 \hspace{.3cm}
 \mathrm{fib}_c(f)
 \;
 \simeq
 \;
 \ast
$
\\
\hline
\hline
{\bf Data type bijection}
&
{\bf Contractible fibers}
\\
\hline
\end{tabular}
\end{equation}

These two notions happen to coincide \cite[\S 4.4]{UFP13}:
\begin{equation}
  \label{TheTwoNotionsOfEquivalencesCoincide}
  \begin{tikzcd}[
    row sep=1pt,
    column sep=0pt
  ]
    \mathrm{isBijection}(f)
    \ar[
      rr,
      shift right=4pt
    ]
    \ar[
      from=rr,
      shift right=4pt
    ]
    &&
    \mathrm{isEquivalence}(f)
    \\
    \bigg(
      (c : C)
        \,\mapsto\,
      \Big(
        \big(
          d,\,
          f(d)
            {\rightsquigarrow}
          c
        \big)
          :
        \mathrm{fib}_c(f)
      \Big)
        \mapsto
      \big(
        d_0(c)
          \overset{
            \mathrm{ctr}_d
          }{
            \rightsquigarrow
          }
        d
        ,\,
        \cdots
      \big)
    \bigg)
    &\longmapsto&
    \left(
    \def\arraystretch{1.2}
    \begin{array}{l}
        \overline{f}_l
        \,:\,
        c \,\mapsto\, d_0(c)
        \,,\;
        d \,\mapsto\,
          \big(
            d_0(c)
              \overset{\mathrm{ctr}_d}{\rightsquigarrow}
            d
          \big)
        \\
        \overline{f}_r
        \,:\,
        c \,\mapsto\, d_0(c)
        ,\,\;
        c \,\mapsto\,
       \mathrm{id}_c
    \end{array}
  \!\!\!  \right)
    \,.
  \end{tikzcd}
\end{equation}

If one reads the ``for all'' quantifier on the right of \eqref{HomotopyEquivalenceOfTypes} naively as in \eqref{LogicalConnectivesAsTypeFormation} (even if $C$ is not 0-truncated), then
in the classical homotopy theory \eqref{InterpretationInTheClassicalModelTopos} of topological spaces this state of affairs is the content of the  {\it classical Whitehead theorem} (e.g. \cite[Cor. 11.14]{Bredon93}\cite[Thm. 6.3.31]{AguilarGitlerPrieto02}) which says that maps between cofibrant spaces (CW-complexes) are homotopy equivalences as soon as they are {\it weak} homotopy equivalences in that the homotopy groups of all their homotopy fibers vanish. For this reason the type-theoretic ``bijections'' above were originally called {\it weak equivalence} in \cite{Voevodsky10}.

However, the beauty of simply  recasting
\eqref{LogicalConnectivesAsTypeFormation}
the naive quantifier ``$\forall_{c : C}$'' as the dependent function constructor $(c : C) \to (-)$ dramatically increases the generality of the statement: While various versions of the Whitehead theorem actually fail in general model toposes (\cite[\S 6.5]{Lurie09}\cite[\S 8.8]{UFP13}), the equivalence \eqref{TheTwoNotionsOfEquivalencesCoincide} holds generally.

\smallskip

We denote the type of equivalence like the function type but equipped with a ``$\sim$''-symbol:
\begin{equation}
  \label{DataTypeOfEquivalences}
  D, C \,:\, \Types
  \;\;\;\;\;\;\;\;
  \vdash
  \;\;\;\;\;\;\;\;
  \big(
    D \xrightarrow{\sim} C
  \big)
  \;\;:\defneq\;\;
  \dsum{
    f : D \to C
  }
  \mathrm{isEquivalence}(f)
\end{equation}

The tautological example of an equivalence is of course the identity function $a \mapsto a$, which is canonically its
own left and right inverse, as certified by its reflexive identification $\mathrm{id}_{(a \mapsto a)}$:
\begin{equation}
  \label{IdentityFunctionAsAnEquivalence}
  \hspace{-2mm}
  A \,:\, \Types
  \;\;\;\;\;
  \vdash
  \;\;\;\;\;
  \bigg(
    (a \,\mapsto\, a)
    ,\,
    \Big(
      \big(\,
        \overline{(a \mapsto a)}^{\,l}
        \,:\defneq\,
        (a \mapsto a)
        ,\,
        \mathrm{id}_{(a \mapsto a)}
      \big)
      ,\;
      \big(\,
        \overline{(a \mapsto a)}^{\,r}
        \,:\defneq\,
        (a \mapsto a)
        ,\,
        \mathrm{id}_{(a \mapsto a)}
      \big)
    \Big)
 \!\! \bigg)
  \,:\;
  \mathrm{Equiv}(A,\, A) \;.
\end{equation}


\medskip
\noindent
{\bf The idea of univalent universes  of data types.}
However, there is a priori {\it another} sensible way to understand identification of data types. Namely, since we regard data
types $A, B$ themselves as being data of type ``data type'', denoted $A, B \,:\, \Types$ \eqref{TypeOfSmallTypes}, there is the notion of their identification
certificates \eqref{IdentificationType} $c \,:\, \mathrm{Id}_{\Types}(A,\, B)$, just as for data of any other type.
Now, the $\mathrm{Id}$-induction principle  \eqref{PathInduction} gives a transformation from such
{\it identifications} of data to operational {\it equivalences} \eqref{DataTypeOfEquivalences} between data, induced from taking
the identical equivalence $a \mapsto a$
\eqref{IdentityFunctionAsAnEquivalence} to be the operationalization of the self-identification $\mathrm{id}_A$ \eqref{CanonicalIdentification}:
\vspace{-3mm}
\begin{equation}
  \label{UnivalenceAxiom}
  \begin{tikzcd}[
    column sep=0pt,
    row sep=-7pt
  ]
  \mathrm{operationalize}
  &[-5pt]:&[-5pt]
  \underset{
    A, B \colon \Types
  }{\prod}
  \bigg(
  &
  \overset{
    \mathclap{
      \raisebox{6pt}{
        \scalebox{.7}{
        \color{darkblue}
        \bf
        \def\arraystretch{.9}
        \begin{tabular}{c}
          identification certificates
          \\
          between data types
        \end{tabular}
        }
      }
    }
  }{
    \mathrm{Id}_{\Types}(A,\,B)
  }
  \ar[rr]
  &[-4pt]&[-4pt]
  \overset{
    \mathclap{
      \raisebox{6pt}{
        \scalebox{.7}{
        \color{darkblue}
        \bf
        \def\arraystretch{.9}
        \begin{tabular}{c}
          operational equivalences
          \\
          between data types
        \end{tabular}
        }
      }
    }
  }{
    \mathrm{Equiv}(A, \,B)
  }
  &\bigg)
  \,,
  \\
  &&&
  \mathrm{id}_A
  &\;\;\longmapsto\;\;&
  (a \,\mapsto\, a)
  \end{tikzcd}
  \;\;\;
  \mathrm{univalence}
  \;:\;
  \mathrm{isEquivalence}
  \big(
    \mathrm{operationalize}
  \big)
  \,.
\end{equation}

One says\footnote{On this point see again footnote \ref{UnivalenceAttribution}.} that a homotopically typed language satisfies {\it type
universe extensionality} \cite[\S 5.4]{HofmannStreicher98} or that it has a {\it univalent type universe} \cite[p. 11]{Voevodsky10}
if this comparison function is itself an equivalence \eqref{HomotopyEquivalenceOfTypes} (cf. \cite[\S 3.11]{Escardo19}\cite[\href{https://1lab.dev/1Lab.Univalence}{\S Univalence}]{1lab}).

\smallskip
Another incarnation of the  Univalence Axiom \eqref{UnivalenceAxiom} says \cite[Theorem 4.8.3]{UFP13} that
the operation of recording homotopy fibers \eqref{HomotopyFiber} for all base points constitutes an equivalence \eqref{HomotopyEquivalenceOfTypes} between functions into a given type $X$ and $X$-dependent types \eqref{TypeOfSmallDependentTypes}. Semantically this means (originally conjectured by \cite{Awodey12}, proven in special cases in \cite{KLV12}\cite{Shulman15} and generally in \cite{Shulman19}, review in \cite{Riehl22}) that the type universes of univalent homotopy-typed languages are ``object classifiers'' \cite[\S 6.1.6]{Lurie09} reflecting the ambient category of types as an ``$\infty$-topos'' \cite{Simpson99}\cite{ToenVezzosi05}\cite{Lurie09}\cite{Rezk10}:
\vspace{0mm}
\begin{equation}
  \label{TypeClassification}
  \adjustbox{}{
  \def\arraystretch{1.5}
  \begin{tabular}{|c|c|}
  \hline
  {\bf Univalent Data Type Universe}
  &
  {\bf Object classifier}
  \\
  \hline
  \hline
  \begin{minipage}{11cm}
  $$
 \begin{tikzcd}[
    decoration=snake,
    column sep=65pt
  ]
      \overset{
        \mathclap{
        \raisebox{2pt}{
          \scalebox{.7}{
            \color{darkblue}
            \bf
            Functions into $X$
          }
        }
        }
      }{
        (Y :\Types)
          \times
        (Y \to X)
      }
    \ar[
      rr,
      bend left=12,
      "{\ }"{name=s},
      "{
        \scalebox{.8}{
           \bf
           \color{greenii}
           fibers
        }
      }"{swap, yshift=0pt},
      "{
        (Y, f)
        \;\mapsto\;
        \left(
          x
            \,\mapsto\,
          \mathrm{fib}_x(f)
        \right)
      }"
    ]
    \ar[
      rr,
      phantom,
      "{ \simeq }"
    ]
    \ar[
      rr,
      <-,
      bend right=12,
      "{\ }"{name=t},
      "{
        \left(
          (x \, :\, X) \times E_x
          ,\;
          \mathrm{pr}_X
        \right)
        \;\mapsfrom\;
        \left(
          x \,\mapsto\, E_x
        \right)
      }"{swap},
      "
        \scalebox{.8}{
          \color{greenii}
          \bf
          total space
        }
      "{yshift=0pt}
    ]
    &&
      \overset{
        \mathclap{
        \raisebox{2pt}{
          \scalebox{.7}{
            \color{darkblue}
            \bf
            $X$-Dependent types
          }
        }
        }
      }{
        \big(
          X \to \Types
        \big)
      }
    \end{tikzcd}
  $$
  \end{minipage}
  &
  \begin{minipage}{4cm}
  $$
    \adjustbox{raise=20pt}{
    \begin{tikzcd}
      Y
      \ar[
        d,
        "{\ }"{name=t}
      ]
      \ar[
        r,
        "{\ }"{swap, name=s}
      ]
      \ar[
        from=s,
        to=t,
        Rightarrow,
        "{ \scalebox{.7}{(pb)} }"
      ]
      &
      \widehat{\Objects}
      \ar[d]
      \\
      X
      \ar[
        r
      ]
      &
      \Objects
    \end{tikzcd}
    }
  $$
  \end{minipage}
  \\
  \hline
  \end{tabular}
  }
\end{equation}

\smallskip

In more detail, the homotopy type-theoretic syntax directly interprets
into 1-categories (cf. Lit. \ref{VerificationLiterature}) and here those understood to ``present'' these $\infty$-toposes (this is the
key point of \cite{Lurie09}), known as {\it model toposes} \cite{Rezk10} or more specifically as  {\it type-theoretic model toposes}
\cite[\S 1.3]{Shulman19} (building on \cite{Shulman15}, review in \cite[\S 6.1]{Riehl22}), in which the object classifier exists as a universal fibration of small objects.

\smallskip
The archetypical example of such model toposes  is the ``classical model topos'' of classical homotopy theory, traditionally known as
the {\it Kan-Quillen model category of simplicial sets}. This is suitably equivalent (namely Quillen equivalent, presenting the same
$\infty$-topos --- for review and references see \cite[\S A]{FSS20Character}) to the {\it Serre-Quillen model category of k-topological spaces}
which tacitly underlies the considerations in \cref{ViaParameterizedPointSetTopology} but which fails to be strictly ``type-theoretic''
in that it is not {\it quite} locally Cartesian closed (only over base spaces which are Hausdorff, cf. \eqref{HausdorffBaseSpace}):
\begin{equation}
  \label{InterpretationInTheClassicalModelTopos}
  \begin{tikzcd}[
    column sep=40pt
  ]
    \underset{
      \mathclap{
        \raisebox{-4pt}{
          \scalebox{.8}{
            \color{gray}
            \cref{ViaDependentHomotopyTypeTheory}
          }
        }
      }
    }{
    \overset{
      \mathclap{
        \raisebox{8pt}{
          \scalebox{.7}{
            \color{darkblue}
            \bf
            \def\araystretch{.9}
            \begin{tabular}{c}
              syntactic category of
              \\
              homotopy data types
            \end{tabular}
          }
        }
      }
    }{
      \mathrm{HoTTSyntax}
    }
    }
    \ar[
      rr,
      "{
        \scalebox{.7}{
          \color{greenii}
          \bf
          semantics
        }
      }"{swap, yshift=-1pt}
    ]
    &&
    \quad
    \overset{
      \mathclap{
        \raisebox{6pt}{
          \scalebox{.7}{
            \color{darkblue}
            \bf
            \def\araystretch{.9}
            \begin{tabular}{c}
              type-theoretic
              \\
              model topos
              \\
              of simplicial sets
            \end{tabular}
          }
        }
      }
    }{
    \SimplicialSets_{\mathrm{Qu}}
    }
    \quad
    \ar[
      rr,
      "{
        \scalebox{.7}{
          \color{greenii}
          \bf
          Quillen equivalence
        }
      }"{swap, yshift=-1pt}
    ]
    &&
    \underset{
      \mathclap{
        \raisebox{-4pt}{
          \scalebox{.8}{
            \color{gray}
            \cref{ViaParameterizedPointSetTopology}
          }
        }
      }
    }{
    \overset{
      \mathclap{
        \raisebox{6pt}{
          \scalebox{.7}{
            \color{darkblue}
            \bf
            \def\araystretch{.9}
            \begin{tabular}{c}
              model topos
              \\
              of topological spaces
            \end{tabular}
          }
        }
      }
    }{
    \kTopologicalSpaces_{\mathrm{Qu}}
    }
    }
  \end{tikzcd}
\end{equation}
(The categorical semantics of dependent type theory on the left is due to \cite{Seely84}\cite{Hofmann97}\cite{Jacobs98}, the homotopy/model-category theoretic aspect due to \cite{AwodeyWarren09}, and finally the construction of the univalent object classifier due \cite{KLV12} and then in generality due to \cite{KLV12}. The Quillen equivalence on the right is classical, going back to \cite{Quillen67}, cf. \cite[\S I.11]{GoerssJardine99})

\medskip

Incidentally, under restriction to propositional types \eqref{DependentSumOfProps}
the equivalence \eqref{TypeClassification}
is a homotopy-theoretic generalization of the classical fact that a proposition about data of type $X$ is equivalently encoded in the sub-type of data satisfying this proposition: Semantically this is the existence of {\it sub-object classifiers} known from topos theory \cite{Lawvere70} (see e.g. \cite[\S 5.1]{BorceuxVol3})

\vspace{-.4cm}
\begin{equation}
  \label{PropositionsEquivalentToSubtypes}
  \adjustbox{}{
  \def\arraystretch{1.5}
  \begin{tabular}{|c|c|}
  \hline
  {\bf Propositional Type Universe}
  &
  {\bf Sub-Object classifier}
  \\
  \hline
  \hline
  \begin{minipage}{11cm}
  $$
  \begin{tikzcd}[
    decoration=snake,
    column sep=65pt
  ]
      \overset{
        \mathclap{
        \raisebox{2pt}{
          \scalebox{.7}{
            \color{darkblue}
            \bf
            Injections into $X$
          }
        }
        }
      }{
        (Y :\Types)
          \times
        (Y \hookrightarrow X)
      }
    \ar[
      rr,
      bend left=12,
      "{\ }"{name=s},
      "{
        \scalebox{.8}{
           \bf
           \color{greenii}
           fibers
        }
      }"{swap, yshift=0pt},
      "{
        (Y, f)
        \;\mapsto\;
        \left(
          x
            \,\mapsto\,
          \mathrm{fib}_x(f)
        \right)
      }"
    ]
    \ar[
      rr,
      phantom,
      "{ \simeq }"
    ]
    \ar[
      rr,
      <-,
      bend right=12,
      "{\ }"{name=t},
      "{
        \left(
          (x \,:\, X) \times P_x
          ,\;
          \mathrm{pr}_X
        \right)
        \;\mapsfrom\;
        \left(
          x \,\mapsto\, P_x
        \right)
      }"{swap},
      "
        \scalebox{.8}{
          \color{greenii}
          \bf
          total space
        }
      "{yshift=0pt}
    ]
    &&
  \;\;\;    \overset{
        \mathclap{
        \raisebox{2pt}{
          \scalebox{.7}{
            \color{darkblue}
            \bf
            Propositions about $X$
          }
        }
        }
      }{
        \big(
          X \to \Propositions
        \big)
      }
    \end{tikzcd}
  $$
  \end{minipage}
  &
  \begin{minipage}{4cm}
  $$
    \adjustbox{raise=20pt}{
    \begin{tikzcd}
      Y
      \ar[
        d,
        hook,
        "{\ }"{name=t}
      ]
      \ar[
        r,
        "{\ }"{swap, name=s}
      ]
      \ar[
        from=s,
        to=t,
        Rightarrow,
        "{ \scalebox{.7}{(pb)} }"
      ]
      &
      \widehat{\Objects_{-1}}
      \ar[d]
      \\
      X
      \ar[
        r
      ]
      &
      \Objects_{-1}
    \end{tikzcd}
    }
  $$
  \end{minipage}
  \\
  \hline
  \end{tabular}
  }
\end{equation}

For example, given a data set $D : \Sets$, then a {\it relation} on such data is given by the sub-type $R \hookrightarrow D \times D$ of those pairs of data which are in relation to each other, which is equivalently the proposition $D \times D \to \Propositions$ asserting about any pair that its data are in relation to each other:
\begin{equation}
  \label{RelationsAsSubtypesAndPropositions}
  \begin{tikzcd}
    \big(
      R : \Types
    \big)
    \times
    \big(
      r : R \to D \times D
    \big)
    \times
    \big(
      (d_1, d_2 : D)
        \to
      \mathrm{isProp}\big(
        \underbrace{
          \mathrm{fib}_{(d_1, d_2)}(r)
        }_{
          \mathclap{
            \scalebox{.7}{
              \color{darkblue}
              \bf
              \begin{tabular}{c}
                proposition that
                \\
                $d_1, d_2$ are in relation
              \end{tabular}
            }
          }
        }
      \big)
    \big)
    \quad
    \ar[
      rr,
      <->,
      "{ \sim }"
    ]
    &&
  \quad   \big(
      D \times D \to \Propositions
    \big)
  \end{tikzcd}
\end{equation}

\medskip

\newpage

\noindent
{\bf The idea of inductive data types.}
\label{InductiveTypes}
While we have seen how to construct new data types from given ones --- by forming dependent function types, dependent pair types \eqref{DependentFunctionNotation} and identification types \eqref{IdentificationType} --- it remains to discuss how to introduce definite data types in the first place.

The archetypical example is the data type $\Bits \,:\, \Types$ of Boolean truth values (often denoted $\mathrm{Bool}$, instead). It is clear that its {\it term introduction rule} should say that there are two pieces of data of this type, namely $0, \,1  : \Bits$. But it remains to introduce a language construct ensuring that there is {\it no other} data of this type. An elegant idea for achieving this is to declare that we obtain a function $\Bits \to P$ to any other data type $D$ {\it as soon} as we have correspondingly two pieces of data $0_D, \, 1_D \,:\, D$. But a subtle point has to be taken care of for this and analogous {\it term elimination rules} to work as expected: Since we are working in the generality of {\it dependent} data types \eqref{DependentData}, we need to declare this in the generality of {\it dependent functions}. But it turns out that it suffices to consider the case when $D$ depends on $\Bit$ itself (e.g. \cite[p. 35-37]{MartinLof84}\cite[\S 3.1]{AwodeyGambinoSojokova12}):

\begin{equation}
\label{BitsType}
\adjustbox{}{
\tabcolsep=5pt
\def\arraystretch{1.2}
\begin{tabular}{|l|l|l|}
\hline
\multicolumn{3}{|l|}{
{\bf Bits.}
}
\\
\hline
\hline
& {\bf Homotopy type-theory}
& {\bf Homotopy theory}
\\
\hline
\rotatebox{90}{
  \clap{
  \bf Formation
  }
}
&
$
\def\arraystretch{1.6}
\begin{array}{c}
  \\[-10pt]
  \\
  \hline
  \vdash \Bits \,:\, \Types
  \\[-13pt]
  {\phantom{A}}
\end{array}
$
&
\\
\cline{1-2}
\rotatebox{90}{
  \clap{
  \bf Introduction
  }
}
&
$
\def\arraystretch{2}
\begin{array}{l}
\def\arraystretch{1.6}
\begin{array}{c}
  \\
  \\
  \hline
  \vdash 0
    \,:\,
  \Bits
  \\
\end{array}
\;\;\;\;\;\;\;
\def\arraystretch{1.6}
\begin{array}{c}
  \\
  \\
  \hline
  \vdash 1
    \,:\,
  \Bits
  \\
\end{array}
\\[-18pt]
\phantom{-}
\end{array}
$
&
\hspace{.6cm}
\begin{minipage}{5cm}

\vspace{-2cm}

$
\begin{tikzcd}[sep=50pt]
  {}
  &
  \ast
  \ar[
    d,
    dashed,
    "{ 0 }"
  ]
  \\
  \ast
  \ar[
    r,
    dashed,
    "{ 1 }"{swap}
  ]
  &
  \Bits
\end{tikzcd}
$
\end{minipage}
\\
\hline
\rotatebox{90}{
  \clap{
    \bf
    Elimination
  }
}
&
$
  \def\arraystretch{1.6}
  \begin{array}{c}
    \begin{array}{l}
    b
      \,\colon\,
    \Bits
    \;\;\vdash\;\;
    P(b)
    \,\colon\,
    \Types
    \;;
    \\
   \vdash 0_P \,:\, P(0)
    ;\;\;\;
  \vdash  1_P \,:\, P(1)
    \end{array}
  \\
  \hline
    \mathclap{\phantom{\vert^{\vert}}}
    b \,\colon\, \Bits \,\vdash\,
    \mathrm{ind}_{(P,\,0_P,\,1_P)}(b)
    \,\colon\,
    P(b)
    \mathclap{\phantom{\vert_{\vert_{\vert_{\vert_{\vert_{\vert}}}}}}}
  \end{array}
$
&
\\
\cline{1-2}
\rotatebox{90}{
  \clap{
    \bf
    Computation
  }
}
&
$
\def\arraystretch{1.4}
\begin{array}{rlc}
  \\[-4pt]
    \mathrm{ind}_{(P,\,0_P,\,1_P)}(0)
    &\defneq&
    0_P
    \\
    \mathrm{ind}_{(P,\,0_P,\,1_P)}(1)
    &\defneq&
    1_P
    \\
    {\phantom{A}}
\end{array}
$
&
\hspace{-8pt}
\begin{minipage}{6.3cm}

\vspace{-2cm}

$
\begin{tikzcd}
  &[-7pt]
  \ast
  \ar[
    dr,
    shorten >=-3pt,
    "{ 0 }"{swap}
  ]
  \ar[
    rrrd,
    bend left=30,
    "{ 0_P }",
    "{\ }"{name=s, swap, pos=.57}
  ]
  &[-7pt]
  &[-8pt]
  &[-8pt]
  \\[-10pt]
  {}
  &&
  \Bits
  \ar[
    ddr,
    equals,
    gray,
    shorten <=-4pt,
    shorten >=-1pt
  ]
  \ar[
    rr,
    shorten <=-2pt,
    dashed,
    "{ \mathrm{ind}_{(P,\, \cdots)} }"
  ]
  &&
  P
  \ar[
    ddl,
    gray,
    shorten >=-2pt
  ]
  \\
  &
  \ast
  \ar[
    ur,
    shorten >=-1pt,
    "{ 1 }"
  ]
  \ar[
    rrru,
    bend right=45,
    "{ 1_P }"{swap, pos=.325},
    "{\ }"{name=t, pos=.35}
  ]
  \\[-40pt]
  && &
  \color{gray}
  \Bits
\end{tikzcd}
$
\end{minipage}
\\
\hline
\end{tabular}
}
\end{equation}

\smallskip
In homotopy theory this defines the space which is the disjoint union of two points -- the {\it 0-sphere} -- , universally characterized as the {\it coproduct} of
two copies of the 1-point space $\ast$. Here the universal map out of the coproduct (the ``term eliminator'') is denoted
``$\mathrm{ind}_{(\cdots)}$'' because this turns out to be an example of the same general notion of {\it induction} which also
controls the classical notion of induction over the natural numbers \eqref{NaturalNumberstype}.
Therefore, one refers to this and analogous data types as {\it inductive types} (\cite{CoquandPaulin90}\cite{Dybjer94}, early review in \cite{PaulinMohring}\cite[\S 9.2.2]{Luo94}, exposition in \cite[\S 5.6]{UFP13}).

\medskip
Concretely, there are two ways to introduce data of natural number type: On the one hand, there is certainly the datum
$0 \,:\, \mathbb{N}$; on the other hand, if data $n \,:\, \mathbb{N}$ is already given, then there is the datum $\mathrm{succ}(n) \,=\, n+1 \,:\, \mathbb{N}$. As before with the type of bits, the idea now is to enforce that these two introduction rules produce {\it all} $\mathbb{N}$-data by declaring that we obtain a function $\mathbb{N} \xrightarrow{\phantom{-}} D$ to any other data type {\it as soon} as that type $D$ is equipped with images $0_D$  and $\mathrm{succ}_D$ of these two constructors. Saying this in the generality that $D$ is a type depending \eqref{DependentData} on $\mathbb{N}$ yields the following inference rules \eqref{NaturalNumberstype} for natural numbers \cite[pp. 38]{MartinLof84}\cite[p. 52-53]{CoquandPaulin90}\cite[\S 3]{Dybjer94}:

\begin{equation}
\label{NaturalNumberstype}
\adjustbox{}{
\def\arraystretch{1.4}
\begin{tabular}{|c||l|l|}
\hline
\multicolumn{3}{|l|}{
  \bf Natural numbers.
}
\\
\hline
&
{\bf Homotopy type-theory}
&
{\bf Homotopy theory}
\\
\hline
\rotatebox{+90}{
  \clap{\bf Formation}
}
&
\; $
  \def\arraystretch{1.4}
  \begin{array}{c}
    \\
    {}
    \\
    \hline
   {\vdash\, \mathbb{N} \,:\, \Types }
    \\
    \\
  \end{array}
$
&
\multicolumn{1}{|c|}{
}
\\
\cline{1-2}
\rotatebox{+90}{
  \clap{\bf Introduction}
}
&
\; $
  \def\arraystretch{1.4}
  \begin{array}{c}
    \\
    {}
    \\
    \hline
    \vdash\, 0 \,:\, \mathbb{N}
    \\
    \\
  \end{array}
  \;\;\;\;\;\;\;\;
  \begin{array}{c}
    \\
   \vdash\,  n \,:\, \mathbb{N}
    \\
    \hline
   \vdash\,  \mathrm{succ}(n) \,:\, \mathbb{N}
    \\
    \\
  \end{array}
$
&
\hspace{1cm}
\begin{minipage}{5cm}

  \vspace{-2.3cm}

  $
    \begin{tikzcd}
    \ast \,\sqcup\, \mathbb{N}
    \ar[
      rr,
      dashed,
      "{
        (0,\, \mathrm{succ})
      }"
    ]
    &&
    \mathbb{N}
    \end{tikzcd}
  $
 \end{minipage}
\\
\hline
\rotatebox{+90}{
  \clap{\bf Elimination}
}
&
$
  \def\arraystretch{1.4}
  \arraycolsep=4pt
  \begin{array}{c}
    \begin{array}{rcl}
      n \,:\, \mathbb{N}
      &\vdash&
      D(n) \,:\, \Types
      \\
      &\vdash& 0_D \,:\, D(0)
      \\
      n \,:\, \mathbb{N} \,,\;\;
      d \,:\, D(n)
      &\vdash&
      \mathrm{succ}_D(n,\,d)
        \,:\,
      D\big(\mathrm{succ}(n)\big)
    \end{array}
    \\
    \hline
    \mathclap{\phantom{\vert^{\vert}}}
    n \,:\, \mathbb{N}
    \;\;\vdash\;\;
    \mathrm{ind}_{(D, 0_D, \mathrm{succ}_D)}(n)
    \,:\,
    D(n)
    \\[-12pt]
    {}
  \end{array}
$
&
\\
\cline{1-2}
\rotatebox{+90}{
  \clap{\bf Computation}
}
&
$
\def\arraystretch{1.6}
\begin{array}{l}
  \\[-14pt]
  \mathrm{ind}_{(D,\, 0_D,\,\mathrm{succ}_D)}
  (0)
  \;\defneq\;
  0
  \;;
  \\
  \mathrm{ind}_{(D,\, 0_D,\,\mathrm{succ}_D)}
  \big(
    \mathrm{succ}(n)
  \big)
  \\
  \;\;\;\;\;\defneq\;
  \mathrm{succ}_D
  \big(
    n
    ,\,
    \mathrm{ind}_{(D,\,0_D,\,\mathrm{succ}_D)}(n)
  \big)
  \\[-12pt]
  {}
\end{array}
$
&
\hspace{-.1cm}
\begin{minipage}{6.6cm}

\vspace{-2.2cm}

$
  \begin{tikzcd}[
    column sep=30pt,
    row sep=25pt
  ]
    &[+34pt]
    &
    [-60pt]
    \ast \,\sqcup\, \mathbb{N}
    \ar[
      ddll,
      "{
        (0,\mathrm{succ})
      }"{sloped}
    ]
    \ar[
      dr,
      equals,
      shorten >=-2pt
     ]
    \ar[
      r,
      dashed,
      shorten=-1pt,
      "{
        \mathrm{id}_\ast
        \,\sqcup\,
        \mathrm{ind}
      }"
    ]
    &[+16pt]
    \ast \,\sqcup\, D
    \ar[
      d,
      "{
        \mathrm{id}_\ast
          \,\sqcup\,
        \pi_D
      }"
    ]
    \\
    &
    &&
    \ast \,\sqcup\, \mathbb{N}
    \ar[
      ddll,
      "{
        (0, \mathrm{succ})
      }"{description, sloped}
    ]
    \\[-20pt]
    \mathbb{N}
    \ar[dr, equals]
    \ar[
      r,
      dashed,
      "{
        \mathrm{ind}
      }"{description}
    ]
    &
    D
    \ar[
      d,
      "{ \pi_D }"
    ]
    \ar[
      from=uurr,
      crossing over,
      "{
        (0_D,\, \mathrm{succ}_D)
      }"{description, sloped, pos=.6}
    ]
    \\
    &
    \mathbb{N}
  \end{tikzcd}
$

\end{minipage}
\\
\hline
\end{tabular}
}
\end{equation}

\medskip

In the denotational semantics on the right we see that $\mathbb{N}$ has the structure of a (homotopy-){\it initial algebra over the endofunctor}
$\ast \,\sqcup\, \mathrm{Id}$ on spaces (i.e. the endofunctor which reflects the domains of a nullary constructor $0$ and of a unary constructor $\mathrm{succ}$). In general, the denotational semantics of ``well-founded'' inductive types (``$\mathcal{W}$-types'' \cite[pp. 43]{MartinLof84})
in homotopy theory is given by (homotopy-)initial algebras of polynomial endofunctors \cite{Dybjer97}\cite{AwodeyGambinoSojokova12}\cite{AwodeyGambinoSojokova15}.

\medskip

The inductive rules for the natural number type capture the classical notions both of {\it proof of propositions by induction} and of
{\it construction of functions by recursion}:

\vspace{1mm}
\begin{itemize}[leftmargin=.5cm]

\item {\bf $\NaturalNumbers$-Induction.}
When  the dependent type in \eqref{NaturalNumberstype} is propositional \eqref{Propositions}, so that
$n \,:\, \mathbb{N} \;\vdash\; D(n) \,:\, \Propositions$ is a {\it proposition about}
natural numbers \eqref{DependentSumOfProps}, then:
\begin{itemize}[leftmargin=.4cm]
\item
the assumption of the elimination rule \eqref{NaturalNumberstype} is that we have a certificate $0_P \,:\, P(0)$ -- hence a {\it proof} of
the proposition about $0$ -- and moreover with each certificate/proof $d \,:\, P(n)$ of the proposition about any $n$ also a a proof
$\mathrm{succ}_P \,:\, P(n+1)$ of the proposition about $n + 1$;
\item in which case the conclusion of the elimination rule is a proof of the proposition for all natural numbers: $n \,:\, \mathbb{N}\;\;\vdash\;\; \mathrm{ind}(n) \,:\, P(n)$ -- thereby recovering the classical induction principle.
\end{itemize}

\medskip

\item {\bf $\NaturalNumbers$-Recursion.}
\label{NaturalNumberRecursion}
When $D$ happens to be independent of $n : \NaturalNumbers$ then the induction principle \eqref{NaturalNumberstype} is that of {\it recursive functions}.
For example, addition and multiplication of natural numbers may be recursively defined as follows:
\begin{equation}
\label{AdditionAndMultiplicationOfNaturalNumbers}
\def\arraystretch{1.3}
\begin{array}{l}
  D \,:\defneq\, \mathbb{N} \to \mathbb{N}
\\
  0_D :\defneq (k \mapsto k)
\\
\mathrm{succ}_{D}\big(
  n, f
  \big) :\defneq
  \big(
    k
    \mapsto
    \mathrm{succ}\big(f(k)\big)
  \big)
\\
\hline
+
\,:\defneq\,
\mathrm{ind}_{(D,0_D, \mathrm{succ}_D)}
\,:\,
\NaturalNumbers
\to
\big(
  \NaturalNumbers
  \to
  \NaturalNumbers
\big)
\end{array}
\hspace{1.4cm}
\def\arraystretch{1.3}
\begin{array}{l}
  D \,:\defneq\, \mathbb{N} \to \mathbb{N}
\\
  0_D :\defneq (k \mapsto 0)
\\
\mathrm{succ}_{D}\big(
  n, f
  \big) :\defneq
  \big(
    k
      \mapsto
    f(k)
    +
    k
  \big)
\\
\hline
\cdot
\,:\defneq\,
\mathrm{ind}_{(D,0_D, \mathrm{succ}_D)}
\,:\,
\NaturalNumbers
\to
\big(
  \NaturalNumbers
  \to
  \NaturalNumbers
\big)
\end{array}
\end{equation}

\end{itemize}

\medskip

\noindent
{\bf Unique and non-existent data.}
Finally, the most simple but important examples of inductive data types:

\noindent
The  {\it singleton type} (often: ``unit type'', e.g. \cite[p. 30]{UFP13}), whose induction rule witnesses it
as a (necessarily contractible) type \eqref{IsContractible} with an unique datum:
\vspace{-1mm}
\begin{equation}
  \label{UnitType}
  \def\arraystretch{1.3}
  \begin{array}{c}
    {}
    \\
    \hline

    \vdash\,\ast \,:\, \Types
  \end{array}
  \;\;\;\;\;\;\;\;\;
  \def\arraystretch{1.3}
  \begin{array}{c}
    {}
    \\
    \hline
    \vdash\,\scalebox{.7}{\raisebox{1pt}{$\bullet$}} \,:\, \ast
  \end{array}
  \;\;\;\;\;\;\;\;\;
  \def\arraystretch{1.3}
  \begin{array}{c}
    x \,:\, \ast
    \;\,\vdash\,\;
    D(x) \,:\, \Types
    \;;
    \;\;\;\;
    \scalebox{.7}{\raisebox{1pt}{$\bullet$}}_D \,:\, D(\scalebox{.7}{\raisebox{1pt}{$\bullet$}})
    \\
    \hline
   x : \ast\, \vdash\,
    \mathrm{ind}_{(D, \scalebox{.7}{\raisebox{1pt}{$\bullet$}}_D)}(x)
    \,:\,
    D(x)
  \end{array}
  \;\;\;\;\;\;\;\;\;\;
    \mathrm{ind}_{(D, \scalebox{.7}{\raisebox{1pt}{$\bullet$}}_D)}
    (\scalebox{.7}{\raisebox{1pt}{$\bullet$}})
    \;=\;
    \scalebox{.7}{\raisebox{1pt}{$\bullet$}}_D
\end{equation}

\begin{equation}
  \label{EmptyType}
\hspace{-.7cm}
\adjustbox{}{
\begin{tabular}{ll}
\begin{minipage}{10.6cm}
The {\it empty type} is inductively generated by {\it no} data (e.g. \cite[\S 1.7, \S A.8]{UFP13}\cite[\S 2.6]{Escardo19}):
\end{minipage}
&
\vspace{-.2cm}
$
  \def\arraystretch{1.3}
  \begin{array}{c}
    {}
    \\
    \hline
    \vdash\,\varnothing : \Types
  \end{array}
  \;\;
  \def\arraystretch{1.3}
  \begin{array}{c}
    x \,:\, \varnothing
    \;\,\vdash\,\;
    D(x) \,:\, \Types
    \\
    \hline
    x : \varnothing
    \vdash\,
    \mathrm{ind}_{D}
    \,:\,
    D(x)
  \end{array}
$
\vspace{-.2cm}
\end{tabular}
}
\end{equation}

\vspace{.4cm}
\hspace{-.8cm}
\begin{tabular}{ll}
\begin{minipage}{11.1cm}
This degenerate case of the general rules for inductive types gives the ancient logical rule of {\it ex falso quodlibet} --- from absurdity we may derive anything --- when we regard propositions as types \eqref{isProposition}. There can be no data of type $\varnothing$ since it would take {\it no} conditions to produce data $\mathrm{ind}(x) : P$ of any type $P$ (which is an absurdity) {\it assuming} $x : \varnothing$ as given (which hence must not exist). In practice, the induction principle of $\varnothing$ is used when doing a case analysis to escape from cases that cannot happen; a case that cannot happen will imply $\varnothing$, and so we may still prove our goal using the induction principle of $\varnothing$ in that case.
\end{minipage}
&
$
  \def\arraystretch{1.6}
  \begin{array}{c}
    P \,:\, \Types
    \\
    \hline
    \mathrm{const}_P
    :\defneq
    \big(
      (x : \varnothing)
      \,\mapsto\,
      D
    \big)
    \,:\,
    \big(
      \varnothing \to \Types
    \big)
    \\
    \hline
    \mathrm{ind}_{\mathrm{const}_P}
    \,:\,
    (x : \varnothing)
    \to
    D
  \end{array}
$
\end{tabular}

\vspace{3mm}
\noindent
{\bf The idea of higher inductive types.}
\label{HigherInductiveTypes}
But in a homotopically typed language, these induction principles for constructing concrete data types are to be generalized  to account for the introduction not just of plain data, but also of (re-)identitications \eqref{IdentificationType} of such data (``higher inductive types'' \cite[\S 6]{UFP13}\cite[\S 2.2.6]{vanDoorn18}\cite[\S 4]{VezzosiMoertbergAbel19}).
For example, in practice, one often considers data that is either of some type $Y$ or of some type $Y'$, except that a datum $y \,:\, Y$ is meant to be identified with data $y' \,:\, Y'$ whenever the pair $(y,y')$ \eqref{PairTypes} arises as the output of a given program
\vspace{-1mm}
$$
(f,f') \,:\, X \longmapsto Y \times Y'.
$$

\vspace{-2mm}
\noindent The type of such combined data is called the homotopy-{\it pushout} (or {\it cofiber coproduct}) of $f$ and (along) $f'$,
denoted $Y {}^{f}\!\!\underset{X}{\sqcup}^{f'} Y'$ or usually just $Y \underset{X}{\sqcup} Y'$, for brevity.
The inference rules
for such homotopy-cofiber/pushout types (e.g. \cite[p. 4]{HouFinsterLicataLumsdaine16}) are an evident expression of their homotopy-theoretic interpretation as homotopy cofiber/pushout spaces (e.g. \cite[\S 7.1]{Strom11}\cite[\S 6]{Arkowitz11}), as shown in \eqref{InferenceRulesForHomotopyPushout}:

\begin{equation}
\label{InferenceRulesForHomotopyPushout}
\adjustbox{}{
\tabcolsep=5pt
\def\arraystretch{1.2}
\begin{tabular}{|l|l|l|}
\hline
\multicolumn{3}{|l|}{
{\bf Homotopy pushout (cofiber coproducts)}
}
\\
\hline
\hline
& {\bf Homotopy type-theory}
& {\bf Homotopy theory}
\\
\hline
\rotatebox{90}{
  \clap{
  \bf Formation
  }
}
&
$
\def\arraystretch{1.6}
\begin{array}{c}
  \\[-10pt]
  \vdash \; X,\, Y,\, Y' \,:\, \Types
  ;\;
  \;\;
  \vdash f \,:\, X \to Y \;;\;\;
  \vdash f' \,:\, X \to Y'
  \\
  \hline
   \vdash \; Y \underset{X}{\sqcup} Y'
  \,:\, \Types
  \\[-13pt]
  {\phantom{A}}
\end{array}
$
&
\\
\cline{1-2}
\rotatebox{90}{
  \clap{
  \bf Term introduction
  }
}
&
$
\def\arraystretch{2}
\begin{array}{l}
\def\arraystretch{1.6}
\begin{array}{c}
   \vdash \; y \,:\, Y
  \\
  \hline
 \vdash \; \mathrm{cpr}(y)
    \,:\,
   Y \underset{X}{\sqcup} Y'
\end{array}
\;\;\;\;\;\;\;
\def\arraystretch{1.6}
\begin{array}{c}
  \vdash \;  y' \,:\, Y'
  \\
  \hline
  \vdash \; \mathrm{cpr}'(y')
    \,:\,
  Y \underset{X}{\sqcup} Y'
\end{array}
\\
\def\arraystretch{1.6}
\begin{array}{c}
 \vdash \; x \,:\, X
  \\
  \hline
  \mathclap{\phantom{\vert^{\vert^{\vert^{\vert^{\vert^{\vert}}}}}}}
 \vdash \; \mathrm{hmt}
    \,:\,
  \mathrm{Id}_{\big(
    Y \underset{X}{\sqcup} Y'
  \big)}
  \Big(
    \mathrm{cpr}\big(f(x)\big)
    ,\,
    \mathrm{cpr}'\big(f'(x)\big)
  \Big)
\end{array}
\\[-18pt]
\phantom{-}
\end{array}
$
&
\hspace{.6cm}
\begin{minipage}{5cm}

\vspace{-2cm}

$
\begin{tikzcd}[sep=50pt]
  X
  \ar[
    r,
    "{ f }"
  ]
  \ar[
    d,
    "{ f' }"{swap}
  ]
  &
  Y
  \ar[
    d,
    dashed,
    "{ \mathrm{cpr} }"
  ]
  \ar[
    dl,
    Rightarrow,
    dashed, color=orangeii,
    shorten=7pt,
    start anchor={[xshift=-8pt]},
    end anchor={[xshift=+8pt]},
    "{ \mathrm{hmt} }"{sloped, swap}
  ]
  \\
  Y'
  \ar[
    r,
    dashed,
    "{ \mathrm{cpr}' }"{swap}
  ]
  &
  Y \underset{X}{\sqcup} Y'
\end{tikzcd}
$
\end{minipage}
\\
\hline
\rotatebox{90}{
  \clap{
    \bf
    Term elimination
  }
}
&
$
  \def\arraystretch{1.6}
  \begin{array}{c}
    \begin{array}{l}
    \hat{y}
      \,\colon\,
    Y \underset{X}{\sqcup} Y'
    \;\vdash\;
    P(\hat{y})
    \,\colon\,
    \Types
    \;;
    \\
    \vdash \; \mathrm{cpr}_P
    \,\colon\,
    \underset{ y \colon Y }{\prod}
    P\big(
      \mathrm{cpr}(y)
    \big)
    \;\;
    \vdash
    \;\;
    \mathrm{cpr}'_P
    \,\colon\,
    \underset{ y' \colon Y' }{\prod}
    P\big(
      \mathrm{cpr}'(y')
    \big)
    \;;
    \\
   \vdash \; \mathrm{hmt}_P
    \,\colon\,
    \underset{x \colon X}{\prod}
    \mathrm{Id}_{
    }
    \Big(
      \mathrm{hmt}(x)_\ast
      \big(
        \mathrm{cpr}_P(
          f(x)
        )
      \big)
      ,\,
        \mathrm{cpr}'_P\big(
          f'(x)
        \big)
    \Big)
    \mathclap{\phantom{\vert_{\vert_{\vert_{\vert_{\vert_{\vert}}}}}}}
    \end{array}
  \\
  \hline
    \mathclap{\phantom{\vert^{\vert}}}
    \hat{y} \,\colon\, Y \underset{X}{\sqcup} Y'
    \,\vdash\;
    \mathrm{ind}_{\big(P,\, \mathrm{cpr}_P,\,\mathrm{cpr}'_P,\,\mathrm{hmt}_P\big)}(y)
    \,\colon\,
    P(\hat{y})
    \\[-15pt]
\phantom{-}
    \end{array}
$
&
\\
\cline{1-2}
\rotatebox{90}{
  \clap{
    \bf
    Computation
  }
}
&
\begin{minipage}{6.3cm}

$
\def\arraystretch{1.6}
\begin{array}{rlc}
  \\[-4pt]
    \mathclap{\phantom{\vert}}
    \mathrm{ind}_{(P,\, \mathrm{cpr}_P,\,\mathrm{cpr}'_P,\,\mathrm{hmt}_P)}
      \,\circ\,
    \mathrm{cpr}
    &\defneq&
    \mathrm{cpr}_P
    \;;
    \\
    \mathclap{\phantom{\vert^{\vert}}}
    \mathrm{ind}_{(P,\, \mathrm{cpr}_P,\,\mathrm{cpr}'_P,\,\mathrm{hmt}_P)}
      \,\circ\,
    \mathrm{cpr}'
    &\defneq&
    \mathrm{cpr}'_P
    \;;
    \\
\end{array}
$
\newline
$$
    x : X
    \;\vdash\;
    \mathrm{comp}
    \;:\; \Id_{}\Big(
    \mathrm{apd}_{
      \hspace{-3pt}
      \raisebox{2pt}{
        \scalebox{.7}{$
        \mathrm{ind}_{}
        $}
      }
      \hspace{-3pt}
   }
    (\mathrm{hmt}(x)),\, \mathrm{hmt}_P(x)
  \Big)
$$
\end{minipage}
&
\hspace{-8pt}
\begin{minipage}{6.3cm}

\vspace{-3.4cm}

$
\begin{tikzcd}
  &[-7pt]
  Y
  \ar[
    dr,
    shorten >=-3pt,
    "{ \mathrm{cpr} }"{description, sloped}
  ]
  \ar[
    dd,
    Rightarrow,
    shorten=11pt, color=orangeii,
    start anchor={[xshift=+6pt]},
    end anchor={[xshift=-6pt]},
    "{ \mathrm{hmt} }"{description}
  ]
  \ar[
    rrrd,
    bend left=30,
    "{ \mathrm{cpr}_P }",
    "{\ }"{name=s, swap, pos=.57}
  ]
  &[-7pt]
  &[-8pt]
  &[-8pt]
  \\[-10pt]
  X
  \ar[
    ur,
    "{ f }"
  ]
  \ar[
    dr,
    "{ f' }"{swap}
  ]
  &&
  Y \underset{X}{\sqcup} Y'
  \ar[
    ddr,
    equals,
    gray,
    shorten <=-4pt,
    shorten >=-1pt
  ]
  \ar[
    rr,
    shorten <=-2pt,
    dashed,
    "{ \mathrm{ind}_{(P,\, \cdots)} }"
  ]
  &&
  P
  \ar[
    ddl,
    gray,
    shorten >=-2pt
  ]
  \\
  &
  Y'
  \ar[
    ur,
    shorten >=-6pt,
    "{ \mathrm{cpr}' }"{description, sloped}
  ]
  \ar[
    rrru,
    bend right=45,
    "{ \mathrm{cpr}'_{P} }"{swap, pos=.325},
    "{\ }"{name=t, pos=.35}
  ]
  \ar[
    from=s,
    to=t,
    shorten=8pt,
    Rightarrow, color=orangeii,
    crossing over,
    "{\mathrm{hmt}_P}"{description, pos=.7}
  ]
  \\[-55pt]
  && &
  \color{gray}
  Y \underset{X}{\sqcup} Y'
\end{tikzcd}
$
\end{minipage}
\\
\hline
\end{tabular}
}
\end{equation}

The computation rule for the pushout is a little different than for the other inductive types. For the generatoring $\mathrm{hmt}$, we do not have a definitional equality but rather an identity $\mathrm{comp}$ which identifies the application of the induction function $\mathrm{ind}$ applies to $\mathrm{hmt}$ with t$\mathrm{hmt}_P$, the homotopy in the codomain. The function $\mathrm{apd}$ is the dependent version of $\mathrm{ap}$ \eqref{FunctionApplicationToIdentifications}, and is defined by path induction in the same way:

\vspace{-4mm}
\begin{equation}
    \label{FunctionDepApplicationToIdentification}
  \hspace{-.2cm}
  \def\arraystretch{1.4}
  \begin{array}{rccc}
    f \,:\, (c : C) \to D(c)
    \;\;\;\;\;\;
    \vdash
    \;\;\;\;\;\;
    \mathrm{apd}_f
    \;:
    \;
    (c_1,\, c_2 \,:\, C) \longrightarrow
        &
    \Big(
    \hspace{1pt}
    (p : \mathrm{Id}_D\big(c_1,\, c_2\big))
    &\to&
    \mathrm{Id}_{D(c_2)}\big(
      p_{\ast}f(c_1)
      ,\,
      f(c_2)
    \big)
    \mathrlap{
         \Big)
    }
    \\[-1pt]
    &
    \mathrm{id}_{c_1}
      &\mapsto&
    \mathrm{id}_{f(c_1)}
    \mathclap{\phantom{
      \vert_{\vert_{\vert_{\vert}}}
    }}
  \end{array}
\end{equation}
\vspace{-2mm}

For more on the computation rules of higher inductive types, see \cite[\S 6]{UFP13}. In other variants of homotopy type theory such as {\it cubical type theory} (\cite{CCHM15} \cite{CHM18}), the computation rules for the pushout and other higher inductive types can be given by definitional equalities like those for other inductive types.

\medskip

\noindent
{\bf The idea of higher homotopy bits.}
As an example of a pushout, the (un-reduced) {\it suspension} of a data type $X$, is the homotopy pushout \eqref{InferenceRulesForHomotopyPushout}
of $X \xrightarrow{ x \,\mapsto\, \scalebox{.6}{\raisebox{.9pt}{$\bullet$}} } \ast$ \eqref{UnitType}
along itself, hence the data type where each datum $x \,\colon\, X$ is promoted to a certificate of identification of a data pair $(\mathrm{nth}, \mathrm{sth})$:

\vspace{-.7cm}
\begin{equation}
  \label{Suspension}
  \adjustbox{}{
  \begin{tabular}{cc}
  \def\arraystretch{1.4}
  \tabcolsep=4pt
  \begin{tabular}{|l|l|}
    \hline
    \multicolumn{2}{|l|}{
      \bf Suspensions.
    }
    \\
    \hline
    {\bf Homotopy type-theory}
    &
    {\bf Homotopy theory}
    \\
    \hline
    $
    \mathrm{S}(X)
    \;:\defneq\;
    \ast \underset{X}{\sqcup} \ast
    \;:\;
    \Types
    $
    &
    \begin{tikzcd}[column sep=large]
      X
        \ar[r]
        \ar[d]
      &
      \ast
      \ar[
        d,
        "{ \mathrm{nth}_{\mathrm{S}X} }"
      ]
      \ar[
        dl,
        Rightarrow,
        "{ \mathrm{mer}_{\mathrm{S} X} }"{sloped}
        shorten=6pt, color=orangeii,
        start anchor={[xshift=-6pt]},
        end anchor={[xshift=+6pt]},
      ]
      \\
      \ast
      \ar[
        r,
        "{ \mathrm{sth}_{\mathrm{S} X} }"{swap}
      ]
      &
      \mathrm{S} X
    \end{tikzcd}
    \\
    \hline
  \end{tabular}
  &
  \hspace{1cm}
\adjustbox{}{
$
\overset{
  \mathclap{
    \raisebox{8pt}{
    \scalebox{.7}{
      \def\arraystretch{.8}
      \begin{tabular}{c}
        \scalebox{.9}{(un-reduced)}
        \\
        \color{darkblue}
        \bf
        suspension of $X$
      \end{tabular}
    }
    }
  }
}{
  \mathrm{S} X
}
\;\;\;\;\;=\;\;\;\;\;
\left\{\!\!\!
\adjustbox{raise=-2.5cm}{
\begin{tikzpicture}[scale=0.7]

\begin{scope}
  \draw (0,0) node{
    \adjustbox{scale={1}{.7}}{
      \includegraphics[width=7cm, angle=-90]{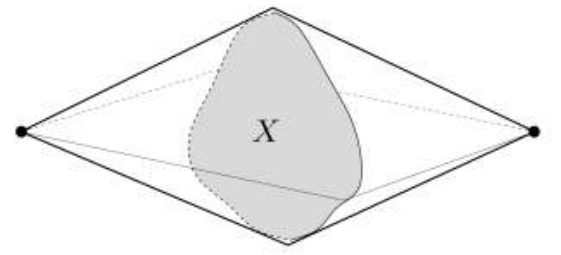}}
  };
  \draw[fill=lightgray, draw=lightgray]
    (0,+.18) circle (.3);
  \draw[fill=gray, draw=gray, draw opacity=.08,fill opacity=.12]
    (0,+.18) circle (.3);
\end{scope}
\node at (0,0) {
  \adjustbox{scale={1}{.65}}{
    \scalebox{1}{$X$}
  }
  };
\node at (1.33,.06) {
  \adjustbox{scale=.9}{
  \adjustbox{scale={1}{.65}}{
    \scalebox{1}{$x$}
  }
  }
  };

\node at (1.52, .07) {
  \adjustbox{scale={.9}{.7}}{$\bullet$}
};

\node at (1,1.1) {
  \rotatebox{-56}{
    \scalebox{.7}{
      $\mathrm{mer}(x)$
    }
  }
};
\node at (.94,-1.1) {
  \rotatebox{-124}{
    \scalebox{.7}{
      $\mathrm{mer}(x)$
    }
  }
};

  \draw (0,2.5) node {
    \scalebox{.7}{$\mathrm{nth}$}
  };
  \draw (0,-2.43) node {
    \scalebox{.7}{$\mathrm{sth}$}
  };
\end{tikzpicture}
}
\!\!\!\right\}
$
}
  \end{tabular}
  }
\end{equation}
\vspace{-.2cm}

\noindent
Indicated on the right\footnote{The graphics on the right of \eqref{Suspension} is adapted from \cite[p. 14]{Muro10}.} is the corresponding semantics as the topological suspension space  (e.g. \cite[p. 41]{Janich84}\cite[p. 8]{Hatcher02}).
For example, the suspension
\eqref{Suspension}
of the empty type \eqref{EmptyType} is inductively generated by two terms with {\it no} homotopy between them --- this is equivalently the type of bits:
$
  \mathrm{S}\varnothing
  \;\;
  \simeq
  \;\;
  \Bits
$
\eqref{BitsType}.

\vspace{-.3cm}
\begin{equation}
  \label{Spheres}
  \hspace{-.55cm}
  \adjustbox{}{
\begin{tabular}{ll}
\begin{minipage}{8.3cm}
Next, the suspension of the type of bits is freely generated by (1.) a pair of data points with (2.) a pair of identifications between them. This is the homotopy type of the circle, cf. the illustration in \eqref{HigherHomotopyBits}.

Inductively, if $X = S^n$ is an $n$-sphere homotopy type, then its suspension is the $n+1$-sphere homotopy type, realized as the union of (1.) a pair of poles $\mathrm{nth}$, $\mathrm{sth}$ and (2.) the {\it meridians} $\mathrm{mer}(s)$ through all points $s$ in the {\it equator} $n$-sphere.
\end{minipage}
  &
  \hspace{-.4cm}
  \begin{tabular}{ll}
  \def\arraystretch{1.4}
  \tabcolsep=4pt
  \begin{tabular}{|l|l|}
    \hline
    \multicolumn{2}{|l|}{
      \bf $n$-Spheres.
    }
    \\
    \hline
    {\bf Homotopy type-theory}
    &
    {\bf Homotopy theory}
    \\
    \hline
    $
      \arraycolsep=3pt
      \def\arraystretch{1.4}
      \begin{array}{lcl}
      \\[-15pt]
      S^{-1} &:\defneq& \varnothing
      \\
      S^0 &:\defneq& \Bits
      \\
      S^{n+1} &:\defneq& \mathrm{S} S^n
      \end{array}
    $
    &
    \begin{tikzcd}[column sep=large]
      S^n
        \ar[r]
        \ar[d]
      &
      \ast
      \ar[
        d,
        "{ \mathrm{nth}_{S^{n+1}} }"
      ]
      \ar[
        dl,
        Rightarrow, color=orangeii,
        "{ \mathrm{mer}_{S^{n+1}} }"{sloped}
        shorten=6pt,
        start anchor={[xshift=-6pt]},
        end anchor={[xshift=+6pt]},
      ]
      \\
      \ast
      \ar[
        r,
        "{ \mathrm{sth}_{S^{n+1}} }"{swap}
      ]
      &
      S^{n+1}
    \end{tikzcd}
    \\
    \hline
  \end{tabular}
  \end{tabular}
  \end{tabular}
  }
\end{equation}
In other words, in homotopy-typed programming languages, the archetypical data type $\Bits$ of bits \eqref{BitsType}
is accompanied by a tower of higher homotopy types, obtained as its iterated suspensions \eqref{Suspension} via
$\mathbb{N}$-induction \eqref{NaturalNumberstype}:

\smallskip
By \eqref{isTruncatedIfLoopingsAreContractible}  below, these higher sphere types serve to detect all the higher homotopy nature of data types.
In this sense, homotopy-typed programming is all about {\it generalizing bits to ``higher homotopy bits''}:
\begin{equation}
\label{HigherHomotopyBits}
\adjustbox{}{
\hspace{-.8cm}
\begin{tabular}{ll}
\begin{minipage}{5.5cm}
While this perspective on homotopy data type-theory is mathematically as compelling as it is intriguing, the practical content of ``computation on higher homotopy bits'' in actual computer science (as opposed to its interpretation in mathematical homotopy theory) has arguably remained somewhat elusive. Our claim in \cref{KZConnectionsInHomotopyTypeTheory} is that
(certification of)
topological quantum computation fills this gap.
\end{minipage}
&
\;\;\;
{\small
\def\arraystretch{1.5}
\tabcolsep=14pt
\begin{tabular}{|c|c|c|}
  \hline
  \multicolumn{3}{|c|}{
    \bf Higher homotopy bits
  }
  \\
  \hline
  {\bf Classical bits}
  &
  {\bf Circle type}
  &
  {\bf Sphere type}
  \\
  \hline
  $S^0 \,=\, \Bits$
  &
  $S^1 \,=\, \mathrm{S} \,\Bits$
  &
  $S^2 \,=\, \mathrm{S}^2 \,\Bits$
  \\
  \hline
  \hline
  $
  \begin{tikzcd}[
    row sep=20pt,
    decoration=snake
  ]
  &
  \color{white}\mathrm{nth}
  \ar[
    dd,
    bend right=80,
    decorate,
    white,
    "{
      \scalebox{1.5}{\color{black}$0$}
    }"{description, xshift=-4pt}
  ]
  \ar[
    dd,
    bend left=80,
    decorate,
    white,
    "{
      \scalebox{1.5}{\color{black}$1$}
    }"{description, xshift=1pt}
  ]
  \\
  \\
  &
  \color{white}\mathrm{sth}
  \end{tikzcd}
  $
  &
  $
  \begin{tikzcd}[
    row sep=20pt,
    decoration=snake
  ]
  &
  \scalebox{1.3}{$0'$}
  \ar[
    dd,
    bend right=80,
    shorten >=-2pt,
    decorate,
    "{
      \scalebox{1.5}{$0$}
    }"{description, xshift=-4pt}
  ]
  \ar[
    dd,
    bend left=80,
    decorate,
    shorten >=-2pt,
    "{
      \scalebox{1.5}{$1$}
    }"{description, xshift=1pt}
  ]
  \\
  \\
  &
  \scalebox{1.3}{$1'$}
  \end{tikzcd}
  $
  &
\adjustbox{raise=-1.4cm}{
\begin{tikzpicture}[decoration=snake]

\draw (-.03,0) node {
  \begin{tikzcd}[
    row sep=20pt,
    decoration=snake
  ]
  &
  \scalebox{1.3}{$0'$}
  \ar[
    dd,
    bend right=80,
    shorten >=-3,
    decorate,
    "{
      \scalebox{1.6}{$0$}
    }"{description, xshift=-4pt},
  ]
  \ar[
    dd,
    bend left=80,
    decorate,
    shorten >=-3,
    "{
      \scalebox{1.6}{$1$}
    }"{description, xshift=2pt},
  ]
  \\
  \\
  &
  \scalebox{1.3}{$1'$}
  \end{tikzcd}
  };

  \begin{scope}[xscale=-1]
  \begin{scope}[xshift=-1,yshift=-10]
  \draw[
    line width=2.1,
    decorate
  ] (0,0) arc (-94:-180:.85 and .3);
  \end{scope}
  \begin{scope}[yshift=0]
  \draw[
    line width=1.2,
    dashed,
  ] (-180:.85 and .3) arc (-180:-258:.85 and .3);
  \draw[
    ->,
    line width=.4
  ] (-.17,.3) to (-.17+.001, .3);
  \end{scope}
  \end{scope}

  \begin{scope}[yshift=0]
  \draw[
    line width=1.2,
    dashed,
  ] (-180:.85 and .3) arc (-180:-258:.85 and .3);
  \node at (0,.3) {
    \colorbox{white}{
    \hspace{-11pt}
    \scalebox{.8}{
      $1''$
    }
    \hspace{-11pt}
    }
  };
  \draw[
    ->,
    line width=.4
  ] (-.17,.3) to (-.17+.001, .3);
  \end{scope}

  \begin{scope}[xshift=-1,yshift=-10]
  \draw[
    line width=2.1,
    decorate
  ] (0,0) arc (-94:-180:.85 and .3);
  \node at (0,0) {
    \hspace{0pt}
    \colorbox{white}{
      \hspace{-11pt}
      \scalebox{1.4}{
        $0''$
      }
    \hspace{-11pt}
    }
  };
  \end{scope}

\end{tikzpicture}
}
  \\
  \hline
\end{tabular}
}
\end{tabular}
}
\end{equation}

\noindent
{\bf Homotopy cofibers, CW-complexes, and sequential colimits.}
Another important special case of homotopy pushouts \eqref{InferenceRulesForHomotopyPushout} are {\it homotopy cofibers} (the notion ``dual'' to homotopy fibers \eqref{HomotopyFiber}),
{\it cell attachments}
(e.g. \cite[\S 3.1]{AguilarGitlerPrieto02})
and the resulting cell complexes (specifically ``CW-complexes'', e.g. \cite[pp. 5]{Hatcher02}\cite[\S 5.1]{AguilarGitlerPrieto02}):

\begin{equation}
  \label{Cofiber}
  \adjustbox{}{
  \begin{tabular}{|l|l|l|}
  \hline
  \begin{minipage}{5.5cm}
The {\bf homotopy cofiber} of a function $f \,\colon\, Y \to X$
is the special case of
the cofiber coproduct type
\eqref{InferenceRulesForHomotopyPushout} where one summand is the singleton type
\eqref{UnitType}:
  \end{minipage}
  &
  \begin{minipage}{5cm}
  $
  \mathrm{cof}(f)
  \;:\defneq\;
  \ast
    \underset{X}{\sqcup}
  Y
  $
  \end{minipage}
  &
  \begin{minipage}{4cm}
  \begin{tikzcd}
    X
    \ar[
      r,
      "{ f }",
      "{\ }"{swap, name=s, pos=.7}
    ]
    \ar[
      d,
      "{\ }"{name=t, pos=.8}
    ]
    &
    Y
    \ar[d]
    \\
    \ast
    \ar[r]
    &
    \mathrm{cof}(f)
    \ar[
      from=s,
      to=t,
      Rightarrow,
      color=orangeii
    ]
  \end{tikzcd}
  \end{minipage}
  \\
  \hline
  \end{tabular}
  }
\end{equation}
\begin{equation}
  \label{CellAttachment}
  \adjustbox{}{
  \begin{tabular}{|l|l|l|}
  \hline
  \begin{minipage}{5.5cm}
    A single $n$-{\bf cell attachment} to a type $X$ is a cofiber \eqref{Cofiber} of a function $f : S^n \to X$ out of the $n$-sphere type \eqref{Spheres}.
  \end{minipage}
  &
  \begin{minipage}{5cm}
  $
  \mathrm{cof}
  (
    S^n
    \xrightarrow{f}
    X
  )
  $
  \end{minipage}
  &
  \begin{minipage}{4cm}
  \begin{tikzcd}
    S^{n}
    \ar[
      r,
      "{ f }",
      "{\ }"{swap, name=s, pos=.7}
    ]
    \ar[
      d,
      "{\ }"{name=t, pos=.8}
    ]
    &
    X
    \ar[d]
    \\
    \ast
    \ar[r]
    &
    X \cup_f D^{n}
    \ar[
      from=s,
      to=t,
      Rightarrow,
      color=orangeii
    ]
  \end{tikzcd}
  \end{minipage}
  \\
  \hline
  \end{tabular}
  }
\end{equation}
\begin{equation}
  \label{ParameterizedCellAttachment}
  \adjustbox{}{
  \begin{tabular}{|l|l|l|}
  \hline
  \begin{minipage}{5.5cm}
    More generally:

    An indexed $n$-{\bf cell attachment} to a type $X$ is
    such a pushout \eqref{InferenceRulesForHomotopyPushout}
    relative to a parameter type $R$.
   \end{minipage}
  &
  \begin{minipage}{5cm}
  $
  \mathrm{po}
  \left(\!\!\!
    \begin{tikzcd}
      R \times S^n
      \ar[r, "{f}"]
      \ar[
        d,
        "{\mathrm{pr}_R}"
      ]
      &
      X
      \\
      R
    \end{tikzcd}
 \!\!\! \right)
  $
  \end{minipage}
  &
  \begin{minipage}{4cm}
  \begin{tikzcd}
    R \times S^{n}
    \ar[
      r,
      "{ f }",
      "{\ }"{swap, name=s, pos=.7}
    ]
    \ar[
      d,
      "{\mathrm{pr}_R}"{swap},
      "{\ }"{name=t, pos=.8}
    ]
    &
    X
    \ar[d]
    \\
    R
    \ar[r]
    &
    X \cup_f D^{n}
    \ar[
      from=s,
      to=t,
      Rightarrow,
      color=orangeii
    ]
  \end{tikzcd}
  \end{minipage}
  \\
  \hline
  \end{tabular}
  }
\end{equation}
\begin{equation}
  \label{FiniteCWComplex}
  \adjustbox{}{
  \begin{tabular}{|l|l|l|}
  \hline
  \begin{minipage}{5.5cm}
   A {\bf finite CW-complex} is  obtained from the empty type
   \eqref{EmptyType}
   by a finite sequence of cell attachments \eqref{CellAttachment} of increasing dimension $n_{k+1} \,\geq\, n_k$
  \end{minipage}
  &
  \begin{minipage}{5cm}
  $
  \def\arraystretch{1.3}
  \arraycolsep=2pt
  \begin{array}{lcl}
  X^{-1}
  &:\defneq&
  \varnothing
  \\
  X^{k+1}
  &:\defneq&
  \mathrm{cof}
  \Big(\!
    { S^{n_{k+1}} }
    \;
    \xrightarrow{
        f^{(k)}
    }
    X^k
    \Big)
  \\
  X
  &:\defneq&
  X^{k_{\mathrm{max}}}
  \end{array}
  $
  \end{minipage}
  &
  \hspace{-4pt}
  \begin{minipage}{4cm}
  $
  \hspace{-4pt}
  \def\arraystretch{1.2}
  \begin{array}{c}
    \varnothing
    \hookrightarrow
    \cdots
    \hookrightarrow
    \\
    X^{k}
    \xhookrightarrow{}
    X^{k} \cup_{f^{(k)}} D^{n_{k+1}}
    =:
    X^{k+1}
    \\
    \hookrightarrow
    \cdots
    \hookrightarrow X
  \end{array}
  $
  \end{minipage}
  \\
  \hline
  \end{tabular}
  }
\end{equation}
\begin{equation}
  \label{FiniteSet}
  \adjustbox{}{
  \begin{tabular}{|l|l|l|}
  \hline
  \begin{minipage}{5.5cm}
   A {\bf finite set} is a finite CW-complex
   \eqref{FiniteCWComplex}
   of dimension 0.
  \end{minipage}
  &
  \begin{minipage}{5cm}
  $
  \def\arraystretch{1.3}
  \arraycolsep=2pt
  \begin{array}{lcl}
  X^{-1}
  &:\defneq&
  \varnothing
  \\
  X^{k+1}
  &:\defneq&
  \mathrm{cof}
  \big(\!
    \varnothing
    \;
    \xrightarrow{
      \;\;
    }
    X^k
    \big)
  \\
  X
  &:\defneq&
  X^{k_{\mathrm{max}}}
  \end{array}
  $
  \end{minipage}
  &
  \hspace{-4pt}
  \begin{minipage}{4cm}
  $
  \hspace{-4pt}
  \def\arraystretch{1.2}
  \begin{array}{c}
    \big\{
      x_1
      ,\,
      x_2
      ,\,
      \cdots
      ,\,
      x_n
    \big\}
  \end{array}
  $
  \end{minipage}
  \\
  \hline
  \end{tabular}
  }
\end{equation}

\smallskip

\begin{equation}
  \label{SequenceColimitPushoutDiagram}
  \adjustbox{}{
  \begin{tabular}{|l|l|l|}
  \hline
  \begin{minipage}{5.5cm}
   $\mathclap{\phantom{\vert^{\vert^{\vert}}}}$
   A {\bf sequential colimit}
   over functions

   $
     \begin{array}{l}
     n : \NaturalNumbers
     \;
     \vdash
     \;
     X_n : \Types
     \,,
     \\
     n : \NaturalNumbers
     \;\vdash\;
     f_n : X_n \to X_{n+1}
     \end{array}
   $

   indexed by the natural numbers
   \eqref{NaturalNumberstype} is
   also obtained by a homotopy pushout
   \eqref{InferenceRulesForHomotopyPushout}
   (e.g. \cite[\S 15]{Rijke18}).
  \end{minipage}
  &
    \begin{minipage}{9.4cm}
    \begin{tikzcd}[
      row sep = 5.1em,
      column sep=110pt
    ]
    \big(\dsum{n : \mathbb{N}} X_n\big) \times S^0 \ar[
      r,
      "{
        \left\{
        \hspace{-4pt}
        \def\arraystretch{1}
        \begin{array}{l}
        \scalebox{1.3}{$($}
          (n, x), 0
        \scalebox{1.3}{$)$}
        \,\mapsto\,
        \scalebox{1.3}{$($}
          n+1,\, f_n(x)
        \scalebox{1.3}{$)$}
        \\
        \scalebox{1.3}{$($}
          (n, x), 1
        \scalebox{1.3}{$)$}
        \,\mapsto\,
        (n, x)
        \end{array}
        \hspace{-4pt}
        \right.
        }"{description},
        "{\ }"{swap, yshift=-7pt, pos=.11, name=s}
       ]
       \ar[
         d,
         "{
           \begin{array}{c}
             \big((n, x), i\big)
             \\
             \rotatebox{270}{
             \hspace{-.5em}$\mapsto$}
              \\
              (n, x)
             \end{array}
          }"',
          "{\ }"{name=t}
         ]
         \ar[
           from=s,
           to=t,
           Rightarrow,
           color=orangeii
         ]
         &
       \dsum{n : \mathbb{N}} X_n
    \ar[
      d,
      dashed
    ]
    \\
    \dsum{n : \mathbb{N}} X_n
    \ar[
      r,
      dashed,
      "{
        \overset{
         \mathclap{
           \raisebox{4pt}{
             \scalebox{.7}{
               \color{greenii}
               colimiting co-cone
             }
           }
         }
        }{
        n
        \;\;\;\mapsto\;\;\;
        \mathrm{cpr}_n : X_n \to X_\infty
        }
      }"
    ]
    &
    X_{\infty}
    \end{tikzcd}
    \end{minipage}
  \\
  \hline
  \end{tabular}
  }
\end{equation}


\medskip

\noindent
{\bf The idea of homotopy $n$-truncation of data types.}
\label{TheIdeaOfNTruncation}
When $X : \Types$ is not necessarily $n$-truncated \eqref{HigherHomotopyType},
we may construct (e.g. as in \cite[\S 7.3, p. 223]{UFP13}) its {\it $n$-truncation} $\Truncation{}{n}{X} : \TruncatedTypes{n}$: which is the ``best approximation'' to $X$ by a type that {\it is} $n$-truncated, in that it comes with a function of the form $\eta_n : X \to \Truncation{}{n}{X}$ which is {\it initial} among functions from $X$ to $n$-truncated types, namely uniquely factoring any such function up to equivalence \eqref{HomotopyEquivalenceOfTypes}. As before with other inductive types, it is natural to state such a {\it universal mapping property} ($\mathrm{ump}$) in the generality of dependent types and dependent functions, where it looks as follows:
\begin{equation}
    \label{UniversalPropertyTrunctation}
    \adjustbox{}{
\def\arraystretch{1.6}
\begin{tabular}{|c|c|}
  \hline
  {\bf Homotopy type theory}
  &
  {\bf Homotopy theory}
  \\
  \hline
  $
  Y :
    \Truncation{}{n}{X}
    \to \TruncatedTypes{n}
    \;\;\;
    \vdash
    \;\;\;
    \mathrm{ump} \,:\,
    \mathrm{isEquiv}
    \Big(
       \big(
         (-) \circ \eta_n
       \big)
         :
       \underset
         {c : \Truncation{}{n}{X}}
         {\prod}
         Y(c)
       \to
       \underset
         {x : X}
         {\prod}
       Y\big(
         \eta_n(x)
       \big)
    \Big)
  $
  &
\(
  \begin{tikzcd}
	& Y \\
	X & {\Truncation{}{n}{X}}
	\arrow[
	  from=2-1,
	  to=2-2,
	  "{ \eta_X }"
	]
	\arrow[from=1-2, to=2-2, "{\mbox{\color{gray} $n$-truncated}}"]
	\arrow["{\exists!}", bend left, dashed, from=2-2, to=1-2]
	\arrow["\forall", bend left, dashed, from=2-1, to=1-2]
\end{tikzcd}
  \)
  \\
  \hline
\end{tabular}
}
\end{equation}

\vspace{1mm}
\noindent
For example, since any $n$-truncated type is also $(n+1)$-truncated, these factorizations yield for each $X : \Types$ a tower of truncations

\vspace{-.9cm}
\begin{equation}
  \label{PostnikovTower}
  \begin{tikzcd}
    X
    \ar[r, ->]
    \ar[
      rrr,
      rounded corners,
      to path={
           ([yshift=1pt]\tikztostart.north)
        -- node {
          \hspace{4.8cm}
          \raisebox{10pt}{\scalebox{.8}{\colorbox{white}{$
            \eta_n
          $}}}
        }
        ([yshift=8pt]\tikztostart.north)
        -- ([yshift=7pt]\tikztotarget.north)
        -- (\tikztotarget.north)
      },
    ]
    &
    \cdots
    \ar[r]
    &
    \Truncation{}{n+1}{X}
    \ar[r]
    &
    \Truncation{}{n}{X}
    \ar[r]
    &
    \cdots
    \ar[r]
    &
    \Truncation{}{0}{X}
    \ar[r]
    &
    \Truncation{}{-1}{X}
    \;\defneq:\;
    \exists X
  \end{tikzcd}
\end{equation}

\vspace{-1mm}
\noindent which in the homotopy theoretic semantics is known as the
{\it Postnikov tower} (\cite[Cor. 3.7]{GoerssJardine99}\cite[\S 5.5.6 \& \S 6.5]{Lurie09}). Classical homotopy theory has been interested mostly in the higher stages of the Postnikov tower, but homotopy type theory brings out that its low stages are of profound {\it logical} relevance:

\begin{itemize}

\item
 (-1)-truncation of a type $X$ -- also called {\it propositional truncation} \cite{AwodeyBauer04}\cite[\S 3.7]{UFP13}\cite{Kraus14} -- may be understood
 as producing the proposition \eqref{isProposition} that
 {\it there exists data of type $X$}  (that $X$ is ``inhabited''), cf. \eqref{LogicalConnectivesAsTypeFormation}:

\begin{equation}
\label{MeaningOfPropositionalTruncation}
\adjustbox{}{
\def\arraystretch{1.6}
\begin{tabular}{|c|c|}
  \hline
  {\bf Type theory}
  &
  {\bf Logic}
  \\
  \hline
  $
    \Gamma
    \;\;\;
    \vdash
    \;\;\;
   \exists X \,:\defneq\, \Truncation{\big}{-1}{X}
    \;:\;
    \Propositions
  $
  &
  $
  \exists_{x \in X}
  $
  \\
  \hline
\end{tabular}
}
\end{equation}

The universal property \eqref{UniversalPropertyTrunctation} of propositional truncation is the usual rule for using an existential quantifier in a proof: to prove a proposition $P$ assuming $\exists X$, we may assume we have an $x : X$. Explicitly, if $P$ is any proposition then to prove $\exists X \to P$ it suffices by the universal property to give a function $X \to P$ which (like any function) may be defined by assuming an element $x : X$ and then proving $P$.

\item
0-truncation of a type $X$ -- also called {\it set truncation} -- may be understood as producing the set \eqref{Sets} of equivalence classes of $X$-data; semantically this is the passage to the set $\pi_0$ of connected components of a space, and for dependent data types all this applies fiber-wise \eqref{TypeClassification}, cf. Lem \ref{FiberwiseTruncationPreservedByBaseChange}:
\begin{equation}
\label{FiberwiseZeroTruncationAndConnectedComponents}
\adjustbox{}{
\def\arraystretch{1.6}
\begin{tabular}{|c|c|}
  \hline
  {\bf Homotopy type theory}
  &
  {\bf Homotopy theory}
  \\
  \hline
  $
    \Gamma
    \;\;\;
    \vdash
    \;\;\;
    \Truncation{\big}{0}{X}
    \;:\;
    \Sets
  $
  &
  $
  \pi_{
    0/\Gamma
  }(X)
  $
  \\
  \hline
\end{tabular}
}
\end{equation}
\end{itemize}

\medskip

\noindent
{\bf Cell-complex construction of $n$-truncation.}
While in homotopy type theory it is popular, following \cite[\S 7.3, p. 223]{UFP13}, to construct $n$-truncation \eqref{UniversalPropertyTrunctation} in one step as a clever higher inductive type construction; we now highlight an alternative construction of $n$-truncation, which is implicit in \cite{Rijke19} and whose semantics is closer to the classical construction of $n$-truncations in homotopy theory.

First notice that from unwinding the definition of {\it $n$-truncation} \eqref{HigherHomotopyType}, {\it looping} \eqref{LoopSpace} and {\it higher spheres} \eqref{Spheres} one finds (\cite[Thm. 7.2.9]{UFP13}\cite[Thm. 3.10]{CherubiniRijke21}), in close analogy to classical homotopy theoretic arguments, that a type is $n$-truncated precisely if  its $(n+1)$-fold loopings are contractible \eqref{IsContractible}, and that an $n+1$-fold looping is equivalently the function type out of the ${n+1}$-sphere preserving the given base datum (\cite[Lem. 6.5.4]{UFP13}):
\begin{equation}
  \label{isTruncatedIfLoopingsAreContractible}
  \def\arraystretch{.9}
  \begin{array}{l}
    n : \NaturalNumbers
    ,\;
    \\
    X : \Types
  \end{array}
  \;\;\;\;
  \vdash
  \;\;\;\;
  \isTruncatedType{n}(X)
  \;\;
  \simeq
  \;\;
  \Big(
    (x : X)
    \to
    \big(
    \exists !
    \;
    \Omega^{n+1}_{x} X
    \big)
  \Big)
  \;\;
  \simeq
  \;\;
  \Big(
  (x : X)
  \to
  (f : S^{N + 1} \to X)
  \times
  \Id
  \big(
    f(\mathrm{nth})
    ,\,
    x
  \big)
  \Big)
  \,.
\end{equation}
Therefore one may expect that the $n$-truncation $\Truncation{}{n}{X}$ is obtained by adjoining trivializations of all $f : S^{n+1} \to X$, via the following indexed cell-attachment \eqref{ParameterizedCellAttachment}:
\begin{equation} \label{StageOneOfTruncationPushout}
\begin{tikzcd}
    (S^{n+1}\!\! \to X) \times S^{n+1}
    \ar[
      d,
      "{ \mathrm{ev} }"{swap},
      "{\ }"{name=t}
    ]
    \ar[
      r,
      "{\mathrm{pr}_1 }",
      "{\ }"{swap, name=s, pos=.7}
    ]
    &
    (S^{n+1}\!\! \to X)
    \ar[d]
    \\
    X
    \ar[r]
    &
    X_1
    \ar[
      from=s,
      to=t,
      Rightarrow,
      color=orangeii
    ]
  \end{tikzcd}
\end{equation}
While this is the right idea, the result $X_1$ may still fail to be $n$-truncated., but it is getting closer: We must
pass to the colimit of iterating this construction (following \cite[\S 7.2]{Rijke19}):

\begin{definition}[Truncation]
\label{TruncationDefn}
 For $X : \Types$ and $n : \NaturalNumbers$ we define $\eta_n : X \to \Truncation{}{n}{X}$ to be the colimiting co-cone of the
 sequential colimit \eqref{SequenceColimitPushoutDiagram}
 \begin{equation}
   \label{SequentialColimitForTruncation}
   \Truncation{}{n}{X}
   \;:\defneq\;
   X_\infty
   \;=\;
   \mathrm{colim}
   \big(
     X :\defneq X_0
     \xrightarrow{i_0}
     X_1
     \xrightarrow{i_1}
     X_2
     \xrightarrow{i_2}
     \cdots
   \big)
   \,,
   \hspace{1cm}
   \eta_n
   \;
   :\defneq
   \;
   \mathrm{cpr}_0
   \;:\;
   X \to [X]_n
 \end{equation}
 over $n+2$-cell attachments $i_k$
 that are indexed
 \eqref{ParameterizedCellAttachment} by the functions $S^{n+1}\!\! \to X_k$ into the previous stage of the sequence\footnote{
   In classical homotopy theory the construction \eqref{TruncationPushoutDiagram} is expressed by explicitly attaching sets of cells of ever higher dimension $n + 2 + k$, thereby iteratively filling up the higher homotopy groups. Here in the type-theoretic formulation this increase in dimension is {\it implicit} in the fact that the cell attachments are indexed not just by a set, but by the higher homotopy type $S^{n+1}\!\! \to X_k$ (i.e. by ``internalizing'' the classical construction). For example, applying the construction for $n = -1$ to $X_0 = \ast$ yields in the first stage $X_1 = S^1$, whence the index type for the second stage is the 2-type $S^0\! \to S^1 \;\simeq\; S^1 \times S^1$.
 }
 \begin{equation} \label{TruncationPushoutDiagram}
    \begin{tikzcd}
    (S^{n + 1}\!\! \to X_k)
      \times
    S^{n + 1}
    \ar[
      d,
      "{ \mathrm{ev} }"{swap},
      "{\ }"{name=t, pos=.5}
    ]
    \ar[
      r,
      "{\mathrm{pr}_1 }"{pos=.5},
      "{\ }"{swap, name=s}
    ]
    &
    (S^{n+1}\!\! \to X_k)
    \ar[d]
    \\
    X_k
    \ar[r, "{i_k}"]
    &
    X_{k+1}
    \mathrlap{\,.}
    \ar[
      from=s,
      to=t,
      Rightarrow,
      color=orangeii
    ]
  \end{tikzcd}
 \end{equation}
 \end{definition}


\begin{proposition}[Universal property for $n$-truncation]
The construction of $\eta_n : X \to \Truncation{}{n}{X}$ from Def. \ref{TruncationDefn} has the universal property
\eqref{UniversalPropertyTrunctation} of the $n$-truncation.
\end{proposition}
\begin{proof}
  This is a special case of the construction of the localization at a class of maps between compact types given in
  \cite[Thm. 7.2.10]{Rijke19}. Specifically, the pushout \eqref{TruncationPushoutDiagram} is a special case of the
  initial quasi-$F$-local extension (\cite[Defn. 7.2.6]{Rijke19}) where the family $F$ is taken to be the terminal map
  $S^{n+1} \to \ast$ to the singleton type \eqref{UnitType}. Then \cite[Thm. 7.2.10]{Rijke19} applies since the spheres
  are sequentially compact by \cite[Cor. 7.1.12]{Rijke19}.

  The universal property given in \cite[Thm. 7.2.10]{Rijke19} is not the dependent one we gave in \eqref{UniversalPropertyTrunctation}.
  However, a variety of similar universal properties which characterize {\it modalities} (\cite[\S 7.7]{UFP13}) are proven equivalent
  in \cite[\S 1]{RSS20}; these include the universal property \eqref{UniversalPropertyTrunctation} as \cite[Defn. 1.2]{RSS20} and the
  universal property appearing in \cite[Thm. 7.2.10]{Rijke19} as \cite[Defn. 1.3]{RSS20} (noting that $n$-truncated types are closed
  under pair types, \cite[Thm. 7.1.8]{UFP13}).
\end{proof}

\noindent
{\bf Quotient sets.} The presence of higher inductive type formation (pp. \pageref{HigherInductiveTypes})
works wonders even when all data types involved are sets \eqref{Sets} as opposed to higher homotopy types \eqref{HigherHomotopyType}; namely it implies the existence of data types encoding ordinary {\it quotient sets} of data. This solves a decade-old problem in type theory (review in \cite[\S 1.1]{Li14}\cite[\S 4.3.2]{Murray22}) which goes back to the roots of the notion of {\it sets} in constructive mathematics \cite[p. 13]{Bishop67}\cite[p. 15]{BishopBridges85} as sets equipped with equivalence relations, also called  ``setoids'' \cite[\S 1.3, \S 5.1]{Hofmann95}\cite{BartheCaprettaPons03}
or ``Bishop sets'' \cite[\S 3.1]{CoquandSpiwack07}.

Concretely, given a data set \eqref{Sets}
equipped with a relation
\eqref{RelationsAsSubtypesAndPropositions}
which is an {\it equivalence relation} in that the following conditions are satisfied:
\begin{equation}
  \label{isEquivalenceRelation}
  \def\arraystretch{1.5}
  \begin{array}{l}
  X \,:\, \Sets
  \\
  P \,:\, X \times X \to \Propositions
  ,\,
  \\
  R \,:\defneq\, (x_0, x_1 : X) \times P(x_0,x_1)
  ,
  \end{array}
  \;\;\;\;\;\;
  \def\arraystretch{1.5}
  \begin{array}{rcl}
  \mathrm{isReflexive}(P)
  &:\defneq&
  (x : X) \to P(x,x)
  \,,
  \\
  \mathrm{isSymmetric}(P)
  &:\defneq&
  (x_0,\, x_1 : P)
  \times
  P(x_0,\, x_1)
  \to
  P(x_1,\, x_0)
  \,,
  \\
  \mathrm{isTransitive}(P)
  &:\defneq&
  (x_0, x_1, x_2 : X)
  \times
  P(x_0,\, x_1)
  \times
  P(x_1,\, x_2)
  \to
  P(x_0,\, x_3)
  \,;
  \end{array}
\end{equation}
then we obtain the {\it quotient type} $X / R$  of $X$ by $R$, defined
(cf. \cite[\S 6.10]{UFP13}\cite[\S 1.5, \S 2.4]{RjkeSpitters})
as the set truncation of the pushout \eqref{InferenceRulesForHomotopyPushout} of
the function which picks the pairs of data of type $X$ that are in relation,
along the function which projects out their distinction in
$S^0  = \Bits$ \eqref{HigherHomotopyBits}:
\begin{equation}
  \label{QuotientSetAsPushout}
  \cdots
  \;\;\;\;\;
  \vdash
  \;\;\;\;\;
  X/R
  \;\;
    :\defneq
  \;\;
  \left[ \mathrm{po}
  \left(\!\!\!\!
  \begin{tikzcd}[
    sep=0pt
  ]
    \big(
      (x_0,\, x_1),\, i
    \big)
    &\mapsto&
    x_i
    \\
    R \times S^0
    \ar[rr]
    \ar[
      dd,
      "{
        \mathrm{pr}_R
      }"
    ]
    &&
    X
    \\
    \phantom{A}
    \\
    R
  \end{tikzcd}
  \!\right)\right]_0
  \;\;
  :
  \;\;
  \Sets
\end{equation}

\noindent
We could also define the quotient $X / R$ as the set of {\it equivalence classes} of the equivalence relation $R$. An equivalence class is a property $E : X \to \mathrm{Prop}$ so that there exists an $x : X$ with $E(y)$ if and only if $P(x, y)$ for all $y$:
\begin{equation}\label{EquivalenceClass}
    \mathrm{isEquivalenceClass}(E) \;:\defneq\; \exists \left(\dsum{x : X} \big(\dprod{y : X} (E(y) \overset{\sim}{\to} R(x, y) \big) \right).
\end{equation}
Then we may define
\begin{equation} \label{QuotientSetAsEquivalenceclasses}
    \cdots\; \vdash\; X / R \;:\defneq\; \{E : X \to \mathrm{Prop} \mid \mathrm{isEquivalenceClass}(E)\}.
\end{equation}

A multitude of examples arises in the construction of the hierarchy of types of {\it number systems} which we come to in \cref{KZConnectionsInHomotopyTypeTheory}.

\medskip

\noindent
{\bf The idea of certified data structures.}\label{DataStructures}
We had motivated, around \eqref{FunctionDeclaration}, the notion of data typing by the promise of software verification; this now becomes nicely manifest:

\vspace{0cm}
\hspace{-.8cm}
\begin{tabular}{ll}
\begin{minipage}{11.2cm}
Via iteration of dependent data pairings
\eqref{DependentSumInference} (called {\it data telescopes} \cite{Zucker75}\cite{deBruijn91}, {\it data records} \cite[p. 2]{CoenTassi08} or type {\it classes} \cite[\S 2]{GGMR09})
of {\it data base} types $B \,:\, \Sets$
\eqref{Sets}
with $B$-dependent functions \eqref{DependentFunctionNotation} constituting read/write/compute-operations on $B$ data (``methods'') and further with identification certificates \eqref{IdentificationType} constituting data consistency statements,
one obtains fully verifiable {\it data structures}, whose denotational semantics is just that of ``mathematical structures'' (see e.g. pointers in \cite{Sakaguchi20}) in the orignal sense \cite{Corry04} of algebra (often referred to as: ``the hierarchy of structures'' or similar).
\end{minipage}
&
\def\arraystretch{2}
\begin{tabular}{|c|c|}
  \hline
  {\bf
  \def\arraystretch{.9}
  \begin{tabular}{c}
    Data
    \\
    structure
  \end{tabular}}
  &
  {\bf
  \def\arraystretch{.9}
  \begin{tabular}{c}
    Mathematical
    \\
    structure
  \end{tabular}}
  \\
  \hline
  Data base
  &
  Underlying set
  \\
  \hline
  \def\arraystretch{.9}
  \begin{tabular}{c}
    Data access
    \\
    methods
  \end{tabular}
  &
  \def\arraystretch{.9}
  \begin{tabular}{c}
    Algebraic
    \\
    structure
  \end{tabular}
  \\
  \hline
  \def\arraystretch{.9}
  \begin{tabular}{c}
    Consistency
    \\
    specification
  \end{tabular}
  &
  \def\arraystretch{.9}
  \begin{tabular}{c}
    Laws/
    \\
    properties
  \end{tabular}
  \\
  \hline
\end{tabular}
\end{tabular}
\vspace{.1cm}

For example, data bases $B$ equipped with methods to consistently read/write given $D$-data (known as well-behaved $D$-``lens''-structure \cite[\S 3]{BPV06}) are of the following form:
\begin{equation}
  \label{DLensStructure}
  \scalebox{.7}{
    \color{purple}
    \bf
    \def\arraystretch{.9}
    \begin{tabular}{c}
      $D$-lens structure
      \\
      on data base
    \end{tabular}
  }
  D\mathrm{Lens}
  \;:\defneq\;
  \left\{
  \def\arraystretch{2}
  \begin{array}{l}
    \left.
    (B \,:\, \Sets)
    \;
    \mathclap{\phantom{\vert^{\vert^{\vert^{\vert}}}}}
    \right\}
    \scalebox{.7}{
      \color{darkblue}
      \bf
      \def\arraystretch{.9}
      \begin{tabular}{c}
        data
        \\
        base
      \end{tabular}
    }
    \\
    \left.
    \begin{array}{ll}
      \;\times\;
      \big(
      \mathrm{read}_D
      \,:\,
      B \to D
      \big)
      \\
      \;\times\;
      \big(
        \mathrm{write}_D
        :
        D \times B \to B
      \big)
    \end{array}
    \;
    \right\}
    \scalebox{.7}{
      \color{darkblue}
      \bf
      \def\arraystretch{.9}
      \begin{tabular}{l}
        access
        \\
        structure
      \end{tabular}
    }
    \\
    \left.
    \begin{array}{l}
      \;\times\;
      \bigg(
      \mathrm{rw}
      :
      \scalebox{1.3}{$($}
      {
        { b \,:\, B }
        \atop
        {d \,:\, D}
      }
      \scalebox{1.3}{$)$}
      \to
      \mathrm{Id}_{D}
      \Big(
        \mathrm{read}_D\big(\mathrm{write}_D(d,b)\big)
        ,\;
        d
      \Big)
      \bigg)
      \\
      \;\times\;
      \bigg(
      \mathrm{wr}
      :
      \dprod{
        { b \,:\, B }
      }
      \mathrm{Id}_{B}
      \Big(
        \mathrm{write}_D\big(\mathrm{read}_D(b),\, b\big)
        ,\;
        b
      \Big)
    \bigg)
    \end{array}
    \right\}
    \scalebox{.7}{
      \color{darkblue}
      \bf
      \def\arraystretch{.9}
      \begin{tabular}{l}
        consistency
        \\
        specification
      \end{tabular}
    }
  \end{array}
  \right.
\end{equation}
Notice that to give a $D$-lensed data structure \eqref{DLensStructure} means, by the pair type introduction rule \eqref{DependentSumInference},
\begin{equation}
  \label{ALensType}
  \big(
    \underbrace{
      B
    }_{
      \mathclap{
        \scalebox{.7}{
          \raisebox{-3pt}{
          \color{darkblue}
          \bf
          data base
        }
        }
      }
    }
    ,\,
    \underbrace{
      \mathrm{read}_D
      ,\,
      \mathrm{write}_D
    }_{
      \mathclap{
        \scalebox{.7}{
          \color{darkblue}
          \bf
          methods
        }
      }
    }
    ,\,
    \underbrace{
      \mathrm{rm}
      ,\
      \mathrm{rw}
    }_{
      \mathclap{
        \scalebox{.7}{
          \color{darkblue}
          \bf
          certificates
        }
      }
    }
  \big)
  \;:\;
  D\mathrm{Lens}
\end{equation}
to instantiate the data base $B$ equipped with its read/write methods  {\it and} with certificates that these work as expected.

It is in this way that fully data-typed programs \eqref{FunctionDeclaration} are automatically {\it certified} and {\it verified}: To produce data of a given structured type
necessarily involves supplying a certificate that correct data behaviour has been verified.
\medskip

\noindent
{\bf Group structure.}
A data structure of fundamental mathematical relevance
is {\it group data structure} (e.g. \cite{Kachapova09}\cite[\S 33.10]{Escardo19}\cite[\S ``\href{https://1lab.dev/\#group-theory}{Group theory}'']{1lab}) in the sense of abstract group theory (e.g. \cite[\S 1.1]{Miller72}\cite{Rotman95}):
\begin{equation}
  \label{TypeOfGroups}
  \scalebox{.7}{
    \color{purple}
    \bf
    \def\arraystretch{.9}
    \begin{tabular}{c}
      group
      \\
      data
      \\
      structure
    \end{tabular}
  }
  \Groups
  \;\;
  :\defneq
  \;\;
  \left\{
  \def\arraystretch{2}
  \begin{array}{l}
    \left.
    \Big.
    \big(
      G \,:\,\Sets
    \big)
    \;\;
    \mathclap{\phantom{\vert^{\vert^{\vert^{\vert}}}}}
    \mathclap{\phantom{\vert_{\vert_{\vert_{\vert}}}}}
    \right\}
    \scalebox{.7}{
      \color{darkblue}
      \bf
      \def\arraystretch{1}
      \begin{tabular}{l}
        data
        \\
        base
      \end{tabular}
    }
    \\
    \left.
    \begin{array}{l}
      \;\times\;
      \big(
        \mathrm{e}
        \,:\,
        G
      \big)
      \\
      \;\times\;
        \big(
          \cdot
          \,:\,
          G \times G \to G
        \big)
        \\
        \;\times\;
        \big(
          (-)^{-1}
          \,:\,
         G \to G
       \big)
    \end{array}
    \;\;
    \right\}
    \scalebox{.7}{
      \color{darkblue}
      \bf
      \def\arraystretch{1}
      \begin{tabular}{l}
        group
        \\
        structure
      \end{tabular}
    }
    \\
    \left.
    \begin{array}{l}
      \;\times\;
        \Big(
        \mathrm{unt}
        \,:\,
        \dprod{
          g \,:\, G
        }
        \mathrm{Id}_G
        \big(
          g \cdot \mathrm{e}
          ,\,
          g
        \big)
        \,\times\,
        \mathrm{Id}_G
        \big(
          \mathrm{e} \cdot g
          ,\,
          g
        \big)
      \Big)
      \\
      \;\times\;
         \Big(
          \mathrm{asc}
          :
          \scalebox{1.3}{$($}
           {
            { g_1 ,\, g_2, }
            \atop
            { g_3 \,:\, G }
           }
          \scalebox{1.3}{$)$}
          \to
          \mathrm{Id}_G
           \big(
                (g_1 \cdot g_2)
                \cdot
                g_3
              ,\,
                g_1
                \cdot
                (
                  g_2 \cdot g_3
                )
            \big)
        \Big)
      \\
      \;\times\;
      \Big(
        \mathrm{inv}
        :
        (
          { g \,:\,G }
        )
        \to
            \mathrm{Id}_G
            \big(
                g
                \cdot
                g^{-1}
              ,\,
              \mathrm{e}
            \big)
            \,\times\,
            \mathrm{Id}_G
            \big(
                g^{-1}
                \cdot
                g
              ,\,
              \mathrm{e}
            \big)
          \Big)
      \end{array}
      \right\}
      \;
      \scalebox{.7}{
        \color{darkblue}
        \bf
        \def\arraystretch{1}
        \begin{tabular}{l}
          group
          \\
          laws
        \end{tabular}
      }
  \end{array}
  \right.
  \,.
\end{equation}
Historically, the mathematical term ``group'' is short for {\it symmetry group} (e.g. \cite{Miller72}) or {\it transformation group} (e.g. \cite{tomDieck87}) in the sense of: {\it group of transformational symmetries of some object} (cf. \cite[(2.3)]{tomDieck87}).
Curiously this meaning is natively brought out by the {\it magic of homotopy type theory} (cf. \cite{BBCDG21}): From the discussion on p \pageref{TheMagicOfHomotopyTypeTheory} it is clear that the self-identifications $d_0 \rightsquigarrow d_0$ (symmetries) of any datum $d_0 : D$ in a groupoid $D : \TruncatedTypes{1}$ \eqref{TypeOfGroupoids} -- hence the data in its loop type \eqref{LoopSpace} at $d_0$ -- form a group \eqref{TypeOfGroups} under concatenation \eqref{ConcatenationOfIdentifications}:
\begin{equation}
 \label{LoopTypesAsGroupTypes}
 \hspace{-2mm}
 \overset{
   \mathclap{
     \raisebox{4pt}{
       \scalebox{.7}{
       \color{orangeii}
       \bf
       \llap{Given}
       \color{darkblue}
       data 1-type
       with base datum
     }
     }
   }
 }{
  D : \TruncatedTypes{1}
  ,\;\;
  d_0 : D
  }
  \overset{
    \mathclap{
      \raisebox{4pt}{
        \scalebox{.7}{
          \color{orangeii}
          \bf
          obtain
        }
      }
    }
  }{
  \qquad \quad
  \vdash
  \qquad \quad
  }
  \big(
    \overset{
      \mathclap{
        \raisebox{4pt}{
        \scalebox{.7}{
          \color{darkblue}
          \bf
          \;\;{self-identifications}
        }
        }
      }
    }{
      \Omega_{d_0}D
    }
    ,\,
    \underset{
     \mathclap{
       \raisebox{-0pt}{
         \scalebox{.7}{
           \llap{
             \color{orangeii}
             \bf
              equipped with
           }
           \color{darkblue}
           \bf
           \def\arraystretch{.9}
           \begin{tabular}{c}
             self-
             \\
             identification
           \end{tabular}
         }
       }
     }
    }{\quad
      \mathrm{e} :\defneq \mathrm{id}_{d_0}
    }
    ,\,
    \underset{
      \mathclap{
        \raisebox{-5pt}{
          \scalebox{.7}{
            \color{darkblue}
            \bf
            \;\;
            concatenation
          }
        }
      }
    }{\quad
    \,\cdot\,
    :\defneq
    \mathrm{conc}
    }
    ,\,
    \underset{
      \mathclap{
        \raisebox{-2pt}{
          \scalebox{.7}{
            \color{darkblue}
            \bf
            reversal
            \;\;\;
          }
        }
      }
    }{\;\;\;
    (-)^{-1}
      :\defneq
    \mathrm{inv}
    }
    \underset{
      \mathrlap{
        \raisebox{-2pt}{
          \scalebox{.7}{
            \color{darkblue}
            \bf
            associativity etc.
          }
        }
      }
    }{
    ,\quad
    \cdots
  \quad   }
  \big)
  \overset{
    \mathclap{
      \raisebox{5pt}{
        \scalebox{.7}{
          \color{orangeii}
          \bf
       \;   forming
          \;\;\;\;\;
        }
      }
    }
  }{
  \quad:\qquad
  }
  \overset{
    \mathclap{
      \raisebox{3pt}{
        \scalebox{.7}{
          \color{purple}
          \bf
          \;\;\;\;\;\;\;\;
          group data.
        }
      }
    }
  }{
    \Groups
  }
\end{equation}

Further in this vein: {\bf subgroup data structure} $H \subset G$ (cf. e.g. \cite[\S 1.2]{Miller72}) of a given group structure $G$ \eqref{TypeOfGroups} may be formulated as the {\it propositions} \eqref{DependentSumOfProps} of the form $P_H : G \to \Propositions$  (which we may think of as) asserting: $P_H : g \mapsto \mbox{``$g$ is in $H \subset G$''}$, equipped with certificates that these propositions do define subgroups (cf. \cite[\S 33.12]{Escardo19}):
\begin{equation}
  \label{SubgroupTypes}
  \hspace{-1mm}
  \big(
    G
    ,\,
    \NeutralElement
    ,\,
    \cdot
    ,\,
    (-)^{-1}
  \big)
  :
  \Groups
  \;\;\;\;\;\;
  \vdash
  \;\;\;\;\;\;
  \overset{
    \mathclap{
      \raisebox{4pt}{
        \scalebox{.7}{
          \color{purple}
          \bf
          \def\arraystretch{.9}
          \begin{tabular}{c}
            subgroup
            \\
            structure
          \end{tabular}
        }
      }
    }
  }{
  \mathrm{SubGrp}(G)
  }
  :\defneq
  \left\{\!\!\!\!
  \def\arraystretch{1.6}
  \begin{array}{l}
    \left.
    \big(
     P : G \to \Propositions
     \big)
     \mathclap{\phantom{\vert^{\vert^{\vert}}}}
     \right\}
     \scalebox{.7}{
       \color{darkblue}
       \bf
       structure
     }
     \\
     \left.
     \begin{array}{l}
     \times
     \;\;
     \big(
       \mathrm{ptd}
       :
       P(\NeutralElement)
     \big)
     \\
     \times
     \;\;
     \Big(
       \mathrm{mld} :
     \big(
       (g_1, g_2 : G)
       \times
       P(g_1) \times P(g_2)
     \big)
     \to
     P(g_1 \times g_2)
     \Big)
     \\
     \times
     \;\;
     \Big(
      \mathrm{ivd}
      :
     \big(
       (g : G)
       \times
       P(g)
     \big)
     \to
     P\big(
       g^{-1}
     \big)
     \Big)
     \end{array}
    \!\!\!\! \right\}
     \scalebox{.7}{
       \color{darkblue}
       \bf
       properties
     }
  \end{array}
  \right.
\end{equation}

From such $P : G \to \Propositions$ the actual subgroup data is recovered \eqref{PropositionsEquivalentToSubtypes}
as the dependent pairings $H \,:\defneq\, (g : G) \times P_H(G)$ (elements of $G$ paired with a certificate that they are in fact in the subgroup), which inherits group structure by using the group operations on $G$ paired with the certificates that on $H$ they do restrict to land again in $H$:
\vspace{-2mm}
\begin{equation}
\hspace{-4mm}
  \begin{tikzcd}[
    row sep = -3pt, column sep=0
  ]
      \big(
      G
      ,\,
      \NeutralElement
      ,\,
      \cdot
      ,\,
      (-)^{-1}
      \big)
    :
    \Groups
    \;\;\;
    \vdash
    &[-11pt]
    \mathrm{SubGrp}(G)
    \ar[
      rr,
      "{
        \scalebox{.7}{
          \color{greenii}
          \bf
          underlying abstract group structure
        }
      }"
    ]
    &&
    \Groups
    \\
    &
    \big(
      P
      ,\,
      \mathrm{ptd}
      ,\,
      \mathrm{mld}
      ,\,
      \mathrm{ivd}
    \big)
    &\! \longmapsto&
   \! \left(
    \def\arraystretch{1.4}
    \arraycolsep=1pt
    \begin{array}{l}
      H \;\;\;:\defneq
      (g : G) \times P(g)
      ,\,
      \\
      \NeutralElement_H
      \;\;
      :\defneq
      (\NeutralElement, \mathrm{ptd})
      ,\,
      \\
      \;\cdot_H
      \;\;
      :\defneq
      \big(
        (g_1, p_1)
        ,
        (g_2, p_2)
        : H
      \big)
      \mapsto
      \big(
        g_1 \cdot g_2,\, \mathrm{mld}(g_1, g_2, p_1, p_2)
      \big)
      ,\,
      \\
      (-)^{-1}_H
      :\defneq
      \big(
        (g, p)
        : H
      \big)
      \mapsto
      \big(
        g^{-1}
        ,\, \mathrm{ivd}(g,p)
      \big)
    \end{array}
  \!\!\! \right)
  \end{tikzcd}
\end{equation}

\smallskip

\noindent
{\bf The idea of data structure identification.}
\label{StructureIdentityPrinciple}
With data structures
(p. \pageref{DataStructures})
defined as telescopes of dependently paired dependent functions and identification, we can make explicit their operational {\it equivalences} \eqref{DataTypeOfEquivalences} or equivalently their identifications \eqref{UnivalenceAxiom} as soon as we have an explicit handle on the identification of any
dependent functions
and
dependent pairs.

Using the univalence axiom \eqref{UnivalenceAxiom}, these work component-wise as expected, via comparison functions readily defined by $\mathrm{Id}$-induction \eqref{PathInduction},
a statement known as {\it function extensionality} \cite[p. 8]{Voevodsky10}\cite[\S 4.9, Thm. 2.9.7]{UFP13} and its analogue for
pairings \cite[Thm 2.7.2]{UFP13}):
\vspace{-3mm}
\begin{equation}
  \label{happlyFormula}
  \hspace{-11mm}
  \scalebox{.7}{
    \color{orangeii}
    \bf
    \def\arraystretch{.9}
    \begin{tabular}{c}
      function
      \\
      extensionality
    \end{tabular}
  }
  \begin{tikzcd}[
    column sep=10pt,
    row sep=-10pt
  ]
  \dprod{
    {
    f,\,g
    }
    \atop
    {
    \,:\,
    (d: D) \,\to\, C_d
    }
  }
    &
  \bigg(\qquad
    \overset{
      \mathclap{
        \raisebox{6pt}{
          \scalebox{.7}{
            \color{darkblue}
            \bf
            \def\arraystretch{.9}
            \begin{tabular}{c}
              identifications
              of
              dependent functions
            \end{tabular}
            \;\;\;
          }
        }
      }
    }{
    \mathrm{Id}_{ (d : D) \,\to\, C_d }
    \big(
      f
      ,\,
      g
    \big)
    }
   \qquad  \ar[
      rrr,
      "{ \sim }",
      "{
        \hspace{4pt}
        \scalebox{.7}{
          \color{greenii}
          \bf
          equivalent to
        }
      }"{yshift=10.5pt}
    ]
    &&&
    \Big(
    \overset{
      \mathclap{
        \;\;\;\;\;\;
        \raisebox{6pt}{
          \scalebox{.7}{
            \color{darkblue}
            \bf
            \def\arraystretch{.9}
            \begin{tabular}{c}
              dependent functions
              of
              identifications
            \end{tabular}
          }
        }
      }
    }{
    \big(
      d : D
    \big)
    \to
    \mathrm{Id}_{C_{d}}
    \big(
      f(d)
      ,\,
      g(d)
    \big)
    }
    \Big)
       \;\; \bigg)
   &
    \\
    &
    \mathrm{id}_{f}
    & \longmapsto&&
  \big(
      d
      \mapsto
      \mathrm{id}_{f(d)}
    \big)
  \end{tikzcd}
\end{equation}
\begin{equation}
  \label{PairExtensionality}
  \hspace{-4mm}
   \scalebox{.7}{
    \color{orangeii}
    \bf
    \def\arraystretch{.9}
    \begin{tabular}{c}
      pair
      \\
      extensionality
    \end{tabular}
  }
 \begin{tikzcd}[
    column sep=10pt,
    row sep=-10pt
  ]
  \dprod{
    {
    (d,c)
    ,\,
    (d',c')
    }
    \atop
    {
    \,:\,
    (d: D) \times C_d
    }
  }
  &
  \bigg(\;
    \overset{
      \mathclap{
        \raisebox{4pt}{
          \scalebox{.7}{
            \color{darkblue}
            \bf
            \def\arraystretch{.9}
            \begin{tabular}{c}
              identifications
              of
              dependent pairs
            \end{tabular}
          }
        }
      }
    }{
    \mathrm{Id}_{ (d : D) \times C_d }
    \big(
      (d,c)
      ,\,
      (d',c')
    \big)
    }
    \ar[
      rrr,
      "{ \sim }",
      "{
        \,
        \scalebox{.7}{
          \color{greenii}
          \bf
          equivalent to
        }
      }"{yshift=8.5pt}
    ]
    &&&
    \overset{
      \mathclap{
        \raisebox{4pt}{
          \scalebox{.7}{
            \color{darkblue}
            \bf
            \def\arraystretch{.9}
            \begin{tabular}{c}
              dependent pairs
              of
              identifications
            \end{tabular}
          }
        }
      }
    }{
    \big(
      p : \mathrm{Id}_D( d, \, d' )
    \big)
    \times
    \mathrm{Id}_{C_{d'}}
    \big(
      p_\ast c
      ,\,
      c'
    \big)
    }
    \bigg)
    &
      \\
    &
    \mathrm{id}_{(d,c)}
    &\;\;\longmapsto\;\;&&
    \big(
      \mathrm{id}_{d}
      ,\,
      \mathrm{id}_c
    \big)
  \end{tikzcd}
\end{equation}

By iteration of these two rules, the equivalences of telescope/record data structures (p. \pageref{DataStructures}) are found to be those equivalences on the underlying base types which are compatible with all the given structure -- hence which are {\it homo-morphisms},  thus recovering the original algebraic notion of {\it iso-morphism}.

\medskip
For example, an equivalence of group data structures \eqref{TypeOfGroups} comes out to be a {\it bijective group homo-morphism} (e.g. \cite[\S 1.3]{Miller72}), hence a {\it group iso-morphism} just as in the traditional algebraic sense, but here {\it emergent} from the principles of homotopy type theory:
\vspace{-3mm}
\begin{equation}
  \label{GroupStructureIsomorphism}
  \overset{
    \mathclap{
      \raisebox{4pt}{
        \hspace{-3.3cm}
        \scalebox{.7}{
          \color{orangeii}
          \bf
          \def\arraystretch{.9}
          \begin{tabular}{c}
            identifications of
            \\
            types with group structure
          \end{tabular}
        }
      }
    }
  }{
  \mathrm{Id}_{
    \Groups
  }
  \Big(
    (G,\mathrm{e}, \mathrm{cmp}, \mathrm{inv})
    ,\,
    (G',\mathrm{e}', \mathrm{cmp}', \mathrm{inv}')
  \Big)
  }
  \;\;
  \xrightarrow{ \;\;\; \sim \;\;\; }
  \;\;
  \left\{
  \def\arraystretch{1.5}
  \begin{array}{ll}
    \left.
    \big(
      p \,:\, \mathrm{Id}_{\Sets}(G,G')
    \big)
    \mathclap{\phantom{\vert^{\vert^{\vert^{\vert}}}}}
    \;\;
    \right\}
    \scalebox{.7}{
      \color{darkblue}
      \bf
      \def\arraystretch{1}
      \begin{tabular}{l}
        underlying
        \\
        bijection
      \end{tabular}
    }
    \\
    \left.
    \begin{array}{l}
      \;\times\;
        \mathrm{Id}_G(p_\ast \mathrm{e},\,\mathrm{e}')
      \\
      \;\times\;
      \mathrm{Id}_{G \times G \to G}
      \big(
        p_\ast \mathrm{cmp}
        ,\,
        \mathrm{cmp}'
      \big)
        \\
        \;\times\;
        \mathrm{Id}_{G \to G}
        \big(
          p_\ast \mathrm{inv}
          ,\,
          \mathrm{inv}'
        \big)
    \end{array}
    \hspace{0cm}
    \right\}
    &
    \scalebox{.7}{
      \hspace{-20pt}
      \color{darkblue}
      \bf
      \def\arraystretch{1}
      \begin{tabular}{l}
        homomorphism
        \\
        property
      \end{tabular}
    }
  \end{array}
  \right.
\end{equation}

\hspace{-.8cm}
\begin{tabular}{cc}
\begin{minipage}{11.2cm}

\smallskip
Essentially this state of affairs \eqref{happlyFormula} - \eqref{GroupStructureIsomorphism} in univalent homotopy type theory was first made explicit in \cite{CoquandDanielsson13}
and has come to be known as the ``{\it structure identity principle}'' (following \cite{Aczel11}), further discussed in
\cite[\S 9.8]{UFP13} \cite[\S \S3.33.1]{Escardo19}\cite{ANST20}
(see also in relation to {\it structuralism} in \cite[\S 5.1]{Tsementzis17}).

Notice that, thereby, dependent data type theory pleasantly resolves a long historical quest (cf. \cite{Corry04}) for a good meta-theory of ``mathematical structures'', along the way unifying it with the notion of ``data structures''.
\end{minipage}
&
\hspace{1mm}
\def\arraystretch{1.6}
\begin{tabular}{|c|c|}
  \hline
  {\bf Type theory} & {\bf Homotopy theory}
  \\
  \hline
  \hline
  Data types & Homotopy types
  \\
  \hline
  Data structures & Math. structures
  \\
  \hline
  Data bases & Carrier types
  \\
  \hline
\end{tabular}
\end{tabular}


\bigskip

\noindent
{\bf The idea of higher structures.}
\label{TheIdeaOfHigherStructures}
The previous examples of data/mathematical structures (pp. \pageref{DataStructures})
and many of those considered further below are based on (data-){\it sets} \eqref{Sets}, this being the classical situation traditionally considered in the literature. But in a homotopy-typed language we may just as well consider {\it higher homotopy types} \eqref{HigherHomotopyType}
as base types, to obtain ``higher structures'' (in the now popular sense, see e.g. \cite{CGX02}\cite{JSSW19}\cite{FSSHigherStruc}).

\medskip
Fundamental examples of higher structures are group {\it deloopings} \eqref{GroupsAsPointedConnectedOneTypes} and higher deloopings \eqref{PointedDeloopingStructure}; these we turn to in \cref{GMConnectionOnTwistedCohomologyData} below.

\vspace{2cm}

\noindent
{\bf In summary}, the remarkable insight of  {homotopy type theory} as a statement in computer science may be expressed as follows:

\begin{center}
{\it
\colorbox{lightgray}{
Any programming language with truly thorough data typing}

\colorbox{lightgray}{is natively a verification language
for constructions}

\colorbox{lightgray}{in algebraic topology and homotopy theory.}
}
\end{center}

\noindent
We next turn to show how this allows for a slick construction of homotopy data types of Gauss-Manin monodromy in general (\cref{GMConnectionOnTwistedCohomologyData}) and then of anyonic topological quantum gates in particular (\cref{KZConnectionsInHomotopyTypeTheory}).

\newpage

\subsection{Monodromy of cohomological data}
\label{GMConnectionOnTwistedCohomologyData}

We now describe the programming language construct
(Def. \ref{TypeTheoreticGaussManinConnection} below)
which, under the dictionary in \cref{HoTTIdea},
encodes the monodromy of
Gauss-Manin connections on twisted Cohomology
groups as developed in \cref{ViaParameterizedPointSetTopology},
and hence (below in \cref{KZConnectionsInHomotopyTypeTheory})
specifically
the operation of anyon braid gates via monodromy of KZ-connections.

\medskip

As a type formation in itself, our main construction in Def. \ref{TypeTheoreticGaussManinConnection} below is rather immediate: It is nothing but the dependent 0-truncation of a doubly iterated dependent function type between (higher) delooping types --- the reader with insider knowledge in homotopy type theory will not need any further introduction to parse Def. \ref{TypeTheoreticGaussManinConnection} and may want to jump ahead.

\medskip

Recall that this type-theoretic simplicity is our main point: Under the dictionary of \cref{HoTTIdea} the transport operation \eqref{TransportRule} in this
readily constructed type of Def. \ref{TypeTheoreticGaussManinConnection} clearly interprets as what in  \cref{ViaParameterizedPointSetTopology} we showed is the parallel transport by Gauss-Manin connections --- whose traditional construction however is rather less immediate (Lit. \ref{LiteratureGaussManinConnections}, Lit. \ref{KZConnectionsOnConformalBlocksReferences}).

\medskip

However, to be self-contained to a broader audience and since the type theoretic literature on the following issues remains thin,
we first proceed now with laying out some type-theoretic foundations regarding what one might call the theory of {\it transformation groups} or {\it abstract Galois theory}. If nothing else, what follows may serve as an illustrative example for how to work concretely (albeit ``informally'' in the style of \cite{UFP13}) with the homotopy data type language of \cref{HoTTIdea} (cf. Rem. \ref{PerspectiveOnProof} below).

\medskip

\noindent
{\bf Groups of self-identifications.}
In \eqref{LoopTypesAsGroupTypes} we saw that a natural source of group structures \eqref{TypeOfGroups} are the ``loopings'' \eqref{LoopSpace} of pointed types, $\Omega_{d_0} D  : \Groups$.
Since this only depends on the single datum $d_0 : D$ on which the self-identifications of these loops are based,
then given any $G \,:\, \Groups$ it makes sense to ask for a 1-type \eqref{TypeOfGroupoids} -- to be denoted
$\mathbf{B}G$ and called a {\it delooping} of $G$ -- for which there is an essentially  unique datum in the first place
and whose self-identifications recover $G$ in this way \eqref{LoopTypesAsGroupTypes}
\vspace{-2mm}
$$
  G \;\simeq\;
  \Omega_\ast \mathbf{B}G
$$

\vspace{-1mm}
\noindent Proposition \ref{TheLoopingDeloopingEquivalence} below asserts (in particular) that all groups
arise this way, up to equivalence,
This leads to an alternative slick definition of group types which is only available in homotopy-typed
languages: the {\it pointed connected 1-types}:
\begin{equation}
  \label{GroupsAsPointedConnectedOneTypes}
  \scalebox{.7}{
    \color{purple}
    \bf
    \def\arraystretch{.9}
    \begin{tabular}{c}
      groups as
      pointed
      \\
      connected 1-types
    \end{tabular}
  }
  \PointedConnectedOneTypes
  \;:\defneq\;
  \left\{
  \def\arraystretch{1.3}
  \begin{array}{ll}
    \mathbf{B}G : \TruncatedTypes{1}
    &
    \big\}
    \;
    \scalebox{.7}{\bf \color{darkblue} higher data}
    \\
    \times
    \;\;
    \mathrm{pt} : \mathbf{B}G
    &
    \big\}
    \;\;
    \scalebox{.7}{\bf \color{darkblue} structure}
    \\
    \times
    \;\;
      (t : \mathbf{B}G)
      \to
      \exists \,
      \Id_{\mathbf{B}G}(\mathrm{pt} ,\, t)
    &
    \big\}
    \;
    \scalebox{.7}{{\bf \color{darkblue} property} (connectivity)}
  \end{array}
  \right.
\end{equation}
Semantically, this alternative homotopy-theoretic conception of groups, and its equivalence (Prop. \ref{TheLoopingDeloopingEquivalence}) to the algebraic definition
is the content of the `'May recognition theorem''
for loop spaces \cite{May72} generalized to groups internal to $\infty$-toposes \cite[Lem. 7.2.2.11]{Lurie09}
and as such much amplified in \cite[p. 7]{NSS12}\cite[Prop.]{SS20OrbifoldCohomology}\cite[Prop. 0.2.1]{SS21EPB}, cf. \eqref{GroupInfinityAction} below. The perspective has been picked up by the type-theoretic literature in \cite[p. 6]{BvDR18}\cite[\S 4]{BBCDG21} (see also \cite{Warn22}).

In order to prove this equivalence in type theoretic detail (Prop. \ref{TheLoopingDeloopingEquivalence} below), we step back and lay out all the ingredients:

\medskip

\noindent
{\bf $G$-Actions and torsors.} For a given group structure $G$ there is a classical notion of its {\it actions} on sets (e.g. \cite[(1.1)]{tomDieck87}):

\begin{definition}[$G$-Sets]
\label{TheTypeOfGSets}
For $G$ a group \eqref{TypeOfGroups}, a
{\it left $G$-action structure on a set} --- or just {\it left $G$-set} for short --- is data of the following type:
\vspace{-2mm}
\begin{equation}
\label{AlgebraicGActions}
\hspace{-2mm}
  \big(
    G,\,
    \NeutralElement
    ,\,
    \cdot
    ,\,
    (-)^{-1}
  \big)
  :
  \Groups
  \;\;\;
    \vdash
  \;
  \scalebox{.7}{
    \def\arraystretch{.9}
    \begin{tabular}{c}
      left
      \\
      \color{purple}
      \bf
      $G$-action
      \\
      /
      \color{purple}
      \bf
      $G$-set
    \end{tabular}
  }
  \Actions{G}_L
  \;:\defneq
  \left\{\!\!\!\!\!\!
  \def\arraystretch{1.6}
  \begin{array}{l}
    \left.
    (S : \Sets)
    \mathclap{\phantom{\vert^{\vert}}}
    \;
    \right\}
    \scalebox{.7}{
      \color{darkblue}
      \bf
      data
    }
    \\
    \left.
    \begin{array}{l}
    \times
    \;\;
    \big(
      \acts : G \times S \to S
    \big)
    \end{array}
    \mathclap{\phantom{\vert^{\vert}}}
    \right\}
    \;
    \scalebox{.7}{
      \color{darkblue}
      \bf
      structure
    }
    \\
    \left.
    \begin{array}{l}
      \times
      \;\;
      \Big(
        \mathrm{unt}
        :
        (s : S)
        \to
        \Id\big(
          \NeutralElement
          \, \acts \,
          s
          ,\,
          s
        \big)
      \Big)
      \\[+4pt]
      \times
      \;\;
      \bigg(
        \mathrm{act}
        :
        (g_1,\, g_2 : G)
        \to
        \Id\Big(
          (g_2 \cdot g_1)
          \acts \,
          s
          ,\,\;
          g_2
          \acts\,
          \big(
            g_1
            \acts
            \,
            s
          \big)
        \Big)
     \!\! \bigg)
    \end{array}
    \;
    \!\!\!\!\!\!\!\right\}
    \scalebox{.7}{
      \color{darkblue}
      \bf
      properties
    }
  \end{array}
  \right.
\end{equation}
Correspondingly there are {\it right} $G$-sets, a mild distinction which does and will matter in some applications:
\begin{equation}
\label{AlgebraicRightGActions}
\hspace{-2mm}
  \big(
    G,\,
    \NeutralElement
    ,\,
    \cdot
    ,\,
    (-)^{-1}
  \big)
  :
  \Groups
  \;\;\;
    \vdash
  \;
  \scalebox{.7}{
    \def\arraystretch{.9}
    \begin{tabular}{c}
      right
      \\
      \color{purple}
      \bf
      $G$-action
      \\
      /
      \color{purple}
      \bf
      $G$-set
    \end{tabular}
  }
  \Actions{G}_R
  \;:\defneq
  \left\{\!\!\!\!\!\!
  \def\arraystretch{1.6}
  \begin{array}{l}
    \left.
    (S : \Sets)
    \mathclap{\phantom{\vert^{\vert}}}
    \;
    \right\}
    \scalebox{.7}{
      \color{darkblue}
      \bf
      data
    }
    \\
    \left.
    \begin{array}{l}
    \times
    \;\;
    \big(
      \rightacts : S \times G  \to S
    \big)
    \end{array}
    \mathclap{\phantom{\vert^{\vert}}}
    \right\}
    \;
    \scalebox{.7}{
      \color{darkblue}
      \bf
      structure
    }
    \\
    \left.
    \begin{array}{l}
      \times
      \;\;
      \Big(
        \mathrm{unt}
        :
        (s : S)
        \to
        \Id\big(
          s
          \, \rightacts \,
          \NeutralElement
          ,\,
          s
        \big)
      \Big)
      \\[+4pt]
      \times
      \;\;
      \bigg(
        \mathrm{act}
        :
        (g_1,\, g_2 : G)
        \to
        \Id\Big(
          s
          \,
          \rightacts
          (g_1 \cdot g_2)
          ,\,\;
          \big(
            s
            \,
            \rightacts
            g_1
          \big)
          \rightacts
          g_2
        \Big)
     \!\! \bigg)
    \end{array}
    \;
    \!\!\!\!\!\!\!\right\}
    \scalebox{.7}{
      \color{darkblue}
      \bf
      properties
    }
  \end{array}
  \right.
\end{equation}

We will follow the usual convention of referring to a $G$-set by the name of its underlying set, often leaving the structure and properties implicit.

\smallskip
A function between $G$-sets \eqref{AlgebraicGActions}
is {\it equivariant} if and only if it commutes with the action of $G$,
and we denote the data set of equivariant maps by $\mathrm{Hom}_G(X, Y)$:
\begin{equation}
\label{EquivarianceTypeDefinition}
\left.
\def\arraystretch{1.4}
\begin{array}{l}
G \,:\, \Groups
,
\\
(S, \acts ),
(T, \acts)
\,:\, \Actions{G}_L
\\
\end{array}
\hspace{-4pt}
\right\}
\def\arraystretch{1.5}
\begin{array}{rccl}
\phi \,:\,  S \to T
\;\;
\vdash
\;\;
&
\mathrm{isEquivariant}(\varphi)
  &
    :\defneq
  &
    (g : G) \times (x : S)
      \to
    \Id_Y
    \big(
      \varphi(g \acts\, s)
      ,\,
      g \acts\, \varphi(s)
    \big)
  \\
  \;\;
  \vdash
  \;\;
  &
  \mathrm{Hom}_G(X, Y)
  &
    :\defneq
  &
  \big\{
    \varphi : S \to T
    \,\mid\,
    \mathrm{isEquivariant}(\varphi)
  \big\}
  \,.
\end{array}
\end{equation}
\end{definition}
Via  homotopy theory, the type of $G$-actions \eqref{TheTypeOfGSets}
has a slick reformulation (Prop. \ref{GSetsAreMapsOutOfBG} below) in line with the delooping equivalence of Prop. \ref{TheLoopingDeloopingEquivalence}.

The following
is the evident type-theoretic formulation of the classical notion of $G$-torsors
(e.g. \cite[\S 3.4]{Milne80}, cf. \cite[Def. 4.8.1]{BBCDG21}):

\begin{definition}[$G$-Torsors]
\label{GTorsorDef}
For $G$ a group \eqref{TypeOfGroups},
a {\it left $G$-torsor structure} $T$ is an inhabited
\eqref{MeaningOfPropositionalTruncation}
left $G$-set $G \acts \, T$
\eqref{AlgebraicGActions}
whose action is regular
(i.e., free and transitive) in that for any two elements $x, y : T$ there is a unique $g : G$ for which $g \acts\, x = y$.
\begin{equation}
\label{AlgebraicGTorsor}
  \big(
    G,\,
    \NeutralElement
    ,\,
    \cdot
    ,\,
    (-)^{-1}
  \big)
  :
  \Groups
  \;\;\;
    \vdash
  \;
  \scalebox{.7}{
    \color{orangeii}
    \bf
    \def\arraystretch{.9}
    \begin{tabular}{c}
      $G$-torsor
    \end{tabular}
  }
    G\Torsors_L
      \;:\defneq\;
      \left\{
      \def\arraystretch{1.5}
      \begin{array}{ll}
      (T, \acts, \mathrm{unt}, \mathrm{act}) : G \mathrm{Act}
      &
      \scalebox{.7}{\bf
        \color{darkblue}
        $G$-action which
      }
      \\
      \times
      \;
      (x, y : T)
        \to
      \exists !
        \big\{
          g : G
          \;\big|\;
          \Id_T(g \acts\, x, y)
        \big\}
      &
      \scalebox{.7}{\bf
        \color{darkblue}
        is regular
      }
      \\
      \times
      \;
      \exists \, T
      &
      \scalebox{.7}{\bf
        \color{darkblue}
        and inhabited.
      }
   \end{array}
  \right.
\end{equation}
\end{definition}

For example, the left/right multiplication action of any group on itself (i.e.: on its own underlying data set) makes a left/right $G$-torsor:
\begin{equation}
 \label{GroupsAsLeftTorsorsOverThemselves}
  \big(
    G
    \,\,
    \NeutralElement
    ,\,
    (-)^{-1}
  \big) : \Groups
  ,\,
  \;\;\;
  \big(
    T \:\defneq G
    ,\,
    \acts :\defneq \cdot
    ,\,
    \mathrm{act} :\defneq \mathrm{asc}
  \big)
  :
  G\mathrm{Tor}_L
\end{equation}
In fact, up to equivalence \eqref{HomotopyEquivalenceOfTypes} this is the {\it only} example of a $G$-torsor:
Every $G$-torsor is isomorphic to the canonical
one \eqref{GroupsAsLeftTorsorsOverThemselves}, and the choice of isomorphism amounts to choosing which of its elements
is identified with the neutral element of $G$. This standard fact is re-proven type-theoretically as Lem. \ref{TorsorsDeloopAGroup} below; it is the main reason we care about $G$-torsors at this point, because it implies that the type of $G$-torsors \eqref{AlgebraicGTorsor} is (up to equivalence) the delooping \eqref{GroupsAsPointedConnectedOneTypes} of $G$ (Prop. \ref{TheLoopingDeloopingEquivalence}).

\medskip
To this end, first we need to see that every equivariant function between $G$ torsors is an equivalence.
The following proof of this statement is a standard argument, but to showcase how the type-theoretic rules surveyed
in \cref{HoTTIdea} are at work, we spell out this proof in more detail. The upshot however is that the formal rules
allow reasoning just as one informally expects, which justifies leaving them more implicit as we proceed (cf. Rem. \ref{PerspectiveOnProof} below).

\begin{lemma}[Equivariant maps between torsors are isomorphisms]
\label{EquivariantMapBetweenTorsorsIsIso}
Let $G$ be a group \eqref{TypeOfGroups} and $S$ and $T$ be left $G$-torsors (\cref{GTorsorDef}). Then any equivariant
map $\varphi : \mathrm{Hom}_G(S, T)$ \eqref{EquivarianceTypeDefinition} is an isomorphism, in that we have an equivalence \eqref{HomotopyEquivalenceOfTypes}
\[
  \mathrm{Hom}_G(S, T)
    \;\simeq\;
  \Id_{G\mathrm{Tors}_L}(S, T)
\]
of equivariant maps \eqref{EquivarianceTypeDefinition} with identifications \eqref{IdentificationType} of $G$-torsors \eqref{AlgebraicGTorsor},
under which composition of homomorphisms corresponds to concatenation \eqref{ConcatenationOfIdentifications} of identifications.
\end{lemma}
\begin{proof}
  We need to construct a dependent term of the following form
  $$
    G : \Groups
    ,\;\;
    S,\, T : G\mathrm{Tor}_L
    ,\;\;
    \phi : \mathrm{Hom}_G(S,T)
    \;\;\;\;
    \vdash
    \;\;\;\;
    \mathrm{proof}(\phi)
    :
    \mathrm{isBijection}(\phi)
    \,.
  $$
  With that in hand the statement will follow fairly readily by the structure-identity-principle and using univalence.

  \medskip

  Unwinding the definition of $\mathrm{isBijection}(-)$
  \eqref{HomotopyEquivalenceOfTypes}, our goal is a term of this form:
  $$
    S,\, T : G\mathrm{Tor}_L
    ,\;\;
    \phi : \mathrm{Hom}_G(S,T)
    \;\;\;\;
    \vdash
    \;\;\;\;
    \mathrm{proof}(\phi)
    \,:\,
    (t : T)
      \to
    \exists !
    \Big(
      (s : S) \times \Id\big( \phi(s),\, t \big)
    \Big)
    \,,
  $$
  where we are now notationally suppressing the assumption of the group $G$ on the left, just for brevity.

  But by the rules for dependent function types \eqref{PiInferenceRules},
  such may be inferred from a term of this form:
  $$
    S,\, T : G\mathrm{Tor}_L
    ,\;\;
    \phi : \mathrm{Hom}_G(S,T)
    ,\;\;
    t : T
    \;\;\;\;
    \vdash
    \;\;\;\;
    \mathrm{proof}(\phi)
    \,:\,
    \exists !
    \Big(
      (s : S) \times \Id\big( \phi(s),\, t \big)
    \Big)
    \,.
  $$

  So far this is closely analogous to classical reasoning. But in a subsequent step we will need to get our hands on an element of $S$, which type-theoretically is a little more subtle than classically: Namely what we nominally have in the assumptions on the left -- as part of the assumption that $S$ is a torsor \eqref{AlgebraicGTorsor} -- is only a term $p : \exists S$. Making this notationally explicit, we are really looking for
  $$
    S,\, T : G\mathrm{Tor}_L
    ,\;\;
    \phi : \mathrm{Hom}_G(S,T)
    ,\;\;
    t : T
    ,\;\;
    p : \exists S
    \;\;\;\;
    \vdash
    \;\;\;\;
    \mathrm{proof}(\phi)
    \,:\,
    \exists !
    \Big(
      (s : S) \times \Id\big( \phi(s),\, t \big)
    \Big)
    \,.
  $$
  By itself, the hypothesis $p : \exists S$ just says that ``there is'' such an element, but does not in generally allow us to actually put such an element into the context.
  What saves the day here is that the {\it right} hand side of the above dependent term is a proposition \eqref{isProposition}. Therefore the universal property \eqref{UniversalPropertyTrunctation} of propositional truncation applies to show that a term of the above form actually is equivalent to a term of this form:
  $$
    S,\, T : G\mathrm{Tor}_L
    ,\;\;
    \phi : \mathrm{Hom}_G(S,T)
    ,\;\;
    t : T
    ,\;\;
    s : S
    \;\;\;\;
    \vdash
    \;\;\;\;
    \mathrm{proof}(\phi)
    \,:\,
    \exists !
    \Big(
      (s : S) \times \Id\big( \phi(s),\, t \big)
    \Big)
    \,.
  $$

  Now unwinding the definition of $\exists ! (-)$ on the right, our goal is finally in the following form (where on the right we are notationally suppressing, for readability, an iterated identification type, as that is contractible anyway in the present case since $S$ is a set \eqref{Sets}):
  \begin{equation}
    \label{StageInProofThatTorsorMapsAreIsos}
    \begin{array}{l}
    S,\, T : G\mathrm{Tor}_L
    ,\;\;
    \phi : \mathrm{Hom}_G(S,T)
    ,\;\;
    t : T
    ,\;\;
    s : S
    \\
    \hspace{4.2cm}
    \;\;\;\;
    \vdash
    \;\;\;\;
    \mathrm{proof}(\phi)
    \,:\,
    \bigg(
      \big(s_t : S \big)
        \times
      \Id\big( \phi(s_t),\, t \big)
        \times
      \Big(
        (s'_t : S)
        \times
        \Id\big( \phi(s'_t), t \big)
        \to
       \Id( s_t,\, s'_t )
      \Big)
    \bigg)
    \,.
    \end{array}
  \end{equation}
  Notice how this expression, when read out aloud, is pretty much the statement that a classical proof of the lemma would {\it start} with: It says that given a $t : T$ we need to show that there is an $s_t$ in its pre-image under $\phi$ --- where we are allowed to assume that we have {\it some} $s : S$ ---, and that any other element $s'_t$ in the preimage of $t$ is identifiable with $s_t$.
  Accordingly, from here the classical proof idea applies essentially verbatim, using the torsor property of $S$ and $T$ (notationally hidden in the assumptions on the left):




  First, by the regularity of the $G$-action on $T$ there is $g_t : G$ such that $g_t \acts\, \varphi(s) = t$; and by equivariance of $\varphi$ this implies $\varphi(g_t \acts\, s) = t$. Hence we may take $s_t$ in \eqref{StageInProofThatTorsorMapsAreIsos} to be
  $$
    t : T
    ,\;\;
    s : S
    \;\;\;\;
    \vdash
    \;\;\;\;
    s_t \,:\defneq\, g_t \acts s \;\;:\;\; S
    \,.
  $$
  Finally, if $s'_t : S$ with $\varphi(s'_t) = t$ is any other pre-image of $t$, then by the same reasoning we have
  $g'_t : G$
  with $g'_t \acts\, s = s'_t$, and hence $g'_t \acts\, \varphi(s) = t$. But since an  equation of this form also
  defined $g_t$,
  regularity of the $G$-action implies that $g'_t = g_t$; and therefore that $s'_t = g_t \acts\, s = s_t$.
\end{proof}
\medskip

\noindent
{\bf Delooping of groups and classification of principal bundles.}
One of the crown jewels of classical homotopy theory is the proof (good review is in \cite[Thm. 3.5.1]{RudolphSchmidt17}),
under mild
conditions, that for a topological group $G$ there exists a topological space $B G$ such that homotopy classes of maps
$X \to B G$ are
in bijection to isomorphism classes of ``principal $G$-bundles'' over $X$, meaning: fiber bundles of $G$-torsors
(Def. \ref{GTorsorDef}) --- whence one speaks of the {\it classifying space} $B G$.

\noindent
A modern way to re-prove this classical theorem is (see \cite[p. 7]{SS21EPB}) to observe:

\smallskip
\begin{itemize}[leftmargin=.55cm]
\item[{\bf (i)}]
\begin{itemize}[leftmargin=.5cm]
\item[{\bf (a)}]
 that before passage to their isomorphism classes, the {\it groupoid} \eqref{TypeOfGroupoids} of $G$-principal bundles is equivalent
 to that of maps of ``topological stacks'' $X \to \mathbf{B}G$, where $\mathbf{B}G$ is the ``delooping stack'' of $G$,

\item[{\bf (b)}] which is thereby identified with the ``moduli stack'' of principal $G$-bundles, hence of $G$-torsors: $\mathbf{B}G \simeq G\mathrm{Tors}$;
\end{itemize}

\item[{\bf (ii)}] that the classifying space $B G$ is the underlying {\it cohesive shape} (cf. p. \ref{CohesiveHoTT}) of $\mathbf{B}G$.\footnote{It is for this reason that we use the boldface $\mathbf{B}(-)$ for ``delooping'' \eqref{GroupsAsPointedConnectedOneTypes}, following \cite[p. 74]{Schreiber13} (cf. \cite[p. 7]{SS21EPB}), to indicate that it is a topological/cohesive {\it enrichment} of what classically is denoted $B G$ -- which notation,
in turn, is a historical memory of its first construction, known as the ``bar construction''.}

\end{itemize}

\smallskip
This modernized re-proof of classifying space theory makes it essentially a formality if only one has a good abstract grasp on
(cohesive) homotopy theory --- as is provided by (cohesive) homotopy type theory. (In particular, the local triviality condition
on fiber bundles does not need to be declared as an explicit condition, but is implied by the topos-theoretic semantics of
univalent HoTT; this is discussed in \cite[\S 4.2]{SS21EPB}).

\vspace{-1mm}
\begin{equation}
\label{SemanticsOfGTorsors}
\adjustbox{}{
\def\arraystretch{1.4}
\begin{tabular}{|c|c|}
\hline
{\bf Type theory}
&
{\bf Geometric homotopy theory}
\\
\hline
\hline
$
  x : X
  \;\;\;\;
  \vdash
  \;\;\;\;
  P_x \,:\, G \Torsors_L
$
&
\hspace{8pt}
$
  \begin{tikzcd}[column sep=large]
    \mathclap{
      \scalebox{.7}{
        \color{darkblue}
        \bf
        \def\arraystretch{1}
        \begin{tabular}{c}
          principal $G$-bundle
          \\
          aka: $G$-torsor over $X$
        \end{tabular}
      }
    }
    &&
    \mathclap{
      \scalebox{.7}{
        \color{darkblue}
        \bf
        \def\arraystretch{1}
        \begin{tabular}{c}
          bundle of
          \\
          pointed $G$-torsors
        \end{tabular}
      }
    }
    &
    \mathclap{
      \scalebox{.7}{
        \color{darkblue}
        \bf
        \def\arraystretch{1}
        \begin{tabular}{c}
          universal
          \\
          principal $G$-bundle
        \end{tabular}
      }
    }
    \\[-20pt]
    P
    \ar[rr]
    \ar[d]
    \ar[
        out=180+38,
        in=180-38,
        decorate,
        looseness=3.5,
        shift right=4pt,
        "{ G }"{description}
    ]
    \ar[
      drr,
      phantom,
      "{ \scalebox{.7}{(pb)} }"{pos=.6}
    ]
    &&
    G \Torsors^{G\!\raisebox{-1pt}{\scalebox{.7}{$/$}}}
    \ar[d]
    \ar[r, phantom, "{\simeq}"]
    &[-11pt]
    \mathbf{E}G
    \ar[r, phantom, "{\simeq}"]
    \ar[d]
    &[-11pt]
    \ast
    \\
    X
    \ar[rr]
    &&
    G \Torsors
    \ar[r, phantom, "{\simeq}"]
    &
    \mathbf{B}G
    \\[-22pt]
    &&
    \mathclap{
      \scalebox{.7}{
        \color{darkblue}
        \bf
        \def\arraystretch{1}
        \begin{tabular}{c}
          moduli stack
          \\
          of $G$-torsors
        \end{tabular}
      }
    }
    &
    \mathclap{
      \scalebox{.7}{
        \color{darkblue}
        \bf
        \def\arraystretch{1}
        \begin{tabular}{c}
          delooping of
          \\
          group stack
        \end{tabular}
      }
    }
  \end{tikzcd}
$
\\
\hline
\end{tabular}
}
\end{equation}

In this spirit we now indicate the formal type-theoretic proof of the first item above (the cohesive aspect in the second item is beyond the scope of the present text, but see the outlook on p. \pageref{CohesiveHoTT}). This is the content of Prop. \ref{TheLoopingDeloopingEquivalence} below, for which we first establish Lemmas \ref{TorsorsDeloopAGroup} and \ref{DeloopingsAreUnique}.
While these should not be surprising to the experts, the result is of profound importance and currently not citable in this form from the literature (though see \cite[\S 4.13]{BBCDG21} for a less structural proof of Prop. \ref{TheLoopingDeloopingEquivalence}).

\smallskip
\begin{lemma}[The type of torsors deloops a group]
\label{TorsorsDeloopAGroup}
For $G$ a group type \eqref{TypeOfGroups}, the type $G\mathrm{Tors}_L$ of $G$-torsors (Def. \ref{GTorsorDef})
is a pointed connected 1-type \eqref{LoopTypesAsGroupTypes}, pointed by $G : G\Torsors$ \eqref{GroupsAsLeftTorsorsOverThemselves}
\vspace{-2mm}
\begin{equation}
  \label{TypeOfTorsorsIsPointedConnectedOneType}
  G \,:\, \Groups
  \;\;\;\;
  \vdash
  \;\;\;\;
  \big(
    G\Torsors_L
    ,\;
    \mathrm{pt} \,:\defneq\, G
  \big)
  \;:\;
  \PointedConnectedOneTypes
  \,,
\end{equation}
whose loop type \eqref{LoopSpace} is $G$:
\begin{equation}
  \label{LoopsOfTypeOfTorsors}
  G : \Groups
  \;\;\;\;
    \vdash
  \;\;\;\;
  p
    \,:\,
  \Id_{\Groups}
  \big(
  \Omega_{G} \big(G \mathrm{Tors}\big)
  ,\,
  G
  \big)
  \,.
\end{equation}
\end{lemma}
\begin{proof}
  First to see that the type of torsors is $0$-connected, namely to provide a certificate of the form
  required in the last line in \eqref{GroupsAsPointedConnectedOneTypes}:
  $$
    T : G\Torsors_L
    \;\;\;\;
    \vdash
    \;\;\;\;
    \exists\,
    \Id_{G \Torsors_L}
    (
      G ,\, T
    )\;.
  $$
  Applying the type equivalence of Lem. \ref{EquivariantMapBetweenTorsorsIsIso}, it suffices to show
  that there exists an equivariant map $\varphi_T : \mathrm{Hom}_G(G, T)$:
  $$
    T : G\Torsors_L
    \;\;\;\;
    \vdash
    \;\;\;\;
    \exists\,
    \mathrm{Hom}_{G}
    (
      G ,\, T
    )\;.
  $$
  Making here explicit the inhabitation certificate \eqref{AlgebraicGTorsor}
  provided by a $G$-torsor structure on the left, this is
  $$
    T : G\Torsors_L
    ,\;\;
    p \,:\, \exists \, T
    \;\;\;\;
    \vdash
    \;\;\;\;
    \exists\,
    \mathrm{Hom}_{G}
    \big(
      G ,\, T
    \big)
    \,.
  $$
  But since on the right we need a proof of the proposition $\exists(-)$ which is given by (-1)-truncation \eqref{MeaningOfPropositionalTruncation}, its universal property \eqref{UniversalPropertyTrunctation} implies that such dependent terms are equivalent to those where we assume a specific element $t : T$ on the left:
  $$
    T : G\Torsors_L
    ,\;\;
    t \,:\, T
    \;\;\;\;
    \vdash
    \;\;\;\;
    \exists\,
    \mathrm{Hom}_{G}
    \big(
      G ,\, T
    \big)
    \,.
  $$
  Moreover, applying the rules of function types \eqref{PiInferenceRules} to the (-1)-truncation unit
  $\eta :
    \mathrm{Hom}_{G}
    \big(
      G ,\, T
    \big)
    \to
    \exists \,
    \mathrm{Hom}_{G}
    \big(
      G ,\, T
    \big)
$ \eqref{PostnikovTower} means that we will infer such an existence proof from constructing one example:
$$
    T : G\Torsors_L
    ,\;\;
    t \,:\, T
    \;\;\;\;
    \vdash
    \;\;\;\;
    \phi_t
    :
    \mathrm{Hom}_{G}
    \big(
      G ,\, T
    \big)
    \,.
  $$
  But such a term we obtained by setting
  \begin{equation}
    \label{HomomorphismOfTorsorsOutOfG}
    \varphi_t(g)
    \;\;
    :\defneq
    \;\;
    g \acts\, t
  \end{equation}
  and providing the following certificate that this function is equivariant \eqref{EquivarianceTypeDefinition},
  where we use concatenation of equalities \eqref{ConcatenationOfEqualities}:
  $$
    g', g \,:\, G
    \;\;\;\;\;\;
      \vdash
    \;\;\;\;\;\;
    \phi_t( g' \cdot g )
    \underset{
     \scalebox{.7}{
       \eqref{HomomorphismOfTorsorsOutOfG}
     }
    }{=}
    (g' \cdot g) \acts t
    \underset{
        \scalebox{.7}{
          \eqref{AlgebraicGActions}
        }
    }{
      \overset{
        \mathrm{act}_T
      }{=}
    }
    g' \acts (g \acts t)
    \underset{
     \scalebox{.7}{
       \eqref{HomomorphismOfTorsorsOutOfG}
     }
    }{=}
    g' \acts \phi_t(g)
    \;\;
    :
    \;\;
    \Id_{T}\Big(
     \phi_t(g' \cdot g)
     ,\,
     g' \acts \phi_t(g)
    \Big)
    \,.
  $$
  In summary, this provides the connectivity certificate required in \eqref{GroupsAsPointedConnectedOneTypes}.
  The 1-truncation certificate also required there will immediately follow from the proof of the second claim \eqref{LoopsOfTypeOfTorsors},
  to which we now turn (since $G$ is 0-truncated and by the recursive definition of $n$-truncation \eqref{TypeOfGroupoids}).

  The proof of \eqref{LoopsOfTypeOfTorsors}, after unwinding the definition of looping \eqref{LoopSpace}, is to be a dependent term of the form
  $$
    G : \Groups
    \;\;\;\;\;
      \vdash
    \;\;\;\;\;
    \mathrm{proof}_G
    \,:\,
    \Id_{ \Groups }
    \big(
     \Id_{G \Torsors}
     (
       G
       ,\,
       G
     )
     ,\,
     G
    \big)
    \,.
  $$
  By applying the type equivalence of Lem. \ref{EquivariantMapBetweenTorsorsIsIso} our goal is equivalently a term of the following type:
  $$
    G : \Groups
    \;\;\;\;\;
      \vdash
    \;\;\;\;\;
    \mathrm{proof}_G
    \,:\,
    \Id_{ \Groups }
    \big(
     \mathrm{Hom}_G
     (
       G
       ,\,
       G
     )
     ,\,
     G
    \big)
    \,.
  $$
  Moreover, by the type equivalence of univalence \eqref{UnivalenceAxiom}, we equivalently need to produce an inverted function between group types
  $$
    G : \Groups
    \;\;\;\;\;
      \vdash
    \;\;\;\;\;
    \mathrm{proof}_G
    \,:\defneq\,
    \big(
      f_G, \, f_G^{-1}, \, \cdots
    \big)
    \,:\,
    \mathrm{Hom}_G
     (
       G
       ,\,
       G
     )
     \overset{\sim}{\longrightarrow}
     G
    \,.
  $$

  Now take this function $f_G$ to be given by evaluation at the neutral element:
  \begin{equation}
    \label{FunctionFromGEndosAsTorsorToG}
    \phi
    \,:\,
    \mathrm{Hom}_G(G,\,G)
    \;\;\;\;
      \vdash
    \;\;\;\;
    f_G(\phi)
    \;\,
      :\defneq
    \;\,
    \phi(\NeutralElement)
    \;\;
      :
    \;\;
    G
  \end{equation}
  and take its reverse $f^{-1}_G$ to be
  \begin{equation}
    \label{InverseOfFunctionFromGEndosAsTorsorToG}
    h : G
    \;\;\;\;
      \vdash
    \;\;\;\;
    f_G^{-1}(h)
    \;\,
      :\defneq
    \;\,
    \big(
      h \mapsto h \cdot g
    \big)
    \;:\;
    G \longrightarrow \mathrm{Hom}_G(G,\,G)
    \,.
  \end{equation}

  The certificates that this does constitute a pair of inverse functions follow, via the composition
  rule for equality certificates \eqref{ConcatenationOfEqualities},
  just the way one would prove this classically:

  $$
    f \circ f^{-1}(g)
    \underset{
      \scalebox{.7}{
        \eqref{InverseOfFunctionFromGEndosAsTorsorToG}
      }
    }{\;\;=\;\;}
    f\big(
      h \mapsto h \cdot g
    \big)
    \underset{
      \scalebox{.7}{
        \eqref{FunctionFromGEndosAsTorsorToG}
      }
    }{\;\;=\;\;}
    \NeutralElement \cdot g
    \underset{
      \scalebox{.7}{
        \eqref{AlgebraicGActions}
      }
    }{
      \overset
        {\mathrm{unt}}
        {\;\;=\;\;}
    }
    g
    \;\;\;\;
    :
    \;\;\;\;
    \Id\big(
      f \circ f^{-1}(g)
      ,\,
      g
    \big)
  $$

  $$
    f^{-1}
      \circ
    f(\phi)
    \underset{
      \scalebox{.7}{\eqref{FunctionFromGEndosAsTorsorToG}}
    }{
    \;\;=\;\;
    }
    f^{-1}\big(
      \phi(\NeutralElement)
    \big)
    \underset{
      \scalebox{.7}{\eqref{InverseOfFunctionFromGEndosAsTorsorToG}}
    }{
    \;\;=\;\;
    }
    \big(
      h
      \,\mapsto\,
      h \acts \, \phi(\NeutralElement)
    \big)
    \underset{
      \scalebox{.7}{
       \eqref{EquivarianceTypeDefinition}
      }
    }{
    \;\;=\;\;
    }
    \big(
      h
      \,\mapsto\,
      \phi(h \acts\,  \NeutralElement)
    \big)
    \underset{
      \scalebox{.7}{
        \eqref{AlgebraicGActions}
      }
    }{
      \overset
        {\mathrm{unt}}
        {\;=\;}
    }
    \big(
      h
      \,\mapsto\,
      \phi(h)
    \big)
    \underset{
      \scalebox{.7}{
        \eqref{PiInferenceRules}
      }
    }{
      \;=\;
    }
    \phi
    \;\;\;\;
      :
    \;\;\;\;
    \Id\big(
      f^{-1} \circ f(\phi)
      ,\,
      \phi
    \big).
  $$
  This completes the proof of \eqref{LoopsOfTypeOfTorsors}
  and with it that of \eqref{TypeOfTorsorsIsPointedConnectedOneType}.
\end{proof}

\begin{remark}[Perspective on proof]
\label{PerspectiveOnProof}
The above proofs of Lem. \ref{EquivariantMapBetweenTorsorsIsIso} and
Lem. \ref{TorsorsDeloopAGroup} should serve to illustrate how type-theoretic constructions proceed (of course, many
more illustrative examples may be found in the literature listed at Lit. \ref{LiteratureHomotopyTypeTheory}).
Therefore, in the following proofs we shall omit the fine-grained manipulation of dependent terms and just indicate
enough of the proof strategy that obtaining a fully formal type-theoretic proof is straightforward, if maybe tedious.

Of course, this is exactly the style in which all rigorous math has been communicated during the last century, it being
understood that in principle all rigorous human-readable proofs could be transformed (straightforwardly, if tediously)
into fully formal proofs in a classical logical foundation like ZFC set theory, if desired. The big difference however is that there
is little reason to desire a formal proof in ZFC set theory,
while here in the new foundations of homotopy type theory, proofs are computer programs whose construction may be highly desirable, such as the construction of a program evaluating topological quantum gate execution that we are headed towards.
\end{remark}

\begin{lemma}[Any delooping is equivalent to the type of torsors]
\label{DeloopingsAreUnique}
For $G$ a group
\eqref{TypeOfGroups}
and  $\big(\mathbf{B}G, \mathrm{pt}\big)$ any delooping \eqref{GroupsAsPointedConnectedOneTypes}, then the function which sends data $t$ of type $\mathbf{B}G$ to its identification type with $\mathrm{pt}$
extends to a type equivalence \eqref{HomotopyEquivalenceOfTypes}
between $\mathbf{B}G$ and the type of $G$-torsors (Def. \ref{GTorsorDef}):

\begin{equation}
  \label{EquivalenceBetweenBGAndGTors}
  \def\arraystretch{1.4}
  \begin{array}{l}
  G : \Groups
  ,\;\;
  \mathbf{B}G : \PointedConnectedOneTypes
  ,
  \\
  \ell
  :
  \big(
    \Omega_{\mathrm{pt}} \mathbf{B}G
    ,\,
    \mathrm{id}_{\mathrm{pt}}
    ,\,
    \mathrm{conc}
  \big)
  \overset{\sim}{\to}
  \big(
    G
    ,\,
    \NeutralElement
    ,\,
    \cdot
  \big)
  \end{array}
  \;\;\;\;\;\;\;
    \vdash
  \;\;\;\;\;\;\;
   \begin{tikzcd}[row sep=-3pt]
    \mathbf{B}G
    \ar[
      rr,
      "{ \sim }"
    ]
    &&
    G \mathrm{Tors}_L
    \\
    t
    &\longmapsto&
    \big(
      \Id_{\mathbf{B}G}
      (
        \mathrm{pt}
        ,\,
        t
      )
      ,\,
      \mathrm{conc}
    \big)
    \mathrlap{\,,}
  \end{tikzcd}
\end{equation}
where the $G$-action on $\Id(\mathrm{pt},t)$ is given by concatenation of identifications, under the given looping equivalence $\ell$:
\vspace{-2mm}
$$
  \begin{tikzcd}[row sep=-2pt, column sep=10pt]
    G
      \times
    \Id_{\mathbf{B}G}
    (
      \mathrm{pt}
      ,\,
      t
    )
    \ar[rr, "{\sim}"]
    &&
    \Id_{\mathbf{B}G}
    \big(
      \mathrm{pt}
      ,\,
      \mathrm{pt}
    \big)
    \times
    \Id_{\mathbf{B}G}
    (
      \mathrm{pt}
      ,\,
      t
    )
    \ar[
      rr,
      "{ \acts }"
    ]
    &&
    \Id_{\mathbf{B}G}
    (
      \mathrm{pt}
      ,\,
      t
    )
    \\
    \big(
      g
      ,\,
      \mathrm{pt}
        \overset{p}{\rightsquigarrow}
      t
    \big)
    &\longmapsto&
    \big(
      \mathrm{pt}
      \overset{
        \ell_g
      }{\rightsquigarrow}
      \mathrm{pt}
      ,\,
      \mathrm{pt}
        \overset{p}{\rightsquigarrow}
      t
    \big)
    &\longmapsto&
      \mathrm{pt}
        \overset{
         \mathrm{conc}(
           \ell_g
           ,\,
           p
         )
        }{\rightsquigarrow}
      t
  \end{tikzcd}
$$
\end{lemma}
\begin{proof}
  First to see that the construction \eqref{EquivalenceBetweenBGAndGTors}
  really produces torsors:

  The  inhabitation condition, $\exists \, \Id(\mathrm{pt},t)$, is part of the connectivity certificate that comes with a pointed connected type \eqref{GroupsAsPointedConnectedOneTypes}.
  Moreover, given a pair of torsor elements, then using their inversion
  \eqref{InversionOfIdentifications}
  and concatenation \eqref{ConcatenationOfIdentifications}
  produces a group element taking one to the other, under the above action:
 \begin{equation}
\label{ActionByTransportDefinition}
\def\arraystretch{1.5}
\begin{array}{rccl}
p,q \,:\, \Id(\mathrm{pt,t})
\;\;
\vdash
\;\;
&
g
  &
    :\longmapsto
  &
     \adjustbox{raise=-14pt}{
    \begin{tikzcd}[
      decoration=snake,
    ]
      \mathrm{pt}
      \ar[
        dr,
        decorate,
        "{ p }"{swap}
      ]
      \ar[
        rr,
        decorate,
        "{
          \ell_g
          \;:\defneq\;
          \mathrm{conc}
          \scalebox{1.2}{$($}
            p
            ,\,
            \mathrm{inv}(q)
          \scalebox{1.2}{$)$}
        }"{yshift=2pt}
      ]
      &&
      \mathrm{pt}
      \ar[
        dl,
        decorate,
        "{ q }"
      ]
      \\
      &
      t
    \end{tikzcd}
    }
    \!\!\!\!
    :
    \Id(\mathrm{pt},\, \mathrm{pt})
  \\
 p,q \,:\, \Id(\mathrm{pt,t})
\;\;
\vdash
\;\;
  &
  g \acts q
    =
    p
    & : &
    \Id\big(
      g \acts q
      ,\,
      p
    \big)
    \,.
\end{array}
\end{equation}
  Finally, this group element is unique, because for  another $g' : G$ with  $g' \acts q = p$, we have
  \begin{equation}
    \label{IntermediateStepInProofThatBGIsGTors}
    \mathrm{conc}(\ell_g, \,q)
    =
    p
    =
    \mathrm{conc}(\ell_{g'}, q)
    \;\;
    :
    \;\;
    \Id\big(
      \mathrm{pt}
      ,\,
      t
    \big)
  \end{equation}
  and hence
$$
  \ell_g
  \underset{
    \scalebox{.7}{
      \eqref{ReversesAreInverses}
    }
  }{
    =
  }
    \mathrm{conc}
    \big(
      \mathrm{conc}(\ell_g, \,q)
      ,\,
      \mathrm{inv}(q)
    \big)
    \underset{
      \scalebox{.7}{
        \eqref{IntermediateStepInProofThatBGIsGTors}
      }
    }{=}
    \mathrm{conc}
    \big(
      \mathrm{conc}(\ell_{g'}, \,q)
      ,\,
      \mathrm{inv}(q)
    \big)
  \underset{
    \scalebox{.7}{
      \eqref{ReversesAreInverses}
    }
  }{
    =
  }
  \ell_{g'}
    \;\;
    :
    \;\;
    \Id\big(
      \mathrm{pt}
      ,\,
      \mathrm{pt}
    \big)
    \,.
$$

Second, that this function \eqref{EquivalenceBetweenBGAndGTors} is an equivalence follows from the fundamental lemma of pointed connected types \cite[Thm. 2.1.1]{Myers22}: If a function between pointed connected types induces an equivalence on loop types, then it is itself an equivalence.
\end{proof}

 We now have the tools in hand to state the inverse equivalence to the looping operation \eqref{LoopTypesAsGroupTypes}:
\begin{proposition}[The looping-delooping equivalence]
\label{TheLoopingDeloopingEquivalence}
There is an equivalence \eqref{HomotopyEquivalenceOfTypes}
between the type $\Groups$ of algebraically defined groups \eqref{TypeOfGroups} and the type $\mathrm{Type}_{1\scalebox{.7}{$\bullet$}}^{>0}$ \eqref{GroupsAsPointedConnectedOneTypes} of pointed connected 1-types:
\vspace{-1mm}
\begin{equation}
  \label{LoopingDeloopingEquivalence}
  \begin{tikzcd}[
    row sep=2pt
  ]
    \PointedConnectedOneTypes
    \ar[
      rr,
      shift right=4pt,
      "{\sim}",
      "{ \Omega_{\mathrm{pt}}(-) }"{swap}
    ]
    \ar[
      from=rr,
      shift right=5pt,
      "{
        \mathbf{B}(-)
        \;:\defneq\;
        (-)\mathrm{Tors}_L
      }"{swap}
    ]
    &&
    \Groups
    \\[5pt]
    (D, \mathrm{pt})
    &\longmapsto&
    \Omega_{\mathrm{pt}} D
  \end{tikzcd}
\end{equation}
In one direction this equivalence is given by the looping construction $\Omega_{\mathrm{pt}}$ \eqref{LoopTypesAsGroupTypes};
in the other direction by sending $G$ to its type $G \mathrm{Tors}_L$ of left $G$-torsors \eqref{GTorsorDef}.

\end{proposition}
Another type-theoretic proof of this statement was earlier given in \cite{LicataFinster14}, using higher inductive types (groupoid quotients \cite{VeltrivdWeide21}). We may now offer a more structural proof, using delooping by the type of torsors:
\begin{proof}
 It is sufficient to show that the two functions are inverses of each other, up to re-identification \eqref{HomotopyEquivalenceOfTypes}, which is now immediate from the previous lemmas:
 In one direction we have, by \cref{TorsorsDeloopAGroup}:
 $$
   \big(
   \Omega_G
   \circ
   (-)\Torsors_L
   \big)
   (G)
   \;\defneq\;
   \Omega_{G}
   \big(
    G \Torsors_L
   \big)
   \underset{
     \scalebox{.7}{
       \eqref{LoopsOfTypeOfTorsors}
     }
   }{\;\simeq\;}
   G
   \,.
 $$
 In the other direction we have, using
 \cref{TorsorsDeloopAGroup}
 and \cref{DeloopingsAreUnique}
 $$
   \big(
     (-)\Torsors_L
     \circ
     \Omega_{\mathrm{pt}}
   \big)
   (\mathbf{B}G)
   \;\defneq\;
   \big(
     \Omega_{\mathrm{pt}}\mathbf{B}G
   \big)
   \Torsors_L
   \underset{
     \scalebox{.7}{
       \eqref{LoopsOfTypeOfTorsors}
     }
   }{
     \;\simeq\;
   }
   G
   \Torsors_L
   \underset{
     \scalebox{.7}{
       \eqref{EquivalenceBetweenBGAndGTors}
     }
   }{
     \;\simeq\;
   }
   \mathbf{B}G
   \,.
 $$

\vspace{-.6cm}
\end{proof}

\medskip

We can now reframe the type of $G$-actions
(Def. \ref{TheTypeOfGSets})
simply as types dependent on a delooping of $G$ -- this is where the right $G$-actions show up.

\begin{proposition}[Functions of $\mathbf{B}G$ are $G$-Actions]
  \label{GSetsAreMapsOutOfBG}
  Given $G : \Groups$ \eqref{TypeOfGroups}, the type of right
  $G$-sets (Def. \ref{TheTypeOfGSets}) is equivalent \eqref{HomotopyEquivalenceOfTypes} to the type of functions from a delooping $\mathbf{B}G$ (Prop. \ref{TheLoopingDeloopingEquivalence}) to $\Sets$ \eqref{Sets}:
  \begin{equation}
    \label{MapsOutOfBGEquivalentToGSets}
    \begin{tikzcd}[
      sep=0pt
    ]
      (\mathbf{B}G \to \Sets)
      \ar[
        rr,
        "{ \sim }"
      ]
      &&
      G\Sets_R
      \\
      f &\longmapsto&
    \quad   \big(
        S :\defneq f(\mathrm{pt})
        ,\;
        s \,\rightacts g
         \,:\defneq\,
        g_\ast(s)
      \big)
    \end{tikzcd}
  \end{equation}
\end{proposition}
\begin{proof}
 By the type equivalence $\mathbf{B}G \simeq G\Torsors_L$ from \cref{DeloopingsAreUnique}, and recalling that this identifies $\mathrm{pt} : \mathbf{B}G$ with $G : G\Torsors_L$,
 we may equivalently show that the following function is an equivalence
 \begin{equation}
    \label{MapsOutOfGTorsEquivalentToGSets}
    \begin{tikzcd}[
      sep=0pt
    ]
      \mathllap{
        F
        \;\;
        :
        \;\;\;
      }
      (G\mathrm{Tors}_L \to \Sets)
      \ar[
        rr
      ]
      &&
      G\Sets_R
      \\
      f &\longmapsto&
    \quad   \big(
        S :\defneq f(G)
        ,\;
        s \,\rightacts g
         \,:\defneq\,
        (h \,\mapsto\,h \cdot g )_\ast (s)
      \big)
      \mathrlap{\,.}
    \end{tikzcd}
  \end{equation}
  We will now do so by constructing an explicit inverse to this function and then showing that it is an inverse:
  \begin{equation}
    \label{ReverseOfMapsOutOfGTorsEquivalentToGSets}
    \begin{tikzcd}[row sep=0pt]
      \mathllap{
        F^{-1}
        \;\;
        :
        \;\;\;
      }
      G\Sets_R
      \ar[rr]
      &&
      \big(
        G\Torsors_L
        \to
        \Sets
      \big)
      \\
      \big(
        X
        ,\,
        \rightacts
      \big)
      &\mapsto&
      \Big(
        (T, \acts)
        \,\mapsto\,
        X \otimes_G T
      \Big)
      \,,
    \end{tikzcd}
  \end{equation}
  where the tensor product ``$\otimes_G$''  (classically denoted ``$\times_G$'', cf. \cite[(1.17)]{SS21EPB} and references given there)
  denotes the following quotient set \eqref{QuotientSetAsPushout}:
 \begin{equation}
   \label{TwistedActionDefinition}
     X : G\Sets_R
     ,\;\;\;
     T : G\Sets_L
     \;\;\;\;\;
      \vdash
     \;\;\;\;\;
     X \otimes_G T
       \;:\defneq\;
     \frac{X \times T}{(x \,\rightacts g,\, t)
       \sim
     (x,\, g\acts\, t)}
     \,.
 \end{equation}
 \noindent
  As usual, we will denote also the images under the quotient map by the tensor product symbol
  \begin{equation}
    \label{TensorProductOfElements}
    \begin{tikzcd}[row sep=0pt]
      \mathllap{
        \mathrm{cpr}
        \;\;
        :
        \;\;\;
      }
      X \times Y
      \ar[rr]
      &&
      X \otimes_G T
      \\
      (x,\,t)
      &\mapsto&
      x \otimes t
    \end{tikzcd}
  \end{equation}
  Notice that/how this implies the
  expected equality certificates \eqref{PropositionalEquality}
  of this form:
  \begin{equation}
    \label{GTensorRelation}
    x : X
    ,\;\;
    t : T
    ,\;\;
    g : G
    \;\;\;\;\;
      \vdash
    \;\;\;\;\;
    (x \,\rightacts g)\otimes t
      \underset{
        \scalebox{.7}{
          \eqref{TensorProductOfElements}
        }
      }{
      =
      }
    \mathrm{cpr}
    \big(
      (x \,\rightacts g)
      ,\,
      t
    \big)
    \underset{
      \scalebox{.7}{
        \eqref{TwistedActionDefinition}
      }
    }{
      =
    }
    \mathrm{cpr}
    \big(
      x
      ,\,
      g
      \acts
      \,
      t
    \big)
      \underset{
        \scalebox{.7}{
          \eqref{TensorProductOfElements}
        }
      }{
      =
      }
    x \otimes (g \acts\, t)
    \;\;
    :
    \;\;
    \Id\big(
      \cdots
      ,
      \cdots
    \big)
  \end{equation}

Now to prove the claim, we construct homotopies \eqref{HomotopyEquivalenceOfTypes}
which exibit $F^{-1}$ \eqref{ReverseOfMapsOutOfGTorsEquivalentToGSets} as a homotopy-inverse to $F$ \eqref{MapsOutOfGTorsEquivalentToGSets}
from both sides.
From one side, we need dependent equivalences of this form:
\begin{equation}
  \label{EquivalenceInProvingThatGSetsAreFunctionsOfBG}
  (X, \rightacts) : G\Torsors_R
  \;\;\;\;
    \vdash
  \;\;\;\;
  F
  \circ
  F^{-1} (X, \rightacts)
  \,\defneq\,
  F
  \big(
    (T, \acts) \,\mapsto\, X \otimes_G T
  \big)
  \,\defneq\,
  \big(
    X \otimes_G G
    ,\rightacts
  \big)
  \;\simeq\;
  (X, \rightacts)
  \,.
\end{equation}
By the structure identity principle, it will suffice to construct a comparison function which is $G$-equivariant \eqref{EquivarianceTypeDefinition} and prove that it is an isomorphism. Consider this one:
\begin{equation}
  \label{ComparingTorsorWithItsTensoringWithGroup}
  \begin{tikzcd}[row sep=0pt]
    \mathllap{
      (X, \rightacts)
      :
      G\Torsors_R
      \;\;\;\;\;\;\;
        \vdash
      \;\;\;\;\;\;\;
    }
    X
    \ar[
      rr
    ]
    &&
    X \otimes_G G
    \\
    x
    &\mapsto&
    x \otimes \NeutralElement
  \end{tikzcd}
\end{equation}
To see that this is indeed equivariant, we first need to describe the $G$-action on the left:
Since $G$ is identified with $\mathrm{Hom}_G(G,G)$ via right multiplication, the induced action on $X \otimes_G G$ is also by right multiplication:
\begin{equation}
  \label{GroupActionOnTensorOfTorsorWithGroup}
  (x \otimes h) \,\rightacts g
  \;=\;
  x \otimes (h \cdot g).
\end{equation}
This can be checked by $\mathrm{Id}$-induction \eqref{PathInduction}, considering how the type family $X \otimes_G (-)$ acts on an identification $\varphi : \Id_{G\mathrm{Tors}_L}(G, T)$ and seeing that it holds for $\id_G : \Id_{G\mathrm{Tors}_L}(G, G)$.
From this we find the certificate that \eqref{ComparingTorsorWithItsTensoringWithGroup} is indeed equivariant,
by composing equality certificates, as follows \eqref{ConcatenationOfEqualities}:
$$
  (x \,\rightacts g) \otimes \NeutralElement
  \underset{
    \scalebox{.7}{
      \eqref{GTensorRelation}
    }
  }{
    =
  }
  x \otimes (g \acts \NeutralElement)
  \underset{
    \scalebox{.7}{
      \eqref{GroupsAsLeftTorsorsOverThemselves}
    }
  }{
    =
  }
  x \otimes (g \cdot \NeutralElement)
  \underset{
    \scalebox{.7}{
      \eqref{TypeOfGroups}
    }
  }{
    =
  }
  x \otimes (\NeutralElement \cdot g)
  \underset{
    \scalebox{.7}{
      \eqref{GTensorRelation}
    }
  }{
    =
  }
  (x \otimes \NeutralElement) \,\rightacts g
  \,.
$$
This establishes that \eqref{ComparingTorsorWithItsTensoringWithGroup} is equivariant and hence invertible (Lem. \ref{EquivariantMapBetweenTorsorsIsIso}). We may also readily define the inverse: it is given by $x \otimes g \mapsto x \,\rightacts g$, which is well defined since $x \,\rightacts (g \cdot h) = (x \,\rightacts g) \,\rightacts h$.

\medskip

From the other side, we need to construct dependent equivalences of this form:
$$
  f : G\Torsors_L \to \Sets
  \;\;\;\;\;\;\;
    \vdash
  \;\;\;\;\;\;\;
  F^{-1}
  \circ
  F
  (f)
  \;\defneq\;
  F^{-1}
  \big(
    f(G)
  \big)
  \;\defneq\;
  \big(
    (T, \acts)
    \,\mapsto\,
    f(G)
      \otimes_G
    T
  \big)
  \;\;
  \simeq
  \;\;
  \big(
    (T, \acts)
    \,\mapsto\,
    f(T, \acts)
  \big)
  \;\defneq\;
  f
  \,.
$$
By (univalence and) function extensionality \eqref{happlyFormula}, this is obtained by constructing $T$-dependent bijections between the arguments of the functions on the right --- consider this one:
$$
  \begin{tikzcd}[row sep=0pt]
    f(G) \otimes_G T
    \ar[rr]
    &&
    f(T)
    \\
    x \otimes t
    &\mapsto&
    (\varphi_t)_\ast(x)
    \mathrlap{\,,}
  \end{tikzcd}
$$
where $\varphi_t : \mathrm{Hom}_G(G, T)$ is the equivariant isomorphism $\varphi_t(h) :\defneq h \acts\, t$ and $(\varphi_t)_{\ast} : f(G) \to f(T)$ is the transport along this identification in $f$. This map is well defined on the tensor product quotient since $(\varphi_{g\acts\, t})_{\ast}(x) = (\varphi_t \circ (h \mapsto hg))_\ast(x) = (\varphi_t)_{\ast}(h \mapsto hg)_{\ast}(x) =: (\varphi_t)_{\ast}(x \,\rightacts g)$, recalling that the right action on $f(G)$ is given by $x \,\rightacts g :\defneq (h \mapsto hg)_{\ast}(x)$ and making use of the functoriality of transport \eqref{FunctorialityOfTransport}. It remains, then, to show that this map is a bijection.

Let $y : f(T)$, seeking to show that it has a unique inverse image in $f(G) \otimes_G T$. Since we are trying to prove the proposition $\exists!\fiber{y}{ x \otimes t
\mapsto
    (\varphi_t)_\ast(x)}$, we may assume we have an element $t : T$ (since by assumption we have $p : \exists T$). We then have that $y = (\varphi_t)_{\ast}((\varphi_t^{-1})_{\ast}(y))$, so that $y$ has an inverse image given by $(\varphi_t^{-1})_{\ast}(y) \otimes t$. Now suppose that $y = (\varphi_s)_{\ast}(x)$, seeking to show that $x \otimes s = (\varphi_t^{-1})_{\ast}(y) \otimes t$. Since $T$ is a torsor, there is a unique $g : G$ for which $g \acts\, t = s$; therefore, $\varphi_s = \varphi_t \circ (h \mapsto hg)$, which rearranges to give us that $(h \mapsto hg) = \varphi_t^{-1} \circ \varphi_s$. This means that $x \,\rightacts g =: (h \mapsto hg)_{\ast}(x) = (\varphi_t^{-1} \circ \varphi_s)_{\ast}(x) = (\varphi_t^{-1})_{\ast}(y)$, so that $x \otimes s = x \otimes (g \acts\, t) = (x \,\rightacts g) \otimes t = (\varphi_t^{-1})_{\ast}(y) \otimes t$.
\end{proof}

\begin{notation}
It will be useful to have generic, concise, and suggestive notation for the
inverse construction to the equivalence  \eqref{MapsOutOfBGEquivalentToGSets}. We will write this as follows:
\begin{equation}
  \label{NotationForGAction}
  \begin{tikzcd}[
    row sep=-1pt
  ]
    G\Sets
    \ar[rr]
    &&
    (\mathbf{B}G \to \Sets)
    \\
    \big(
      S,\;
      {
        \color{purple}
        \acts
      }
      \,
    \big)
    &\longmapsto&
    \big(
      (G \acts \,S)
      \,:\;
      t
      \mapsto
      {\color{purple}
        \acts_{\,{t}}
      }
      \, S
    \big)
  \end{tikzcd}
\end{equation}
\end{notation}

\begin{remark}[Choice of delooping]
  Since (by Prop. \ref{TheTypeOfGSets}) an action of $G$ is equivalently a function $\mathbf{B}G \to \Sets$,
  and since (by Prop. \ref{TheLoopingDeloopingEquivalence}) all deloopings of $G$ are equivalent, it often
  pays to tailor the choice of delooping to the sorts of actions relevant for a given construction.
  We will see this in action when we construct the delooping $\mathbf{B}R^{\times}$ of the group of units
  \eqref{GroupOfUnits} of a ring $R$ as the type of $R$-modules \eqref{ModuleDataStructure} identifiable with $R$ -- in \eqref{TypeOfRLinesDeloopsGroupOfUnits} below.
\end{remark}

\medskip

\noindent
{\bf $G$-$\infty$-Actions.}
The equivalent characterization of group actions
from Prop. \ref{GSetsAreMapsOutOfBG} has the striking advantage that it
immediately generalizes to a notion of $G$-actions on any (higher) homotopy type $\mathcal{A}$: simply as
those functions $\mathbf{B}G \to \Types$ \eqref{TypeOfSmallTypes}
which take $\mathrm{pt} : \mathbf{B}G$ to $\mathcal{A}$. Under the pertinent identifications
\eqref{DependentDataTypesAsFunctions} this are nothing but the $\mathbf{B}G$-dependent data types
(cf. \cite[\S 4.2]{BvDR18}):

\begin{equation}
\label{GroupInfinityAction}
\adjustbox{}{
\def\arraystretch{1}
\begin{tabular}{|c|c|}
\hline
{\bf Delooping-dependent Types}
&
{\bf Group $\infty$-Actions}
\\
\hline
\hline
\begin{minipage}{4cm}
$$
  t
  :
  \mathbf{B}G
  \;\;\;\;
  \vdash
  \;\;\;
  \acts_{\, {t}} \, \mathcal{A}
  :
  \Types
$$
\end{minipage}
&
\begin{minipage}{6cm}
$$
\begin{tikzcd}[column sep=huge]
  \mathcal{A}
  \ar[
    d,
    "{\ }"{name=t}
  ]
  \ar[
    r,
    "{\ }"{swap, name=s}
  ]
  \ar[
    from=s,
    to=t,
    Rightarrow,
    "{ \scalebox{.7}{(pb)} }"
  ]
  &
  \overset{
    \mathclap{
      \scalebox{.7}{
        \color{darkblue}
        \bf
        \def\arraystretch{.9}
        \begin{tabular}{c}
          Universal
          $G$-associated
          \\
          $\mathcal{A}$-fiber
          $\infty$-bundle
        \end{tabular}
      }
    }
  }{
  \HomotopyQuotient
    { \mathcal{A} }
    { G }
  }
  \ar[
    d,
    "{\  }"{name=t2}
  ]
  \ar[
    r,
    "{\ }"{swap, name=s2}
  ]
  \ar[
    from=s2,
    to=t2,
    Rightarrow,
    "{ \scalebox{.7}{(pb)} }"{}
  ]
  &
  \widehat{\Objects}
  \ar[d]
  \\
  \ast
  \ar[
    r,
    "{ \mathrm{pt} }"
  ]
  &
  \mathbf{B}G
  \ar[
    r,
    "{ G \acts \, \mathcal{A}  }",
    "{
      \scalebox{.7}{
        \color{greenii}
        \bf
        \scalebox{1.3}{$G$-$\infty$}-action
      }
    }"{swap}
  ]
  &
  \Objects
\end{tikzcd}
$$
\end{minipage}
\\
\hline
\adjustbox{raise=7pt}{
\begin{minipage}{4cm}
$$
  \vdash
  \;\;
  (t : \mathbf{B}G)
  \times
  \big(\!
    \acts_t \mathcal{A}
  \big)
  \;:\;
  \Types
$$
\end{minipage}
}
&
\begin{minipage}{6cm}
$$
\begin{tikzcd}
  \HomotopyQuotient
    { \mathcal{A} }
    { G }
    \;\simeq\;
    \mathrlap{
      \scalebox{.7}{
        \color{darkblue}
        \bf
        \def\arraystretch{.9}
        \begin{tabular}{c}
          homomotopy
          \\
          quotient
        \end{tabular}
      }
    }
\end{tikzcd}
$$
\end{minipage}
\\
\hline
\adjustbox{raise=7pt}{
\begin{minipage}{4cm}
$$
  \vdash
  \;\;\;
  (t : \mathbf{B}G)
  \to
  \big(\!
    \acts_t \mathcal{A}
  \big)
  \;:\;
  \Types
$$
\end{minipage}
}
&
\begin{minipage}{6cm}
$$
\begin{tikzcd}
    \mathcal{A}^G
    \mathrlap{
      \scalebox{.7}{
        \color{darkblue}
        \bf
        \def\arraystretch{.9}
        \begin{tabular}{c}
          homomotopy
          \\
          fixed locus
        \end{tabular}
      }
    }
\end{tikzcd}
$$
\end{minipage}
\\
\hline
\end{tabular}
}
\end{equation}

\smallskip
What is essentially a syntactic triviality on the left translates semantically, as indicated on the right, to the rather profound notion of ``infinitely homotopy coherent actions'' or ``$\infty$-actions'' of (sheaves of) groups on $\infty$-stacks, see \cite[\S 4]{NSS12}\cite[\S 2.2]{SS20OrbifoldCohomology}\cite[\S 3.23]{SS21EPB} for details and further references. In particular, this shows that the type of dependent pairs  which constitutes the ``total space'' \eqref{TypeClassification} of a $\mathbf{B}G$-dependent type is semantically the {\it homotopy quotient} of the corresponding $\infty$-action (while the analogous dependent function type is semantically the {\it homotopy fixed locus}).

\medskip
Concretely, the canonical model for the $G$-homotopy quotient $\HomotopyQuotient{(-)}{G}$ in topological spaces \eqref{ViaParameterizedPointSetTopology} is given
(under mild conditions, certainly when $G$ is discrete) by the {\it Borel construction} $(-) \times_G E G$, which manifestly forms the topological $\mathcal{A}$-fiber bundle which is {\it associated} to the universal $G$-principal bundle:
\begin{equation}
\label{DeloopingDependentTypeAsBorelConstruction}
\adjustbox{}{
\def\arraystretch{1.7}
\begin{tabular}{|c|c|}
  \hline
  {\bf Homotopy type theory}
  &
  {\bf Homotopy theory}
  \\
  \hline
  \hline
  $
  \vdash
  \;\;\;
  (t : \mathbf{B}G) \times
  \big(\!
    \acts_t \mathcal{A}
 \big)
  \;:\;
  \Types
  $
  &
  $\mathcal{A} \times_G E G$
  \scalebox{.7}{
    \color{darkblue}
    \bf
      Borel construction
  }
  \\
  \hline
\end{tabular}
}
\end{equation}

\noindent
{\bf Delooping of group homomorphisms.}
A similar construction as in the proof of Prop. \ref{GSetsAreMapsOutOfBG} shows that group homomorphisms
\vspace{-4mm}
\begin{equation}
  \label{TypeOfGroupHomomorphisms}
  \hspace{-4mm}
  \def\arraystretch{2.2}
  \begin{array}{l}
  \big(
    G
    ,\,
    \NeutralElement_G
    ,\,
    \cdot_G
    ,\,
    (-)^{-1}_G
  \big)
  ,\;
  \big(
    H
    ,\,
    \NeutralElement_H
    ,\,
    \cdot_H
    ,\,
    (-)^{-1}_H
  \big)
  \,:\,
  \Groups
  \;\;\;
    \vdash
  \;\;\;
  \\
  \scalebox{.7}{
    \color{orangeii}
    \bf
    \def\arraystretch{.9}
    \begin{tabular}{c}
      Group
      \\
      homomorphisms
    \end{tabular}
  }
  \big(
    G \underset{\mathrm{hom}}{\to} H
  \big)
  \;\;
  :\defneq
  \;\;
  \left\{\!\!
  \def\arraystretch{1.6}
  \begin{array}{l}
  \left.
  \big(
    \phi : G \to H
  \big)
  \mathclap{\phantom{\vert^{\vert^{\vert^{\vert}}}}}
  \right\}
  \scalebox{.7}{
    \color{darkblue}
    \bf
    \def\arraystretch{.9}
    \begin{tabular}{c}
      map of
      \\
      underlying data
    \end{tabular}
  }
  \\
  \left.
  \begin{array}{l}
  \times \;\;
  \Big(
    \mathrm{unt}
    :
    \Id
    \big(
      \phi(\NeutralElement_G)
      ,\,
      \NeutralElement_H
    \big)
  \Big)
  \\
  \times \;\;
  \Big(
    \mathrm{hom}
    :
    (g_1, g_2 : G)
    \to
    \Id
    \big(
      \,
      \phi(g_1 \cdot_G g_2)
      ,\;
      \phi(g_1) \cdot_H \phi(g_2)
    \big)
    \,
  \Big)
  \end{array}
 \!\!\! \right\}
\!\!  \scalebox{.7}{
    \color{darkblue}
    \bf
    \def\arraystretch{.9}
    \begin{tabular}{c}
      respecting
      \\
      group structure
    \end{tabular}
  }
  \end{array}
  \right.
  \end{array}
\end{equation}
are equivalently pointed maps between the deloopings:
\vspace{-2mm}
\begin{equation}
  \label{DeloopingOfGroupHomomorphisms}
  \begin{tikzcd}[
    sep=-2pt,
    decoration={snake, segment length=4.5pt, amplitude=1pt}
  ]
  \Big(
    \phi \; :
    &
    G
    &
  \overset{
    \mathclap{
      \raisebox{9pt}{
        \scalebox{.7}{
          \color{darkblue}
          \bf
          group homomorphisms
        }
      }
      \;\;\;\;\;\;
    }
  }{
      \underset{\mathrm{hom}}{\longrightarrow}
  }
    &
    H
    &
    \hspace{-8pt}
  \Big)
  \ar[
    rr,
    "{ \sim }"
  ]
  &&
  \big(
    \mathbf{B}\phi
    \;:
    &
    \overset{
      \mathclap{
        \raisebox{6pt}{
          \scalebox{.7}{
            \color{darkblue}
            \bf
            maps of deloopings
          }
        }
      }
    }{
    \mathbf{B}G
    }
    &
    \longrightarrow
    &
    \mathbf{B}H
    &
  \!\!\!\!\!\big)
  \times
  \overset{
    \mathclap{
      \raisebox{5pt}{
        \scalebox{.7}{
          \color{darkblue}
          \bf
          preserving the base point
          \;\;\;\;\;
        }
      }
    }
  }{
  \Id\big(
    (\mathbf{B}\phi)(\mathrm{pt})
    ,\;
    \mathrm{pt}
  \big)
  }
  \\[-3pt]
  &
  g
  &\longmapsto&
  \phi(g)
  &
  &
  \phantom{---}
  &
  &
  T
  &\longmapsto&
  H \otimes_G T
    &
    \raisebox{0pt}{$
     \,\,\, (h \otimes g) \mapsto h\phi(g)
    $}
  \end{tikzcd}
\end{equation}
\noindent
where the tensor product $H \otimes_G T$ is defined as the following quotient set \eqref{QuotientSetAsPushout} with the evident left action by $H$:
\begin{equation}
    H \otimes_G T \;:\defneq\; \frac{H \times T}{(h\phi(g),\, t) \sim (h,\, g \acts\, t)}.
\end{equation}

\newpage

\noindent
{\bf Free groups.}
The perspective on groups as loop group of pointed connected 1-types \eqref{GroupsAsPointedConnectedOneTypes} is suggestive of a slick construction of the {\it free group} on a set of generators (e.g. \cite[\S 1]{Johnson90}) as being simply the loop group \eqref{LoopTypesAsGroupTypes} of the suspension \eqref{Suspension} of the generating set together with a freely adjoined base element $\NeutralElement$ (cf. \cite[\S 6.2]{BBCDG21}):

\vspace{-2mm}
\begin{equation}
  \label{TheFreeGroup}
  \mathrm{Generators} : \Sets
  \;\;\;\;\;\;
  \vdash
  \;\;\;\;\;\;
  \mathrm{FreeGrp}(\mathrm{Generators})
  \;\;
  :\defneq
  \;\;
    \underset{
      \mathclap{
        \raisebox{-4pt}{
          \hspace{4pt}
          \rotatebox{-40}{
            \hspace{-16pt}
            \rlap{
              \color{darkblue}
              \bf
              \scalebox{.7}{
                looping
              }
            }
          }
        }
      }
    }{
  \Omega_{\mathrm{nth}}
  }
  \Big(
    \underset{
      \mathclap{
        \raisebox{-4pt}{
          \hspace{4pt}
          \rotatebox{-40}{
            \hspace{-16pt}
            \rlap{
              \color{darkblue}
              \bf
              \scalebox{.7}{
                suspension
              }
            }
          }
        }
      }
    }{[{
      \mathrm{S}
    }
  \big(
   \{\NeutralElement\} \sqcup \mathrm{Generators}
  \big)]_1}
  \Big)
  \;:\;
  \Groups
\end{equation}
\vspace{+.4cm}

\noindent
Here each $g : \mathrm{Generators}$ may be identified with the loop obtained by concatenating \eqref{ConcatenationOfIdentifications} the corresponding meridian \eqref{Suspension}
with the reverse \eqref{InversionOfIdentifications} of the merididan indexed by the base generator $\NeutralElement$:
\begin{equation}
  \label{LoopsInSuspension}
  \raisebox{-3pt}{
  \scalebox{.7}{
    \color{darkblue}
    \bf
    \def\arraystretch{.9}
    \begin{tabular}{c}
      meridian
      \\
      corresponding
      \\
      to $g : \mathrm{Generators}$
    \end{tabular}
  }
  }
  \hspace{-.3cm}
  \begin{tikzcd}[
    decoration=snake,
    row sep=30pt
  ]
    \mathrm{nth}
    \ar[
      d,
      bend left=60,
      decorate,
      "{
        \mathrm{mer}_{\NeutralElement}
      }"{xshift=3pt},
      end anchor={[yshift=-2pt]}
    ]
    \ar[
      d,
      bend right=60,
      decorate,
      "{
         \mathrm{mer}_{g}
       }"{swap, xshift=-3pt},
      end anchor={[yshift=-2pt]}
    ]
    \\
    \mathrm{sth}
  \end{tikzcd}
  \hspace{-.48cm}
  \raisebox{-4pt}{
  \scalebox{.7}{
    \color{darkblue}
    \bf
    \def\arraystretch{.9}
    \begin{tabular}{c}
      meridian
      \\
      corresponding
      \\
      to base generator $\NeutralElement$
    \end{tabular}
  }}
  \hspace{1cm}
  \scalebox{.7}{
    \color{darkblue}
    \bf
    \def\arraystretch{.9}
    \begin{tabular}{c}
      loop
      \\
      corresponding
      \\
      to $g : \mathrm{Generators}$
    \end{tabular}
  }
  \hspace{-2cm}
  \adjustbox{raise=20pt}{
  \begin{tikzcd}[
    decoration=snake
  ]
    \mathrm{nth}
      \ar[
        out=-180+40,
        in=-40,
        decorate,
        looseness=15,
        "\scalebox{.77}{$\mathclap{
          \mathrm{conc}\big(
    \mathrm{mer}_{g}
    ,\,
    \mathrm{inv}(\mathrm{mer}_{
      \NeutralElement
    })
  \big)
      }$}"{swap, yshift=-2pt},
      shift left=1]
  \end{tikzcd}
  }
  \hspace{-.6cm}
  \scalebox{1}{
    \begin{tabular}{c}
    in
    the suspension
    \\
    $\mathrm{S}(\mathrm{Generators})$
    \end{tabular}
  }
\end{equation}
Hence in a free group \eqref{TheFreeGroup}
we may compose any finite sequence of generators by forming the concatenation \eqref{ConcatenationOfIdentifications} of the corresponding self-equivalences \eqref{LoopsInSuspension}:
$$
  \begin{tikzcd}[
    sep=0pt
  ]
  \overbrace{
      \mathrm{Generators}
        \times \cdots \times
      \mathrm{Generators}
  }^{
    \mathclap{
      \scalebox{.7}{
        $n$ factors
      }
    }
  }
  \ar[rr]
  &&
  \mathrm{FreeGrp}(S)
  \\
  \big(
    g_1, \cdots, g_n
  \big)
  &\quad \longmapsto&
 \qquad  \mathrm{conc}
  \big(
    \mathrm{mer}_{g_1}
    ,\,
    \mathrm{inv}({\mathrm{mer}_{\NeutralElement}})
    ,\,
    \cdots
    ,\,
    \mathrm{mer}_{g_n}
    ,\,
    \mathrm{inv}(\mathrm{mer}_{\NeutralElement})
  \big)
  \,.
  \end{tikzcd}
$$

For example, the free group on a single non-trivial generator, namely the loop type of the suspension of the pointed set with a single non-base element $\mathrm{Generators} :\defneq \big\{1 \big\}$, is the loop type of the circle \eqref{Spheres}, which is equivalently the group of integers \cite[\S 8.1]{UFP13} (these integers being the ``winding numbers'' of loops around the circle):
\begin{equation}
  \label{FreeGroupOnSingleNontrivialGenerator}
  \mathrm{FreeGrp}
  \big(
     \{\NeutralElement\} \sqcup\{
      1
    \}
  \big)
  \;\simeq\;
  \Omega
  \,
  {[\mathrm{S}
  \big(
    \Bits
  \big)]_1}
  \;\simeq\;
  \Omega S^1
  \;\simeq\;
  \mathbb{Z}
  \,.
\end{equation}

\medskip

\noindent
{\bf Presentations of groups.}
\label{PresentationsOfGroups}
Further, the perspective on groups as pointed connected 1-types \eqref{GroupsAsPointedConnectedOneTypes} allows for a slick construction of groups by {\it presentations} in terms of {\it generators and relations} (e.g. \cite{MKS66}\cite{Johnson90}):
Given $\mathrm{Generators} : \Sets$ and
a set $\mathrm{Relations} : \Sets$ of relations
between the elements of the corresponding free group \eqref{TheFreeGroup}
$$
p_0\, p_1 \;:\; \mathrm{Relations} \to
  \mathrm{FreeGrp}(\mathrm{Generators})
  \,,
$$
where we interpret relation $r : \mathrm{Relations}$ as stating that $p_0(r)$ should equal $p_1(r)$,
then
the loop group \eqref{LoopTypesAsGroupTypes}
of the 1-truncation $\Truncation{}{1}{-}$ \eqref{PostnikovTower}
of the CW-complex \eqref{FiniteCWComplex}
whose 1-cells are the $\mathrm{Generators}$
and whose 2-cell attachment \eqref{ParameterizedCellAttachment}
are the pairs of free group elements which are in $\mathrm{Relations}$,
is the group {\it presented} by $\mathrm{Generators}$ and the $\mathrm{Relations}$:
\begin{equation}
  \label{FinitelyPresentedGroupType}
  \mathrlap{
  \hspace{-.2cm}
  \raisebox{40pt}{$
  \mathrm{Gen} : \Sets
  ,\;\;
  \mathrm{Rel} :  \Sets
  ,\;\; p_1,\, p_2 : \mathrm{Rel} \to \mathrm{FreeGroup}(\mathrm{Gen})
  \;\;\;\;\;\;
  \vdash
  \;\;\;\;\;\;
  \mathbf{B}G(\mathrm{Gen}, \mathrm{Rel})
  :\Groups
  $}
  }
  \hspace{1cm}
  \mathbf{B}G(\mathrm{Gen}, \mathrm{Rel})
  \;\;
    :\defneq
  \;\;
  \left[
  \mathrm{po}
  \left(\!\!\!\!\!
  \begin{tikzcd}[
    row sep=6pt,
    column sep=40pt
  ]
    \Big(
        \mathrm{Rel}
    \times
    {S^1}
    \ar[
      rr,
     "{
      \left(
        r
        ,\,
        \mathrm{mer}_i
      \right)
      \;\mapsto\;
      p_i(r)
    }"
    ]
    \ar[
      d,
      shorten=-3pt
    ]
    &&
    \mathrm{S}
    \big(
      \{\NeutralElement\}\sqcup \mathrm{Gen}
    \big)
    \\
    \mathrm{Rel}
    \end{tikzcd}
  \!\!\!\right)
  \right]_1
\end{equation}

The free group $G(\mathrm{Gen},\mathrm{Rel})$ itself may be defined as $\Omega_{\mathrm{nth}}\mathbf{B}G(\mathrm{Gen},\mathrm{Rel})$. By the uniqueness of deloopings (Lem. \ref{DeloopingsAreUnique}), $\mathbf{B}G(\mathrm{Gen},\mathrm{Rel})$ is then equivalent to the type of left $G(\mathrm{Gen},\mathrm{Rel})$-torsors.

\begin{remark}[Comparison to the literature]
\label{2HitsAs1HITS}
A version of this construction \eqref{FinitelyPresentedGroupType}
is briefly indicated in \cite[Ex. 8.7.17]{UFP13}.
Alternatively, it is popular in homotopy type theory (following \cite[Def. 3.1]{LicataFinster14}) to write the intent of
\eqref{FinitelyPresentedGroupType} as a single-step higher-inductive type whose generators include second-order-identifications
(a ``2-HIT'')  instead of an iterated pushout as above. But a general theory and semantics of $n$-HITs for $n \geq 2$ does not
seem to be available in the literature --- except via reformulation as iterated 1-HITs \cite[\S 3.2]{vanDoorn18}, which is what we are using here.

Last but not least, it is in the form of (fundamental groups of loops in) CW-complexes \eqref{FinitelyPresentedGroupType} that
group presentations are known in classical algebraic topology, see \cite[Thm. 7.34]{Rotman88}\cite{MashayekhMirebrahimi10}.
\end{remark}

\medskip

\newpage

\noindent
{\bf Algebra.} Besides this basic group theory, we need some most basic concepts of algebra (e.g. \cite{Knapp06}, cf. \cite[\href{https://1lab.dev/\#algebra}{\S algebra}]{1lab}):

\medskip
A group structure \eqref{TypeOfGroups} is {\it abelian} (e.g. \cite{Fuchs15}) if the group  operation is
{\it commutative} (then written additively as ``+''), i.e. invariant under permutation of the
arguments (cf. \cite[\S 4.12]{BBCDG21}).

\begin{equation}
  \label{StructureOfAbelianGroups}
  \overset{
    \mathclap{
     \raisebox{3pt}{
  \scalebox{.7}{
    \color{purple}
    \bf
    \def\arraystretch{.9}
    \begin{tabular}{c}
      abelian
      \\
      group
    \end{tabular}
  }
     }
    }
  }{
    \AbelianGroups
  }
  \;\;:\defneq\;\;
  \left\{
  \def\arraystretch{1.3}
  \begin{array}{l}
    \big(
      (
        A,\, 0,\, \,+\, ,\, -
      ) : \Groups
    \big)
    \;\;
    \Big\}
    \scalebox{.7}{
      \color{darkblue}
      \bf
      \def\arraystretch{.9}
      \begin{tabular}{l}
        underlying
        \\
        group
      \end{tabular}
    }
    \\
    \times
    \;\;
    (a_1, a_2 : A)
    \to
    \underset{
      \mathclap{
    \scalebox{.7}{
      \color{purple}
      \bf
      \def\arraystretch{.9}
      \begin{tabular}{l}
        abelian
        property
      \end{tabular}
    }
      }
    }{
    \mathrm{Id}_A( a_1 + a_2 ,\, a_2 + a_1 )}
    \Big\}
  \end{array}
  \right.
\end{equation}
\vspace{2pt}

Given $G : \Groups$ we may think ---
in view of Prop. \ref{GSetsAreMapsOutOfBG} and the structure identity principle \eqref{GroupStructureIsomorphism} ---
of functions from $\mathbf{B}G$ to this type
\eqref{StructureOfAbelianGroups}
of abelian groups as
{\it linear actions} of $G$ (i.e., respecting the zero-element and the additive operation):
\vspace{-2mm}
\begin{equation}
  \label{TypeOfLinearGroupActions}
  G\mathrm{LinAction}
  \;:\defneq\;
  (
  \mathbf{B}G
  \to
  \AbelianGroups
  )
\end{equation}

\vspace{-2mm}
A (commutative) {\it ring structure} (e.g. \cite[\S 1]{AndersonFuller92}\cite[Def. 1.1.1]{Bland11}) is an abelian group structure \eqref{StructureOfAbelianGroups} equipped with a further product operation ``distributing'' over the abelian addition (cf. \cite[\S 33.13]{Escardo19} \cite[\href{https://1lab.dev/\#ring-theory}{\S ring-theory}]{1lab},
examples in \cref{KZConnectionsInHomotopyTypeTheory}):
\begin{equation}
  \label{RingStructure}
  \scalebox{.7}{
    \color{purple}
    \bf
    \def\arraystretch{.9}
    \begin{tabular}{c}
      unital
      \\
      ring
    \end{tabular}
  }
  \mathrm{Ring}
  \;:\defneq\;
  \left\{
  \def\arraystretch{1.2}
  \begin{array}{l}
    R : \Sets
    \;\;\;
    \Big\}
    \scalebox{.7}{
      \color{darkblue}
      \bf
      data base
    }
    \\[-5pt]
    \\
    \scalebox{.7}{
      \rotatebox{+90}{\clap{
        \color{orangeii}
        \bf
        abelian group
      }
      }
    }
    \left\{.
    \hspace{-10pt}
    \begin{array}{l}
    \left.
    \def\arraystretch{1.2}
    \begin{array}{l}
      \times
      \;\;
      (0 : R)
      \\
      \times
      \;\;
      (- : R \to R)
      \\
      \times
      \;\;
      (+ : R \times R \to R)
    \end{array}
    \right\}
    \;\scalebox{.7}{
      \color{darkblue}
      \bf
      structure
    }
    \\
    \left.
    \def\arraystretch{1.2}
    \begin{array}{l}
      \times
      \;\;
      (r : R)
        \to
      \mathrm{Id}_R\big(
        0 + r
        ,\,
        r
      \big)
      \\
      \times
      \;\;
      (r : R)
        \to
      \mathrm{Id}_R\big(
        -r + r
        ,\,
        0
      \big)
      \\
      \times
      \;\;
      (r_1, r_2 : R)
        \to
      \mathrm{Id}_R\big(
        r_1 + r_2
        ,\,
        r_2 + r_1
      \big)
      \\
      \times
      \;\;
      (r_1, r_2, r_3 : R)
        \to
      \mathrm{Id}_R\big(
        (r_1 + r_2) + r_3
        ,\,
        r_1 + ( r_2 + r_3)
      \big)
    \end{array}
    \right\}
    \;\scalebox{.7}{
      \color{darkblue}
      \bf
      properties
    }
    \end{array}
    \right.
    \\
    \\
    \scalebox{.7}{
      \rotatebox{+90}{\clap{
        \color{orangeii}
        \bf
        abelian monoid\phantom{g}
      }
      }
    }
    \left\{
    \hspace{-7pt}
    \begin{array}{l}
    \left.
    \begin{array}{l}
      \times
      \;\;
      (1 : R )
      \\
      \times
      \;\;
      (\cdot : R \times R \to R)
    \end{array}
    \right\}
    \scalebox{.7}{
      \color{darkblue}
      \bf
      structure
    }
    \\
    \left.
    \begin{array}{l}
      \times
      \;\;
      (r : R) \to
      \mathrm{Id}_R(1 \cdot r ,\, r)
      \\
      \times
      \;\;
      (r_1, r_2 : R)
        \to
      \mathrm{Id}_R( r_1 \cdot r_2,\, r_2 \cdot r_1 )
      \\
      \times
      \;\;
      (r_1, r_2, r_3 : R)
        \to
      \mathrm{Id}_R\big(
        r_1 \cdot (r_2 \cdot r_3)
        ,\,
        (r_1 \cdot r_2) \cdot r_3
      \big)
    \end{array}
    \right\}
    \scalebox{.7}{
      \color{darkblue}
      \bf
      properties
    }
    \end{array}
    \right.
    \\[-6pt]
    \\
    \hspace{11pt}
    \left.
    \begin{array}{l}
    \left.
    \begin{array}{l}
    \times
    \;\;
    (r, r_1, r_2 : R)
    \to
    \mathrm{Id}_R
    \big(
      r \cdot (r_1 + r_2)
      ,\,
      r \cdot r_1
      +
      r \cdot r_2
    \big)
    \end{array}
    \right.
    \end{array}
    \hspace{-4pt}
    \right\}
    \hspace{-6pt}
    \scalebox{.7}{
      \color{purple}
      \bf
      \def\arraystretch{.7}
      \begin{tabular}{c}
        distributivity
        \\
        property
      \end{tabular}
    }
  \end{array}
  \right.
\end{equation}
Given $R : \Rings$ \eqref{RingStructure},
then {\it $R$-module structure} (e.g. \cite[\S 2]{AndersonFuller92}\cite[Def. 1.4.1]{Bland11}) is given by the following data type:
\begin{equation}
  \label{ModuleDataStructure}
  R : \Rings
  \;\;\;\;
  \vdash
  \;\;
  \scalebox{.8}{
    \bf
    \color{purple}
    \def\arraystretch{.9}
    \begin{tabular}{c}
      $R$-module
    \end{tabular}
  }
  \mathrm{Mod}_R
  \;\;:\defneq\;
  \left\{
  \def\arraystretch{1.4}
  \begin{array}{l}
  \big( N : \Sets \big)
  \hspace{1cm}
  \Big\}
  \scalebox{.8}{
    \color{darkblue}
    \bf
    data
  }
  \\[-12pt]
  \\
  \scalebox{.8}{
    \color{orangeii}
    \bf
    \rotatebox{+90}{\clap{
      abelian group
    }
    }
  }
  \left\{
  \hspace{-12pt}
  \begin{array}{l}
  \begin{array}{l}
  \left.
  \def\arraystretch{1.4}
  \begin{array}{l}
    \times
    \;\;
    \big(0 : N \big)
    \\
    \times
    \;\;
    \big( - : N \to N\big)
    \\
    \times
    \;\;
    \big(+ : N \times N \to N \big)
  \end{array}
  \hspace{1cm}
  \right\}
  \scalebox{.8}{
    \color{darkblue}
    \bf
      structure
  }
  \\
  \left.
  \def\arraystretch{1.4}
  \begin{array}{l}
     \times
     \;\;
     \dprod{n : N}
     \Id_N(0 + n,\, n)
     \\
     \times
     \;\;
     \dprod{n : N}
     \Id_N(-n + n,\, 0)
     \\
     \times
     \;\;
     \dprod{n_1, n_2 : N}
     \Id_N(n_1 + n_2, n_2 + n_1)
     \\
     \times
     \;\;
     \dprod{n_1,n_2,n_3 : N}
     \Id_N\big(
       (n_1 +n_2) +n_3
       ,\,
       n_1 + (n_2 + n_3)
     \big)
  \end{array}
  \right\}
  \scalebox{.8}{
    \color{darkblue}
    \bf
    \rotatebox{-90}{\clap{properties}}
  }
  \end{array}
  \end{array}
  \right.
  \\[-6pt]
  \\
  \scalebox{.8}{
    \bf
    \color{orangeii}
    \rotatebox{+90}{\clap{linear $R$-action}}
  }
  \left\{
  \begin{array}{l}
    \begin{array}{l}
    \times
    \;\;
    \big(
    \cdot \,:\, R \times N  \to N
    \big)
    \hspace{1.5cm}
    \Big\}
    \scalebox{.8}{
        \color{darkblue}
        \bf
        structure
    }
    \end{array}
     \\
     \left.
    \begin{array}{l}
    \times
    \;\;
    \dprod{n : N}
    \Id_N(0 \cdot n,\, 0)
    \\
    \times
    \;\;
    \dprod{n : N}
    \Id_N(1 \cdot n,\, n)
    \\
    \times
    \;\;
    \dprod{c_1, c_2 : R, n:
        N}
    \Id_N\big(
      c_1 \cdot (c_2 \cdot n)
      ,\,
      (c_1 \cdot c_2) \cdot n
    \big)
    \\
    \times
    \;\;
    \dprod{c : R,\, n_1, n_2 : N}
    \Id_N\big(
      c \cdot (n_1 + n_2)
      ,\,
      c \cdot n_1 + c \cdot n_2
    \big)
    \\
    \times
    \;\;
    \dprod{c_1, c_2 : R,\, n : N}
    \Id_N\big(
      (c_1 + c_2)n
      ,\,
      c_1 \cdot n + c_2 \cdot n
    \big)
    \end{array}
    \hspace{.2cm}
    \right\}
    \scalebox{.8}{
      \rotatebox{-90}{
        \clap{
          \color{darkblue}
          \bf
          \hspace{-8pt}properties
        }
      }
    }
  \end{array}
  \right.
\end{array}
\right.
\end{equation}

Here we will be concerned with those $R$-modules that are $R$-lines \eqref{TypeOfRLines}:

\medskip

\noindent
{\bf Linear action of groups of units.}
Given a ring \eqref{RingStructure}
we obtain:
\begin{itemize}[leftmargin=.4cm]
\setlength\itemsep{-3pt}
\item
the  {\bf underlying abelian group}
 \eqref{StructureOfAbelianGroups} whose group operation is the addition operation of the ring
 and which the remaining monoid structure makes into an $R$-module \eqref{ModuleDataStructure}, which we will denote by the same symbols:
\begin{equation}
  \label{UnderlyingAbelianGroupOfType}
  \begin{tikzcd}[
    sep = -1pt,
  ]
    (-)_{\mathrm{udl}}
    \;:
    &
    \Rings
    \ar[rr]
    &&
    R \Modules
    \ar[rr]
    &&
    \AbelianGroups
    \\
    &
    (R ,\, 0 ,\, - ,\, + ,\, 1 ,\, \cdot)
    &\longmapsto&
    \Big(
      R ,\,
      \big(0 ,\, - ,\, +, \cdots\big)
      ,\,
      \big(\cdot, \cdots\big)
    \Big)
    &\longmapsto&
    (R ,\, 0 ,\, - ,\, + )
    \mathrlap{\,,}
  \end{tikzcd}
\end{equation}

\item the  {\bf group of units}\footnote{
A group of units need not be abelian if the corresponding ring is non-commutative. However, below in \cref{KZConnectionsInHomotopyTypeTheory} we specialize to the ring of complex numbers, whose group of units is, of course,  abelian.},
namely of multiplicatively invertible elements in the ring
(e.g. \cite[p. 143]{Knapp06})
whose group operation is the multiplication operation of the ring:
\vspace{-2mm}
\begin{equation}
  \label{GroupOfUnits}
  \begin{tikzcd}[
    sep=-1pt
  ]
    (-)^\times
    :
    &
    \Rings
    \ar[rr]
    &&
    \Groups
    \\
    &
    \big(
      R
      ,\,
      0
      ,\,
      -
      ,\,
      +
      ,\,
      1
      ,\,
      \cdot
    \big)
    &\longmapsto&
    \left(\!\!\!\!
      \def\arraystretch{.9}
      \begin{array}{l}
        (r, r^{-1} : R)
        \\
        \times
        \;
        \Id_R( r \cdot r^{-1},\, 1 )
      \end{array}
      ,\,
      (1,1) ,\, (\cdot, \cdot) ,\, (-)^{-1}
    \right)
    \mathrlap{\,.}
  \end{tikzcd}
\end{equation}

\item the {\bf $R$-lines} over a ring $R$ (e.g. \cite{ABGHR14}), are those $R$-modules \eqref{ModuleDataStructure} which are isomorphic to $R_{\mathrm{udl}} : R \Modules$ \eqref{UnderlyingAbelianGroupOfType} itself (also called the ``free cyclic $R$-modules''):
\begin{equation}
  \label{TypeOfRLines}
  R : \Rings
  \;\;\;\;\;\;
    \vdash
  \;\;\;\;\;\;
  R \Lines
  \;:\defneq\;
  \Big(
  \big\{
    N \,:\, R\Modules
    \,\big\vert\;
    \exists
    (N \simeq R_{\mathrm{udl}})
  \big\}
  \,,\;
  \mathrm{pt}
    :\defneq
  R
  \Big)
  \;\;
  :
  \;\;
  \PointedConnectedOneTypes
  \,.
\end{equation}
Noticing the close analogy of the lines over a ring to torsors over a group, essentially the same proof as of Lem. \ref{TorsorsDeloopAGroup} shows that the type of $R$-lines deloops the group of units \eqref{GroupOfUnits}. By Thm. \ref{DeloopingsAreUnique}, we therefore have a pointed equivalence:
\begin{equation}
  \label{TypeOfRLinesDeloopsGroupOfUnits}
  \begin{tikzcd}[
    sep=0pt
  ]
    \mathllap{
      R : \Rings
      \;\;\;\;\;\;\;
        \vdash
      \;\;\;\;\;\;\;
    }
    R \Lines
    \ar[rr, "{ \sim }"]
    &&
    R^{\times}\mathrm{Tors}_L
    \\
    L
    &\longmapsto&
    \Id_{R \Lines}(R, L)
    \mathrlap{\,.}
  \end{tikzcd}
\end{equation}

For this reason, we are justified in defining $\mathbf{B}R^{\times} :\defneq R \Lines$ to be the type of $R$-lines.\\

\item
the {\bf action of the group of units on the underlying abelian group}:
Via the ring multiplication \eqref{RingStructure}, the group of units \eqref{GroupOfUnits} acts \eqref{AlgebraicGActions}
on the underlying abelian group
\eqref{UnderlyingAbelianGroupOfType}
and the distributivity property makes this a linear action \eqref{TypeOfLinearGroupActions}:
$$
  R : \Rings
  \;\;\;\;\;\;
    \vdash
  \;\;\;\;\;\;
  R^\times \acts R_{\mathrm{udl}}
  \;:\defneq\;
  \big(
    R_{\mathrm{udl}}
    ,\;
    \mathrm{act} :\defneq \cdot
    \;,\,
    \cdots
  \big)
  \,:\,
  R^\times \mathrm{LinAct}_L
  \,.
$$
But under Prop. \ref{GSetsAreMapsOutOfBG} and with \eqref{TypeOfRLinesDeloopsGroupOfUnits} we may equivalently express this more elegantly under delooping, simply as:
\begin{equation}
  \label{GroupOfUnitsActingOnUnderlyingAbelianGroup}
  \hspace{-1cm}
  \begin{tikzcd}[
    sep=-2pt
  ]
  R : \Rings
  &
  \;\;\;\;\;\;
  \vdash
  \;\;\;\;\;\;
  &
  R^\times \! \acts \, R_{\mathrm{udl}}
  &:&
  \mathbf{B} R^\times
  \ar[rr]
  &&
  \AbelianGroups
  \\
  && &&
  {\big(N,\, (0,\, +,\, -),\, \cdot \big)}
  &\longmapsto&
  \big(N,\, (0,\,+,\,-)\big)
  \mathrlap{\,.}
  \end{tikzcd}
\end{equation}

\end{itemize}

\medskip

\noindent
{\bf Iterated delooping and Eilenberg-MacLane types.}
\label{IteratedDelooping}
In evident generalization of the notion of delooping of groups
(Prop. \ref{TheLoopingDeloopingEquivalence})
the  {\it delooping of a pointed type}
is (cf. \cite[\S 2]{Warn22}\cite{BCTFR23})
another pointed {\it and connected} type whose loop type \eqref{LoopSpace} recovers the given pointed type, up to specified equivalence \eqref{HomotopyEquivalenceOfTypes}:
\begin{equation}
  \label{PointedDeloopingStructure}
  \mathcal{A} : \Types
  ,\,
  a : \mathcal{A}
  \;\;\;\;\;\;\;\;\;
    \vdash
  \;\;\;\;\;\;\;\;\;
  \overset{
    \mathclap{
      \raisebox{4pt}{
        \scalebox{.7}{
          \color{purple}
          \bf
          delooping of $(\mathcal{A},a)$
        }
      }
    }
  }{
    \mathrm{DeLpg}(\mathcal{A}, a)
  }
  \;:\defneq\;
  \left\{
  \def\arraystretch{1.6}
  \begin{array}{l}
  \left.
  \big(
    \mathbf{B}\mathcal{A}
    :
    \Types
  \big)
  \mathclap{\phantom{\vert^{\vert^{\vert^{\vert}}}}}
  \right\}
  \scalebox{.7}{
    \color{darkblue}
    \bf
    higher data base
  }
  \\
  \left.
  \begin{array}{l}
  \times
  \;\;
  (\mathrm{pt} : \mathbf{B} \mathcal{A})
  \\
  \times
  \;\;
  \big(
    \mathrm{equ}
    :
    \Omega_{\mathrm{pt}} \mathbf{B}\mathcal{A}
    \overset{\sim}{\to}
    \mathcal{A}
  \big)
  \end{array}
  \right\}
  \scalebox{.7}{
    \color{darkblue}
    \bf
    higher structure
  }
  \\
  \left.
  \begin{array}{l}
  \times
  \;\;
  \big(
    \mathrm{cnd}
    :
    \exists !
    \Truncation{\big}{0}{\mathbf{B}\mathcal{A}}
  \big)
  \\[+2pt]
  \times
  \;
  \Big(
    \mathrm{ptd}
    :
    \Id
    \big(
      \mathrm{equ}(
        \mathrm{id}_{\mathrm{pt}}
      )
      ,\,
      a
    \big)
  \!\Big)
  \end{array}
  \right\}
  \scalebox{.7}{
    \color{darkblue}
    \bf
    properties
  }
  \end{array}
  \right.
\end{equation}
Since the delooping of a pointed type has itself an underlying pointed type
$$
  \begin{tikzcd}[
    sep = 0pt
  ]
    \mathcal{A} : \Types
    ,\,
    a : \mathcal{A}
    \;\;\;\;\;\;
      \vdash
    \;\;\;\;\;\;
    &
    \mathrm{DeLpg}(\mathcal{A},a)
    \ar[rr]
    &&
    \big(
      \mathbf{B}\mathcal{A}
      :
      \Types
    \big)
    \times
    (\mathrm{pt} : \mathbf{B}\mathcal{A})
    \\
    &
   \!\!\!\!\! \big(
      \mathbf{B}\mathcal{A}
      ,\,
      \mathrm{pt}
      ,\,
      \mathrm{equ}
      ,\,
      \mathrm{cnd}
      ,\,
      \mathrm{ptd}
    \big)
    &\longmapsto&
    \big(
      \mathbf{B}\mathcal{A}
      ,\,
      \mathrm{pt}
    \big)
  \end{tikzcd}
$$
it makes sense to ask for iterated deloopings:
\begin{equation}
  \label{IteratedDeloopingStructure}
  \mathcal{A} : \Types
  ,\,
  a : \mathcal{A}
  ,\,
  n : \mathbb{N}
  \;\;\;\;\;\;\;\;\;
    \vdash
  \;\;\;\;\;\;\;\;\;
  \overset{
    \mathclap{
      \raisebox{6pt}{
        \scalebox{.7}{
          \color{purple}
          \bf
          \def\arraystretch{.9}
          \begin{tabular}{c}
            $n$-fold
            \\
            delooping of $(\mathcal{A},a)$
          \end{tabular}
        }
      }
    }
  }{
    {n}\mathrm{DeLpg}(\mathcal{A}, a)
  }
  \;:\defneq\;
  \left\{
  \def\arraystretch{1.6}
  \begin{array}{l}
  \left.
  \big(
    \mathbf{B}^n \mathcal{A}
    :
    \Types
  \big)
  \mathclap{\phantom{\vert^{\vert^{\vert^{\vert}}}}}
  \right\}
  \scalebox{.7}{
    \color{darkblue}
    \bf
    higher data base
  }
  \\
  \left.
  \begin{array}{l}
  \times
  \;\;
  (\mathrm{pt} : \mathbf{B}^n \mathcal{A})
  \\
  \times
  \;\;
  \big(
    \mathrm{equ}
    :
    \Omega_{\mathrm{pt}}^n \mathbf{B}^n\mathcal{A}
    \overset{\sim}{\to}
    \mathcal{A}
  \big)
  \end{array}
  \right\}
  \scalebox{.7}{
    \color{darkblue}
    \bf
    higher structure
  }
  \\
  \left.
  \begin{array}{l}
  \times
  \;\;
  \big(
    \mathrm{cnt}
    :
    \exists !
    \Truncation{\big}{0}{\mathbf{B}^n\mathcal{A}}
  \big)
  \\[+2pt]
  \times
  \;
  \Big(
    \mathrm{ptd}
    :
    \Id
    \big(
      \mathrm{equ}(
        \mathrm{id}_{\mathrm{pt}}
      )
      ,\,
      a
    \big)
 \! \Big)
  \end{array}
  \right\}
  \scalebox{.7}{
    \color{darkblue}
    \bf
    properties
  }
  \end{array}
  \right.
\end{equation}
Here $\Omega^n_{\mathrm{pt}}$ denotes the $n$-fold iteration of the looping operation \eqref{LoopSpace} regarded as an endo-function on pointed types.

In the denotational semantics of homotopy theory one refers to {\it iterated loop space} structure:\footnote{
  Here we present iterated (de/)looping as extra structure on $\mathbf{B}^n \mathcal{A}$; alternatively one could rewrite these definitions to regard it as extra structure on $\mathcal{A}$. The distinction, which traditional literature glosses over anyway, is a matter of convention rather than of practical content.
} \cite{May72}\cite{Segal73}\cite[\S 5.2.6]{LurieHigherAlgebra}. If a tower of $n$-fold deloopings are compatibly given for all values of $n : \mathbb{N}$ one speaks of a connective {\it spectrum} of spaces exhibiting  {\it infinite loop space} structure \cite{May77}\cite{Adams78}\cite[\S 1.4]{LurieHigherAlgebra} (type-theoretic discussion includes \cite[\S 3.2]{Cavallo15}\cite[\S 5.3]{vanDoorn18}). This is the structure that leads over to the notion of {\it linear} homotopy types, which we turn to in the companion article \cite{QPinLHOTT}.
For example:

\begin{proposition}[{\cite[p. 3 \& Thm. 5.4]{LicataFinster14}}]
The $(n+1)$-fold iterated delooping
\eqref{IteratedDeloopingStructure}
of $A : \AbelianGroups$ \eqref{StructureOfAbelianGroups}
can be constructed
as the $n+1$-truncation
(Def. \ref{TruncationDefn}) of
the $n$-fold iterated suspension \eqref{Suspension}
of the first delooping \eqref{LoopingDeloopingEquivalence}:
\begin{equation}
  \label{HigherDeloopingByHigherSuspension}
  A
  :
  \AbelianGroups
  ,\;\;
  n : \mathbb{N}
  \;\;\;\;\;\;
  \vdash
  \;\;\;\;\;\;
  \big(
    \mathbf{B}^{n+1}A
    \;:\defneq\;
    \Truncation{\big}{n+1}{
      \Sigma^n \mathbf{B} A
    }
    ,\,
    \mathrm{nth}
    ,\,
    \mathrm{equ}
  \big)
  \,:\,
  n\mathrm{DeLpg}(A,\NeutralElement)
\end{equation}
\end{proposition}

In the denotational semantics of homotopy theory, these iterated deloopings of abelian groups are known as {\it Eilenberg-MacLane spaces} (Lit. \ref{LiteratureEilenbergMacLanSpaces}) which are the classifying types/spaces for ordinary cohomology --- see \eqref{TypeTheoreticOrdinaryCohomology} below.

\medskip

\noindent
{\bf Twisted higher deloopings and associated bundles of EM-Types.}
A key point for the definition of twisted cohomology below \eqref{TwistedCohomologySecondVersion} is that type-theoretic constructions such as the higher delooping in \eqref{HigherDeloopingByHigherSuspension} apply in the generality where the data involved may {\it depend} on any given context: Specifically, placing a delooping type $\mathbf{B}G$ \eqref{LoopingDeloopingEquivalence} into the context immediatley gives a notion of higher delooping types that is coherently acted on \eqref{GroupInfinityAction}
by a given group $G$:

\begin{definition}[$G$-Equivariant higher deloopings]
Given $R : \Rings$ \eqref{RingStructure}
---
considered with the action
$R^\times \! \acts \, R_{\mathrm{udl}}$
\eqref{GroupOfUnitsActingOnUnderlyingAbelianGroup}
of $R^\times$ \eqref{GroupOfUnits}
on its underlying abelian group
$R_{\mathrm{udl}}$ \eqref{UnderlyingAbelianGroupOfType}
---,
we obtain twisted higher deloopings of the underlying abelian groups
of rings by applying \eqref{HigherDeloopingByHigherSuspension} in the context of $\mathbf{B}G R^\times$:

\begin{equation}
  \label{GEquivariantHigherDelooping}
  \begin{tikzcd}[
    sep = 0
  ]
  R : \Rings
  ,\;\;
  n : \mathbb{N}
  &
  \;\;\;\;\;
  \vdash
  \;\;\;\;\;
  &
  R^\times \!\acts \; \mathbf{B}^{n} R_{\mathrm{udl}}
  &:&
  \mathbf{B}G
  &
  \longrightarrow
  &
  \Types
  \\
  &&&&
  t
    &\longmapsto&
    \underset{
      \mathclap{
        \raisebox{-3pt}{
          \scalebox{.7}{
            \color{gray}
            \eqref{HigherDeloopingByHigherSuspension}
          }
        }
      }
    }{
    \mathbf{B}^{n}
  }
  (
    \underset{
      \mathclap{
        \raisebox{-3pt}{
          \scalebox{.7}{
            \color{gray}
            \eqref{NotationForGAction}
          }
        }
      }
    }{
      \acts_{\, {t}} \, R_{\mathrm{udl}}
    }
  )
  \end{tikzcd}
\end{equation}
\end{definition}
The semantics of this $\mathbf{B} R^\times$-dependent type
\eqref{GEquivariantHigherDelooping}
is given, as a special case of \eqref{GroupInfinityAction}, by bundles of Eilenberg-MacLane spaces associated to the universal $R^\times$-principal bundle:

\begin{equation}
\label{}
\adjustbox{}{
\def\arraystretch{1.6}
\begin{tabular}{|c|c|}
\hline
{\bf $\mathbf{B} R^\times$-dependent higher $R$-delooping}
&
{\bf Universal $R^\times$-Associated $K(R_{\mathrm{udl}},n)$-Bundle}
\\
\hline
\hline
\begin{minipage}{4cm}
$$
  t
  :
  \mathbf{B} R^\times
  \;\;\;\;
  \vdash
  \;\;\;
  \acts_{\,{t}} \, \mathbf{B}^n R_{\mathrm{udl}}
  :
  \Types
$$
\end{minipage}
&
\begin{minipage}{7cm}
$$
\begin{tikzcd}[column sep=large]
  \mathbf{B}^n R
  \ar[
    d,
    "{\ }"{name=t}
  ]
  \ar[
    r,
    "{\ }"{swap, name=s}
  ]
  \ar[
    from=s,
    to=t,
    Rightarrow,
    "{ \scalebox{.7}{(pb)} }"
  ]
  &
  \HomotopyQuotient
    { \mathbf{B}^n R_{\mathrm{udl}} }
    { R^\times }
  \ar[
    d,
    "{\  }"{name=t2}
  ]
  \ar[
    r,
    "{\ }"{swap, name=s2}
  ]
  \ar[
    from=s2,
    to=t2,
    Rightarrow,
    "{ \scalebox{.7}{(pb)} }"{}
  ]
  &
  \widehat{\Objects}
  \ar[d]
  \\
  \ast
  \ar[
    r,
    "{ \mathrm{pt} }"
  ]
  &
  \mathbf{B} R^\times
  \ar[
    r,
    "{
      R^\times
        \!\acts \,
      \mathbf{B}^n R_{\mathrm{udl}}  }"
  ]
  &
  \Objects
\end{tikzcd}
$$
\end{minipage}
\\
\hline
\end{tabular}
}
\end{equation}

\medskip

With these preliminaries in hand, we come to the main type construction of this section, in Def. \ref{TypeTheoreticGaussManinConnection} below.

\newpage

\noindent
{\bf Twisted cohomology and Gauss-Manin connections.}
Given $R : \Rings$ \eqref{RingStructure} and $n : \mathbb{N}$ \eqref{NaturalNumberstype}, the {\bf ordinary cohomology} with coefficients $R$ in degree $n$ is (specified by) the 0-truncation $\Truncation{}{0}{-}$ \eqref{PostnikovTower} of the type of functions \eqref{DependentFunctionNotation} into the $n$-fold delooping $\mathbf{B}^n(-)$ \eqref{HigherDeloopingByHigherSuspension} of the underlying abelian group \eqref{UnderlyingAbelianGroupOfType} of $R$ (cf. \cite[\S 3. 2]{Cavallo15}\cite[\S 4.1]{Warn22} with implementation in {\tt Agda}: \cite{BLM22}):
\begin{equation}
  \label{TypeTheoreticOrdinaryCohomology}
  \begin{tikzcd}
  \left.
  \def\arraystretch{1.3}
  \begin{array}{l}
  \overset{
    \mathclap{
    \raisebox{4pt}{
    \scalebox{.7}{
      \color{darkblue}
      \bf
      coefficients
      \;\;\;
    }
    }
    }
  }{
  R : \Rings
  ,\;\;
  }
  \overset{
    \mathclap{
      \raisebox{+4pt}{
        \scalebox{.7}{
          \color{darkblue}
          \bf
          \;\;degree
        }
      }
    }
  }{
  n : \mathbb{N}
  }
  ,
  \\
  \underset{
    \mathclap{
      \raisebox{-3pt}{
        \scalebox{.7}{
          \color{darkblue}
          \bf
          domain
        }
      }
    }
  }{
  X : \Types
  }
  \end{array}
  \right\}
  \;\;\;\;\;\;\;
  \vdash
  \;\;\;\;\;\;\;
  \overset{
    \mathclap{
      \raisebox{6pt}{
        \scalebox{.7}{
          \color{purple}
          \bf
          \def\arraystretch{.9}
          \begin{tabular}{c}
            ordinary
            \\
            cohomology
          \end{tabular}
        }
      }
    }
  }{
  H^n\big(
    X
    ;\,
    R
  \big)
  }
  \;:\defneq\;
  \Truncation{\big}{0}{
    X \to \mathbf{B}^n R_{\mathrm{udl}}
  }
  \;:\;
  \Types
  \end{tikzcd}
\end{equation}
Under the dictionary of \cref{HoTTIdea}, this type construction clearly interprets as the traditional notion of ordinary
cohomology (Lit. \ref{LiteratureCohomology}), cf. discussion and references in \cite[Ex. 1.0.2]{FSS20Character}.

\medskip

Given furthermore a {\it twist} $\tau : X \to \mathbf{B} R^\times$ \eqref{GroupOfUnits}, we say that the $\tau$-{\bf twisted ordinary cohomology} of $X$ with coefficients in
$R$ is (for the second version \eqref{TwistedCohomologySecondVersion} cf. \cite[Def. 5.4.2]{vanDoorn18}):
\begin{align}
  \label{TwistedCohomologyFirstVersion}
  \hspace{-1.5cm}
  \begin{array}{l}
  \left.
  \def\arraystretch{1.3}
  \begin{array}{l}
  \overset{
    \mathclap{
    \raisebox{3pt}{
    \scalebox{.7}{
      \color{darkblue}
      \bf
      coefficients
    }
    }
    }
  }{
  R : \Rings
  }
  ,\;\;
  \overset{
    \mathclap{
      \raisebox{+4pt}{
        \scalebox{.7}{
          \color{darkblue}
          \bf
          degree
        }
      }
    }
  }{
  n : \mathbb{N}
  }
  ,
  \\
  \underset{
    \mathclap{
      \raisebox{-2pt}{
        \scalebox{.7}{
          \color{darkblue}
          \bf
          domain
        }
      }
    }
  }{
  X : \Types
  }
  ,\;\;
  \underset{
    \mathllap{
      \raisebox{-4pt}{
        \scalebox{.7}{
          \color{orangeii}
          \bf
          twist
        }
      }
    }
  }{
    \tau : X \to \mathbf{B} R^\times
  }
  \end{array}
  \!\!\!\right\}
  \qquad
  \vdash
  \qquad
  \overset{
    \mathclap{
      \raisebox{+6pt}{
        \scalebox{.7}{
          \bf
          \def\arraystretch{.9}
          \begin{tabular}{c}
            \color{purple}
            twisted
            cohomology
          \end{tabular}
        }
      }
    }
  }{
  H^{n + {\color{orangeii}\tau}}
  (X;\, R)
  }
  \;:\defneq\;
  \Truncation{\bigg}{0}{
    (t : \mathbf{B} R^\times)
    \to
    \Big(
    \overset{
      \mathclap{
        \raisebox{3pt}{
          \scalebox{.7}{
            \color{gray}
            \eqref{HomotopyFiber}
          }
        }
      }
    }{
      \mathrm{fib}_t({\color{orangeii}\tau})
    }
    \to
    \overset{
      \mathclap{
        \raisebox{3pt}{
          \scalebox{.7}{
            \color{gray}
            \eqref{GEquivariantHigherDelooping}
          }
        }
      }
    }{
    \mathbf{B}^n
    (
      \acts_{\, {t}} \, R
    )
    }
    \Big)
  }
  \;\;
  :
  \;\;
  \Types
  \end{array}
  \\[-10pt]
  \label{TwistedCohomologySecondVersion}
 \phantom{aaaaaaa} \hspace{5cm}
   \simeq
  \Truncation{\bigg}{0}{
    \underbrace{
    (t : \mathbf{B} R^\times) \times \big(
      x_t
      :
      \mathrm{fib}_t({\color{orangeii}\tau}))
    \big)
    }_{
      (x \,:\, X)
      \mathrlap{
        \;\;
        \scalebox{.7}{
          \color{gray}
          \eqref{TypeClassification}
        }
      }
    }
   \to
    \underbrace{
    \mathbf{B}^n\big(
      \acts_{\, {t}} \, R
    \big)
    }_{
      \mathclap{
      \scalebox{.7}{$
      \big(
      \tau^\ast
      \mathbf{B}(
        R^\times \!\acts \, R
      )
      \big)(x)
      $}
      }
    }
    }
  :
  \Types
\end{align}

Under the dictionary of \cref{HoTTIdea}, this type construction interprets as the traditional notion of twisted ordinary cohomology (Lit. \ref{LiteratureCohomology}), see the discussion and references in \cite[Rem. 4.22]{NSS12}
and \cite[Rem. 2.94]{SS20OrbifoldCohomology}\cite[Ex. 2.0.5]{FSS20Character}:

\begin{equation}
  \label{TwistedCohomologyDictionary}
\adjustbox{}{
\def\arraystretch{1.5}
\begin{tabular}{|c|c|}
 \hline
  \multicolumn{2}{|l|}{
    \bf Twisted ordinary cohomology
    \;
    $
      H^{n + \tau}
      \big(
        X;\,
        R
      \big)
    $
  }
  \\
  \hline
  {\bf Homotopy type theory}
  &
  {\bf Homotopy theory}
  \\
  \hline
  \hline
  \begin{minipage}{7cm}
    $
      \Truncation{\Big}{0}{
        (t : \mathbf{B} R^\times)
        \to
        \Big(
          \mathrm{fib}_t(\tau)
          \to
          \mathbf{B}^n\big(
            \acts_{{}_{t}} \, R_{\mathrm{uld}}
          \big)
        \Big)
      }
    $
    \hfill
    \eqref{TwistedCohomologyFirstVersion}
  \end{minipage}
  &
  \begin{minipage}{6cm}
    $$
    \begin{array}{c}
     \left\{\!\!\!\!
     \adjustbox{raise=4pt}{
     $
      \begin{tikzcd}[
        column sep=33pt
      ]
        &&
        \HomotopyQuotient
          { \mathbf{B}^n R_{\mathrm{udl}} }
          { R^\times }
        \ar[d]
        \\
        X
        \ar[
          rr,
          "{ \tau }"
        ]
        \ar[
          urr,
          dashed,
        ]
        &&
        \mathbf{B} R^\times
      \end{tikzcd}
      $
      }
    \!\!\!\!\!  \right\}_{\!\!\!\big/\sim_{\mathrm{hmt}}}
      \\[-10pt]
      {}
      \end{array}
    $$
  \end{minipage}
  \\
  \hline
  \begin{minipage}{6.5cm}
    $
    \Truncation{\Big}{0}{
      (x : X)
      \to
      \big(
        \tau^\ast
        \mathbf{B}
        (R^\times \acts \, R_{\mathrm{uld}})
      \big)
      (x)
    }
    $
    \hfill
    \eqref{TwistedCohomologySecondVersion}
  \end{minipage}
  &
  \begin{minipage}{7cm}
    $$
    \begin{array}{c}
     \left\{\!\!\!\!
     \adjustbox{raise=4pt}{
     $
      \begin{tikzcd}[
        column sep=7pt
      ]
        &
        \tau^\ast
        \big(
          \HomotopyQuotient
            { \mathbf{B}^n R_{\mathrm{udl}} }
            { R^\times }
        \big)
        \ar[
          r,
          shorten=-2pt
        ]
        \ar[d]
        &
        \HomotopyQuotient
          { \mathbf{B}^n R_{\mathrm{uld}} }
          { R^\times }
        \ar[d]
        \\
        X
        \ar[
          r,
          equals
        ]
        \ar[
          ur,
          dashed
        ]
        &
        X
        \ar[
          r,
          "{ \tau }"
        ]
        &
        \mathbf{B} R^\times
      \end{tikzcd}
      $
      }
    \!\!\!\!\!\!  \right\}_{\!\!\!\big/\sim_{\mathrm{hmt}}}
      \\[-10pt]
      {}
      \end{array}
    $$
  \end{minipage}
  \\
  \hline
\end{tabular}
}
\end{equation}

For the construction of Gauss-Manin connections from \cref{GaussManinForTwistedGeneralizedCohomology},
we are to consider $B$-indexed families of such twisted cohomology types \eqref{TwistedCohomologyDictionary}. Remarkably, in the syntax of homotopy data types this is a triviality that amounts to introducing a type $B$ into the context and having the terms $X$ and $\tau$ depend on it:

\begin{definition}[Gauss-Manin transport on fibrations of twisted cohomology groups]
\label{TypeTheoreticGaussManinConnection}
We say that the data type of {\it fibrations of twisted ordinary cohomology sets} is:
\vspace{-2mm}
\begin{equation}
  \label{TypeTheoreticGausManinFibration}
  \hspace{-6mm}
  \begin{array}{l}
  \left.
  \def\arraystretch{1.3}
  \begin{array}{l}
  \overset{
    \mathclap{
    \raisebox{3pt}{
    \scalebox{.7}{
      \color{darkblue}
      \bf
      coefficients
    }
    }
    }
  }{
  R : \Rings
  }
  ,\;\;
  \overset{
    \mathclap{
      \raisebox{+2pt}{
        \scalebox{.7}{
          \color{darkblue}
          \bf
          degree
        }
      }
    }
  }{
  n : \mathbb{N}
  }
  ,\,
    \overset{
      \mathclap{
      \raisebox{+3pt}{
       \scalebox{.7}{
         \;\;
         \color{orangeii}
          \bf
          \def\arraystretch{.9}
          \begin{tabular}{c}
            parameter
            base
           \end{tabular}
         }
      }
      }
    }
    {
      B : \Types
      ,\;\;
    }
  \\
  \underset{
    \mathclap{
      \raisebox{-3pt}{
        \scalebox{.7}{
          \color{darkblue}
          \bf
          {\color{orangeii}fibration of} domains
        }
      }
    }
  }{
  X_{(-)} : B \to \Types
  }
  ,\;\;
  \underset{
    \mathllap{
      \raisebox{-2pt}{
        \scalebox{.7}{
          \color{darkblue}
          \bf
          {\color{orangeii}family of} twists
        }
      }
    }
  }{
    \tau_{(-)}
      :
    (b : B)
      \to
    \big(
      X_b \to \mathbf{B} R^\times
    \big)
  }
  \end{array}
  \!\!\!\!\!\right\}
  \;\;
  \vdash
  \;\;
  \left.\!\!\!
  \def\arraystretch{.9}
  \begin{array}{l}
  \overset{
    \mathclap{
      \raisebox{+6pt}{
        \scalebox{.7}{
          \bf
          \def\arraystretch{.9}
          \begin{tabular}{c}
            \color{purple}
            twisted
            cohomology
          \end{tabular}
        }
      }
    }
  }{
  H^{n + \tau_{\color{orangeii}(-)}}
    \big(
      X_{\color{orangeii}(-)}
      ;\,
      R
    \big)
  }
  \;:\defneq\;
  \\
  \Truncation{\bigg}{0}{
    (t : \mathbf{B}R^\times)
    \to
    \Big(
    \underbrace{
      \mathrm{fib}_t
      \big(\tau_{\color{orangeii}(-)}\big)
    }_{
      \mathclap{
        \mathrm{fib}_{(-,\,t)}
        (
          \mathrm{pr}_X
            ,\,
          \tau
        )
        \mathrlap{
          \;
          \scalebox{.7}{
            \color{gray}
            \eqref{FibersOfGlobalTwistPairedWithDomainProjection}
          }
        }
      }
    }
    \to
    \mathbf{B}^n
    (
      \acts_{\, {t}} \, R_{\mathrm{udl}}
    )
    \Big)
  }
  \;\;
  :
  \;\;
  \overset{
    \mathclap{
      \raisebox{5pt}{
        \scalebox{.7}{
          \color{orangeii}
          \bf
          \def\arraystretch{.9}
          \begin{tabular}{c}
            fibered
            \\
            over base
          \end{tabular}
        }
      }
    }
  }{
  B \to
  }
  \Types
  \end{array}
  \right.
  \end{array}
\end{equation}

Given such, its {\it Gauss-Manin monodromy} is the corresponding transport \eqref{TransportRule}
over the base type $B$:
\begin{equation}
  \label{GaussManinMonodromyPathLifting}
  \begin{tikzcd}[
    sep=0pt
  ]
  \mathrm{GMTransport}
  \;:
  \underset{
    {\color{orangeii} b_1, b_2 : B}
  }{\prod}
  &
  \!\!\!
  \bigg(
    \Id_{B}(
    {\color{orangeii}b_1}
    ,\,
    {\color{orangeii}b_2}
  )
  \ar[rr]
  &&
  \Big(
    H^{n + \tau_{\color{orangeii}b_1}}
    \big(
      X_{\color{orangeii}b_1}
      ;\,
      R
    \big)
    \;
    \overset{\sim}{\to}
    \;
    H^{n + \tau_{\color{orangeii}b_2}}
    \big(
      X_{\color{orangeii}b_2}
      ;\,
      R
    \big)
  \Big)
 \!\!    \bigg).
  \\[-4pt]
  &
  \big(
    b_1 \overset{\;
      p_{12}\;
    }{\rightsquigarrow} b_2
  \big)
  &
  \longmapsto
  &
  (p_{12})_\ast
  \end{tikzcd}
\end{equation}
\end{definition}
 Under the dictionary in \cref{HoTTIdea} and using Thm. \ref{GaussManinConnectionInTwistedGeneralizedCohomologyViaMappingSpaces}, this is the type-theoretic construction whose denotational semantics is the parallel transport/monodromy of Gauss-Manin connections on fibrations of twisted ordinary cohomology sets.

 \medskip

 Next, in \cref{KZConnectionsInHomotopyTypeTheory}, we turn to the specialization of this general construction to the case corresponding to Knizhnik-Zamolodchikov connections and hence to the operation of topological quantum gates (Thm. \ref{TheTheorem} below).

\medskip

 In closing this subsection, we comment on the equivalence under the brace in \eqref{TypeTheoreticGaussManinConnection}:

\begin{remark}[Between fiberwise and global twists]
Given a family of domain types equipped with a fiberwise system of twists as assumed in \eqref{TypeTheoreticGausManinFibration},
$$
  X_{(-)} \,:\, B \longrightarrow \Types
  ,\;\;\;\;\;\;\;
  \tau_{(-)}
  \,:\,
  B \longrightarrow \big(
    X_b
    \to
    \mathbf{B} R^\times
  \big)
$$
consider the corresponding total space \eqref{TypeClassification}
\vspace{-2mm}
 \begin{equation}
   \label{TotalSpaceOfParaeterizedDomainTypes}
   \begin{tikzcd}[sep=-2pt]
   X
   \;:\defneq\;
   (b : B) \times X_b
   \,,
   \\
   \phantom{b}
   \end{tikzcd}
   \;\;\;\;\;\;\;\;\;\;
   \begin{tikzcd}[sep=0]
     \mathrm{pr}_X
     \;:
     &
     X
     \ar[rr]
     &&
     (b : X) \times X_b
     \ar[rr]
     &&
     B
     \\
     &
     x
     &\longmapsto&
     (b,\, x_b)
     &\longmapsto&
     b
   \end{tikzcd}
 \end{equation}

 \vspace{-2mm}
\noindent  equipped with the corresponding ``global twist'':
 \vspace{-2mm}
 \begin{equation}
   \label{GLobalTwist}
   \begin{tikzcd}[sep=-2pt]
   \tau :
   &
   X
   \ar[
     rr,
     equals
   ]
   &&
   (b : B) \times X_b
   \ar[rr]
   &&
   \mathbf{B} R^\times
   \\
   &
   x
   &
   \longmapsto
   &
   (b,\, x_b)
   &\longmapsto&
   \tau_b\big(
     x_b
   \big)
   \end{tikzcd}
 \end{equation}

  \vspace{-2mm}
\noindent
Then the fiber type \eqref{HomotopyFiber} of the pairing
$$
  \big(
    \mathrm{pr}_X
    ,\,
    \tau
  \big)
  \;:\;
  X
    \longrightarrow
  B \times \mathbf{B}R^\times
$$
at any $(b,\,t) \,:\, B \times \mathbf{B}R^\times$ is equivalently the fiber type of $\tau_b$ at $t$:
\begin{equation}
  \label{FibersOfGlobalTwistPairedWithDomainProjection}
  \def\arraystretch{1.6}
  \begin{array}{lll}
  \mathrm{fib}_{(b ,\,t)}\big( \mathrm{pr}_X, \, \tau \big)
   &
    =\;
  (x : X) \times
  \Id\Big(
    \big(
      \mathrm{pr}_X(x),\, \tau(x)
    \big)
    ,\,
    \big(
      b,\,t
    \big)
  \Big)
  &
  \proofstep{
    by \eqref{HomotopyFiber}
  }
  \\
  &
  \simeq\;
  (x : X)
    \times
  \Id\big( \mathrm{pr}_X(x) ,\, b \big)
  \times
  \Id( \tau(x),\, t  )
  &
  \proofstep{
    by
    \eqref{happlyFormula}
  }
  \\
  &
 \simeq \;
  (x_b : X_b)
  \times
  \Id\big(
    \tau_b(x_b),\, t
  \big)
  &
  \proofstep{
    by
    \eqref{HomotopyFiber}
    \&
    \eqref{GLobalTwist}
  }
  \\
  &
  =\;
  \mathrm{fib}_t\big(
    \tau_b
  \big)
  &
  \proofstep{
    by
    \eqref{HomotopyFiber}.
  }
  \end{array}
\end{equation}

\end{remark}

\newpage

\section{The homotopy data type of anyon braid gates}
\label{KZConnectionsInHomotopyTypeTheory}

Finally, we discuss the specialization of the data type construction of Gauss-Manin monodromy in
Def. \ref{TypeTheoreticGaussManinConnection}
to that of KZ-monodromy on    $\suTwoAffine{\ShiftedLevel-2}$-conformal blocks,
along the lines of
Ex. \ref{TheKZConnection}, concluding with
Def. \ref{DataStructureOfConformalBlocks} and Thm. \ref{TheTheorem} below.

\medskip

This requires encoding the three specializations
\eqref{DomainFibrationForKZConnectionCase}
-
\eqref{TwistForKZConnectionCase}
invoked in
Ex. \ref{TheKZConnection}:

\medskip

\begin{enumerate}

\item[{\bf (i)}]
 The domain fibration is to be specialized
 to that of delooped {\bf braid groups}
 $\mathbf{B} \, \mathrm{PBr}(n + N) \to \mathbf{B} \, \mathrm{PBr}(N)$;

 \medskip

 \item[{\bf (ii)}] the local coefficients are specialized to
  the EM-type
  $\ComplexNumbers^\times \!\acts\, K(\ComplexNumbers, n)$
  of the {\bf complex numbers};

  \medskip

 \item[{\bf (iii)}] the twist is specialized to
  what on pure Artin generators is a list of
  {\bf complex exponentials} $\exp(2 \pi \ImaginaryUnit - )
  :
  \mathbb{Q} \to \ComplexNumbers^\times$.

\end{enumerate}

\medskip

All three of these are, of course, classical mathematical constructions, and their encodings in typed programming languages are in principle
well-understood. Nonetheless, at the time of this writing,
implementations specifically for a univalently computing and inductively homotopy-typed programming languages as needed here
(such as cubical {\tt Agda}, Lit. \ref{LiteratureAgda}) is  not readily available for import from standard libraries.
Therefore we shall dwell a little on how to get one's hands on these data structures.

\medskip

\noindent
{\bf The fibration of delooped pure braid groups.} The construction of the delooping type $\mathbf{B} \, \mathrm{PBr}(N + 1)$ of the pure
braid group is a straightforward consequence of our previous discussion:
Given that the pure braid group has a finite presentation \eqref{ArtinPureBraidGroup} by pure braid generators \eqref{ArtinPureBraidGeneratorsGraphically},
we may use the construction formula \eqref{FinitelyPresentedGroupType} (or, in {\tt Agda}, the 2-HIT equivalent to it):
First form the suspension type of the set of pure braid generators \eqref{ArtinPureBraidGeneratorsGraphically}
and then pushout-out \eqref{ParameterizedCellAttachment} each loop formed by a pair of free group elements in relation:

\vspace{-.4cm}
$$
  \mathbf{B} \mathrm{PBr}(N+1)
  :\defneq
  \left[
  \mathrm{po}
  \left(\!\!
  \begin{tikzcd}[
    row sep=2pt,
    column sep=9pt
  ]
  \overset{
    \raisebox{3pt}{
    \;\;\;\;
    \scalebox{.7}{
      \color{gray}
      \eqref{FirstPureArtinRelationGraphically}
      \eqref{ThePureArtinLeeRelations}
    }
    }
  }{
    \mbox{(Artin-Lee relations)}
  }
  \times
  S^1
  \quad
  \ar[rr]
  \ar[
    d,
    shorten=-3pt
  ]
  &&
  \quad
  \mathrm{S}
  \left(
      \{\NeutralElement\}
      \sqcup
      {
      \left\{
        b_{i j}
        \,=\,
\color{orange}
\left[
\color{black}
\adjustbox{raise=+6pt, scale=.8}{
\begin{tikzpicture}[line width=1.2, yscale=.8]

\begin{scope}[shift={(-1,0)}, gray]
\draw (-.75,-1) to (-.75,1);
\draw (-.25,-1) to (-.25,1);
\draw (-.5,-.6) node {\scalebox{.7}{$\cdots$}};
\end{scope}

\begin{scope}[shift={(0,0)}, gray]
\draw (-.75,-1) to (-.75,1);
\draw (-.25,-1) to (-.25,1);
\draw (-.5,-.6) node {\scalebox{.7}{$\cdots$}};
\end{scope}

\draw[line width=1.4]
  (0,-1) to (0,1);

\begin{scope}[shift={(+1,0)}, gray]
\draw (-.75,-1) to (-.75,1);
\draw (-.25,-1) to (-.25,1);
\draw (-.5,-.6) node {\scalebox{.7}{$\cdots$}};
\end{scope}

\draw[line width=4pt, white]
  (-1, -1)
    --
  (-1, -.8)
    .. controls (-1,-.4) and (0.15, -.4) ..
  (0.15, 0)
    .. controls (0.15, +.4) and (-1, +.4) ..
  (-1,.8)
    --
  (-1,1);
\draw[line width=1.4]
  (-1, -1)
    --
  (-1, -.8)
    .. controls (-1,-.4) and (0.15, -.4) ..
  (0.15, 0)
    .. controls (0.15, +.4) and (-1, +.4) ..
  (-1,.8)
    --
  (-1,1);

\draw[line width=4, white]
  (0,0) to (0,1);
\draw[line width=1.4]
  (0,0) to (0,1);

\node
  at (-1,-1.3) {
    \scalebox{.7}{
      \color{blue}
      $i$
    }
  };

\node
  at (0,-1.3) {
    \scalebox{.7}{
      \color{blue}
      $j$
    }
  };

\end{tikzpicture}
}
\color{orange}
\right]
\color{black}
\right\}
}_{1 \leq i < j \leq N+1}
 \right)
 \\
 \mbox{(Artin-Lee relations)}
 \end{tikzcd}
 \!\!\!\!\right)
 \right]_1
$$

By the classical homotopy equivalence \eqref{OrderedConfigurationSpaceIsEMSpaceOfPureBraidGroup} we may regard this as our type-theoretic
model of the homotopy type of ordered configuration spaces: In {\it cohesive homotopy type theory} (which we do not discuss here, but see the Outlook on p. \pageref{Outlook}) we would (or will) have a type equivalence of the form
$
  \shape
  \ConfigurationSpace{N+1}
  (\mathbb{R}^2)
  \;\;
  \simeq
  \;\;
  \mathbf{B}\, \mathrm{PBr}(N+1)
  \,.
$

Now observe, by the nature of the pure braid generators $b_{i j}$ \eqref{ArtinPureBraidGroup},
that the canonical fibrations of ordered configuration spaces \eqref{FibrationOfOrderedConfigurationSpaces} which forget
the last  point(s) in a configuration clearly induce on pure braid groups the homomorphisms which trivialize those
generators that carry the label of a discarded strand as an index:
\begin{equation}
  \label{FibrationOfPureBraidGroupsInTermsOfGenerators}
  \begin{tikzcd}
    \ConfigurationSpace{N + 2}(\mathbb{R}^2)
    \ar[
      d,
      ->>
    ]
    \ar[
      r,
      phantom,
      "{
        \underset{
          \mathrm{whe}
        }{\simeq}
      }"{yshift=-3pt}
    ]
    &
    B \, \mathrm{PBr}(N+2)
    \ar[
      d,
      ->>
    ]
    &[-10pt]
    b_{I,\, J}
    \ar[
      d,
      |->,
      shorten=8pt
    ]
    &[-16pt]
    b_{I ,\, N + 1}
    \ar[
      d,
      |->,
      shorten=8pt
    ]
    &[-16pt]
    b_{I ,\, N + 2}
    \ar[
      d,
      |->,
      shorten=8pt
    ]
    \\
    \ConfigurationSpace{N + 1}(\mathbb{R}^2)
    \ar[
      d,
      ->>
    ]
    \ar[
      r,
      phantom,
      "{
        \underset{
          \mathrm{whe}
        }{\simeq}
      }"{yshift=-3pt}
    ]
    &
    B \, \mathrm{PBr}(N+1)
    \ar[
      d,
      ->>
    ]
    &[-10pt]
    b_{I,\, J }
    \ar[
      d,
      |->,
      shorten=8pt
    ]
    &[-16pt]
    b_{I,\, N + 1}
    \ar[
      d,
      |->,
      shorten=8pt
    ]
    &[-16pt]
    \NeutralElement
    \ar[
      d,
      |->,
      shorten=8pt
    ]
    \\
    \ConfigurationSpace{N}(\mathbb{R}^2)
    \ar[
      r,
      phantom,
      "{
        \underset{
          \mathrm{whe}
        }{\simeq}
      }"{yshift=-3pt}
    ]
    &
    B \, \mathrm{PBr}(N)
    &
    b_{I,\, J}
    &
    \NeutralElement
    &
    \NeutralElement
  \end{tikzcd}
  \hspace{.6cm}
  \proofstep{
    for $1 \leq I < J \leq N$
  }
\end{equation}
The delooping \eqref{DeloopingOfGroupHomomorphisms} of this group homomorphism yields the desired fibration:
\begin{equation}
  \label{TheFibrationOfDeloopedBraidGroups}
 \begin{tikzcd}[
    sep=0pt,
    decoration={snake, segment length=4.5pt, amplitude=1pt},
  ]
  \mathrm{pr}
  \,:\;
  &
  \mathbf{B} \, \mathrm{PBr}(N + n)
  \ar[rr]
  &&
  \mathbf{B} \, \mathrm{PBr}(N)
    \\
    &
    \mathrm{pt}
      \ar[orangeii,
        out=-180+55,
        in=-55,
        decorate,
        looseness=3,
        shift left=8pt,
        "{ b_{I \, J} }"{swap, yshift=-2pt}
      ]
    &\mapsto&
    \mathrm{pt}
      \ar[orangeii,
        out=-180+55,
        in=-55,
        decorate,
        looseness=3,
        shift left=8pt,
        "{
          b_{I J}
        }"{swap, yshift=-2pt}
      ]
  \\[18pt]
  &
    \mathrm{pt}
      \ar[orangeii,
        out=-180+55,
        in=-55,
        decorate,
        looseness=3,
        shift left=8pt,
        "{ b_{I \, i} }"{swap, yshift=-2pt}
      ]
    &\mapsto&
    \mathrm{pt}
      \ar[orangeii,
        out=-180+55,
        in=-55,
        decorate,
        looseness=3,
        shift left=8pt,
        "{
          \mathrm{e}
        }"{swap, yshift=-2pt}
      ]
  \\[18pt]
  &
    \mathrm{pt}
      \ar[orangeii,
        out=-180+55,
        in=-55,
        decorate,
        looseness=3,
        shift left=8pt,
        "{ b_{i \, j} }"{swap, yshift=-2pt}
      ]
    &\mapsto&
    \mathrm{pt}
      \ar[orangeii,
        out=-180+55,
        in=-55,
        decorate,
        looseness=3,
        shift left=8pt,
        "{
          \mathrm{e}
        }"{swap, yshift=-2pt}
      ]
  \end{tikzcd}
  \hspace{1cm}
  \proofstep{
    for
    $
    \def\arraystretch{1.5}
    \begin{array}{l}
      1 \leq I < J \leq N
      \\
      N < i < j \leq N + n
    \end{array}
    $
  }
\end{equation}

\newpage

\noindent
{\bf Constructing the continuum.}
\label{ConstructingTheContinuum}
For encoding physical reality in general (and in particular for encoding topological quantum gates, via Def. \ref{DataStructureOfConformalBlocks} below)
we need the data structure of {\it real numbers} (``{\it the continuum}''  $\mathbb{R}$, e.g. \cite{Rudin64}). Elementary as this appears to any
practicing mathematician today, when reconsidered from the bare logical foundations of typed programming languages one is reminded that there is
a fair bit of work and some subtleties involved in constructing the real numbers and verifying their expected properties starting from just the
type of natural numbers.
This invokes non-trivial insights fully developed only in the 19th century (cf. \cite{Corry15}), fully understood
in its ``constructive'' refinement only late in the 20th century (\cite{Bishop67}\cite{BishopBridges85}\cite{Bridges99}),
which only now in the 21st century is being appreciated as the theory of real numbers pertinent to certified computing (e.g. \cite{OConnor07}\cite{GNSW07}, cf. Lit. \ref{ExactRealComputerAnalysis}).

\medskip
This may serve to explain that an actual library for real number arithmetic in the programming language {\tt Agda} (Lit. \ref{VerificationLiterature})
has been started only recently \cite{Murray22} (see also \cite{Lundfall15}) and is still under development. However, its  basic principles of
constructions and proofs are those originally developed already, in full detail,  for the {\it constructive analysis} of
\cite{Bishop67}\cite{BishopBridges85} and are well-understood. Better yet, implemented in homotopy data type theory this seminal historical
program of constructive analysis arguably finds its conceptual conclusion, in that here the required quotient sets \eqref{QuotientSetAsPushout}
actually exist, thus solving the remaining problems (cf. \cite[\S 1.1]{Li14}\cite[\S 4.3.2]{Murray22}) with previous ``setoid'' models.
(A yet better but currently more hypothetical approach may be that of \cite[\S 11]{UFP13}, see Rem. \ref{CauchyCompletenessOfRealNumbers} below).

\medskip

We now briefly list the incremental ring data structure constructions that produce the type of complex numbers, in this fashion, starting
from the type of natural numbers \eqref{NaturalNumberstype} and proceeding through the types of integer numbers \eqref{TypeOfIntegers},
rational numbers \eqref{DataStructureOfRationalNumbers}
and -- which requires the most care: --- the type of real numbers \eqref{TheTypeOfRealNumbers}.

\begin{remark}[Conventions for displaying ring data structures]
In doing so in the following, beware that:

\begin{itemize}[leftmargin=.7cm]

\item[{\bf (i)}]
Throughout we omit displaying the construction of the properties-certificates (such as for associativity etc.): These are all  either straightforward or, in the case of the convergence-certificates for the operations on real numbers \eqref{TheTypeOfRealNumbers}, the required ideas are all given in \cite{Bishop67}\cite{BishopBridges85} and have been coded into {\tt Agda} in \cite{Lundfall15}\cite{Murray22}.

\item[{\bf (ii)}] Consequently, we display functions on quotient types $X/R$ \eqref{QuotientSetAsPushout}
as functions on the quotient{\it ed} types $X$, leaving implicit the proof certificates that these functions respect the pertinent equivalence relations and hence descend to the quotient, much as spelled out in the proof of Prop. \ref{GSetsAreMapsOutOfBG}.

\item[{\bf (iii)}] We systematically ``overload'' notation for operations on number systems, as usual: For instance in the definition of the addition operation on integer numbers \eqref{TypeOfIntegers}
$$
  (n_1 ,\, m_1)
  +
  (n_2 ,\, m_2)
  \;:\defneq\;
  (n_1 + n_2 ,\, m_1 + m_2)
$$
it is understood that the operation on the left goes $ + : \Integers \times \Integers \to \Integers$ (as per the previous item) and is defined by the operation on the right which instead goes $+ : \NaturalNumbers \times \NaturalNumbers \to \NaturalNumbers$ and was constructed previously  \eqref{AdditionAndMultiplicationOfNaturalNumbers}.

\item[{\bf (iv)}] On our use of quotient types of real and complex numbers in the following, see Rem. \ref{CauchyCompletenessOfRealNumbers} below.
\end{itemize}
\end{remark}

\medskip

\noindent
{\bf The integer numbers.}
The archetypical (namely: initial) example of ring data structure
\eqref{RingStructure}
is the data type of {\it integers}: Its underlying set may be given (cf. \cite[Rem. 6.10.7]{UFP13}\cite[\href{https://1lab.dev/Data.Int.html}{\S Data.Int}]{1lab}) as the quotient type \eqref{QuotientSetAsPushout} of that of pairs \eqref{PairTypes} of natural numbers \eqref{NaturalNumberstype} by the equivalence relation \eqref{isEquivalenceRelation} which identifies those pairs of pairs $(n,\,m)$ whose cross-sums \eqref{AdditionAndMultiplicationOfNaturalNumbers} are equal, hence which serve as stand-ins for their difference $n - m$:
\begin{equation}
  \label{TypeOfIntegers}
  \def\arraystretch{1.2}
  \begin{array}{l}
  \mathllap{
    \vdash
    \;\;\;\;\;\;
  }
  \big(
    \Integers
    ,\,
    0
    ,\,
    +
    ,\
    -
    ,\
    1
    ,\,
    \cdot
    ,\,
  \big)
  \;:\; \Rings
  \\
  \mbox{where}
  \\
  \def\arraycolsep{2pt}
  \def\arraystretch{1.4}
  \begin{array}{ccl}
  \Integers
  &:\defneq&
  \NaturalNumbers
  \times
  \NaturalNumbers
  \;
  \big/
  \;
  \big(
    (n_1,\, m_1)
    ,\,
    (n_2,\, m_2)
    \,:\,
    \NaturalNumbers
      \times
    \NaturalNumbers
  \big)
  \times
    \Id_{\NaturalNumbers}
    \big(
      n_1 + m_2
      ,\,
      n_2 + m_1
    \big)
  \\
  0 &:\defneq& (0,0)
  \\
    +
    &:&
    \big(
    (n_1,\,m_1)
    ,\,
    (n_2,\,m_2)
    \big)
    \;\mapsto\;
    \big(
      n_1 + n_2
      ,\,
      m_1 + m_2
    \big)
   \\
    -
    &:&
    (n,\, m)
    \;\mapsto\;
    (m,\, n)
  \\
    1
    &:\defneq&
    \big(
      \mathrm{succ}(0)
      ,\,
      0
    \big)
  \\
    \cdot
    &:&
    \big(
      (n_1,\,m_1)
      ,\,
      (n_1,\,m_2)
    \big)
    \;\mapsto\;
    \big(
      n_1 \cdot n_2
      +
      m_1 \cdot m_2
     ,\;\;
     m_1 \cdot n_2
     +
     n_1 \cdot m_2
    \big)
  \end{array}
  \end{array}
\end{equation}
Using the canonical inclusion of the underlying types\footnote{The inclusion \eqref{InclusionOfNaturalNumbersIntoIntegers} is of course a homomorphism of monoid (semi-group) data structures, but here we do not dwell on monoid structure, for brevity.} of natural numbers \eqref{NaturalNumberstype}
\begin{equation}
  \label{InclusionOfNaturalNumbersIntoIntegers}
  \begin{tikzcd}[sep=-2pt]
    \iota
    \;: &
    \NaturalNumbers
    \ar[rr]
    &&
    \Integers
    \\
    &
    n &\longmapsto& (n,\,0)
  \end{tikzcd}
\end{equation}
we obtain the {\it ordering relation} on the integer numbers
\begin{equation}
  \label{OrderingOnIntegers}
  \begin{tikzcd}[sep=0pt]
    \leq
    \;\;:
    &
    \Integers \times \Integers
    \ar[rr]
    &&
    \Propositions
    \\
    &
    (n_1, n_2)
    &\longmapsto&
    (k : \NaturalNumbers)
    \times
    \Id\big(
      n_2
      ,\,
      n_1 + \iota(k)
    \big)
  \end{tikzcd}
\end{equation}
and hence the sub-type of {\it positive} integers (equivalent to that of poisitive natural numbers)
\begin{equation}
  \label{TypeOfPositiveIntegers}
  \Integers_+
  \;:\defneq\;
  (n : \Integers) \times (1 \leq n)
  \,.
\end{equation}

\medskip

\noindent
{\bf The rational numbers.} The ring data structure \eqref{RingStructure} of {\it rational numbers} may be given by the quotient set \eqref{isEquivalenceRelation} of pairs consisting of a {\it numerator} $p : \Integers$ \eqref{TypeOfIntegers} and a {\it denominator} $q : \Integers_+$ \eqref{TypeOfPositiveIntegers} subject to the usual identification of fractional arithmetic which makes the pair $(p,\,q)$ be a stand-in for the fraction $p/q$:
\begin{equation}
  \label{DataStructureOfRationalNumbers}
  \def\arraystretch{1.3}
  \begin{array}{l}
    \mathllap{
      \vdash
      \;\;\;\;\;
    }
    \big(
      \RationalNumbers
      ,\;
      0
      ,\,
      +
      ,\,
      -
      ,\,
      1
      ,\,
      \cdot
    \big)
    \;:\;
    \Rings
    \\
    \mbox{where}
    \\
    \def\arraycolsep{2pt}
    \def\arraystretch{1.4}
    \begin{array}{ccl}
      \RationalNumbers
      &:\defneq&
     \Integers
       \times
     \Integers_+
     \;\big/\;
     \big(
       (p_1 ,\, q_1)
       ,\,
       (p_1 ,\, q_2)
       \,:\,
       \Integers
       \times
       \Integers_+
     \big)
     \times
     \Id_{\Integers}
     \big(
       p_1 \cdot q_2
       ,\;
       q_1 \cdot p_2
     \big)
    \\
    0 &:\defneq& (0,1)
    \\
      +
      &:&
      \big(
        (p_1 ,\, q_1)
        ,\,
        (p_2 ,\,q_2)
      \big)
      \;\;\mapsto\;\;
      \big(
        p_1 \cdot q_2
        \,+\,
        p_2 \cdot q_1
        ,\;\;
        q_1 \cdot q_2
      \big)
    \\
      -
      &:&
      (p,\,q)
        \;\mapsto\;
      (-p,\, q)
    \\
    1
    &:\defneq&
    (1,\, 1)
    \\
      \cdot
      &:&
      \big(
      (p_1,\, q_1)
      ,\,
      (p_,\,q_2)
      \big)
      \;\mapsto\;
      \big(
        p_1 \cdot p_2
        ,\;
        q_1 \cdot q_2
      \big)
    \end{array}
  \end{array}
\end{equation}
The ordering relation \eqref{OrderingOnIntegers} on the integer numbers induces the ordering on the rational numbers
(remembering that we are forcing the denominators $q$ to be positive):
\vspace{-2mm}
\begin{equation}
  \label{OrderingOnRationalNumbers}
  \begin{tikzcd}[sep=0pt]
    \leq \;:
    &
    \RationalNumbers
    \times
    \RationalNumbers
    \ar[rr]
    &&
    \Propositions
    \\
    &
    \big(
      (q_1,\, p_1)
      ,\,
      (q_2 ,\, p_2)
    \big)
    &\longmapsto&
    q_1 \cdot p_2
    \;\leq\;
    q_2 \cdot p_1
    \mathrlap{\,.}
  \end{tikzcd}
\end{equation}

\vspace{-2mm}
\noindent For the following construction of the type of real numbers
below in \eqref{TheTypeOfRealNumbers} we introduce common notational abbreviations for the multiplicative {\it inverse} of a positive number
and for the {\it square} of any rational number
\vspace{-2mm}
\begin{equation}
  \label{SquareOfARationalNumber}
  \begin{tikzcd}[sep=-2pt]
    \frac{1}{(-)}
    \;:
    &
    \Integers_+
    \ar[rr]
    &&
    \RationalNumbers
    \\
    &
    n &\longmapsto& (1,\,n)
  \end{tikzcd}
  \hspace{1.5cm}
  \begin{tikzcd}[sep=-2pt]
    (-)^2 \;:
    &
    \RationalNumbers
    \ar[rr]
    &&
    \RationalNumbers
    \\
    &
    r &\longmapsto& r \cdot r
  \end{tikzcd}
\end{equation}
Notice that bounds \eqref{OrderingOnRationalNumbers} on a square $r^2$ \eqref{SquareOfARationalNumber} equivalently serve
as bounds on the {\it absolute value} $\vert r \vert$ (which we do not introduce separately).

\medskip
\noindent
{\bf The real numbers.} The following is the construction of the ring data structure \eqref{RingStructure} of {\it real numbers}
as {\it regular sequences} $x_{(-)} : \Integers_+ \to \RationalNumbers$  of rational numbers \eqref{DataStructureOfRationalNumbers}
indexed by positive integers \eqref{TypeOfPositiveIntegers}, serving as stand-ins for the real number to which these converge.
This definition and the verification of its intended properties is due to \cite[pp. 15]{Bishop67}\cite[pp. 18]{BishopBridges85}
and has been implemented in {\tt Agda} by \cite{Lundfall15}\cite[Def. 3.3.1]{Murray22} (also in {\tt Coq} \cite{KrebbersSpitters13}
following a monadic re-formulation due to \cite{OConnor07}) --- there as data sets equipped with equivalence relations
\eqref{isEquivalenceRelation}, while we now display the corresponding quotient type \eqref{QuotientSetAsPushout}, bewaring of Rem. \ref{CauchyCompletenessOfRealNumbers} below:
\begin{equation}
\label{TheTypeOfRealNumbers}
\hspace{3mm}
\def\arraystretch{1.4}
\begin{array}{l}
  \mathllap{
    \vdash
    \;\;\;
  }
  \big(
    \RealNumbers
    ,\,
    0
    ,\,
    +
    ,\,
    -
    ,\,
    1
    ,\,
    \cdot
  \big)
  \,:\,
  \Rings
  \\
  \mbox{where}
  \\[-5pt]
  \def\arraystretch{1.4}
  \def\arraycolsep{2pt}
  \begin{array}{ccl}
    \RealNumbers
    &:\defneq&
    \overset{
      \mathclap{
        \raisebox{6pt}{
          \scalebox{.7}{
            \color{darkblue}
            \bf
            \def\arraystretch{.7}
            \begin{tabular}{c}
              sequences of
              \\
              rational numbers
              \color{gray}
              \normalfont
              \eqref{DataStructureOfRationalNumbers}
            \end{tabular}
          }
        }
      }
    }{
    \big(
      x_{(-)}
      :
      \Integers_+
      \to
      \RationalNumbers
    \big)
    }
    \times
    \bigg(\!\!
    \overset{
      \mathclap{
        \raisebox{4pt}{
          \scalebox{.7}{
            \color{darkblue}
            \bf
            \def\arraystretch{.7}
            \begin{tabular}{c}
              which converge regularly
            \end{tabular}
          }
        }
      }
    }{
      (
        n,\,m : \Integers_+
      )
      \to
      \Big(
        (x_n - x_m)^2
        \,\leq\,
        \big(
          \frac{1}{n} + \frac{1}{m}
        \big)^2
      \Big)
    }
    \!\!\bigg)
    \!\!
    \overset{
      \mathclap{
        \raisebox{2pt}{
          \scalebox{.7}{
            \color{orangeii}
            \bf
            modulo
          }
        }
      }
    }{
      \bigg/
    }
    \!\!
    \overset{
      \mathclap{
        \raisebox{8.5pt}{
          \scalebox{.7}{
            \color{darkblue}
            \bf
            those
          }
        }
      }
    }{
    \big(
      x_{(-)} ,\, y_{(-)}
    \big)
    }
    \times
    \bigg(\!\!
      \overset{
        \mathclap{
          \raisebox{6pt}{
            \scalebox{.7}{
              \color{darkblue}
              \bf
              that converge to zero
            }
          }
        }
      }{
      (n : \Integers_{+})
      \to
      \Big(
        (x_n - y_b)^2
        \,\leq\,
        \big(\frac{2}{n}\big)^2
      \Big)
      }
    \!\! \bigg)
    \\
    0 &:& n \mapsto 0
    \\
    +
    &:&
    \big(
      x_{(-)}
      ,\,
      y_{(-)}
    \big)
    \;\mapsto\;
    \big(
      n \,\mapsto\, x_n + y_n
    \big)
    \\
    -
    &:&
    x_{(-)}
    \,\mapsto\,
    \big(
      n \,\mapsto\, - x_n
    \big)
    \\
    1
    &:&
    n \,\mapsto\, 1
    \\
    \cdot
    &:&
    \big(
      x_{(-)}
      ,\,
      y_{(-)}
    \big)
    \;\mapsto\;
    \big(
      n
      \,\mapsto\,
      x_n
      \cdot
      y_n
    \big)
  \end{array}
\end{array}
\end{equation}

\begin{remark}[\bf Computational content of real numbers]
\label{ComputationalContentOfRealNumbers}
In the data type \eqref{TheTypeOfRealNumbers} any real number is {\it represented} by a sequence of rational numbers; in practice these representing sequences
\begin{equation}
  \label{RealNumberAsProgram}
  \frac{
    \vdash
    \;\;\;
    x_{(-)} : \Integers_+ \to \RationalNumbers
  }
  {
    n : \Integers_+
    \;\;\;
    \vdash
    \;\;\;
    x_n : \RationalNumbers
  }
\end{equation}
are little programs \eqref{FunctionDeclaration}
which for any prescribed but finite precision $\sim \frac{1}{n}$ compute the intended number to within that precision.
That this is really what it means to {\it know} a real number was the conviction of the founding fathers of {\it constructive analysis} \cite{Bishop67}\cite{BishopBridges85}; in any case this is what it means for a finite computing machine to store an exact real number (as opposed to some finite floating-point approximation) in its memory (see \cite[\S 3]{Vuillemin88}).

For example, the circle number $\pi$ may be encoded by the BBP-formula \cite{BaileyBorweinPlouffe97}, for which the required proof certificates of convergence are evident:
\begin{equation}
  \label{Pi}
  \pi
  \;:\defneq\;
  \bigg(
    n
    \;\mapsto\;
    \sum_{k = 0}^n
    \frac{1}{16^k}
    \Big(
      \frac{4}{6 k + 1}
      -
      \frac{2}{8 k + 4}
      -
      \frac{1}{8 k + 5}
      -
      \frac{1}{8 k + 6}
    \Big)
  \bigg)
  \;:\;
  \RealNumbers
\end{equation}
\end{remark}

\begin{remark}[\bf Cauchy completeness and/or its constructive failure for Bishop reals]
\label{CauchyCompletenessOfRealNumbers}
A desirable or even constituting property of the real numbers is their {\it Cauchy completeness} (e.g. \cite[Def. 3.12]{Rudin64}), meaning that any Cauchy sequence of real numbers converges to a real number: Intuitively this is the requirement that real numbers really form a {\it continuum} in that no ``gaps'' remain on the real line. But there is a difference in whether one asks for convergence of {\it constructible} sequences of real numbers, or more generally:

On the one hand, given a Cauchy sequence of {\it operational encodings} of real numbers \eqref{RealNumberAsProgram}, namely of, in turn, Cauchy sequences of rational numbers, hence given a Cauchy sequence of Cauchy sequences
\begin{equation}
  \label{CauchySequenceOfCauchySequenceOfRationals}
  x_{(-)}^{(-)}
  \,:\,
  \Integers_+ \longrightarrow
  \big(
    \Integers_+
    \to
    \RationalNumbers
  \big)
  \,,
\end{equation}
then its convergence is indeed provable (\cite[p. 27, Thm. 2]{Bishop67}\cite[p. 29, Thm. 3.3]{BishopBridges85}, cf. \cite[Thm. 3.4.4]{Murray22}): this is the fully constructive notion of Cauchy completeness of the real numbers.

On the other hand, such Cauchy sequences of Cauchy sequences of rational numbers \eqref{CauchySequenceOfCauchySequenceOfRationals} represent single sequences of real numbers
\begin{equation}
  \label{SequenceOfRealNumbersAsQuotients}
  x^{(-)} \,:\, \Integers_+ \longrightarrow \RealNumbers
\end{equation}
when we regard the latter inside the quotient type \eqref{TheTypeOfRealNumbers}. But given {\it just} such quotiented Cauchy sequence data of type \eqref{SequenceOfRealNumbersAsQuotients}, there is no proof that it converges to an element of $\mathbb{R}$ \cite{Lubarsky07}. By the constructive proof of \cite[p. 27]{Bishop67}\cite[p. 29]{BishopBridges85} it {\it would} be possible if one could {\it lift} the sequence \eqref{SequenceOfRealNumbersAsQuotients} back to one of the ``constructive'' form \eqref{CauchySequenceOfCauchySequenceOfRationals} --- but this is in turn not possible in general. (The relevant ``axiom of countable choice'' does not hold syntactically in the type theory. While it does hold semantically in the classical model topos \eqref{InterpretationInTheClassicalModelTopos}, it does not hold in the more general model toposes that we are eventually interested in, as per p. \pageref{Outlook}.)

\medskip

Luckily, it looks like for our intended application (in Def. \ref{DataStructureOfConformalBlocks} below) we never (need to) care about sequences of real number data
\eqref{SequenceOfRealNumbersAsQuotients}
for which we do not have constructive lifts \eqref{CauchySequenceOfCauchySequenceOfRationals}, whence the operational Cauchy completeness proven by \cite[p. 27]{Bishop67}\cite[p. 29]{BishopBridges85}\cite[Thm. 3.4.4]{Murray22} ensures that real number data of type \eqref{TheTypeOfRealNumbers} is satisfactory in applications.

\medskip

\end{remark}

\begin{remark}[Other descriptions of the reals]
There are other possible constructions of the real numbers which do not share the Cauchy completeness deficiency
of the Bishop real numbers.

\noindent {\bf (i)} The most straightforward is the type of Dedekind real numbers, or two-sided Dedekind
cuts (\cite[Defn. 11.2.1]{UFP13}):
\begin{equation}\label{DedekindRealNumbers}
    \hspace{-2mm}
  \mathbb{R}_D
  \;:\defneq
  \left\{\!\!\!\!\!\!
  \def\arraystretch{1.6}
  \begin{array}{l}
    \left.
    (L,\, U : \mathbb{Q} \to \Propositions)
    \mathclap{\phantom{\vert^{\vert}}}
    \;
    \right\}
    \scalebox{.7}{
      \color{darkblue}
      \bf
      structure
    }
    \\
    \left.
    \begin{array}{l}
      \times
      \;\;
      \Big(
        \exists\{q : \RationalNumbers \mid L(q)\} \times \exists\{r : \RationalNumbers \mid U(r)\}
      \Big)
      \\[+4pt]
      \times
      \;\;
      \bigg(
      (q : \mathbb{Q}) \to
        L(q) \simeq \exists\{r : \RationalNumbers \mid (q < r) \times L(r)\}
        \Big)
     \!\! \bigg)
     \\
     \times
      \;\;\bigg(
      (r : \mathbb{Q}) \to
        R(r) \simeq \exists\{q : \RationalNumbers \mid (q < r) \times U(q)\}
        \Big)
     \!\! \bigg)
     \\
     \times
      \;\;\bigg(
        (q:\mathbb{Q}) \to (L(q) \times U(q)) \to \varnothing
     \bigg)
     \\
     \times
      \;\;\bigg(
            (q, r : \RationalNumbers) \to \exists(L(q) + R(r))
     \bigg)
    \end{array}
    \;
    \!\!\!\!\!\!\!\right\}
    \scalebox{.7}{
      \color{darkblue}
      \bf
      properties
    }
  \end{array}
  \right.
\end{equation}
The Dedekind real numbers have the benefit of being provably Cauchy complete without any non-constructive principles.
Since they evidently contain the rationals by sending $x : \mathbb{Q}$ to the cut $(q \mapsto (q < x),\, r \mapsto (x < r))$,
this implies that the Bishop reals admit a map to the Dedekind reals which respects the quotient relation on the regular sequences.

\vspace{1mm}
\noindent {\bf (ii)} Finally, another construction principle for the type of Cauchy real numbers as a more
intricate ``higher inductive-inductive type'' (further generalizing the notion of higher inductive types, pp.
\pageref{HigherInductiveTypes}) has been laid out in \cite[\S 11.3]{UFP13} and proven to be genuinely Cauchy
complete (\cite[Thm. 11.3.49]{UFP13}). This novel construction principle has attracted some type-theoretic
attention and may well be the way to go forward in the future; but at the time of this writing, it is not
available in practice.
\end{remark}

\medskip

\noindent
{\bf The complex numbers.} With a type \eqref{TheTypeOfRealNumbers} of real numbers in hand, it is straightforward
to construct the corresponding ring data structure \eqref{RingStructure} of complex numbers (applying the classical
formulas; see e.g. \cite[Def. 1.24]{Rudin64}):
\vspace{-1mm}
\begin{equation}
  \label{TheRingOfComplexNumbers}
  \def\arraystretch{1.6}
  \begin{array}{l}
\phantom{AA}  \mathllap{
   \vdash
    \;\;\;\;
  }
  \big(
    \ComplexNumbers
    ,\,
    +
    ,\,
    -
    ,\,
    0
    ,\,
    \cdot
    ,\,
    1
  \big)
  \,:\,
  \Rings
  \\
  \mbox{where}
  \\
  \hspace{-2mm}
  \def\arraystretch{1.4}
  \def\arraycolsep{2pt}
  \begin{array}{ccl}
    \big(
      z : \ComplexNumbers
    \big)
    &:\defneq&
    \big(
      \mathrm{Re}(z)
      ,\,
      \mathrm{Im}(z)
    \big)
    \,:\,
    \RealNumbers \times \RealNumbers
   \\
   0
    &:\defneq&
   (0,\,0)
   \\
    +
    &:&
    \Big(
    \big(
      \mathrm{Re}(z_1)
      ,\,
      \mathrm{Im}(z_1)
    \big)
    ,\,
    \big(
      \mathrm{Re}(z_2)
      ,\,
      \mathrm{Im}(z_2)
    \big)
    \Big)
    \;\mapsto\;
    \big(
      \mathrm{Re}(z_1) + \mathrm{Re}(z_2)
      ,\,
      \mathrm{Re}(z_1) + \mathrm{Re}(z_2)
    \big)
    \\
    -
    &:&
    \big(
      \mathrm{Re}(z)
      ,\,
      \mathrm{Im}(z)
    \big)
    \;\mapsto\;
    \big(
      - \mathrm{Re}(z)
      ,\,
      - \mathrm{Im}(z)
    \big)
   \\
    \cdot
    &:&
    \Big(
      \big(
        \mathrm{Re}(z_1)
        ,\,
        \mathrm{Im}(z_1)
      \big)
      ,\,
      \big(
        \mathrm{Re}(z_2)
        ,\,
        \mathrm{Im}(z_2)
      \big)
    \Big)
    \mapsto
    \Big(
    \mathrm{Re}(z_1)
    \cdot
    \mathrm{Re}(z_2)
    -
    \mathrm{Im}(z_1)
    \cdot
    \mathrm{Im}(z_2)
    ,\;\;
    \mathrm{Re}(z_1) \cdot \mathrm{Im}(z_2)
    +
    \mathrm{Im}(z_1) \cdot \mathrm{Re}(z_2)
    \Big)
\\
       1
   &:\defneq&
   (1,\, 0)
  \end{array}
  \end{array}
\end{equation}

So the imaginary unit is

\vspace{-.5cm}
\begin{equation}
  \ImaginaryUnit
  \;:\defneq\;
  (0,\,1) \,:\,
  \ComplexNumbers
\end{equation}

We now obtain the complex exponential function on rational arguments by composing its classical series expansion $n \,\mapsto\, \sum_{k=0}^n \tfrac{1}{k!} (-)^k$ (e.g. \cite[p. 178]{Rudin64}) with the series representation of $\pi$ \eqref{Pi}; and by classical computations, the result is readily certified to be a group homomorphism \eqref{TypeOfGroupHomomorphisms}
from the underlying abelian group \eqref{UnderlyingAbelianGroupOfType} of the ring of rational numbers \eqref{DataStructureOfRationalNumbers}
to the group of units \eqref{GroupOfUnits} of the ring of complex numbers \eqref{TheRingOfComplexNumbers}:

\begin{equation}
\begin{tikzcd}
  \exp\big(
    2 \pi \ImaginaryUnit
    \cdot (-)
  \big)
  \;:\;
  \RationalNumbers_{\mathrm{udl}}
  \ar[r, "{\mathrm{hom}}"{swap}]
  &
  \ComplexNumbers^\times
  \,.
\end{tikzcd}
\end{equation}

Similarly, we readily equip $\ComplexNumbers^\times$ with a certificate that it is abelian, $\ComplexNumbers^{\times} : \AbelianGroups$ \eqref{StructureOfAbelianGroups}.

\begin{lemma}[Assigning phases to pure Artin generators]
\label{AssigningPhasesToPureArtinGenerators}
Any list of rational numbers, one for each pure braid generator, defines a group homomorphism:
\begin{equation}
  \label{APhaseForEachArtinGenerator}
\hspace{-5mm}
\begin{tikzcd}[
    sep=0pt,
    decoration={snake, segment length=4.5pt, amplitude=1pt}
  ]
    N \,:\, \NaturalNumbers_+
    ,\;\;
    n \,:\, \NaturalNumbers
    \;\;\;\;\;\;\;
    \vdash
    \;\;\;\;\;\;\;
    &
    \left(
      r_{(--)}
      :
      \def\arraystretch{1}
      \begin{array}{l}
      (k_1, k_2 : \Integers_+)
      \\
        \; \times
        (k_1 + 1 \leq k_2)
      \\
        \; \times
        (k_2 \leq N + n)
      \end{array}
      \to
      \RationalNumbers
    \right)
    \ar[rr]
    &&
    \Big(
    \mathbf{B}\, \mathrm{PBr}(N + n)
    \ar[rr]
    &&
    \mathbf{B} \, \ComplexNumbers^\times
    \Big)
    \\
    &
    r_{(--)}
    &\longmapsto&
    \mathllap{
      \big(\;\;\;\;
    }
    \mathrm{pt}
      \ar[orangeii,
        out=-180+55,
        in=-55,
        decorate,
        looseness=3,
        shift left=8pt,
        "{
           \underset{
             \mathclap{
               \raisebox{-4pt}{
                 \scalebox{.7}{
                   \color{darkblue}
                   \bf
                   \def\arraystretch{.9}
                   \begin{tabular}{c}
                     pure braid
                     \\
                     generators
                   \end{tabular}
                 }
               }
             }
           }{
             b_{k_1 \, k_2}
           }
        }"{swap, yshift=-2pt}
      ]
      &\mapsto&
      \mathrm{pt}
      \ar[orangeii,
        out=-180+55,
        in=-55,
        decorate,
        looseness=3,
        shift left=8pt,
        "{
           \exp
           \big(
             2 \pi \ImaginaryUnit
             r_{(k_1 \, k_2)}
           \big)
         }"{swap, yshift=-2pt}
      ]
    \mathrlap{
      \;\;\;\;\big)
    }
  \end{tikzcd}
\end{equation}
\end{lemma}
\begin{proof}
Since the relations
\eqref{ThePureArtinLeeRelations}
on the pure Artin generators
\eqref{ArtinPureBraidGeneratorsGraphically}
are all group commutator relations,
and since in the abelian target group every group commutator is canonically witnessed as an identity.
\end{proof}

\begin{remark}[Special cases for applications]
  There is much room to replace this general construction with optimized special-purpose constructions
  in special cases. For example, if in applications we are to focus on rational numbers with a numerator
  equal to $q = 4$ (which is the case of Majorana anyons!, Lit. \ref{LiteratureAnyonSpecies}), then the
  corresponding exponential is an integer complex number and may be defined directly:
  $$
    \exp\big(
      2 \pi \ImaginaryUnit
      \,
      {p}/{4}
    \big)
    \;\;=\;\;
    \ImaginaryUnit^p
    \,.
  $$
\end{remark}

\newpage

\noindent
{\bf The homotopy data structure of topological quantum gates.}
With all these data structures in hand, we may conclude.

\begin{definition}[Homotopy data structure of conformal blocks]
\label{DataStructureOfConformalBlocks}
In specialization of Def. \ref{TypeTheoreticGaussManinConnection}, we obtain this data type:
\vspace{-1mm}
\begin{equation}
  \label{TypeTheoreticKZFibration}
  \hspace{-3cm}
  \begin{array}{rcl}
  \left.
  \def\arraystretch{1.6}
  \begin{array}{l}
  \overset{
    \mathclap{
      \raisebox{4pt}{
        \scalebox{.7}{
          \color{darkblue}
          \bf
          punctures
        }
      }
    }
  }{
    N : \mathbb{N}_+
  }
  ,\;\;
  \overset{
    \mathclap{
      \raisebox{4pt}{
        \scalebox{.7}{
          \color{darkblue}
          \bf
          degree
        }
      }
    }
  }{
    n : \mathbb{N}
  }
  ,\;\;\;
  \overset{
    \mathclap{
      \raisebox{4pt}{
        \scalebox{.7}{
          \color{darkblue}
          \bf
          \def\arraystretch{.9}
          \begin{tabular}{c}
            shifted
            level
          \end{tabular}
        }
      }
    }
  }{
    \ShiftedLevel : \mathbb{N}_{\geq 2}
  }
  \\
  \underset{
    \mathclap{
      \raisebox{-4pt}{
        \color{darkblue}
        \bf
        \scalebox{.7}{
          weights
        }
      }
    }
  }{
    \weight_{(-)}
  }
  :
    N \to
    \{0, \cdots, \ShiftedLevel-2\}
  \end{array}
  \right\}
  \;\;
  \vdash
  &
 \!\!\! \Bigg(
  \vec z
    \,\mapsto\,
  \def\arraystretch{1.2}
  \Truncation{\bigg}{0}{
    (t : \mathbf{B}\ComplexNumbers^\times)
    \to
    \Big(
    \mathrm{fib}_{(t,\vec z)}
    \big(
      \mathrm{pr}^{N+n}_N
      ,\,
      \tau_{(\ShiftedLevel, \weight_\bullet)}
    \big)
    \to
    \mathbf{B}^n
    (
      \acts_{\, {t}} \, \ComplexNumbers_{\mathrm{udl}}
    )
    \Big)
  }
  \Bigg)\mathrlap{
  :
  \;
  \overset{
    \mathclap{
      \raisebox{5pt}{
        \scalebox{.7}{
          \color{orangeii}
          \bf
          \def\arraystretch{.9}
          \begin{tabular}{c}
          \end{tabular}
        }
      }
    }
  }{
  \mathbf{B} \, \mathrm{PBr}(N)
  \to
  }
  \Types
  }
  \\
  {}
  \\[-5pt]
  \mbox{where}
  &&
  \\[-7pt]
 &
 \hspace{-3cm}
 \begin{tikzcd}[
    sep=0pt,
    decoration={snake, segment length=4.5pt, amplitude=1pt},
  ]
  \mathllap{
    \scalebox{.7}{
      \color{gray}
      \eqref{TheFibrationOfDeloopedBraidGroups}
      \;\;
    }
  }
  \mathrm{pr}^{N + n}_N
  \,:\;
  &
  \mathbf{B} \, \mathrm{PBr}(N + n)
  \ar[rr]
  &&
  \mathbf{B} \, \mathrm{PBr}(N)
  \\
  &
    \mathrm{pt}
      \ar[orangeii,
        out=-180+55,
        in=-55,
        decorate,
        looseness=3,
        shift left=8pt,
        "{ b_{I \, i} }"{swap, yshift=-2pt}
      ]
    &\mapsto&
    \mathrm{pt}
      \ar[orangeii,
        out=-180+55,
        in=-55,
        decorate,
        looseness=3,
        shift left=8pt,
        "{
          \mathrm{e}
        }"{swap, yshift=-2pt}
      ]
    \\[18pt]
    &
    \mathrm{pt}
      \ar[orangeii,
        out=-180+55,
        in=-55,
        decorate,
        looseness=3,
        shift left=8pt,
        "{ b_{I \, J} }"{swap, yshift=-2pt}
      ]
    &\mapsto&
    \mathrm{pt}
      \ar[orangeii,
        out=-180+55,
        in=-55,
        decorate,
        looseness=3,
        shift left=8pt,
        "{
          b_{I J}
        }"{swap, yshift=-2pt}
      ]
  \end{tikzcd}
  \hspace{5mm}
  \begin{tikzcd}[
   row sep=-4pt, column sep=0pt,
    decoration={snake, segment length=4.5pt, amplitude=1pt}
  ]
    \tau_{(\ShiftedLevel,\,\weight_\bullet)}
    \,:
    &
    \mathbf{B} \, \mathrm{PBr}(N + n)
    \ar[rrr]
    &
    &
    &
    \mathbf{B} \, \ComplexNumbers^\times
    \\
    \scalebox{.7}{
      \color{gray}
      \eqref{TheTwistParameters}
      \eqref{APhaseForEachArtinGenerator}
    }
    &
    \mathrm{pt}
      \ar[orangeii,
        out=-180+55,
        in=-55,
        decorate,
        looseness=3,
        shift left=8pt,
        "{ b_{I \, i} }"{swap, yshift=-2pt}
      ]
    &\mapsto&&
     \mathrm{pt}
      \ar[orangeii,
        out=-180+55,
        in=-55,
        decorate,
        looseness=3,
        shift left=8pt,
        "{
          \exp(
            2 \pi \ImaginaryUnit
            \frac{\weight_I}{\ShiftedLevel}
          )
        }"{swap}
      ]
    \\[20pt]
    &
    \mathrm{pt}
      \ar[orangeii,
        out=-180+55,
        in=-55,
        decorate,
        looseness=3,
        shift left=8pt,
        "{ b_{i \, j} }"{swap, yshift=-2pt}
      ]
    &\mapsto&&
    \mathrm{pt}
      \ar[orangeii,
        out=-180+55,
        in=-55,
        decorate,
        looseness=3,
        shift left=8pt,
        "{
          \exp(
            2 \pi \ImaginaryUnit
            \frac{2}{\ShiftedLevel}
          )
        }"{swap}
      ]
    \\[20pt]
    &
    \mathrm{pt}
      \ar[orangeii,
        out=-180+55,
        in=-55,
        decorate,
        looseness=3,
        shift left=8pt,
        "{ b_{I \, J} }"{swap, yshift=-2pt}
      ]
    &\mapsto&&
    \mathrm{pt}
      \ar[orangeii,
        out=-180+55,
        in=-55,
        decorate,
        looseness=3,
        shift left=8pt,
        "{
          \exp(
            2 \pi \ImaginaryUnit
            \frac{\weight_I \weight_J}{2\ShiftedLevel}
          )
        }"{swap}
      ]
  \end{tikzcd}
  \end{array}
\end{equation}

\end{definition}
\begin{theorem}[{\bf Topological quantum  gates as homotopy data structure}]
  \label{TheTheorem}
  The semantics in the classical model topos \eqref{InterpretationInTheClassicalModelTopos}
  of the transport operation \eqref{GaussManinMonodromyPathLifting} in this data type \eqref{TypeTheoreticKZFibration} is
  given by the monodromy of the Knizhnik-Zamolodchikov connection, on $\suTwoAffine{\ShiftedLevel-2}$-conformal blocks
  (on the Riemann sphere with $N + 1$ punctures weighted by $(\weight_I)_{I=1}^N$ and $\weight_{N + 1} = n + \sum_I \weight_I $).
\end{theorem}
\begin{proof}
  By
  Example \ref{TheKZConnection}
  of Theorem
  \ref{GaussManinConnectionInTwistedGeneralizedCohomologyViaMappingSpaces},
  we are reduced to showing that the semantics
  of the type formation \eqref{TypeTheoreticKZFibration} equals the topological construction expressed by the formula \eqref{KZConnectionByFiberwiseMapping}.
    This follows by applying the syntax/semantics dictionary \cref{HoTTIdea} iteratively to the sub-terms
    of \eqref{TypeTheoreticKZFibration}, as shown in the following steps:

\vspace{3mm}
  \hspace{-.85cm}
  \hypertarget{TableT}{}
  \def\arraystretch{2.2}
  \def\tabcolsep{2pt}
  \begin{tabular}{|l|c|c|}
   \hline
   \multicolumn{1}{|c|}{
    {\bf Syntax}
    }
    &
    $\xleftrightarrow{
      \scalebox{.7}{
        \cref{HoTTIdea}
      }
    }$
    &
    {\bf Semantics}
    \\
    \hline
    \hline
    $
    \def\arraystretch{1.6}
    \begin{array}{l}
    \vec z : \mathbf{B}\,\mathrm{PBr}(N)
    ,\;
    t : \mathbf{B}\ComplexNumbers^\times
    \;\;\;
    \vdash
    \;\;\;
    \\
    \hspace{.6cm}
    \mathrm{fib}_{(t,\,\vec z)}
     (
       \mathrm{pr}^{N + n}_N
       ,\,
       \tau_{(\ShiftedLevel, \weight_\bullet)}
     )
     \,:\,
     \Types
     \end{array}
    $
    &
    \def\arraystretch{.9}
    \begin{tabular}{c}
      \eqref{TypeClassification}
      \\
      \eqref{OrderedConfigurationSpaceIsEMSpaceOfPureBraidGroup}
    \end{tabular}
    &
    $
      \begin{tikzcd}
        \ConfigurationSpace{N+n}(\ComplexPlane)
        \ar[
          d,
          shorten=-2pt,
          "{
            (
            \mathrm{pr}^{N+n}_N
            ,\,
            \tau_{(\ShiftedLevel,\weight_\bullet)}
            )
          }"
          {swap}
        ]
        \\
        \ConfigurationSpace{N}(\ComplexPlane)
        \times
        B \ComplexNumbers^\times
      \end{tikzcd}
    $
    \\
    \hline
    $
    \def\arraystretch{1.6}
    \begin{array}{l}
    \vec z : \mathbf{B}\,\mathrm{PBr}(N)
    ,\;
    t : \mathbf{B}\ComplexNumbers^\times
    \;\;\;
    \vdash
    \;\;\;
    \\
    \hspace{.6cm}
    \mathbf{B}^n\big(
      \acts_{t}
      \ComplexNumbers_{\mathrm{udl}}
    \big)
    \,:\,
    \Types
    \end{array}
    $
    &
    \def\arraystretch{.9}
    \begin{tabular}{c}
      \eqref{LCCRules}
      \\
      \eqref{DeloopingDependentTypeAsBorelConstruction}
    \end{tabular}
    &
    $
    p_{
      \!\!\!
      \scalebox{.7}{$
        \ConfigurationSpace{N}
      $}
    }^\ast
    \begin{tikzcd}
      \mathrm{K}(\ComplexNumbers ,\, n)
      \times_{\ComplexNumbers^\times} E \ComplexNumbers^\times
      \ar[d]
      \\
      B \ComplexNumbers^\times
    \end{tikzcd}
    $
    \\
    \hline
    $
    \def\arraystretch{1.7}
    \begin{array}{l}
    \vec z : \mathbf{B}\,\mathrm{PBr}(N)
    ,\;
    t : \mathbf{B}\ComplexNumbers^\times
    \;\;\;
    \vdash
    \;\;\;
    \\
    \hspace{.6cm}
    \mathrm{fib}_{(t,\,\vec z)}
     (
       \mathrm{pr}^{N + n}_N
       ,\,
       \tau_{(\ShiftedLevel, \weight_\bullet)}
     )
    \to
    \mathbf{B}^n\big(
      \acts_{t}
      \ComplexNumbers_{\mathrm{udl}}
    \big)
    \;:\;
    \Types
    \end{array}
    $
    &
    \def\arraystretch{1}
    \begin{tabular}{c}
      \eqref{FunctionTypesAndMappingSpaces}
    \end{tabular}
    &
    $
      \mathrm{Map}\!
      \left(\!\!\!\!
      \adjustbox{raise=8pt}{$
      \adjustbox{raise=-6pt}{$
      \begin{tikzcd}
        \ConfigurationSpace{N+n}(\ComplexPlane)
        \ar[
          d,
          shorten=-2pt,
          "{
            (
            \mathrm{pr}^{N+n}_N
            ,\,
            \tau_{(\ShiftedLevel,\weight_\bullet)}
            )
          }"{swap}
        ]
        \\
        \ConfigurationSpace{N}(\ComplexPlane)
        \times
        B \ComplexNumbers^\times
      \end{tikzcd}
      $}
      \;,\;\;
      p_{
        \!\!\!
        \scalebox{.7}{$
          \ConfigurationSpace{N}
        $}
      }^\ast
      \begin{tikzcd}
        \mathrm{K}(\ComplexNumbers ,\, n)
        \times_{\ComplexNumbers^\times} E \ComplexNumbers^\times
        \ar[d]
        \\
        B \ComplexNumbers^\times
      \end{tikzcd}
      $}
    \!\!\!\!\!  \right)
    $
    \\
    \hline
    $
    \def\arraystretch{1.6}
    \begin{array}{l}
    \vec z : \mathbf{B}\,\mathrm{PBr}(N)
    \;\;\;
    \vdash
    \;\;\;
    \\
    \hspace{.6cm}
      (t : \mathbf{B}\ComplexNumbers^\times)
      \to
      T_{\vec z, \, t}
      \;:\;
      \Types
    \end{array}
    $
    &
    \def\arraystretch{.9}
    \begin{tabular}{c}
      \eqref{LCCRules}
      \\
      \eqref{BaseChangeAdjunction}
    \end{tabular}
    &
    $
      \big(
        \mathrm{id}_{
          \!\!\!
          \ConfigurationSpace{N}
        }
        \times
        P_{B \ComplexNumbers^\times}
      \big)_\ast
      \;
      \begin{tikzcd}
        T
        \ar[d, shorten=-2pt]
        \\
        \ConfigurationSpace{N}(\ComplexPlane)
        \times
        B \ComplexNumbers^\times
      \end{tikzcd}
    $
    \\
    \hline
    $
    \def\arraystretch{1.7}
    \begin{array}{l}
    \vec z : \mathbf{B}\,\mathrm{PBr}(N)
    ,\;
    t : \mathbf{B}\ComplexNumbers^\times
    \;\;\;
    \vdash
    \;\;\;
    \\
    \hspace{.6cm}
    \Truncation{\big}{0}{
      (t : \mathbf{B}\ComplexNumbers^\times)
      \to
      T_{\vec z, \, t}
      \;:\;
      \Types
    }
    \end{array}
    $
    &
    \def\arraystretch{.9}
    \begin{tabular}{c}
      \eqref{FiberwiseZeroTruncationAndConnectedComponents}
    \end{tabular}
    &
    $
      \pi_{
        \scalebox{.7}{$
        0/
        \ConfigurationSpace{N}(\ComplexPlane)
        $}
      }
      \left(\!\!
      \big(
        \mathrm{id}_{
          \!\!\!
          \ConfigurationSpace{N}
        }
        \times
        P_{B \ComplexNumbers^\times}
      \big)_\ast
      \;
      \begin{tikzcd}
        T
        \ar[d, shorten=-2pt]
        \\
        \ConfigurationSpace{N}(\ComplexPlane)
        \times
        B \ComplexNumbers^\times
      \end{tikzcd}
     \!\!\!\! \right)
    $
    \\
    \hline
    \hline
    \def\arraystretch{1}
    \begin{tabular}{l}
      Homotopy data type structure
      \\
      of Def. \ref{DataStructureOfConformalBlocks}
      specializing
      Def. \ref{TypeTheoreticGaussManinConnection}
    \end{tabular}
    &
    $\leftrightarrow$&
    \def\arraystretch{1}
    \begin{tabular}{l}
    Fibration of conformal blocks
    \eqref{KZConnectionByFiberwiseMapping}
    \\
    via
    Thm. \ref{GaussManinConnectionInTwistedGeneralizedCohomologyViaMappingSpaces}
    \&
    Ex. \ref{TheKZConnection}
    \end{tabular}
    \\
    \hline
  \end{tabular}

\vspace{-.24cm}
\end{proof}

\newpage


\end{document}